\documentclass[11pt, oneside]{Thesis} 
\usepackage{amsmath,systeme}
\usepackage{cancel}
\usepackage{stackengine,graphicx}
\graphicspath{{Pictures/}} 

\usepackage[colorlinks]{hyperref}
\hypersetup{urlcolor=blue, colorlinks=true} 

\hypersetup{pdfsubject=\subjectname}
\hypersetup{pdfauthor=\authornames}
\hypersetup{pdfkeywords=\keywordnames}
 
\usepackage[style=numeric,sorting=none]{biblatex}
\bibliography{Bibliography.bib}

\usepackage[normalem]{ulem}

\usepackage[dvipsnames]{xcolor}
\definecolor{mygray}{gray}{0.2}

\usepackage[english]{babel}
\usepackage[autostyle]{csquotes}

\title{Study of General Relativity and \texorpdfstring{$f(R)$}{frtitle} Modified Gravity with Cosmological Constant for Einstein and Jordan Frames} 

\begin{document}

\frontmatter 

\setstretch{1.3} 

\fancyhead{} 
\rhead{\thepage} 
\lhead{} 

\pagestyle{fancy} 

\newcommand{\HRule}{\rule{\linewidth}{0.5mm}} 


\begin{titlepage}
\begin{center}
\includegraphics[width=5cm]{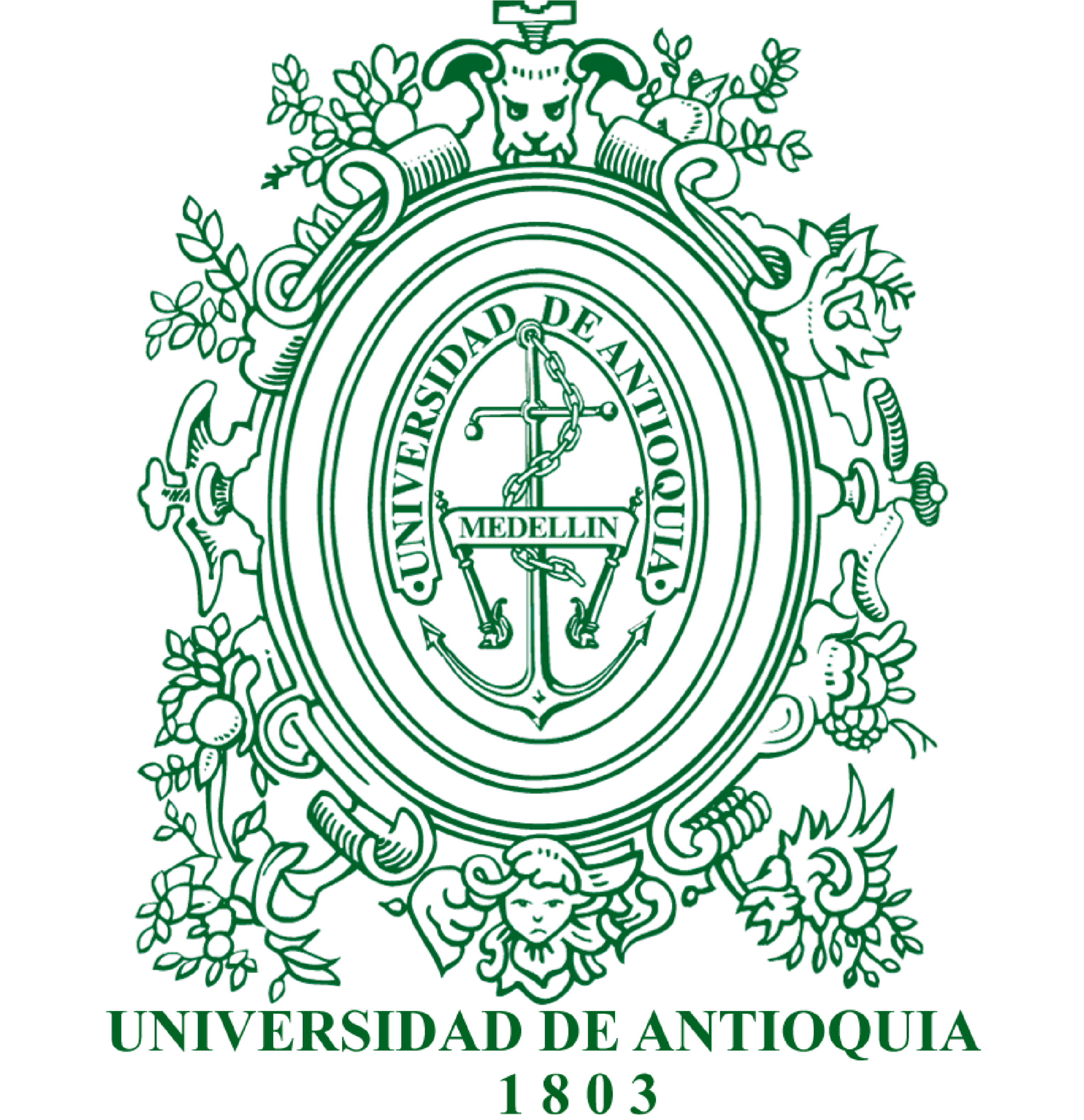} 
 
\textsc{\LARGE \univname}\\[0.4cm]
\deptname\\[1cm] 
\textsc{\Large Doctoral Thesis}\\[0.2cm] 

\HRule \\[0.4cm] 
{\huge \bfseries \ttitle}\\[0.3cm] 
\HRule \\[1cm] 
 
\begin{minipage}{0.4\textwidth}
\begin{flushleft} \large
\emph{Author:}\\
\href{https://orcid.org/0000-0001-6142-7305}{\authornames} 
\end{flushleft}
\end{minipage}
\begin{minipage}{0.4\textwidth}
\begin{flushright} \large
\emph{Supervisor:} \\
\href{http://www.if.ufrj.br/~joras}
{\supname} 
\end{flushright}
\end{minipage}
\\[1cm]
\begin{minipage}{0.8\textwidth}
\begin{flushright} \large
\emph{Co-Supervisor:} \\
\href{https://orcid.org/0000-0001-6455-5564}{\examname} 
\end{flushright}
\end{minipage}\\[2cm]
 
\large \textit{A thesis submitted in fulfilment of the requirements\\ for the degree of \degreename}\\[0.3cm] %

{\large September 19, 2020}\\[1cm] 

\vfill
\end{center}

\end{titlepage}


\Declaration{

\addtocontents{toc}{\vspace{1em}} 

I, C\'esar Daniel Peralta Gonz\'alez, declare that this thesis titled, {\it \ttitle} and the work presented in it is my own. I confirm that this work submitted for assessment is my own and is
  expressed in my own words. Any uses made within it of the works of
  other authors in any form (e.g., ideas, equations, figures, text,
  tables, programs) are properly acknowledged at any point of their
  use. A list of the references employed is included.

 \vspace{2cm} 
Signed:\\
\rule[1em]{25em}{0.5pt} 
 
Date:\\
\rule[1em]{25em}{0.5pt} 
}

\clearpage 


\pagestyle{empty} 

\null\vfill 

\begin{flushright}
\textit{``In the beginning God created the Heauen, and the Earth."} \textbf{Gen. 1}: 1 \cite{bible}.
\end{flushright}
\vfill\vfill\vfill\null

\begin{flushright}
\textit{So shall I have wherewith to answer him that reproacheth me:\\
for I trust in thy word.} \textbf{Psaml. 119}: 42 \cite{bible}
\end{flushright}

\vfill\vfill\vfill\null

\begin{flushright}
\textit{Mama, ay mama.\\
Que bello sue\~no tuve ayer.\\
\ldots \\
\'Ibamos los dos, \textbf{en un gran barco de papel}.\\
Donde yo era el capit\'an \textbf{en el pa\'is de la ilusi\'on}.\\
Y que orgullosa estabas tú.\\
Pasó el tiempo, mucho tiempo.} \\
\textbf{Mama}, \textbf{Joe Arroyo.}\\
\vfill
\textit{El sue\~no del Joe es mi realidad.}
\end{flushright}

\vfill\vfill\vfill\vfill\vfill\vfill\null 

\clearpage 


\addtotoc{Abstract} 


 {\huge{\textit{Abstract}}\par}{\addtocontents{toc}{\vspace{1em}}} 
This thesis investigates a toy model for inflation in a class of modified theories of gravity in the metric formalism. Instead of the standard procedure --- assuming a non-linear Lagrangian $f(R)$ in the Jordan frame --- we start from a simple $\phi^2$ potential in the Einstein frame and investigate the corresponding $f(R)$ in the former picture. Such approach yields plenty of new pieces of information, namely a self-terminating inflationary solution with a linear Lagrangian, a robust criterion for stability of such theories, a dynamical effective potential for the Ricci scalar $R$, the addition of an ad-hoc Cosmological Constant in the Einstein frame leads to a Thermodynamical interpretation of this physical system, which allows further insight on its (meta)stability and evolution.

\vspace{2cm}

\clearpage 


\setstretch{1.3} 

\acknowledgements{\addtocontents{toc}{\vspace{1em}} 

I want to express all my grateful, glory and honor to God, who fulfilled all my dreams, and exceeded all my expectations with his Word of truth and comfort in my life and through the lives of many more.

I would like to thank my advisor, Dr. Sergio Jor\'as, not only for his guidance and fruitful discussions but also for his confidence in giving me the opportunity to carry on this thesis. I would like the express my gratitude to Dr. Diego Restrepo for
giving me the opportunity, support and patience, during the realization of this work.

I would also like to express my special gratitude to Dr. Yeinzon Rodr\'iguez and Dr. Leonardo Casta\~neda whom gave me the basis and motivation to study cosmology.

I would like to express my gratitude to all the academic partners with whom I have shared these 5 years of
work, in particular to: Sergio, Alejandro, Amalia, Juan David, Alexander, Andr\'es, Anyeres, Sheryl and Calambre.

I am very grateful for the invaluable help, and humility received from David Felipe Tamayo, Omar Alberto Rold\'an and Arthur Luna, there are no words to describe such human worth of this persons and their special families. I would also like to express my
special gratitude to all, former and current members, of the Astrophysics, Relativity and Cosmology (ARCOS) group with whom I have shared during my travel to Rio de Janeiro.

This thesis reached its completion with the financial support from COLFUTURO/COLCIENCIAS, Colombia, under the program “Becas Doctorados Nacionales 647” and the administrative and academic management of Instituto de F\'isica from U. Antioquia. 

Finally, I would like to express my deepest gratitude to my parents, Roger and Marina, my brothers Carlos, Roger, Iv\'an and Sa\'id, and my beloved wife and son, Jenny and Daniel for their unconditional love, encouragement and support, without them, there would not either be BSc, nor MSc, nor PhD thesis. Likewise, the support and affection from all my family, friends and people in Sincelejo, Bucaramanga, Bogot\'a, and Medell\'in have been fundamental to
me. I love them all.
}
\clearpage 


\pagestyle{fancy} 

\lhead{\emph{Contents}} 
\tableofcontents 

\lhead{\emph{List of Figures}} 
\listoffigures 

\setstretch{1.3} 

\pagestyle{empty} 

\dedicatory{\begin{flushright} \Large Dedicated to\ldots \\ My good Lord, \textit{\textbf{Jesus}}. \end{flushright}} 

\addtocontents{toc}{\vspace{2em}} 


\mainmatter 

\pagestyle{fancy} 


\part{Preliminars}

\chapter*{Introduction} 
\label{Introduction}
\lhead{Introduction \emph{Introduction}} 
Einstein's general relativity theory is constituted as the most important scientific paradigm in the field of gravitation, presenting a coherent description of space, time, and matter at a macroscopic level. Its conceptual contributions, very high precision experimental measurements, and, the recently computational simulations field research, present to general relativity as one of the most accurate theories ever studied in the history of science together with quantum mechanics \cite{will2014confrontation}. The greatest impact of this theory happens in the field of cosmology, since it is the natural area of the gravity's domain, opening up new possibilities for measurement and modeling. 

Since he began his work, Einstein wondered if general relativity was the definitive theory able to describe the gravitational interactions; from this moment it began the race to test and verify this theory. At the same time, Einstein proposed a static universe model \cite{einstein1922kosmologische}, for this reason he introduced a modification that includes a universal constant whose action was to counteract the possible collapse of the universe, known as cosmological constant. To his surprise, years later, the observations and measurements made at the Monte Palomar observatory by the American astronomer Edwin Hubble, allowed Einstein to contrast his model by concluding that his idea of the cosmological constant was one of his worst mistakes. 

Nevertheless, the scientific community did not discard the idea of introducing a cosmological constant in the theory given that it presented a more complete perspective of it. The proposals that best fit the observations were the cosmological models of the Russian mathematician Alexander Friedmann \cite{friedmann1922125} and the Belgian physicist Georges Lema\^{i}tre \cite{lemaitre1931contributions}. After a while, a good collection of verifiable results such as the existence of the cosmic microwave background (CMB) \cite{penzias1965measurement}, the prediction of observed abundances of light elements \cite{gamow1948origin}, a thermal description, and an estimate of the age and evolution of the universe, led to consolidate these proposals as what is now well known as a standard cosmological model \cite{weinberg2008cosmology,mukhanov2005physical}. 

In recent decades, however, different problems were evident that seriously challenged and questioned one of the fundamental principles of the model: The principle of homogeneity and isotropy of the universe. This set of problems was called cosmological problems. From the 80's, an ingenious solution was proposed by the American physicist Alan Guth. He said that these problems can be resolved simultaneously in a primordial stage of the universe called: Inflation. Guth's main idea is that the universe in its very early stages of evolution had an accelerated exponential expansion, dominated by some primordial $\phi$ scalar field with an inflationary potential $V(\phi)$ known as Inflaton. The rapid expansion makes the universe flat, homogeneous and isotropic. The method used by Guth presented inconsistencies in estimating the amount of time for inflation \cite{guth1981inflationary}. 

A new inflationary scenario known as \textit{slow-roll inflation} was proposed by Russian physicist Andrei Linde in which the problems presented in the previous one are solved. According to Linde, the period of inflation occurs when the scalar field ``rolls very slowly'' through the potential while the universe experiences an exponential expansion. Before reaching the minimum of the potential, inflation ends and a period of reheating occurs \cite{linde1982new}. The inflationary period became perhaps more important due to its ability to extend the quantum fluctuations of the field that filled the universe by making the classics shortly after leaving the horizon. This corresponds to small inhomogeneities in the primary energy density and is responsible, via gravitational interaction, for large-scale universe structures. In fact, the 1992 space satellite COBE (Cosmic Background Explorer) found small inhomogeneities in the CMB temperature of the order of $10^{-5}$K, with an average temperature of $T_0 = 2,725 \pm 0.002$ K, on scales of the order of $10^3$ Mpc \cite{smoot1992structure}. After this, the WMAP satellite (Wilkinson Microwave Anisotropy Probe), which has a better angular resolution and a sensitivity 30 times better than that of COBE, was put into orbit in 2003, confirming and improving the measurements of the previous satellite \cite{spergel2003first}. Thus, inflation begins to establish itself as a consistent and verifiable theory by providing a new scientific paradigm. After inflation, the universe evolves as described in the standard model.
 
Along with these new discoveries, the American astronomers Adam Riess and Saul Perlmutter made the surprising discovery that the universe is currently in a new stage of cosmic acceleration \cite{riess1998observational, perlmutter1999measurements}. The models that sought to explain this phenomenon did not wait. The simplest explanation, which has best fit the data is that of the cosmological constant $\Lambda$, reappearing Einstein's old idea of an universal constant but this time in favor of the expansion. The observations suggest that there is a kind of unknown fluid of negative pressure, in the density of matter, known as dark energy, which dominates the entire universe (about $68\%$ according to its more recent report \cite{Planck2018}), to which the accelerated expansion is associated. However, the cosmological constant model presents a fundamental problem: the inferred value of the vacuum energy density in the universe is too small compared to the value estimated in the Standard Model of particle physics (approximately 120 orders of magnitude below \cite{RevModPhys.61.1, martin2012everything}) which qualifies as a catastrophe in the theory. One way of approaching this problem is to assume that the acceleration is not directly associated with this unknown fluid but is a pure action of the dynamics of spacetime. From this point of view, we try to modify Einstein's equations from its geometry to try to adjust the reported data without the need to include exotic fluids in the model. 

Modified theories of gravity are either used to replace the cosmological constant or the inflaton field -- explaining, respectively, the current and the early accelerated phase of expansion of the universe. Nevertheless, they were introduced (see Ref. \cite{Bicknell_1974} and references therein) long before any experimental data on either subject were available, just for the sake of completeness/diversity. 

In this study we focus on $f(R)$ theories \cite{DeFelice:2010aj,Capozziello:2011et,Nojiri:2017ncd} --- nonlinear functions of the Ricci scalar $R$ defined, as usual, in the Jordan Frame (JF) --- in the metric formalism, which features an extra degree of freedom (d.o.f). Upon a suitable conformal transformation (see below), the modified gravitational Lagrangian assumes the usual Einstein-Hilbert form and the extra d.o.f. is materialized as a scalar field --- for obvious reasons, this is the so-called Einstein Frame (EF). 

We start from a standard potential $V_E(\phi)=\frac{1}{2}m_\phi^2\phi^2$ in the EF (with also standard slow-roll initial conditions) and investigate the corresponding $f(R)$ in the JF. The evolution of the system is then carefully followed in both frames. Our purpose is twofold: to discuss the stability of the initial inflationary phase (and subsequent ones as well), comparing our results to the usual definitions in the current literature, and to point out the phase transition that takes place when the system is driven towards a non-trivial global minimum by a dynamical effective potential $V_J(R)$.  We will then be lead to an unexpected analogy to a non-ideal gas.

\section*{Outline of the Thesis}

In Chapter \ref{Chapter1} we derive the Einstein's field equations in the metric formalism using elementary variational principles with a boundary term in the action. Also we briefly review the main elements in the standard cosmology.

In Chapter \ref{Chapter2} we study briefly the theoretical foundations of inflation, its motivations and definition. We present its main tools and equations that will be used in the rest of this thesis. We consider the slow-roll analysis and numerical solutions for the inflationary models of interest.

In Chapter \ref{Chapter3} we derive the $f(R)$ gravity field equations in the metric formalism, including a boundary term in the total action, and we derive its dynamical equations considering a universe with a Friedman-Lema\^{i}tre-Robertson-Walker (FLRW) metric.

In Chapter \ref{Chapter5} we present the Legendre transformation, through which we can re write the Lagrangian in the JF and obtain the Hilbert-Einstein Lagrangian with a scalar field. After that, we consider the inverse problem and about this is going to be presented the main results of this study. 

In Chapter \ref{Chapter6} we recall the thermodynamics properties of the van der Waals (vdW) fluid, we show the most important quantities that describe a thermodynamics system. The van der Waals models is the most simple model that present a phase transitions, we show the equilibrium conditions and plot the coexistence and metastability regions for this phases.

In Chapter \ref{Chapter7} we showing the main results of this thesis: we explore three different criteria of stability in $f(R)$ gravity theory. At first, we start focus on a particularly simple toy-model potential for inflation in the EF and analyze the usual criteria of stability (effective squared masses) of the corresponding $f(R)$ theory in the JF. We point out where they do not agree (and why). We then introduce a novel interpretation of the physics in JF. 
In section~\ref{catastrophesection} we investigate the behaviour of the effective masses previously defined in the literature and the coalescence of the extrema (maxima and minima) of the effective potential $V_J(R)$, via Catastrophe Theory\cite{poston1996catastrophe}. In section~\ref{numerical} we present the numerical solutions in a concrete example, using standard slow-roll initial conditions in the EF. As we will show, the effective potential for the Ricci scalar $R$ presents a dynamical character. 
In Section~\ref{ThermoAnalogySection} we show an unexpected analogy of such effective potential to the well-known van der Waals gas.

\chapter{General Relativity: Field Equations}
\label{Chapter1}
\lhead{Chapter 1. \emph{General Relativity: Field Equations}} 
General relativity constitutes the current gravitational paradigm and establishes the basis for the standard cosmological model. As a result, our current understanding of the universe at large scale is thanks to this theory. In this chapter, we show the main mathematical tools used in the framework of Einstein's general relativity to describe a universe with a four-dimensional space, FLRW metric, matter and cosmological constant.

\section{FLRW Universe with Einstein's Gravity}

A space-time is described by a pair $(M, g)$, where $M$ is a smooth, connected, orientable and Hausdorff-type four-dimensional manifold; $g$ is a Lorentzian metric over $M$. $g$ is physically interpreted as the generalization of the gravitational potential. For a torsion-free manifold there is a unique symmetric connection $\Gamma^{\alpha}_{\beta\gamma}$ which is compatible with the $g$ metric called the Levi-Civita connection. In this way, the connection is interpreted as the generalization of the gravitational fields generated by the metric potentials of $g$.
 
Einstein's field equations can be obtained from a Lagrangian and the variational principle  $\delta S_{HE} = 0$, where $S_{HE}$ is the action of the gravitational field proposed by Albert Einstein and David Hilbert in 1915 \cite{poisson2004relativist}:
\begin{equation}\label{actionEH}
S_{EH} = \frac{1}{2 k}\int_{V} d^4x \sqrt{-g} R.
\end{equation}
The total action $S$ also includes an action associated with the matter fields $S_M$ and a Gibbons-York-Hawking bound term $S_{GYH}$ \cite{Gibbons}:
\begin{equation}\label{totalaction}
S = S_{EH} + S_{GYH} + S_M.
\end{equation}
with, 
\begin{equation}
    S_{GYH} = 2 \oint_{\partial V} d^3 y \epsilon \sqrt{|h|} K
\end{equation}
where $V$ is a hypervolume on the manifold $M$, $\partial V$ its boundary, $h$ the determinant of the induced metric, $K$ is the trace of the extrinsic curvature of the boundary $\partial V$, and $ \varepsilon$ takes the values $+1$ or $-1$ if $\partial V$ is timelike or spacelike, respectively (it is assumed that $\partial V$ is nowhere null). Coordinates $x^{\alpha}$ are used for the finite region $V$ and $y^{\alpha}$ for the boundary $\partial V$. Below we show how to obtain the Einstein field equations, varying the total action (\ref{totalaction}) respect to the metric $g^{\mu\nu}$ . Such variation keeps fixed the boundary terms $\delta g_{\mu\nu} \Big|_{\partial V}$ \cite{poisson2004relativist}. That is, the variation of the metric is null on the boundary $\partial V$. So, for the Einstein-Hilbert action, the variation is
\begin{equation}\label{EHvariation}
    \delta S_{EH} = \int_{V} d^4 x (R \delta \sqrt{-g} + \sqrt{-g}\delta R).
\end{equation}
The following results are useful. First, let us consider a $(n \times n)$-square matrix ${\cal A} = \{a_{ij}\}$ and its inverse ${\cal B}=\{b^{ij}\}$, whose elements are given by \cite{inverno}
\begin{eqnarray}
b^{ij} = \frac{1}{a}(A^{ij})^T = \frac{1}{a} A^{ji},
\end{eqnarray}
where $a\equiv \det({\cal A})$ and $A^{ij}$ is the cofactor of $a_{ij}$. Setting a row $i$, the determinant can be written as
\begin{eqnarray}
a = \sum_{j=1}^n a_{ij}A^{ij}.
\end{eqnarray}

Partially deriving the previous equation with respect to $a_{ij}$ we obtain
\begin{eqnarray}
A^{ij} = \frac{\partial a}{\partial a_{ij}}.
\end{eqnarray}
Since the determinant is a functional of $a_{ij}$ that depends on the coordinates $x^{\sigma}$, such that $a = a \left(a_{ij}(x^{\sigma})\right)$, then 
\begin{eqnarray}
\partial_\sigma  a \equiv \frac{\partial a}{\partial x^{\sigma}} &=& \frac{\partial a}{\partial a_{ij}}\frac{\partial a_{ij}}{\partial x^{\sigma}}\nonumber\\
&=& A^{ij}\frac{\partial a_{ij}}{\partial x^{\sigma}}\nonumber\\
&=& a \, b^{ji} \frac{\partial a_{ij}}{\partial x^{\sigma}} \equiv a \, b^{ji} \partial_\sigma  a_{ij}.
\end{eqnarray}
Applying this result to the determinant of the metric and recalling that $g^{\mu\nu}$ is symmetrical, then
\begin{eqnarray}\label{variaciondeterminanteg}
\partial_{\sigma} g &=& gg^{\mu\nu} \partial_{\sigma}g_{\mu\nu} ,\nonumber\\
 \text{or} \quad  \delta g &=& gg^{\mu\nu} \delta g_{\mu\nu}.
\end{eqnarray}
In this manner,
\begin{eqnarray}
\delta(\sqrt{-g}) &=& \frac12 (-g)^{-1/2}(-g)g^{\mu\nu}\delta g_{\mu\nu},\nonumber\\
&=& \frac12\sqrt{-g} g^{\mu\nu} \delta g_{\mu\nu}.
\label{eq:delta}
\end{eqnarray}
It is convenient to leave the variation in terms of $\delta g^{\mu\nu}$. For this purpose, we substitute $\delta g_{\alpha\beta} = - g_{\alpha\mu}g_{\beta\nu}\delta g^{\mu\nu}$ in \eqref{eq:delta}, which yields:
\begin{eqnarray} \label{resultado1}
\delta (\sqrt{-g}) &=& -\frac12\sqrt{-g}g^{\alpha\beta} g_{\alpha\mu}g_{\beta\nu}\delta g^{\mu\nu} \nonumber\\
&=& -\frac12\sqrt{-g} \delta^{\beta}_{\mu}g_{\beta\nu}\delta g^{\mu\nu} \nonumber\\
&=& -\frac12\sqrt{-g} g_{\mu\nu} \delta g^{\mu\nu}.
\end{eqnarray}
Second, the variation of the Ricci scalar $\delta R$ is calculated as follows: Since
\begin{equation}
R^{\rho}_{\sigma\mu\nu} \equiv \partial_{\mu}\Gamma^{\rho}_{\sigma\nu} - \partial_{\nu}\Gamma^{\rho}_{\sigma\mu} + \Gamma^{\lambda}_{\sigma\nu}\Gamma^{\rho}_{\lambda\mu} - \Gamma^{\lambda}_{\sigma\mu}\Gamma^{\rho}_{\lambda\nu} \, ,
\end{equation}
then
\begin{equation}
\delta R^{\rho}_{\sigma\mu\nu} = \partial_{\mu}\delta\Gamma^{\rho}_{\sigma\nu} - \partial_{\nu}\delta\Gamma^{\rho}_{\sigma\mu} + \delta\Gamma^{\lambda}_{\sigma\nu}\Gamma^{\rho}_{\lambda\mu} + \Gamma^{\lambda}_{\sigma\nu}\delta\Gamma^{\rho}_{\lambda\mu} - \delta\Gamma^{\lambda}_{\sigma\mu}\Gamma^{\rho}_{\lambda\nu} - \Gamma^{\lambda}_{\sigma\mu}\delta\Gamma^{\rho}_{\lambda\nu}\, .
\end{equation}
We point out that
\begin{eqnarray}
\Gamma^{\lambda'}_{\mu'\nu'} &=& \partial_{\rho}x^{\lambda'}\partial_{\mu'}x^{\sigma}\partial_{\nu'}x^{ \tau}\Gamma^{\rho}_{\sigma\tau} - \partial^2_{\sigma\tau}x^{\lambda'}\partial_{\mu'}x^{\sigma}\partial_{\nu'}x^{ \tau} ,\nonumber\\
\delta \Gamma^{\lambda'}_{\mu'\nu'} &=& \partial_{\rho}x^{\lambda'}\partial_{\mu'}x^{\sigma}\partial_{\nu'}x^{\tau}\delta\Gamma^{\rho}_{\sigma\tau}, 
\end{eqnarray}
i.e, the variation of the metric connection transforms as a tensor. 

On the other hand, one can obtain the variation of the Ricci tensor using the variations of the metric connection,
\begin{eqnarray}
\nabla_{\lambda}(\delta\Gamma^{\rho}_{\mu\nu}) &=& \partial_{\lambda}\delta\Gamma^{\rho}_{\mu\nu} + \Gamma^{\rho}_{\sigma\lambda}\delta\Gamma^{\sigma}_{\nu\mu} - \Gamma^{\sigma}_{\nu\lambda}\delta\Gamma^{\rho}_{\sigma\mu} - \Gamma^{\sigma}_{\mu\lambda}\delta\Gamma^{\rho}_{\nu\sigma}, \nonumber\\
 \nabla_{\nu}(\delta\Gamma^{\rho}_{\mu\sigma}) - \nabla_{\mu} (\delta\Gamma^{\rho}_{\nu\sigma}) &=& \partial_{\nu}\delta\Gamma^{\rho}_{\mu\sigma} + \Gamma^{\rho}_{\lambda\nu}\delta\Gamma^{\lambda}_{\mu\sigma} - \Gamma^{\lambda}_{\mu\nu}\delta\Gamma^{\rho}_{\lambda\sigma} - \Gamma^{\lambda}_{\sigma\nu}\delta\Gamma^{\rho}_{\mu\lambda} \nonumber\\
 &-& \partial_{\mu}\delta\Gamma^{\rho}_{\nu\sigma} - \Gamma^{\rho}_{\lambda\mu}\delta\Gamma^{\lambda}_{\nu\sigma} + \Gamma^{\lambda}_{\nu\mu}\delta\Gamma^{\rho}_{\lambda\sigma} + \Gamma^{\lambda}_{\sigma\mu}\delta\Gamma^{\rho}_{\nu\lambda}.
\end{eqnarray}
from which we get the well-known {\it Palatini's identity}:
\begin{eqnarray} \label{palatini}
\delta R_{\mu\nu} &=& \delta R^{\rho}_{\mu\rho\nu} = \nabla_{\nu}(\delta\Gamma^{\rho}_{\mu\rho}) - \nabla_{\mu} (\delta\Gamma^{\rho}_{\mu\nu}).
\end{eqnarray}
Since $R = g^{\mu\nu} R_{\mu\nu}$, then
\begin{eqnarray} \label{varricci}
\delta R = \delta g^{\mu\nu} R_{\mu\nu} + g^{\mu\nu}\delta R_{\mu\nu},
\end{eqnarray}
Substituting Palatini's identity (\ref{palatini}) into the previous equation we get:
\begin{eqnarray}\label{resultado2}
\delta R &=& \delta g^{\mu\nu} R_{\mu\nu} + g^{\mu\nu} \left(\nabla_{\gamma}(\delta\Gamma^{\gamma}_{\mu\nu}) - \nabla_{\nu} (\delta\Gamma^{\gamma}_{\mu\gamma}) \right) \nonumber\\
&=& \delta g^{\mu\nu} R_{\mu\nu} + \nabla_{\sigma}\left( g^{\mu\nu}(\delta\Gamma^{\sigma}_{\mu\nu}) - g^{\alpha\sigma}(\delta\Gamma^{\gamma}_{\alpha\gamma}) \right).
\end{eqnarray}
Replacing \eqref{resultado1} and \eqref{resultado2} in \eqref{EHvariation} yields:
\small
\begin{eqnarray}\label{variaction}
 \delta S_{EH} &=& \int_{V} d^4 x (R \delta \sqrt{-g} + \sqrt{-g}\delta R) \nonumber\\
 &=& \int_{V} d^4 x \left(-\frac12 R\,g_{\mu\nu}  \sqrt{-g} \delta g^{\mu\nu} + R_{\mu\nu}  \sqrt{-g}  \delta g^{\mu\nu} +  \sqrt{-g}\nabla_{\sigma}\left( g^{\mu\nu}(\delta\Gamma^{\sigma}_{\mu\nu}) - g^{\alpha\sigma}(\delta\Gamma^{\gamma}_{\alpha\gamma}) \right)  \right)\nonumber\\
 &=& \int_{V} d^4 x \sqrt{-g}\left(R_{\mu\nu} -\frac12 R\,g_{\mu\nu} \right) +\int_{V} d^4 x \sqrt{-g} \nabla_{\sigma}\left( g^{\mu\nu}(\delta\Gamma^{\sigma}_{\mu\nu}) - g^{\alpha\sigma}(\delta\Gamma^{\gamma}_{\alpha\gamma}) \right).
\end{eqnarray} 
\normalsize
The divergence term corresponding to the integral on the right of previous equation is defined as the \textit{boundary term}:
\begin{align}
    \delta S_{B} = &\int_{V} d^4 x \sqrt{-g} \nabla_{\sigma}\left( g^{\mu\nu}(\delta\Gamma^{\sigma}_{\mu\nu}) - g^{\alpha\sigma}(\delta\Gamma^{\gamma}_{\alpha\gamma}) \right)\\
    \delta S_{B} = &\int_{V} d^4 x \sqrt{-g} \nabla_{\sigma} V^{\sigma}.
\end{align}
where $V^{\sigma}$ is the vector defined as
\begin{equation}
    V^{\sigma} \equiv \nabla_{\sigma}\left( g^{\mu\nu}(\delta\Gamma^{\sigma}_{\mu\nu}) - g^{\alpha\sigma}(\delta\Gamma^{\gamma}_{\alpha\gamma})\right).
\end{equation}
Using the Gauss-Stokes theorem 
\begin{equation}
     \int_{V} d^n x \sqrt{|g|} \nabla_{\mu} A^{\mu} = \oint_{\partial V} d^{n-1} y \varepsilon \sqrt{|h|}n_{\mu} A^{\mu},
\end{equation}
where $n_{\mu}$ is a vector normal to $\partial V$, the boundary term can be written as:
\begin{equation}
     \delta S_{B} = \oint_{\partial V} d^{3} y \varepsilon \sqrt{|h|}n_{\sigma} V^{\sigma}.
\end{equation}
For the variation of $\delta\Gamma^{\sigma}_{\mu\nu}$ we have:
\begin{eqnarray}
    \delta\Gamma^{\sigma}_{\mu\nu} &=& \delta \left(\frac12 g^{\sigma\gamma}\left[\partial_{\mu}g_{\gamma\nu} +\partial_{\nu}g_{\gamma\mu}-\partial_{\gamma}g_{\mu\nu} \right] \right)\nonumber\\
    &=& \frac12 \delta g^{\sigma\gamma}\left[\partial_{\mu}g_{\gamma\nu} + \partial_{\nu}g_{\gamma\mu}-\partial_{\gamma}g_{\mu\nu} \right] + \frac12 g^{\sigma\gamma} \left[\partial_{\mu}(\delta g_{\gamma\nu}) +\partial_{\nu}(\delta g_{\gamma\mu}) -\partial_{\gamma}(\delta g_{\mu\nu}) \right]\nonumber\\
    &=& \frac12 g^{\sigma\gamma} \left[\partial_{\mu}(\delta g_{\gamma\nu}) +\partial_{\nu}(\delta g_{\gamma\mu}) -\partial_{\gamma}(\delta g_{\mu\nu}) \right].
\end{eqnarray}
Recalling that $\delta g_{\mu\nu} = \delta g^{\mu\nu} = 0$ evaluated on the boundary, then
\begin{equation}
    V^{\sigma}\vert_{\partial V} = g^{\mu\nu}\left[\frac12 g^{\sigma\gamma} \left[\partial_{\mu}(\delta g_{\gamma\nu}) +\partial_{\nu}(\delta g_{\gamma\mu}) -\partial_{\gamma}(\delta g_{\mu\nu}) \right] \right] -  g^{\alpha\sigma}\left[\frac12 g^{\tau\gamma} \partial_{\alpha}(\delta g_{\tau\gamma})\right]. 
\end{equation}
Additionally,
\small
\begin{eqnarray}
    V_{\sigma}\vert_{\partial V} &=& g_{\sigma \xi} V^{\xi}\vert_{\partial V} \nonumber\\
    &=& g_{\sigma \xi} \left\{g^{\mu\nu}\Bigg[\frac12 g^{\xi\gamma} \bigg(\partial_{\mu}(\delta g_{\gamma\nu}) +\partial_{\nu}(\delta g_{\gamma\mu}) -\partial_{\gamma}(\delta g_{\mu\nu}) \bigg) \Bigg] \right\} - g_{\sigma \xi}\left\{ g^{\alpha\xi}\left[\frac12 g^{\tau\gamma} \partial_{\alpha}(\delta g_{\tau\gamma})\right]\right\} \nonumber\\
    &=& \frac12 \delta^{\gamma}_{\sigma}g^{\mu\nu}\bigg[\partial_{\mu}(\delta g_{\nu\gamma}) + \partial_{\nu}(\delta g_{\gamma\mu}) - \partial_{\gamma}(\delta g_{\mu\nu})\bigg]- \frac12 \delta^{\mu}_{\sigma}g^{\tau\gamma}\bigg[\partial_{\mu}(\delta g_{\tau\gamma})\bigg]\nonumber\\
    &=& g^{\mu\nu}\bigg[\partial_{\nu}(\delta g_{\sigma\mu})-\partial_{\sigma}(\delta g_{\mu\nu})\bigg].
\end{eqnarray}
\normalsize
Replacing $g^{\mu\nu}= h^{\mu\nu} + \varepsilon \,  n^{\mu}n^{\nu}$ into the previous equation, the term $n^{\sigma}V_{\sigma}$ can be written as
\begin{eqnarray}
n^{\sigma}V_{\sigma}\vert_{\partial V} &=& n^{\sigma}\left(h^{\mu\nu} + \varepsilon n^{\mu}n^{\nu}\right)\left[\partial_{\nu}(\delta g_{\sigma\mu})-\partial_{\sigma}(\delta g_{\mu\nu})\right]\nonumber\\
&=& n^{\sigma}h^{\mu\nu}\left[\partial_{\nu}(\delta g_{\sigma\mu})-\partial_{\sigma}(\delta g_{\mu\nu})\right],\label{eq:sigmaV}
\end{eqnarray}
where the antisymmetric part of $\varepsilon \, n^{\mu}n^{\nu}$ with $\varepsilon = n^{\nu}n_{\nu}$ is $\pm 1$. Since $\delta g_{\mu\nu}=0$ on the boundary, its tangencial derivatives must also vanish: $\partial_{\sigma} \delta g_{\mu\nu} = 0$ \cite{poisson2004relativist}. Therefore, equation \eqref{eq:sigmaV} simplifies to:
\begin{equation}
    n^{\sigma}V_{\sigma}\bigg\vert_{\partial V} = - n^{\sigma}h^{\mu\nu}\partial_{\sigma}(\delta g_{\mu\nu}).
\end{equation}
This result is non zero because the \textit{normal} derivative of $\partial g_{\mu\nu}$ is not required to vanish on the hypersurface. Collecting the above results we obtain:
\begin{equation}
\delta S_{HE} = \int_{V} d^4 x (R \delta \sqrt{-g} + \sqrt{-g}\delta R)\delta g^{\mu\nu} - \oint_{\partial V} d^3y \varepsilon \sqrt{|h|} h^{\mu\nu}\partial_{\sigma}(\delta g_{\mu\nu})n^{\sigma}.
\end{equation}
\subsection{Variation of the boundary term}
Since the induced metric is fixed on $\partial V$, the trace of the extrinsic curvature $K$ is the only quantity to be varied \cite{poisson2004relativist}:
\begin{eqnarray}
    K &=& \nabla_{\mu}n^{\mu}\nonumber\\
    &=& g^{\mu\nu} \nabla_{\nu}n^{\mu}\nonumber\\
    &=& \left(h^{\mu\nu} + \varepsilon n^{\mu}n^{\nu}\right)\nabla_{\nu}n_{\mu}\nonumber\\
    &=& h^{\mu\nu} \nabla_{\nu}n_{\mu}\nonumber\\
    &=& h^{\mu\nu} \left(\partial_{\nu}n_{\mu} - \Gamma^{\gamma}_{\mu\nu} n_{\gamma} \right),
\end{eqnarray}
The variation of $K$ is derived from the variation of $\delta\Gamma^{\gamma}_{\mu\nu}$ evaluated on the boundary, and the fact that $h^{\mu\nu}\partial_{\nu}(\delta g_{\sigma\mu}) = 0$ and $h^{\mu\nu}\partial_{\mu}(\delta g_{\sigma\nu}) = 0$:
\begin{eqnarray}
\delta K &=& -h^{\mu\nu}\delta \Gamma^{\gamma}_{\mu\nu}n_{\gamma}  \nonumber\\
&=& -\frac12 h^{\mu\nu}(\partial_{\nu}(\delta g_{\mu\sigma}) +\partial_{\mu}(\delta g_{\nu\sigma}) -\partial_{\sigma}(\delta g_{\mu\nu})) \nonumber\\
&=& \frac12 h^{\mu\nu}\partial_{\sigma}(\delta_{\mu\nu}n^{\sigma}).
\end{eqnarray}
 
Using previous equation, the variation of the boundary term is given by
\begin{equation}
    \delta S_{GYH} = \oint_{\partial V} d^3y \varepsilon \sqrt{|h|}h^{\mu\nu}\partial_{\sigma}(\delta_{\mu\nu}n^{\sigma}).
\end{equation}
This last result cancels out exactly the boundary contribution from the Hilbert-Einstein action. 
\subsection{Variation of the Matter Action}
The variation of the matter action yields
\begin{eqnarray}
\delta S_{M} &=& \int_{V}d^4x \delta\left(\sqrt{-g} L_{M} \right),\nonumber\\
&=& \int_{V}d^4x\sqrt{-g} \left(\frac{\partial L_M}{\partial g^{\mu\nu}}\delta g^{\mu\nu} + L_M \delta \sqrt{-g} \right) \nonumber\\ 
&=& \int_{V}d^4x\sqrt{-g} \left(\frac{\partial L_M}{\partial g^{\mu\nu}} - \frac12 L_M g_{\mu\nu} \right)\delta g^{\mu\nu}\nonumber\\
&=&-\frac12 \int_{V}d^4x T_{\mu\nu} \delta g^{\mu\nu},
\end{eqnarray}
where we define $T_{\mu\nu}$ as the momentum-energy tensor given by
\begin{eqnarray}
T_{\mu\nu} \equiv -2\frac{\partial L_M}{\partial g^{\mu\nu}} + L_M g_{\mu\nu}.
\end{eqnarray}

The variation of the total action respect to the metric $g_{\mu\nu}$, with fixed ends\footnote{That is, the variations of the metric are annulled at the boundary.}, produces the Einstein field equations \cite{Wald:1984rg}:
\begin{equation}\label{ecucampoeinstein}
 G_{\mu\nu} = 8\pi G T_{\mu \nu}.
\end{equation}

The Einstein tensor on the left-hand side is responsible for describing the geometry on the manifold. This tensor is defined by the expression
\begin{eqnarray}\label{tensoreinstein}
G_{\mu\nu} \equiv R_{\mu\nu} -\frac12 R g_{\mu\nu},
\end{eqnarray}
where $R_{\mu\nu}$ is the Ricci tensor and $R$ is the Ricci scalar defined as the trace of the Ricci tensor $R \equiv R^{\mu}_{\mu} = g^ {\mu\nu}R_{\mu\nu}$. The material content of the universe is characterized by the energy-momentum tensor $T_{\mu\nu}$ on the right-hand side. 
\subsection{Standard Cosmology}
The standard cosmological model is based on the principle of homogeneity and isotropy of the universe at large scales\footnote{Scales greater than 100 Mpc.}. In order to satisfy this principle, a perfect fluid is assumed. A perfect fluid is characterized by three physical quantities: A quadrivelocity $u^{\mu} = dx^{\mu}/d\tau$; a field of proper density $\rho$; and a scalar field of pressure $p$. At the limit where $p$ is zero (i.e, its kinetic energy is much smaller than its rest energy), the perfect fluid is reduced to a dust model. This suggests that the perfect fluid is of the form
\begin{eqnarray}
T^{\mu\nu}=(\rho + p)u^{\mu}u^{\nu} + p g^{\mu\nu}. 
\end{eqnarray}
Since the perfect-fluid tensor conforms to the description of an isotropic and homogeneous universe, this global characteristic makes its components independent of spatial coordinates\footnote{Actually, it means that it is possible to find a coordinate system such that those quantities do not depend on the spatial coordinates.}. Additionally, we will use a metric of Friedman-Lema\^{i}tre-Robertson-Walker under the same assumption of homogeneity and isotropy. This metric makes it possible to expand or contract the tri-space by means of a scale factor $a = a(t)$ that depends on the cosmic time; this factor describes the dynamics of the evolution of the universe. This metric is characterized by having constant curvature $K$ that determines the spacial geometries according to three possible values assigned $K = \{1,0, -1\}$ from which either a closed, or a flat or an open universe is obtained, respectively. In addition, for this metric and for the rest of the work, the signature $(-, +, +, +)$ is taken. Accordingly, the line element acquires the following form:\cite{weinberg2008cosmology}
\begin{equation}\label{metric}
{\rm d} s^2 = -{\rm d} t^2 + a^2(t)\left(\frac{dr^2}{1-Kr^2} + r^2d\Omega^2 \right),
\end{equation}
This metric allows to solve the Einstein field equations exactly. With the above elements, the non-zero components of the Ricci tensor are
\begin{eqnarray}
R_{00} &=& 3\frac{\ddot{a}}{a},\nonumber\\
R_{ij} &=& \left(\frac{\ddot{a}}{a^2} + 2\frac{\dot{a}^2}{a^2} + \frac{2 K}{a^2} \right)g_{ij},\nonumber\\
R_{0j} &=& 0 \, \forall \, j,
\end{eqnarray}
The Ricci scalar can be obtained through the expression $R = R^{\mu}_{\mu} \equiv g^{\mu\alpha}R_{\alpha\nu}$ so,
\begin{eqnarray}
R &=& 6\left[\frac{\ddot{a}}{a} + \left(\frac{\dot{a}}{a}\right)^2 + \frac{K}{a^2} \right].
\end{eqnarray}
The non-zero components of the Einstein tensor with mixed indices are
\begin{eqnarray}
G^0_0 &=& 3\left[ \left(\frac{\dot{a}}{a}\right)^2 + \frac{K}{a^2}\right]\nonumber\\
G^i_j &=& \left[2\frac{\ddot{a}}{a} + \left(\frac{\dot{a}}{a}\right)^2 + \frac{K}{a^2}\right]\delta^i_j.
\end{eqnarray}
Taken together, these tools make it possible to generate the equations that govern the dynamics of the universe known as the Friedmann and Raychaudhuri equations, respectively:
\begin{eqnarray} 
\left(\frac{\dot{a}}{a}\right)^2 + \frac{K}{a^2} &=& \frac{8\pi G}{3}\rho,  \label{friedmann}\\ 
2\frac{\ddot{a}}{a} + \left(\frac{\dot{a}}{a}\right)^2 + \frac{K}{a^2} &=& -8\pi G p.\label{raychaudury}
\end{eqnarray}
From the above equation, an expression for the cosmic acceleration is obtained by subtracting the equation (\ref{friedmann}) from the equation (\ref{raychaudury}):
\begin{eqnarray}\label{acceleration}
\frac{\ddot{a}}{a} = -\frac43 \pi G (\rho + 3p).
\end{eqnarray}
To complement this set of equations, an equivalent equation for the conservation of the energy known as continuity equation is obtained by taking the derivative with respect to the cosmological time of the equation (\ref{friedmann}) and then using the result in equation (\ref{raychaudury}):
\begin{eqnarray} \label{continuidad1}
\dot{\rho} + 3H(\rho + p)= 0.
\end{eqnarray}
Henceforth the Hubble parameter is defined as $H \equiv \dot{a}/a$ and the reduced mass Planck as $M^2_P = 1/8\pi G$. 
\subsection{The Cosmological Constant}

The possibility to introduce an universal constant in the field equations (\ref{ecucampoeinstein}) led Einstein in 1917 to make the first modification of general relativity in order to justify the idea of a static universe. In other words, the role of the cosmological constant in this model was to compensate for the action of gravity so that the universe does not collapse \cite{Einstein1917}. Later on, in 1929, the observations and measurements made by American astronomer Edwin Hubble --- concerning the linear relationship between the redshift and the radial distance of the galaxies --- led Einstein to abandon the idea of the cosmological constant and to reproduce an expanding universe model, consistent with the observations to that date. Einstein also recognized that introducing a cosmological constant into a static universe meant an unstable universe. However, some of his colleagues agree to retain the cosmological constant for reasons of mathematical generality and for the use of other possible physical purposes associated with cosmic expansion, as expressed by Richard Tolman in a letter to Einstein in 1931 \cite{Tolman1931}:

\blockquote{``...since the introduction of the $\Lambda$–term provides the most general possible expression of the second order which would have the right properties for the energy-momentum tensor, a definite assignment of $\Lambda = 0$, in the absence of experimental determination of its magnitude, seems arbitrary and not necessarily correct.''}

Since then, two relevant concerns have emerged regarding the cosmological constant of both theoretical and experimental nature: what is the meaning of the cosmological constant and how to obtain direct observational evidence from it. In the 90's a new era of space exploration carried out by satellites and telescopes such as COBE and the Hubble space telescope brought in new data on fluctuations in the temperature of the cosmic background radiation of the order of $\Delta T/T \approx 10^{-5}$ and an observational value of the Hubble constant of $(80 \pm 17)\, {\rm km}\, {\rm s}^{-1}\,{\rm MPc}^{-1}$ began to suggest a universe in which the contribution of the energy density associated with the cosmology constant was dominant --- the new model was known as $\Lambda$CDM model. Between 1998 and 1999, the Supernova Cosmology Project and High-Z Supernova programs were aimed at measuring the decceleration parameter $q_0$ using a class of supernovae known as type Ia, which could serve as standardizable candles for the measurement of the distant galaxies and yield a value of the Hubble constant over great distances. These pieces of information provided strong evidence of a period of accelerated expansion over the last billion years. The measurements suggest that $q_0$ has a value of about $-1.2037 \pm 0.175$ were recently obtained in \cite{10.1093/mnras/stz176}.  It was then estimated that the contribution from matter was close to $\Omega_M \approx 0.3$, and a large part of the contribution of the total energy density was from the cosmological constant with $\Omega_{\Lambda} \approx 0.7$. The last observations by satellites Wilkinson Microwave Anisotropy Probe (WMAP) \cite{Wmap2013ApJS}, PLANCK \cite{Planck2015}, the Sloan Digital Sky Survey \cite{sloan2015ApJ}, the Hubble Space Telescope \cite{hubble2018}, and the Chandra X-ray Observatory \cite{ Chandra2018reu}, equipped with very high precision measurement technologies have corroborated this data. However, despite having such an excellent collection of data, the question about the meaning of the cosmological constant is still quite elusive because, to date, nothing is known about its physical nature or why the value of the cosmological constant is extremely small and nonzero.

Let us now revise the field equations with cosmological constant. Einstein's proposal was to modify the equation
 (\ref{ecucampoeinstein}) to
\begin{eqnarray}\label{ecucampoeinsteinC}
 R_{\mu\nu} -\frac12 R g_{\mu\nu} + g_{\mu\nu}\Lambda  = 8\pi G T_{\mu \nu}.
\end{eqnarray}
without loss of of the general covariance. 

With this modification, the Friedmann (\ref{friedmann}) and Raychaudury (\ref{raychaudury}) equations become 
\begin{eqnarray} 
\left(\frac{\dot{a}}{a}\right)^2 + \frac{K}{a^2} + \frac{\Lambda}{3} &=& \frac{8\pi G}{3}\rho,  \label{friedmannC}\\ 
2\frac{\ddot{a}}{a} + \left(\frac{\dot{a}}{a}\right)^2 + \frac{K}{a^2} + \frac{\Lambda}{3} &=& -8\pi G p. \label{raychauduryC}
\end{eqnarray}
From these equations it is possible to obtain a solution for a static universe. Classically, the vacuum has no energy density, but according to the quantum field theory the vacuum is the state of the lowest possible energy density. The idea of associating the cosmological constant with a non-zero vacuum energy density is therefore assumed. Einstein defined the vacuum energy density as
\begin{equation}\label{vac}
    \rho_{\Lambda} = \frac{\Lambda}{8 \pi G}.
\end{equation}
Incorporating the continuity equation (\ref{continuidad1}) into the previous vacuum density energy (\ref{vac}) yields
\begin{equation}
    \dot{\rho_{\Lambda}} = -3H (\rho_{\Lambda} + p_{\Lambda}) = 0,
\end{equation}
which reduces to
\begin{equation}
    p_{\Lambda} = -\rho_{\Lambda}.
\end{equation}
The pressure associated with the vacuum energy density is negative. Consider the density and total energy pressure of the universe composed of a contribution of ordinary matter and a contribution associated with vacuum energy, that is
\begin{eqnarray}
\rho_{tot} &=& \rho_m + \rho_{\Lambda},\\
p_{tot} &=& p_m + p_{\Lambda}.
\end{eqnarray}
The cosmic acceleration, given by equation (\ref{acceleration}), is
\begin{equation}
    \frac{\ddot{a}}{a} = -\frac43 \pi G (\rho_m + 3p_m - 2\rho_{\Lambda}),
\end{equation}
from which one can see that the  vacuum energy density can drive the acceleration of the universe.

Considering now the Friedmann equation (\ref{friedmann}) when the energy density of the vacuum dominates:
\begin{eqnarray}
    H_{\Lambda} = \left(\frac{\dot{a}}{a}\right)^2 &=& \frac{8\pi G}{3}\rho_{\Lambda}, 
\end{eqnarray}
we get that
\begin{equation}
    a(t) \propto e^{H_{\Lambda}t}.
\end{equation}
It is also notable that adding the cosmological constant to Einstein's field equations will increase the age of the universe . 

Next we calculate the age of the universe from observable cosmological parameters such as the densities of radiation, matter and cosmological constant. From the Friedmann equation (\ref{friedmann}) we get
\begin{eqnarray}
     \left(\frac{\dot{a}}{a}\right)^2 &=& \frac{8\pi G}{3}\left( \rho_m + \rho_r + \rho_{\Lambda}\right) - \frac{K}{a^2}.
\end{eqnarray}
For matter we have that
\begin{equation}
    \rho_m(t) = \left(\frac{a(t_0)}{a(t)}\right)^3 \rho_{m,0}.
\end{equation}
The subscript ``\textit{0}'' indicates the present moment. For radiation we have
\begin{equation}
    \rho_r(t) = \left(\frac{a(t_0)}{a(t)}\right)^4 \rho_{r,0}\, ,
\end{equation}
and, for the cosmological constant
\begin{equation}
    \rho_{\Lambda}(t) = \rho_{\Lambda,0}.
\end{equation}
In cosmology it is usual to express mass densities in terms of the fraction $\Omega$ of the \textit{critical density} $\rho_c$, where the critical density is defined as that for which the total mass density makes the universe flat:
\begin{equation}
    \rho_c = \frac{3H^2}{8\pi G}.
\end{equation}
So, by definition,
\begin{equation}
    \Omega_x \equiv \frac{\rho_x}{\rho_c}\,,
\end{equation}
and defining
\begin{equation}
    \Omega_K \equiv - \frac{k}{a^2_0 H^2_0}.
\end{equation}
We can write
\begin{eqnarray}
 \rho_m(t) &=& \frac{3H^2_0}{8\pi G}\left(\frac{a(t_0)}{a(t)}\right)^3 \Omega_{m,0},\\
 \rho_r(t) &=& \frac{3H^2_0}{8\pi G}\left(\frac{a(t_0)}{a(t)}\right)^4 \Omega_{r,0},\\
 \rho_{\Lambda}(t) &=& \frac{3H^2_0}{8\pi G} \Omega_{\Lambda,0},
\end{eqnarray}
such that Friedmann's equation becomes
\begin{equation}\label{friedmannomega}
    H^2(t) = H^2_0\left[\Omega_{r,0}\left(\frac{a_0}{a(t)}\right)^4 + \Omega_{m,0} \frac{a_0}{a(t)} +\Omega_{K,0}\left(\frac{a_0}{a(t)}\right)^2 + \Omega_{\Lambda,0} \right].
\end{equation}
For $t=t_0$, this reduces to
\begin{equation}
    1 = \Omega_{r,0} + \Omega_{m,0} +\Omega_{K,0} + \Omega_{\Lambda,0},
\end{equation}
so that
\begin{equation}\label{omegaK}
    \Omega_{K,0} \equiv 1 - \Omega_{r,0} -\Omega_{m,0} - \Omega_{\Lambda,0}.
\end{equation}
In equation (\ref{friedmannomega}), we add the constraint\footnote{which can be seen as a mere normalization procedure if $K=0$.} $a_0 = 1$:
\begin{eqnarray}
     \frac{\dot{a}}{a} &=& H_0\sqrt{\Omega_{r,0}a^{-4}+ \Omega_{m,0} a^{-3} +\Omega_{K,0}a^{-2} + \Omega_{\Lambda,0}},\nonumber\\
     a \frac{da}{dt}&=& H_0\sqrt{\Omega_{r,0}a^{-4}+ \Omega_{m,0} a^{-3} +\Omega_{K,0}a^{-2} + \Omega_{\Lambda,0}},\nonumber\\
     dt &=& \frac{da}{H_0\, a\sqrt{\Omega_{r,0}a^{-4}+ \Omega_{m,0} a^{-3} +\Omega_{K,0}a^{-2} + \Omega_{\Lambda,0}}}.
\end{eqnarray}
Integrating the previous equation, we get
\begin{equation}
    t_0 = \int_0^1 \frac{da}{H_0\, a\sqrt{\Omega_{r,0}a^{-4}+ \Omega_{m,0} a^{-3} +\Omega_{K,0}a^{-2} + \Omega_{\Lambda,0}}}.
\end{equation}
This is the equation used to calculate the age of the universe. The only possibility to integrate this equation is numerically and the result is approximately equal to $13.8 \times10^9$ years \cite{Planck2018}. Therefore, including the cosmological constant to Einstein's equations generates an age consistent with the estimated age of the oldest stars\footnote{But $t_0 = 9.6\times10^9$ years \cite{Planck2018} if $\Lambda=0$, which is somewhat younger than the oldest objects in the galaxy, though not by many standard deviations \cite{weinberg2008cosmology}.}. A particular possibility of analytical integration results when we take both $ \Omega_K = 0 $ and $ \Omega_r = 0 $ and considering only the contribution of matter density and vacuum density, that is
\begin{equation}
    1 = \Omega_{m} + \Omega_{\Lambda}.
\end{equation}
In this case, the possible solutions are
\begin{equation}
t_0 = \begin{cases}
&\frac{2}{3H_0}\frac{\tan^{-1}\sqrt{\Omega_{m,0}-1}}{\sqrt{\Omega_{m,0}-1}}, \;\; \text{for}\;\;  \Omega_{m, 0} > 1, \;\; \Omega_{\Lambda} <0, \\
&\frac{2}{3H_0}, \;\;\;\;\;\;\;\;\;\;\;\;\;\;\;\;\;\;\;\;\;\;\;\, \text{for}\;\; \Omega_{m, 0} = 1, \;\; \Omega_{\Lambda} =  0, \\
&\frac{2}{3H_0}\frac{\tan^{-1}\sqrt{1 - \Omega_{m,0}}}{\sqrt{1 - \Omega_{m,0}}},  \;\; \text{for}\;\; \Omega_{m, 0} < 1, \;\; \Omega_{\Lambda}> 0.
\end{cases}
\end{equation}
We use (\ref{raychaudury}) to define the deceleration parameter as $q_0 \equiv -\ddot{a}(t_0)a(t_0)/\dot{a}^2(t_0)$, since the state parameter $\omega = p/\rho$ has the values of $-1,0$ and $-1/3$ for vacuum, matter and radiation respectively, the pressure in the present is
\begin{equation}
    p_0 =  \frac{3H_0^2}{8\pi G}\left(-\Omega_{\Lambda,0} + \frac13 \Omega_{r,0} \right),
\end{equation}
so we can write $q_0$ in terms of the $\Omega$s as   
\begin{eqnarray}
    q_0 &\equiv& -\frac{\ddot{a}(t_0)a(t_0)}{\dot{a}^2(t_0)}\nonumber\\
    &=& \frac{4\pi G (\rho_0 + 3p_0)}{3 H^2_0}\nonumber\\
    &=& \frac12\left(\Omega_{m,0} - 2\Omega_{\Lambda,0} + 2\Omega_{r,0}\right).
\end{eqnarray}
Considering only the contribution of matter density and vacuum density with good approximation we estimated that $q_{0} = -0.5275 < 0$ \cite{Planck2018}, which means that the expansion of the universe is accelerating. 

Finally, according to the last analysis of results of the Planck mission measurements that reported that \cite{Planck2018}:

\blockquote{
 ... the Universe is spatially flat to high accuracy ($\Omega_K = 0.0007 \pm 0.0019$), matter density parameter $\Omega_m = 0.315 \pm 0.007$, a Hubble constant $H_0 = (67.4 \pm 0.5) km s^{-1} Mpc^{-1}$, in substantial $3.6\sigma$ tension with the latest local determination by Riess et al. \cite{Riess:2018byc}, ..., None of the extended models that we have studied in this paper convincingly resolves the tension with the Riess et al. \cite{Riess:2018byc} value of $H_0$. The $\Lambda$CDM model provides an astonishingly accurate description of the Universe from times prior to $380000$ years after the Big Bang, defining the last-scattering surface observed via the CMB, to the present day at an age of 13.8 billion years.}

It is very exciting that such a simple modification of general relativity will achieve such extraordinary advances, but one important thing is that the nature and meaning of the cosmological constant is still unknown and therefore it is necessary to investigate and explore new theoretical alternatives on the problems associated with this. 

\section{Conclusions}
The variation of the total action with respect to the metric $g_{\mu\nu}$ fixed on the boundary produces Einstein field equations. The boundary term cancels exactly the boundary contribution from the Hilbert-Einstein action. 

The FLRW metric and the perfect fluid tensor, valid under the condition of homogeneity and isotropy of the universe, allow to solve the Einstein field equations (\ref{ecucampoeinstein}) in an exact way. The dynamics of the FLRW universe is determined by the scale factor $a= a(t)$ by the Friedmann (\ref{friedmann}) and Raychauduri (\ref{raychaudury}) equations. On the other hand, the dynamic of the whole material content obeys the equation of continuity (\ref{continuidad1}). 

The addition of the cosmological constant to the Einstein field equations generates the $\Lambda$CDM model. As a result of this modification, we found that the age of the universe is consistent with the age of the oldest stars, the estimation shows an age of $13.8$ billions, and a universe with accelerated expansion in the present.
\chapter{Inflation} 
\label{Chapter2}
\lhead{Chapter 2. \emph{Inflation}}
In this chapter we study the inflation scenario, its motivation and definition. We also investigate the inflationary scalar field known as inflaton, its equations of movement, and the slow-roll approximation that leads to an analytical solution. The numerical solution for the full field equation is also presented under the initial slow-roll conditions. Finally, we analyze the quadratic inflationary potential and the same case with the contribution of the cosmological constant.
\section{Motivations for Inflation}
The successes of the standard cosmological model are as intriguing as their problems. In particular, the standard cosmological model points to an inevitable singularity at the beginning. On the other hand, the principle of homogeneity and isotropy of the universe is corroborated with observations on large scales but there is never a satisfactory explanation of the origin of this cosmological condition. This in turn leads to what it is known as standard cosmology problems. We will briefly describe three of these problems.
\subsection{Horizon Problem}
\begin{figure}[t]
\begin{center}
\includegraphics[width=0.9\textwidth]{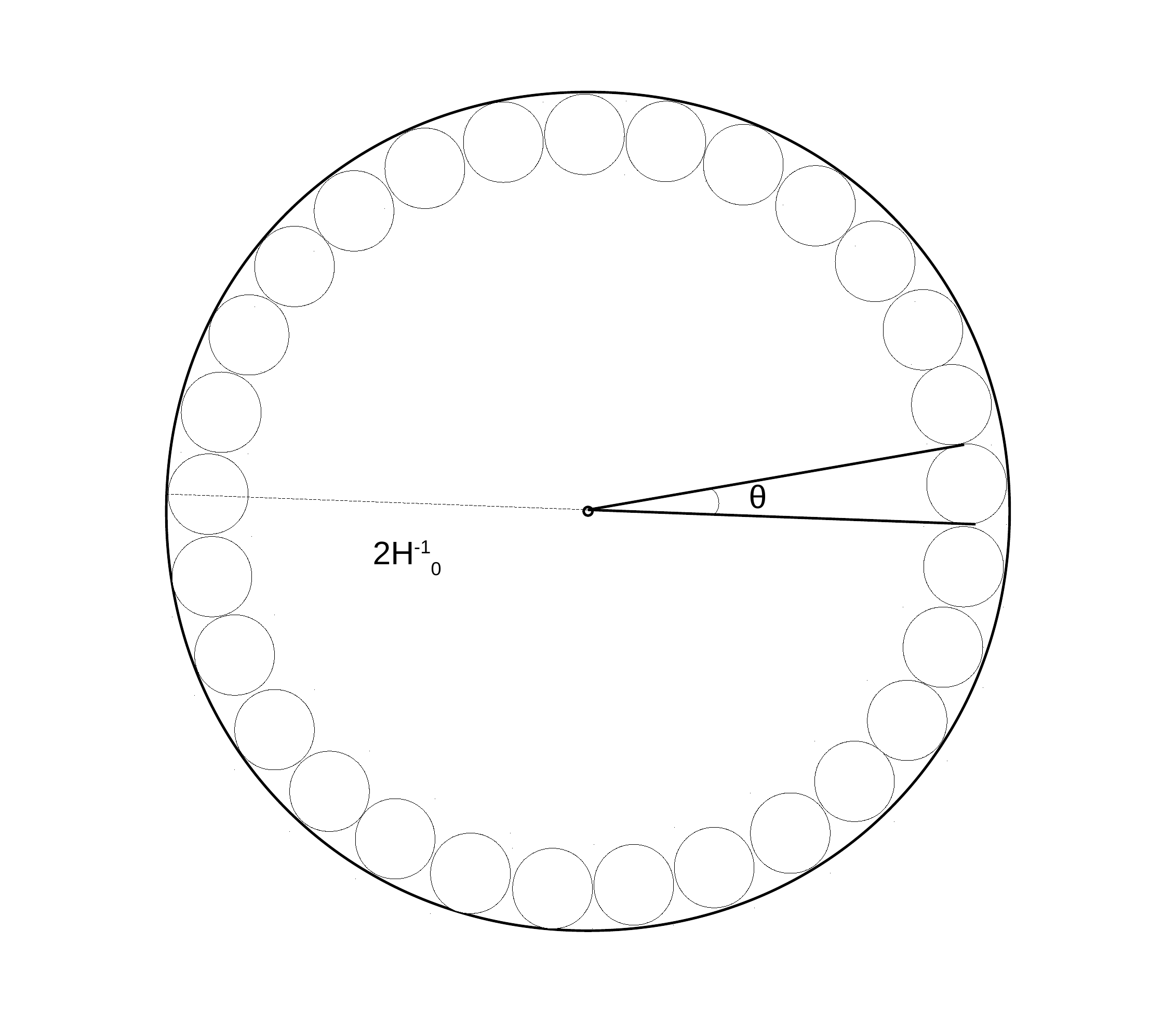}    
\end{center}
\caption[Particle horizon.]{Angles defined by each region causally connected to our current universe. The dotted line defines the current particle horizon $2H_0^{-1}$.}
\label{conos}
\end{figure}

Concerning the uniformity of the universe and by this we mean its physical properties, it is surprisingly uniform. Recent measurements of the CMB show that temperature variations are of the order of $\frac{\delta T}{T} \sim 10^{-5}$. Recall that this cosmic radiation originates during the time of recombination, for $t_{rec} \approx 380,000$ yr. On the other hand, let us point out that the spatial region that corresponds to our current observed universe is obtained by means of the particulate horizon, the maximum distance over which a causal signal can propagate in the age of the universe. By definition, the coordinate distance between a luminous source to the origin\footnote{We considered a FLRW coordinate system in which we are at the origin of the coordinates.} is 
\begin{equation}
    d_c \equiv \frac{2 a_0}{H_0}.
\end{equation}
For the recombination period, we have the same coordinate distance $d_c$ between a luminous source to the origin, but not the same physical distance ($d_{phy}$), that  is,
\begin{eqnarray}\label{distphisica}
    d_{phy,rec} &=& a_{rec}\, d_c\nonumber\\
            &=& 2 H^{-1}_0 \frac{a_{rec}}{a_0}\nonumber\\
            &=& 2 H^{-1}_0 \frac{T_0}{T_{rec}}.
\end{eqnarray}
The particle horizon during recombination is
\begin{eqnarray}
2 H^{-1} &=& 2(H^2)^{-1/2} \propto (\rho_m)^{-1/2} \propto (a^{-3})^{-1/2} \propto (T^3)^{-1/2},\nonumber\\
        &\propto& T^{-3/2}.
\end{eqnarray}
Then,
\begin{equation}
    2 H^{-1}_{rec}T^{3/2}_{rec} = 2 H^{-1}_{0} T^{3/2}_{0},
\end{equation}
and thus,
\begin{equation}
    2 H^{-1}_{rec}  = 2 H^{-1}_{0}\left(\frac{T_0}{T_{rec}} \right)^{3/2}.
\end{equation}
This spatial region contains the following amount of particle horizons:
\begin{eqnarray}
\frac{d^3_{phy,rec}}{\left(2 H^{-1}_{rec}\right)^3} &=& \frac{\left(2 H^{-1}_0 \frac{T_0}{T_{rec}}\right)^3}{\left(2 H^{-1}_0 \frac{T_0}{T_{rec}}\right)^{9/2}} \nonumber\\
&=& \left(\frac{T_{rec}}{T_0}\right)^{3/2}\nonumber\\
&=& \left(\frac{0.26 eV}{2.725 K} \frac{1.1605 \times 10^2 K}{1eV}\right)^{3/2}\nonumber\\
&\sim& 10^4.
\end{eqnarray}
Each of these regions is expected to have a completely different temperature than the temperatures of the other self-causally-connected regions. However, today the same temperature is detected in all of them. Additionally, assume that each one of these regions emits photons with different temperatures, whose light cones are expanding according to the expansion of the universe, see Figure (\ref{conos}).

Then the angle for each region is approximately
\begin{eqnarray}
\theta &=& \frac{2 H_{rec}^{-1}\frac{a_0}{a_{rec}}}{2H_0^{-1}}\nonumber\\
        &=& \frac{ H_{0}^{-1}\left(\frac{T_0}{T_{rec}}\right)^{3/2}\frac{T_{rec}}{T_0}}{H_0^{-1}}\nonumber\\
        &=& \left(\frac{T_0}{T_{rec}}\right)^{1/2}\nonumber\\
        &=& \left(\frac{2.725 K}{0.26 eV} \frac{1eV}{1.1605 \times 10^2 K}\right)^{1/2}\nonumber\\
        &=& 0.03005 \text{~rad}\nonumber\\
        &\approx& 1.72^{\circ}.
\end{eqnarray}
This tells us that if we change the direction of observation in more than in $2^{\circ}$ the sky, the measured temperature should be different, but the background cosmic radiation is highly uniform. This problem of standard cosmology is known as the horizon problem \cite{ weinberg2008cosmology,liddle2000cosmological}.
\subsection{Flatness Problem}
Starting from the definition (\ref{omegaK}) we have
\begin{equation}
    \Omega - 1 = \frac{K}{a^2 H^2}.
\end{equation}
Therefore, if the universe is spatially flat ($K=0$), then $\Omega = 1$ always. But, if $K \neq 0$, $\Omega$ will evolve over time as follows
\begin{equation}
|\Omega -1| \propto \frac{1}{a^2H^2}\begin{cases}
&\frac{1}{a^2 a^{-4}} = a^2, \;\; \text{for radiation domination,} \\
&\frac{1}{a^2 a^{-3}} = a, \;\; \text{for matter domination.}
\end{cases}
\end{equation}
Hence, $|\Omega -1| $ will increase over time, either as $a^2$ or as $a$. However, the current value for $|\Omega -1| $ is \cite{Planck2018}
\begin{equation}
    \Omega_0 - 1 = 0.0007 \pm 0.0019.
\end{equation}
This tells us that if the current value of $\Omega$ is 1, with an accuracy of 1 in $10^4$, then in early times $\Omega$ it was incredibly more accurate closer to $1$. Take for example the value at the start of the nucleosynthesis:
\begin{eqnarray}
\frac{|\Omega -1|_{T = T_N} }{|\Omega -1|_{ T = T_{0}}} &=& \frac{a_N}{a_0}\nonumber\\
        &=& \frac{T_0}{T_N}\nonumber\\
        &=& \frac{2.725 K}{1 MeV} \frac{1 MeV}{1.1605 \times 10^{10} K}\nonumber\\
        &\approx& 0 (10^{-10}).
\end{eqnarray}
This problem is known as a flatness problem \cite{weinberg2008cosmology,liddle2000cosmological}.
\subsection{The Relic Problem}
Some of these relics are the gravitino, the module, and the topological defects\footnote{from expected GUT phase transitions \cite{georgi1974unity, PhysRevD.10.275, linder1990particle}}, such as domain walls, cosmic strings, and magnetic monopoles. We present a brief analysis of this last relic based on some results of particle physics. 

A magnetic monopole is a local monopole possessing a magnetic moment. The work made by 't Hooft and  Polyakov  \cite{tHooft1974, Polyakov} are the first that predicted the magnetic monopoles. For GUT scale of $10^{16}$ GeV, these monopoles would have a mass around $10^{17} - 10^{18}$ GeV. On the other hand, E. Parker obtained an upper bound on the flux of monopoles in the galaxy from astrophysical considerations \cite{Parker:1970xv}. If we take into account this bound, the fraction of the total energy density for the present-day limit for monopoles $\Omega_{mono,0}$
\begin{equation}
    \Omega_{mono,0} = \frac{\rho_{mono,0}}{3 M_p^2 H^2} < 10^{-6}.
\end{equation}
Therefore, the density of magnetic monopoles currently has the following upper bound
\begin{equation}
    \rho_{mono,0} < (3\times 10^{-6})(2.436\times 10^{18}GeV)^2(2.13\times 10^{-42}GeV)^2 h^2 = 3.7 \times 10^{-53} GeV^4.
\end{equation}
The GUT scale temperature is about $10^{28}$ K and, therefore, their density at the time of formation was
\begin{equation}
    \frac{\rho_{mono,form}}{\rho_{mono,0}}= \left(\frac{a_0}{a_{for}}\right)^3 = \frac{g_*(T_{form})}{g_*(T_{0})}\left(\frac{T_{for}}{T_0}\right)^3
\end{equation}
where $g_*(T)$ are the degrees of freedom of the particles for a given temperature $T$. So that
\begin{eqnarray}
\rho_{mono,form} &=& \rho_{mono,0}\frac{g_*(T_{form})}{g_*(T_{0})}\left(\frac{T_{for}}{T_0}\right)^3\nonumber\\
        &=& (3.7\times 10^{-53} GeV^4)\left(\frac{500}{3.36}\right)\left(\frac{10^{28} K}{2.725 K} \right)^3\nonumber\\
        &=& 2.72\times10^{32} GeV^4.
\end{eqnarray}
Nevertheless, according to the theory, one monopole is made within the horizon distance at $t_{GUT} \approx 10^{-36}$, so the density of magnetic monopoles is \cite{ryden2017}
\begin{eqnarray}
n_{mono}(t_{GUT}) = \frac{1 \text{monopole}}{(2 c\, t_{GUT})^3} \approx 10^{82} m^{-3},
\end{eqnarray}
\begin{eqnarray}
\rho_{mono,form} &=& 10^{82} m^{-3}(10^{16} GeV)\nonumber\\
        &=& 10^{98} GeV m^{-3}\left(6.582\times 10^{-25} GeV s\right)^3 \left(2.998\times 10^{8} m s^{-1}\right)^3\nonumber\\
        &=& 7.68 \times 10^{50} GeV^4.
\end{eqnarray}
We observe that the abundance predicted by the models of formation of monopoles is much greater than that observed experimentally. This is the problem of magnetic monopoles \cite{weinberg2008cosmology}.
\subsection{Solution to Cosmological Problems}
To properly solve the problems outlined above, a period of strongly accelerated expansion ---  known as inflation --- is required at the beginning of the universe. 

\begin{figure}[t]
\begin{center}
\includegraphics[width=\textwidth]{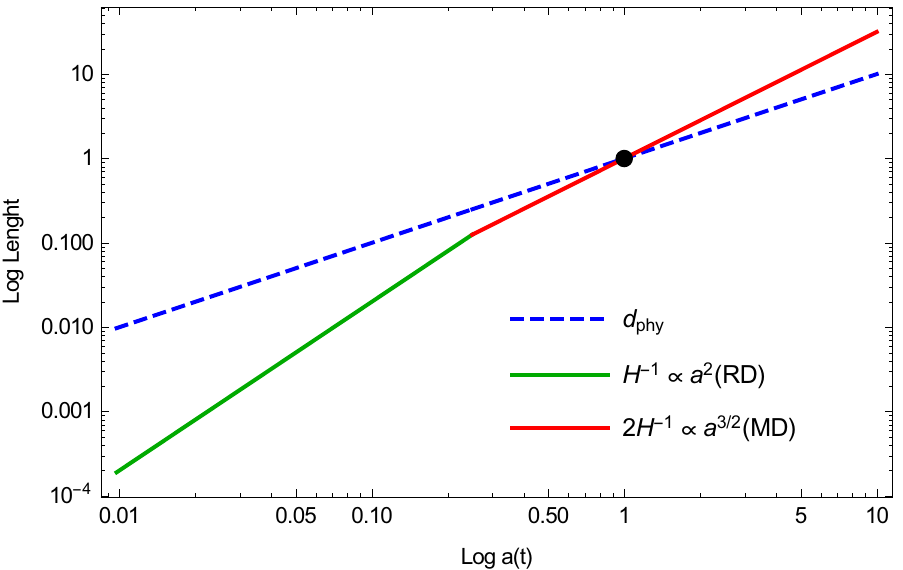}    
\end{center}
\caption[Particle horizon evolution]{Logarithmic plot of particle horizon evolution (green-red/solid) and physical distance (blue/dashed). The intersection (black dot) of both lines corresponds to the current moment of the observed universe. Going back in time the regions are causally disconnected.}
\label{problemahorizonte}
\end{figure}

\begin{figure}[t]
\begin{center}
\includegraphics[width=\textwidth]{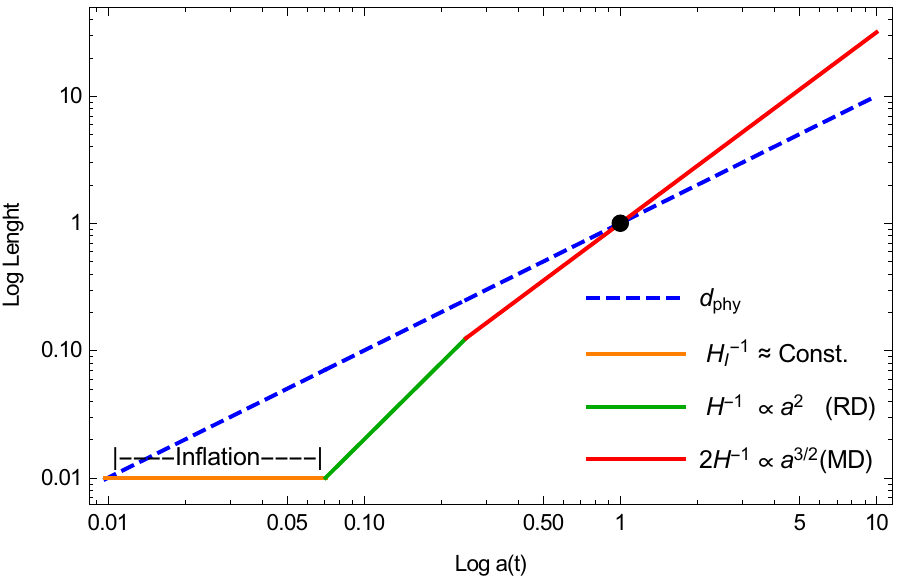}   
\end{center}
\caption[Particle horizon with quasi-constant evolution.]{Logarithmic Plot of Particle Horizon evolution (orange-green-red/solid) and physical distance (blue/dashed). It is considered a period of evolution of the quasi-constant particle horizon at the beginning of the universe (orange/solid) in which the physical distance was within the region of the particle horizon.}
\label{solucionhorizonte}
\end{figure}

Consider, for example, the problem of the horizon. We know that the physical distance is $d_{phy} \propto a$ while the particle horizon evolves according to the cosmological era as $H^{-1} \propto a^2$ during the radiation dominated era (RD) and $2H^{-1} \propto a^{3/2}$ during the matter dominated era (MD). We can see the evolution of each region in the Figure (\ref{problemahorizonte})
We can observe, going back in time, that the region corresponding to our current observed universe will never be in complete causal contact. The condition to solve the horizon problem is that the region corresponding to our current observed universe is reduced during inflation to a value $d_{phy,i}$ smaller than the value of the causally-connected region ($1/2$ of the particle horizon $H_I^{-1}$) at the beginning of inflation (see Figure \ref{solucionhorizonte}).
So that
\begin{eqnarray}
    d_{phy,i} &=& 2 H_0^{-1}\frac{a_i}{a_0}\nonumber\\ 
              &=& 2 H_0^{-1}\frac{a_f}{a_0}\frac{a_i}{a_f}\nonumber\\
              &\simeq& H_I^{-1}.
\end{eqnarray}
Defining the amount of inflation (known as the number of e-folds) as
\begin{equation}
    N = \log\left(\frac{a_f}{a_i}\right),
\end{equation}
we get that
\begin{eqnarray}
    d_{phy,i} &=& 2 H_0^{-1}\frac{T_0}{T_f}e^{-N}\nonumber\\ 
             &\simeq& H_I^{-1}.
\end{eqnarray}
Then,
\begin{eqnarray}
    N &\simeq& \log\left(\frac{2T_0}{T_f}\frac{H_I}{H_0}\right)\nonumber\\ 
    &=& \log\left(\frac{2T_0}{H_0}\right) -\log\left(\frac{T_f}{H_I}\right)\nonumber\\ 
    &=& \log\left(\frac{2 (2.725 K) }{2.1332 h \times 10^{-42}Gev} \frac{1 GeV}{1.1605 \times 10^{13}K}\right) + \log\left(\frac{H_I}{T_f}\right)\nonumber\\ 
    &\approx& 68 + \log\left(\frac{H_I}{T_f}\right).
\end{eqnarray}
Since
\begin{equation}
    \frac{H_I}{T_f} = \sqrt{\frac{\pi^2}{90}g_*(t_f)}\frac{T_f}{M_P},
\end{equation}
and assuming that $g_*(T_f) \approx 500$ and $T_f < M_P$ (to avoid the effects of quantum gravity), we get to relation
\begin{eqnarray}
\frac{H_I}{H_f} &\simeq& 7\nonumber\\
\log\left(\frac{H_I}{H_f}\right) &\simeq& 2.
\end{eqnarray}
Therefore, the amount of inflation required is $N \approx 70$ e-folds of inflation to solve the horizon problem. This solution consecutively solves the problem of flatness and the problem of magnetic monopoles. By definition, during inflation $\Omega$ is driven toward 1. At the end of inflation, $\Omega$ is supposed to be so close to 1 that it remains very close to 1 up to the present \cite{liddle2000cosmological}. A period of exponential expansion that occurred after the production of monopoles (but before photons were created in a period of reheating) would have greatly reduced the monopole to photon ratio. In order to reduce the monopole/photon ratio by a factor $10^{-30}$, it must have increased the horizon size (at some time before the reheating that creates photons) by a factor $10^{10}$, this requires the number $N$ of e-foldings to be greater than $23$ \cite{weinberg2008cosmology}.
%
\section{Inflation}
Formally, inflation is defined as a period when
\begin{equation}\label{inflationdef}
\ddot{a} > 0.
\end{equation}
An important condition for inflation is then obtained from equation (\ref{acceleration}):
\begin{equation} \label{condicion}
\rho + 3p < 0 .
\end{equation}
Thus,
\begin{equation}
p < -\frac{\rho}{3}\;,
\end{equation}
which implies that the necessary condition to generate an inflationary period is that there are negative pressures. This condition is satisfied through a scalar field that evolves with a practically flat potential.

According to the the continuity equation (\ref{continuidad1}) if we consider $p = -\rho$, then the following properties for the inflation period are obtained:
\begin{equation}
\dot{\rho}=0\;;\; H \equiv H_I =\text{Constant}\;,
\end{equation}
and
\begin{equation}
a(t)=a_i e^{H_I(t-t_i)}\;,
\end{equation}
where $H_{I}$ is the Hubble parameter during inflation.
\section{The Inflaton Field}
The scalar field that satisfies the inflation condition is called inflaton, and is denoted by the Greek letter $\phi$. We obtain the density and pressure expressions for a scalar field and we will see what conditions arise for this field according to the definition of inflation. The equations of motion for a scalar field in an universe with a Friedmann-Lema\^{i}tre-Robertson-Walker metric are obtained from the action
\begin{equation}
S=\int d^{4}x\mathcal{L}\;,
\end{equation}
with a Lagrangian density
\begin{equation}
\mathcal{L}=\frac{1}{2}\partial_{\mu}\phi\partial^{\mu}\phi + V(\phi)\;.
\end{equation}
Applying the Euler-Lagrange equations, the equations of field motion known also as Klein-Gordon equations are obtained:
\begin{equation}
\ddot{\phi}+3 H \dot{\phi}-\frac{\nabla^{2}\phi}{a^2}+\frac{\partial V}{\partial \phi}=0\;.
\end{equation}
Since the energy-momentum tensor for a scalar field is defined as
\begin{equation}
T_{\mu\nu}=\partial_{\mu}\phi\,\partial_{\nu}\phi-g_{\mu\nu}\mathcal{L}\;,
\end{equation}
the energy density is
\begin{equation}
\rho_{\phi}\equiv T_{00}= \frac{\dot{\phi}^{2}}{2}+V(\phi)+\frac{(\nabla
\phi)^{2}}{2 a^{2}}\;,
\end{equation}
while the pressure is
\begin{equation}
p_{\phi} \equiv 
\frac{1}{3} T_{\mu\nu} h^{\mu\nu} \equiv 
\frac{1}{3} T_{\mu\nu} \left( g_{\mu\nu} - \delta^0_\mu \delta^0_\nu \right) =
\frac{\dot{\phi}^{2}}{2}-V(\phi)+\frac{(\nabla
\phi)^{2}}{6 a^{2}}\;.
\end{equation}
Considering a homogeneous field, in order to satisfy the cosmological principle, i.e,
\begin{equation} \label{nabla0}
\nabla \phi = 0 ,
\end{equation}
we obtain
\begin{equation} \label{kleingordon}
\ddot{\phi}+3 H \dot{\phi} + \frac{\partial V}{\partial \phi}=0\;,
\end{equation}
\begin{equation}
\rho_{\phi}=  \frac{\dot{\phi}^{2}}{2}+V(\phi)\;,
\end{equation}
\begin{equation}
p_{\phi} = \frac{\dot{\phi}^{2}}{2}-V(\phi)\;.
\end{equation}
For future references, we call equation (\ref{kleingordon}) as the equation of motion for a scalar field in an expanding homogeneous spacetime. In this way, from equation (\ref{condicion}) we obtain
\begin{equation} \label{condicion2}
\dot{\phi}^{2}< V(\phi)\;.
\end{equation}
Equation (\ref{condicion2}) is the only condition of the field during inflation. The condition (\ref{inflationdef}) for inflation can also be written 
\begin{equation}
    -\frac{\dot{H}}{H^2} < 1.
\end{equation}
On the usual assumption that $H$ decreases with time, inflation is an era when $H$ varies slowly on the Hubble timescale. We can begin by assuming that,
\begin{equation}\label{slowasuming}
   |\dot{H}| \ll H^2, 
\end{equation}
then $H$ is practically constant over many Hubble times and we have almost-exponential expansion, $a \propto e^{Ht}$. A universe with H exactly constant is called a de Sitter universe \cite{weinberg2008cosmology}. Then from the equations (\ref{kleingordon}) and (\ref{condicion2}) the following conditions must be satisfied \cite{Lyth:1993eu}:
\begin{equation} \label{slow1}
\dot{\phi}^{2}\ll V(\phi),
\end{equation}
\begin{equation} \label{slow2}
|\ddot{\phi}| \ll |3H\dot{\phi}|.
\end{equation}
The conditions (\ref{slow1}) and (\ref{slow2}) are sufficient but not necessary for the existence of inflation and are known as \textit{slow-roll conditions}.  Applying the slow-roll conditions to equation (\ref{kleingordon}) yields:
\begin{equation}\label{KGsr}
3 H \dot{\phi}\simeq-\frac{\partial V}{\partial \phi}\;.
\end{equation}
Also, using the slow-roll conditions on the Friedmann equation (\ref{friedmann}) approaches
\begin{equation} \label{friedmannsr}
H^{2}\simeq\frac{V(\phi)}{3M^2_P}\;.
\end{equation}
With the pieces of information above, two important parameters for slow-roll conditions, called slow-roll parameters, can be defined:
\begin{equation} \label{parametrosr1}
\epsilon \equiv \frac{M^{2}_{P}}{2}\left( \frac{V'}{V}\right)^{2} \ll 1\;,
\end{equation}
and
\begin{equation} \label{parametrosr2}
\eta \equiv M^{2}_{P}\left|\frac{V''}{V}\right| \ll 1 .
\end{equation}
The primes indicate the derivative with respect to $\phi$. The assumption (\ref{slowasuming}) and the approximations (\ref{KGsr}) and (\ref{friedmannsr}) constitute the slow-roll approximation. The slow-roll approximation implies the conditions (\ref{parametrosr1}) and (\ref{parametrosr2}) on the potential, which we will call flatness conditions \cite{weinberg2008cosmology}.

An additional parameter, already defined, is the number of e-folds $N$. This quantity is calculated using the following expression
\begin{equation}
N(t)\equiv\ln\frac{a(t_{\mathrm{final}})}{a(t_{\mathrm{ini}})}\;,
\end{equation}
which can be expressed in terms of the potential $V(\phi)$ and in terms of the slow-roll parameter $\epsilon$
\begin{equation} \label{efolds}
N = \frac{1}{M^{2}_{P}}\int^{\phi}_{\phi_{\mathrm{end}}}\frac{V(\phi)}{V'(\phi)} d\phi= \frac{1}{M_{P}}\int^{\phi}_{\phi_{\mathrm{end}}}
\frac{1}{\sqrt{2\epsilon}}d\phi\;.
\end{equation}

\section{Spectrum of the Perturbations of the Curvature \texorpdfstring{$\zeta$}{z} in the Inflationary Scenario}
The anisotropies in the temperature of the CMB, $\delta T/T_0$, are directly related to the perturbations in the energy density $\delta \rho/\rho_0$ at the recombination epoch (Sachs-Wolfe effect) \cite{weinberg2008cosmology}, whose primary origin is the extension of the quantum fluctuations of one or more scalar fields $\phi_i$ that fill the universe during inflation \cite{Lyth:1993eu, Rodriguez:2005ru}
\begin{equation} \label{efectosw}
\left(\frac{\delta T}{T_0}\right)_k = -\frac{1}{2}\left(\frac{aH}{k}\right)^2 \left(\frac{\delta \rho}{\rho_0}\right)_k \,. 
\end{equation}
Perturbations in the energy density in the recombination epoch can in turn be quantified by an invariant gauge called \textit{primordial perturbation in the curvature} $\zeta$:
\begin{equation} \label{pert_densidad}
\left(\frac{\delta \rho}{\rho_0}\right)_k = \frac{2}{5}\left(\frac{k}{aH}\right)^2 \zeta_k \,, 
\end{equation}
Geometrically, $\zeta$ measures the spatial curvature of constant-density hypersurfaces, $R^{(3)} = 4\nabla^2 \Psi/a^2$, where $\Psi$ is a scalar metric perturbations \cite{Baumann:2009ds}. It can represent too the difference in the number of e-folds between the two sets of hypersurfaces \cite{kinney2009tasi}. It can be expressed in terms of the fluctuations of the fields $\phi_ {i}$ only. For example, in the case of a single scalar field $\phi$ present during inflation it is given by
\begin{equation}
\zeta = -H_{I} \frac{\delta \phi}{\dot{\phi}_0} \,,
\end{equation}
where $H_{I}$ is the Hubble parameter during inflation. It is convenient to use $\zeta$ to describe the primordial perturbations because it is conserved in superhorizon scales ($k \ll aH_{\rm inf}$), as long as the pressure is a function only of the energy density \cite{Lyth:2004gb, Rigopoulos:2003ak, Wands:2000dp} and it is well defined even after the scalar fields $\phi_i$ have decayed. 
Until recently, the best known and accepted scenario for the origin of density perturbations, the inflationary scenario, identifies inflation as the scalar field whose fluctuations were responsible for density perturbations. This scenario, known as \textit{the inflaton scenario} describes very well the properties of $\zeta$ leading to a nearly scale-invariant power spectrum
\begin{equation} \label{spectrumzeta}
\mathcal{P}_\zeta (k) \equiv A_\zeta^2 \left(\frac{k}{aH_{\rm inf}}\right)^{n_\zeta}\,.
\end{equation}
The amplitude $A_\zeta$ and the spectral index $n_\zeta$ in the equation (\ref{spectrumzeta}) are:
\begin{align} 
\label{AmInflation} 
A^2_\zeta &\simeq \left (\frac{H_*}{\sqrt{8 \varepsilon} \pi M_P} \right )^2 \,\simeq \frac{1}{24\pi^{2} M^4_P}\frac{V}{\epsilon}, \\
n_\zeta &= 2\eta - 6\epsilon,
\end{align}
where we have used equation (\ref{friedmannsr}). The
star $\ast$ denoting the global Hubble parameter evaluated a few Hubble times after horizon exit. It is common to define the spectral index of scalar perturbations as
\begin{equation}\label{ns}
     n_s = 1 + n_{\zeta} = 1 + 2\eta - 6\epsilon,
\end{equation}
in terms of the slow-roll parameters. 

According to last report analysis from PLANCK mission \cite{Planck2018} the value of the amplitude of the spectrum $A_\zeta$ is  $\left( 2 \times 10^{-9} \right)$. In the case of a single scalar field, the bound obtained is
\begin{equation} \label{ncobe}
\frac{V^{1/4}}{\epsilon^{1/4}} = 0.000175 M_P = 4.139 \times 10^{18} \mathrm{GeV} \;,
\end{equation}
that will be known as \textit{the COBE normalization} \cite{COBE1992}. During inflation, fluctuations in the vacuum generate a primordial tensor perturbations, which leads to the corresponding nearly scale-invariant power spectrum
\begin{equation}
\mathcal{P}_T (k) \equiv A_T^2 \left(\frac{k}{aH_{\rm inf}}\right)^{n_T} \,.
\end{equation}
Initially the amplitude of gravitational waves $A_T$ oscillate after entering the horizon with amplitude
\begin{equation}
A^2_T \simeq \left (\frac{\sqrt{2}H_*}{\pi M_P}\right )^2\;.
\end{equation}
It is convenient to specify the relation known as the tensor-to-scalar ratio $r$:
\begin{equation} \label{ratiots}
r \equiv \frac{A^2_T}{A^2_\zeta} = 16\epsilon \;.
\end{equation}

Primordial tensor perturbations have not been observed so far. The current upper limit from the PLANCK observations analysis \cite{Planck2018} on the tensor-to-scalar ratio, $r < 0.10$  $95\%$ of confidence level (CL). Combined with BICEP2/Keck Array BK15/PLANCK data shows that $r < 0.056$ $95\%$ CL \cite{Planck2018}. So, it should be noted that the PLANCK report does not have a lower bound of $r$, and the spectral index $n_s = 0.9649 \pm 0.0042$ $68\%$ CL.
\section{Inflationary Models}
In this section, we will show the inflationary models of interest, the possibility of obtaining inflation of the slow-roll approximation for each proposed potential and the corresponding numerical solution for the full scalar field equation of motion. The models are well known in the literature for their simplicity.
\subsection{The \texorpdfstring{$\phi^2$}{phi2} Potential}
The $\phi^2$ potential is the simplest inflationary potential in the literature: 
\begin{equation}\label{squarepot}
V_{sq}(\phi) = \frac12 m^2_{\phi} \phi^2.
\end{equation}
\subsubsection{Slow-Roll Analysis}
The slow-roll parameters (\ref{parametrosr1}) and (\ref{parametrosr2}) can be written as
\begin{eqnarray}\label{epsilonsq}
\epsilon &=&  \frac{M^{2}_{P}}{2}\left (\frac{V_{sq}'(\phi)}{V_{sq}(\phi)}\right )^{2}\nonumber\\
		 &=&  \frac{M^{2}_{P}}{2}\left (\frac{m^{2}\phi}{\frac12 m^{2}\phi^{2}}\right )^{2} \nonumber\\
		 &=&  \frac{2 M^{2}_{P}}{\phi^2}\;,
\end{eqnarray}
and
\begin{eqnarray}\label{etasq}
\eta &=& M^{2}_{P}\left |\frac{V_{sq}''(\phi)}{V_{sq}(\phi)} \right |\; \nonumber\\
	 &=& M^{2}_{P}\left |\frac{m^2}{\frac12 m^{2}\phi^{2}} \right|\;\nonumber\\
	 &=& \frac{2 M^{2}_{P}}{\phi^2}\;,
\end{eqnarray}
where ($'$) indicates the derivative with respect to $\phi$ ($ \equiv d/d\phi$). slow-roll ends when $\epsilon \simeq 1$, so the scalar field value at the end of inflation $\phi_{\mathrm{end}}$, according to (\ref{epsilonsq}), is
\begin{equation} \label{phiendsq}
\phi_\mathrm{end} \simeq \sqrt{2}M_P\;.
\end{equation}
In order to obtain the initial value of the scalar field ($\phi_{\mathrm{*}}$), equation (\ref{efolds}) can be used; once the integral is solved and replaced in (\ref{phiendsq}), one obtains
\begin{eqnarray} \label{Nsq}
N   &=& \frac{1}{M^2_{P}}\int^{\phi_\mathrm{*}}_{\phi_\mathrm{end}}\frac{V_{sq}(\phi)}{V'_{sq}(\phi)}d\phi \nonumber\\
    &=& \frac{1}{M^2_P}\left(\frac{\phi^2_\mathrm{*}}{4}- \frac{\phi^2_\mathrm{end}}{4}\right)\;.
\end{eqnarray}
Solving for $\phi_{\mathrm{*}}$ from (\ref{Nsq}) we obtain
\begin{eqnarray}\label{phiestrellasq}
\phi_\mathrm{*} = -\sqrt{4N + 2}M_P\;.
\end{eqnarray}
An analytical expression of the field $\phi_{sq} = \phi_{sq}(t)$ can be obtained by solving the system (\ref{KGsr}) and (\ref{friedmannsr}) for the value of the field (\ref{phiestrellasq}), with $N = 60$ efolds, is
\begin{eqnarray}\label{phitsq}
\phi_{sq}(t) = -11 \sqrt{2}\, M_P + \sqrt{\frac{2}{3}}\, m_{\phi}\,t,
\end{eqnarray}
assuming the slow-roll approximation.
\subsubsection{Numerical Solution}
\begin{figure}
\begin{center}
\includegraphics[width=\textwidth]{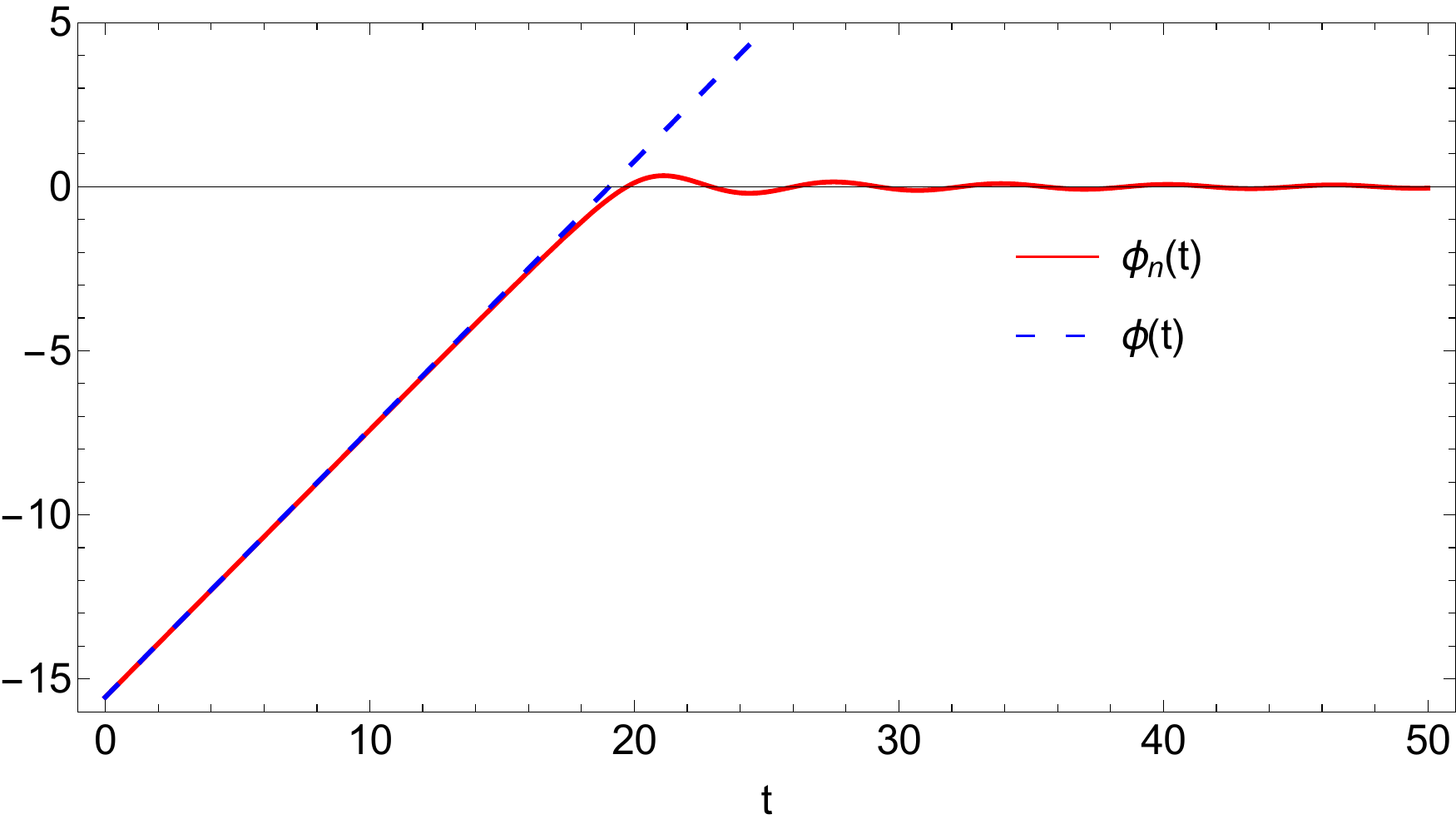}    
\end{center}
\caption[Numerical and analytic solutions for $\phi_{sq}(t)$]{Numerical solution (red/solid) and analytic solutions (blue/dashed) for $\phi_{sq}(t)$ given by Eq.~(\ref{phitsq}), with $N = 60$ efolds, using $m_\phi = 1$ and $M_P = 1$. Note that, before $t\sim 20$, the curves are very close.}
\label{phiplot}
\end{figure}

We used \verb|Mathematica|\textsuperscript{\textcopyright} software \cite{Mathematica} to obtain a numerical solution of the full system equation of movement for the scalar field (\ref{kleingordon})
\begin{equation}  \label{eqphisq}
 \ddot{\phi}(t) + 3 H(t) \dot{\phi}(t) + \partial_\phi V_{sq}\left[\phi(t)\right]=0,
\end{equation}
where
\begin{equation}
H^2(t) = \frac13 \left\{ \frac12 \dot{\phi}(t)^2 + V_{sq}\left[\phi(t)\right] \right\},
\end{equation}
and the initial conditions are the standard ones from the analytic slow-roll solution set for $N=60$ efolds:
\begin{equation}
\phi_{sq}(0) = -11 \sqrt{2}\, M_P \approx - 15.5 M_P \, , \quad
\dot \phi_{sq}(0) = \sqrt{\frac{2}{3}}\, m_{\phi}\approx 0.81 \, m_{\phi}.
\end{equation}
The numerical and analytical solutions are shown in Figure \ref{phiplot}.

The spectral index (\ref{ns}), calculated for $N=60$ (corresponding to the scale $k=0.002/\mathrm{Mpc}$), is
\begin{eqnarray}
n_s &=& 1 - 6 \frac{2 M^{2}_{P}}{(\sqrt{4N + 2}M_P)^2}  + 2 \frac{2 M^{2}_{P}}{(\sqrt{4N + 2}M_P)^2},\nonumber\\
    &=& 1 - \frac{4}{2N + 1},\nonumber\\
    &\approx& 0.96694
\end{eqnarray}
and the tensor-to-scalar ratio (\ref{ratiots}) is
\begin{eqnarray}
r &=& 16\epsilon\nonumber\\
    &=& 16\frac{2 M^{2}_{P}}{(\sqrt{4N + 2}M_P)^2}\nonumber\\
    &=&\frac{16}{2N + 1}\nonumber\\
    &\approx& 0.1322.
\end{eqnarray}
For $N=60$, one gets $n_s = 0.960396$ and $r = 0.158416$. 

The values of the spectral index and the tensor-to-scalar ratio obtained by the quadratic inflationary potential are not favorable according to the last PLANCK's report \cite{Planck2018} --- see Figure ~\ref{fig:planck2018}.    
\subsection{Modifications to \texorpdfstring{$\phi^2$}{phi2mod} Potential}
For the purposes of this study, we will make the following modification to the quadratic potential (\ref{squarepot}):
\begin{equation}\label{potentialsq}
V_{sqm}(\phi) = \frac12 m_{\phi}^2 \left(\phi - a \right)^2 + \Lambda,
\end{equation}
where $a$ is free parameter $\phi$-field Vacuum Expectation Value -- (VEV) and $\Lambda$ is the cosmological constant.
\subsubsection{Slow-Roll Analysis}
For this case, the slow-roll parameters (\ref{parametrosr1}) and (\ref{parametrosr2}) are
\begin{eqnarray}\label{epsilonsqm}
\epsilon &=&  \frac{M^{2}_{P}}{2}\left (\frac{\partial_\phi V_{sqm}(\phi,a,\Lambda)}{V_{sqm}(\phi,a,\Lambda)}\right )^{2} \nonumber\\
	 &=&\frac{2 (a-\phi )^2 M_P^2}{\left(a^2-2 a \phi +2 \Lambda +\phi ^2\right)^2},
\end{eqnarray}
and
\begin{eqnarray}\label{etasqm}
\eta &=& M^{2}_{P}\left |\frac{\partial_{\phi,\,\phi} V_{sqm}(\phi,a,\Lambda)}{V_{sqm}(\phi,a,\Lambda)} \right |\; \nonumber\\
	 &=& \frac{M_P^2}{\frac{1}{2} (a-\phi )^2+\Lambda }.
\end{eqnarray}
Then, the scalar field value at the end of inflation $\phi_{\mathrm{end}}$ according to (\ref{epsilonsqm}) has four roots from which we pick the following one:
\begin{equation} \label{phiendsqm}
\phi_\mathrm{end} \simeq a+\sqrt{-2 \Lambda +M_P^2+\sqrt{M_P^4-4 \Lambda  M_P^2}}\;.
\end{equation}
The initial value of the scalar field ($\phi_{\mathrm{*}}$) is fixed by the minimum number of e-folds $N$:
\begin{eqnarray} \label{Nsqm}
N   &=& \frac{1}{M^2_{P}}\int^{\phi_\mathrm{*}}_{\phi_\mathrm{end}}\frac{V_{sqm}(\phi,a,\Lambda)}{\partial_\phi V_{sqm}(\phi,a,\Lambda)}d\phi \nonumber\\
    &=& \frac{1}{4M^2_P} \left((a-\phi_\mathrm{*} )^2+\Lambda \log(a-\phi_\mathrm{*})-((a-\phi_{end} )^2+\Lambda \log(a-\phi_{end})) \right).
\end{eqnarray}
which yields
\footnotesize
\begin{eqnarray}\label{phiestrellasqm}
\phi_\mathrm{*} = a-\sqrt{2\Lambda}\sqrt{ W_0\left[\frac{\left(-2 \Lambda +M_P^2+\sqrt{M_P^4-4 \Lambda M_P^2}\right) e^{(-2 \Lambda +4 N M_P^2+M_P^2+\sqrt{M_P^4-4 \Lambda  M_P^2}) / 2 \Lambda}}{2 \Lambda }\right]}\;,
\end{eqnarray}
\normalsize
where $W_0$ is Lambert function. 

As we can see, for $N=60$ efolds, the range of $\Lambda$ for which  $\phi_\mathrm{*}$ is Real is $0<\Lambda<\frac{1}{4M_P^2}$. Since $a$ is just a shift, the range of $\phi_\mathrm{*}$ in this case is practically the same as that presented in the case of the quadratic potential. For this reason, we make the following approximation: we turn off the value of $\Lambda$ in the potential (\ref{potentialsq}) and repeating the above procedures again we get
\begin{equation}\label{inisqm}
    \phi_\mathrm{*} = a- \sqrt{2 + 4N}M_P,
\end{equation}
which corresponds to the slow-roll analytical solution
\begin{equation}\label{phisqm}
    \phi_{sqm}(t) = a - 11 \sqrt{2}\, M_P + \sqrt{\frac{2}{3}}\, m_{\phi}\,t.
\end{equation}
\subsubsection{Numerical Analysis}
\begin{figure}
\begin{center}
\includegraphics[width=\textwidth]{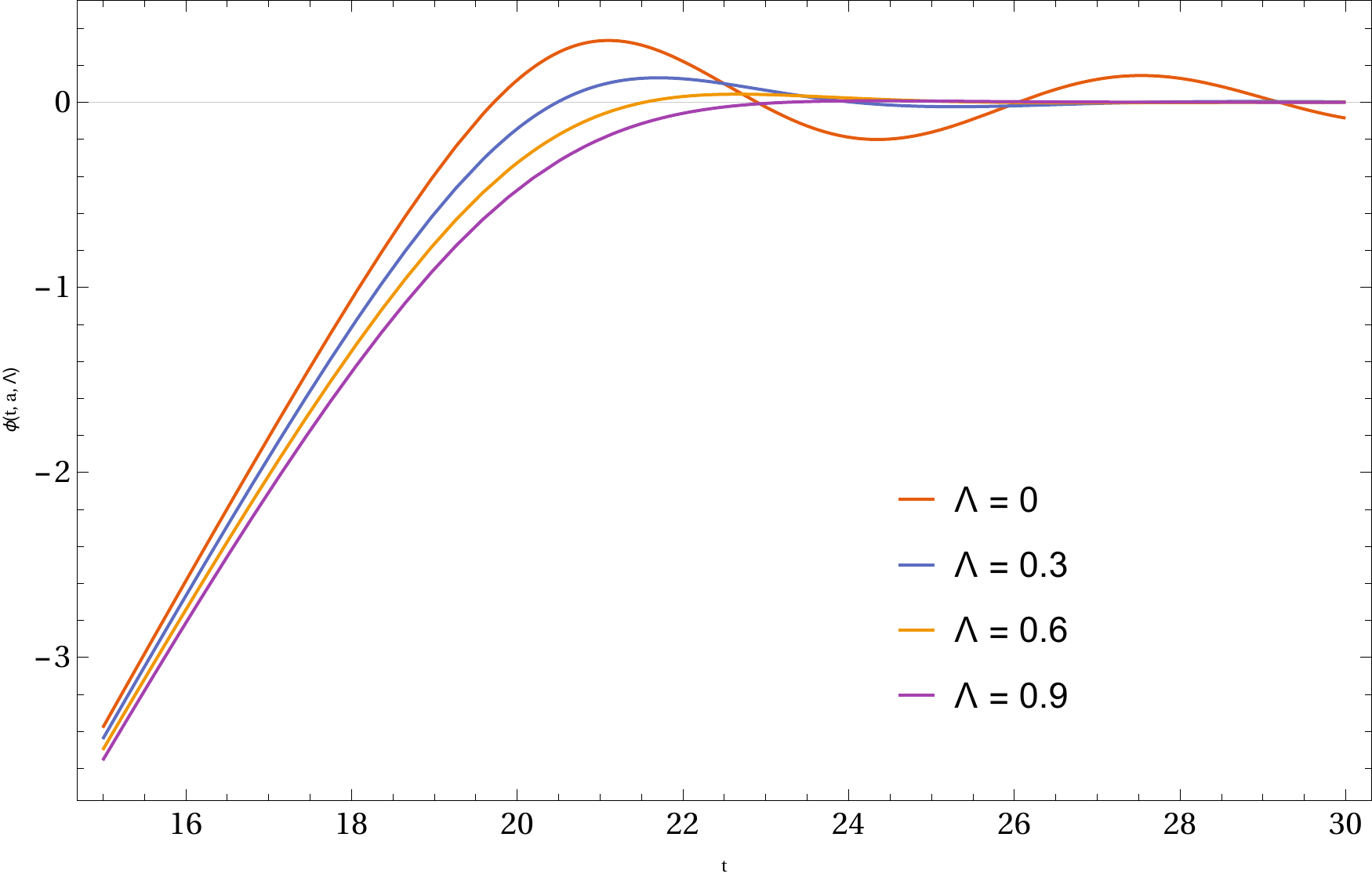}    
\end{center}
\caption[Numerical solutions for Eq.~(\ref{kleingordon})]{Numerical solutions for Eq.~(\ref{kleingordon}), with $N = 60$ efolds and $a=0$, using $m_\phi = 1$ and $M_P = 1$. Note that the curves are smoothed as $\Lambda$ increases. Also, the curves shift up if $a>0$ or shift down if $a<0$.}
\label{smooth}
\end{figure}

\begin{figure}
\center
\includegraphics[width=\textwidth]{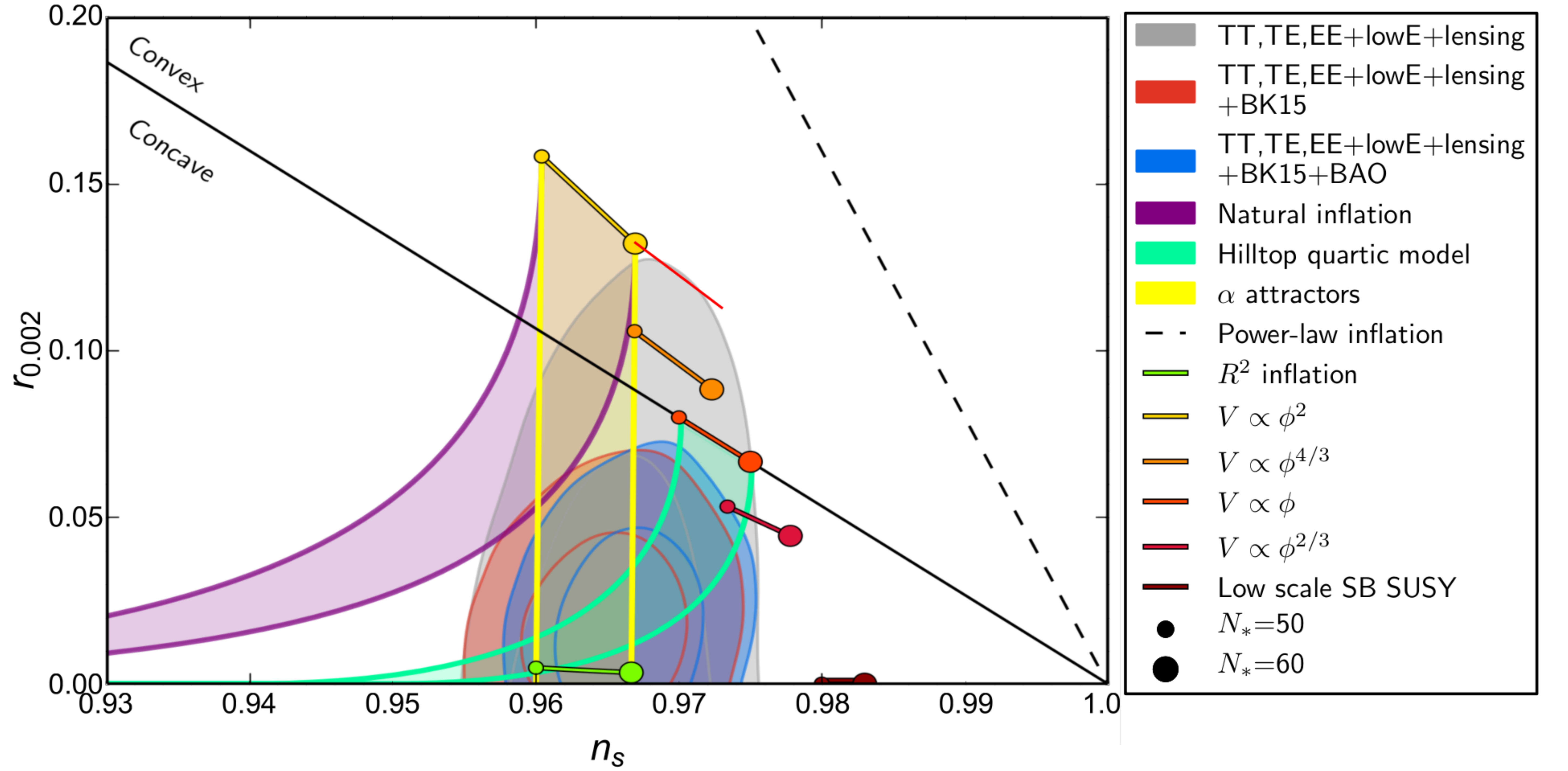} \caption[Planck data 2018]{Marginalized joint $68\%$ and $95\%$ CL regions for $n_s$ and $r$ at $k = 0.002 \text{Mpc}^{-1}$ from Planck alone and in combination with BK15 or BK15+BAO data, compared to the theoretical predictions of selected inflationary models. The modified quadratic potential model (red/solid) for $\Lambda$ range $\{0, 10\}$ $M_P^4$, $N=60$ efolds, and $m_{\phi}=M_P=1$. The $\Lambda$ range $\{2.5,7\}$ $M_P^4$ are in the $95\%$ CL region from Planck alone data. Image adapted from: ESA and the Planck Collaboration \cite{Planck2018}.} 
\label{fig:planck2018} 
\end{figure}

We solved again the full equation system of movement for the scalar field for this case with initial conditions set from the analytic slow-roll above (\ref{phisqm}), namely:
\begin{equation}
\phi_{sqm}(0) \approx a - 15.5 M_P, \quad
\dot \phi_{sqm}(0) = \sqrt{\frac23}\,m_{\phi}\approx 0.81\,m_{\phi}.
\end{equation}
Numerical solutions are shown in Figure \ref{smooth}.
\\ The spectral index (\ref{ns}) for this case is 
\begin{eqnarray}
n_s &=& \frac{\Lambda^2 + M_P^2 (2N - 3)(2N + 1) + 4 \Lambda M_P^2 (N+1)}{\left(\Lambda + M_P^2 (2N+1)\right)^2},
\end{eqnarray}
and the tensor-to-scalar ratio (\ref{ratiots}) is
\begin{eqnarray}
r &=& \frac{16 M_P^4 (1 + 2 N)}{(M_P^2 (1 + 2 N) + \Lambda)^2},
\end{eqnarray}
which do not depend on the $a$ shift. For those, one estimates favorable values with $N=60$ efolds and $\Lambda$ in the range $(2.5,7)$ $M_P^4$ that are in the $95\%$ CL region from PLANCK data alone, but in combination with BK15 or BK15+BAO data, are not preferred \cite{Planck2018}, see Figure ~\ref{fig:planck2018}.

\section{Conclusions}

Inflation is a cosmological scenario that presents an early accelerated expansion. It is generated by a scalar field $\phi$, the inflaton, which solves the problems of standard cosmology and provides a general mechanism for generating (almost scale-free) scalar (density) and tensor perturbations. Additionally, inflaton potential requires a slow-roll region, which leads the slow-roll approximation and the definition of the slow-roll parameters. The minimum inflation duration $N$ to solve the cosmological problems is around 50 to 60 e-folds. The perturbation in the curvature $\zeta$ allows one to quantify the primordial energy-density inhomogeneities produced during inflation. In the Inflation scenario, the properties of $\zeta$ are described by the parameters of the almost scale-invariant power spectrum: the spectral index $n_s$ and the tensor-to-scalar ratio $r$. The measure of this two cosmological parameters is used to discriminate whether an inflationary model is viable.

We presented an analysis of the simple quadratic potential and obtained the analytical solution for the slow-roll approximation and numerical solution for the full field equations. We calculated the slow-roll parameters and the cosmological parameters spectral index $n_s$ and the tensor-to-scalar ratio $r$.  We compared the results with the last-reported data for full analysis from the PLANCK mission that shows a ill-favored evidence for its predictions. A modification for the quadratic potential with a shift $a$ and a cosmological constant $\Lambda$ shows an interesting smooth behavior for non-zero values of the cosmological constant and a better fit in the $95\%$ CL from PLANCK alone data for the spectral index $n_s$ and the tensor-to-scalar ratio $r$, for the following range of $\Lambda$: $\{2.5,7\}$ $M_P^4$.  
\chapter{\texorpdfstring{$f(R)$}{frchapter} Gravity }
\label{Chapter3}
\lhead{Chapter 3. \emph{\texorpdfstring{$f(R)$}{f(R)} Gravity }}
In this chapter we will obtain the field equations of $f(R)$ gravity theory from a modification of the Hilbert-Einstein action using the metric formalism. Additionally, following the calculations in \cite{Guarnizo:2010xr}, we will include a boundary term $S_ {bound}$ in the action which naturally cancels the boundary terms that arise from the aforementioned modification. We derive the dynamical equations (Friedman, Raychauduri and continuity) for a flat FLRW universe with a perfect fluid, from the field equation of $f(R)$ gravity.
\section{Action in \texorpdfstring{$f(R)$}{fRsection} Gravity}
The total action includes a generalized Lagrangian of the Hilbert-Einstein action
\begin{eqnarray}
S_{met} = \frac{1}{2k} \int_{V} d^4x f(R)\sqrt{-g},
\end{eqnarray}
where $f(R)$ is a function of Ricci's scalar $R$, and $k= 8\pi G$. From now on we will use $f \equiv f(R)$ and $f' \equiv df/dR$. A boundary term which is defined by \cite{Dyer:2008hb, Madsen:1989rz} as
\begin{eqnarray}
S_{bound} = 2\oint_{\partial V} d^3y  \, \varepsilon \sqrt{|h|}f' K ,
\end{eqnarray}
and the Lagrangian associated with the fields of matter
\begin{equation}
S_{M} = \int_{V}d^4x \sqrt{-g} L_{M}.
\end{equation}
Now we will obtain the modified Einstein field equations always varying the total action 
\begin{equation}\label{totalactionfr}
   S_{tot} = S_{met} + S_{bound} + S_M 
\end{equation}
with respect to the metric\footnote{An alternative way to obtain the Einstein's field equations was introduced by Palatini in 1919. His formulation consists in treating the $g_{\alpha\beta}$ metric and the $\Gamma^{\alpha}_{\beta\gamma}$ connection as two independent fields \cite{Wald:1984rg, misner2017gravitation}. So, instead the standard metric variation of the Hilbert-Einstein action, a variation respect to the independent connection of an action with gravitational Lagrangian $\mathcal{R} \equiv g^{\alpha\beta}\mathcal{R}_{\alpha\beta}$ where $ \mathcal{R}_{\alpha\beta}$ is the Ricci tensor constructed with the independent connection, and the matter action independent of the connection. In the same way, a generalization of the Hilbert-Einstein action, one can take a Palatini version taken the variation respect to the independent connection in order to obtain the modified Einstein's equations. This version is known as Palatini formalism \cite{Sotiriou:2007yd}. Another alternative is the metric-affine formalism. This formalism is similar to Palatini's, except that now the fields of matter depend on the connections $\Gamma^{\alpha}_{\beta,\gamma}$. In addition, this gives the freedom for the manifold $M$ in this formalism to have torsion ($\Gamma^{\alpha}_{\beta\gamma} = \Gamma^{\alpha}_{\gamma\beta}$). This freedom allows the spin of the particles in some energy regimes to interact with the geometry, which directly gives us the existence of torsion in the theory \cite{papapetrou1978new,Kunstatter1980,tsamparlis1978palatini,Sotiriou:2006qn}.} $g_{\alpha\beta}$, fixing the metric at the boundary $\partial V$, i.e,\cite{poisson2004relativist, Wald:1984rg}
\begin{equation}
\delta g_{\alpha\beta} \bigg\rvert_{\partial V} = 0,
\end{equation}
and then obtaining
\footnotesize
\begin{eqnarray}
\delta S_{tot} &=& \int_{V} d^4x \left(\frac{1}{2k}\frac{\delta(f\sqrt{-g})}{\delta g^{\alpha\beta}}  +  \frac{\delta(L_{M}\sqrt{-g})}{\delta g^{\alpha\beta}} \right)\delta g^{\alpha\beta} + 2\oint_{\partial V} d^3y \, \varepsilon \sqrt{|h|} \frac{\delta \left(f' K \right)}{\delta g^{\alpha\beta}}\, \delta g^{\alpha\beta}  = 0.
\end{eqnarray} \normalsize
The first integral results in
\begin{equation} \label{integral1}
\delta S_{met} = \int_{V} d^4x \frac{1}{2k}\delta(f\sqrt{-g}) = \frac{1}{2k}\int d^4x \left(f \delta(\sqrt{-g}) + \sqrt{-g}\delta f \right).
\end{equation}
On the other hand, from equation (\ref{integral1}) one note that
\begin{equation}
\delta f = f'\delta R.
\end{equation}
In Appendix \ref{AppendixA} we obtain 
\begin{equation} \label{evaluacion1}
g^{\alpha\beta}(\delta\Gamma^{\sigma}_{\beta\alpha}) - g^{\alpha\sigma} (\delta\Gamma^{\gamma}_{\alpha\gamma}) = g_{\mu\nu}\nabla^{\sigma}\delta g^{\mu\nu} + \nabla_{\gamma}\delta g^{\sigma\gamma}.
\end{equation} 
We replace (\ref{evaluacion1}) in (\ref{resultado2})
\begin{eqnarray} \label{deltaR2}
 \delta R &=& \delta g^{\alpha\beta} R_{\alpha\beta} + \nabla_{\sigma} \left( g_{\mu\nu}\nabla^{\sigma}\delta g^{\mu\nu} + \nabla_{\gamma}\delta g^{\sigma\gamma} \right)\nonumber\\
&=& \delta g^{\alpha\beta} R_{\alpha\beta} + g_{\mu\nu}\nabla_{\sigma} \nabla^{\sigma} \delta g^{\mu\nu} - \nabla_{\sigma}\nabla_{\gamma}\delta g^{\sigma\gamma} \nonumber\\
&=& \delta g^{\alpha\beta} R_{\alpha\beta} + g_{\alpha\beta} \square \delta g^{\alpha\beta} - \nabla_{\alpha} \nabla_{\beta}\delta g^{\alpha\beta}.
\end{eqnarray}
We replace (\ref{resultado1}) and (\ref{deltaR2}) in (\ref{integral1}) and arrive at
\footnotesize
\begin{align}
 \delta S_{met} &= \frac{1}{2k}\int_{V}d^4x \left( f \delta \sqrt{-g} + \sqrt{-g} f' \delta R \right)\nonumber\\
&=  \frac{1}{2k}\int_{V}d^4x \Big[ -\frac12 f \sqrt{-g}g_{\alpha\beta}\delta g^{\alpha\beta} + \sqrt{-g} f' \left(\delta g^{\alpha\beta} R_{\alpha\beta} + g_{\alpha\beta} \square \delta g^{\alpha\beta} - \nabla_{\alpha} \nabla_{\beta}\delta g^{\alpha\beta} \right) \Big]\nonumber\\
&=\frac{1}{2k} \int_{V}d^4x \sqrt{-g} \Big[- \frac12 f g_{\alpha\beta}\delta g^{\alpha\beta} + f' \left( \delta g^{\alpha\beta} R_{\alpha\beta} + g_{\alpha\beta} \square \delta g^{\alpha\beta} - \nabla_{\alpha} \nabla_{\beta}\delta g^{\alpha\beta}  \right)  \Big].
\end{align} \normalsize
Let us consider the following integrals
\begin{eqnarray}
\int_{V}d^4x\sqrt{-g} f'g_{\alpha\beta}\square\delta g^{\alpha\beta} \quad {\rm and}\\
\int_{V}d^4x\sqrt{-g} f'\nabla_{\alpha} \nabla_{\beta}\delta g^{\alpha\beta},
\end{eqnarray}
which can be integrated by parts. For this we define the following quantities
\begin{eqnarray}
M_{\tau} &=& f' g_{\alpha\beta} \nabla_{\tau}\delta g^{\alpha\beta} - g_{\alpha\beta} \delta g^{\alpha\beta}\nabla_{\tau} f' \quad {\rm and} \label{Mtau}\\
N^{\sigma} &=& f' \nabla_{\gamma}f'\label{Nsigma}.
\end{eqnarray}
We then take the covariant derivative of $M_{\tau}$:
\begin{eqnarray} \label{cantidad1}
\nabla^{\tau}M_{\tau} &=& \nabla^{\tau} \left( f' g_{\alpha\beta}\nabla_{\tau}\delta g^{\alpha\beta}\right) - \nabla^{\tau}\left(g_{\alpha\beta} \delta g^{\alpha\beta}\nabla_{\tau}f'\right) \nonumber\\
&=& \nabla^{\tau}f' g_{\alpha\beta}\nabla_{\tau}\delta g^{\alpha\beta} + f' g_{\alpha\beta}\square\delta g^{\alpha\beta} - g_{\alpha\beta}\nabla^{\tau}\delta g^{\alpha\beta} \nabla_{\tau}f' - g_{\alpha\beta}\delta g^{\alpha\beta}\square f'\nonumber\\
&=& f'g_{\alpha\beta}\square\delta g^{\alpha\beta} - g_{\alpha\beta}\delta g^{\alpha\beta}\square f'.
\end{eqnarray}
Now, integrating (\ref{cantidad1}) we get
\begin{equation}\label{integralderivadacovarianteM}
\int_{V}d^4x\sqrt{-g} \nabla^{\tau} M_{\tau} = \int_{V}d^4x\sqrt{-g} f' g_{\alpha\beta}\square\delta g^{\alpha\beta} - \int_{V}d^4x\sqrt{-g} g_{\alpha\beta}\delta g^{\alpha\beta}\square f'.
\end{equation}
By the Gauss-Stokes theorem, the left-hand part of (\ref{integralderivadacovarianteM}) is written as
\begin{equation}
\oint_{\partial V}d^3y \varepsilon \sqrt{|h|}n^{\tau}M_{\tau} = \int_{V}d^4x\sqrt{-g} f' g_{\alpha\beta} \square \delta g^{\alpha\beta}- \int_{V}d^4x \delta g^{\alpha\beta}g_{\alpha\beta}\square f' ,
\end{equation}
which can be expressed as
\begin{equation}
\int_{V}d^4x\sqrt{-g} f' g_{\alpha\beta} \square\delta g^{\alpha\beta} =  \int_{V}d^4x \delta g^{\alpha\beta}g_{\alpha\beta}\square f' + \oint_{\partial V}d^3y \varepsilon \sqrt{|h|}n^{\tau}M_{\tau}.
\end{equation}
We follow the same procedure for $N^{\sigma}$:
\begin{eqnarray} \label{cantidad2}
\nabla_{\sigma} N^{\sigma} &=& \nabla_{\sigma}\left(f'\nabla_{\gamma}\delta g^{\sigma\gamma}\right) - \nabla_{\sigma}\left(\delta g^{\sigma\gamma} \nabla_{\gamma}f' \right)\nonumber\\
&=& \nabla_{\sigma}f'\nabla_{\gamma}\delta g^{\sigma\gamma} + f'\nabla_{\sigma}\nabla_{\gamma}\delta g^{\sigma\gamma} - \nabla_{\sigma}\delta g^{\sigma\gamma}\nabla_{\gamma}f' - \delta g^{\sigma\gamma} \nabla_{\sigma}\nabla_{\gamma}f'\nonumber\\
&=& f'\nabla_{\sigma}\nabla_{\beta}\delta g^{\sigma\beta} - \delta g^{\sigma\beta}\nabla_{\sigma}\nabla_{\beta}f',
\end{eqnarray}
which yields, upon integration:
\begin{eqnarray}\label{integralderivadacovarianteN}
\int_{V}d^4x\sqrt{-g} \nabla_{\sigma} N^{\sigma} = \int_{V}d^4x\sqrt{-g} f'\nabla_{\sigma}\nabla_{\beta}\delta g^{\sigma\beta} - \int_{V}d^4x\sqrt{-g} \delta g^{\sigma\beta}\nabla_{\sigma}\nabla_{\beta}f'.
\end{eqnarray}
By the Gauss-Stokes theorem, the left-hand part of (\ref{integralderivadacovarianteN}) is written as
\begin{eqnarray}
\oint_{\partial V}d^3y \varepsilon \sqrt{|h|}n_{\sigma}N^{\sigma}  = \int_{V}d^4x\sqrt{-g} \delta g^{\sigma\beta}\nabla_{\sigma}\nabla_{\beta}f' - \int_{V}d^4x\sqrt{-g} f'\nabla_{\sigma}\nabla_{\beta}\delta g^{\sigma\beta},
\end{eqnarray}
therefore,
\begin{equation}
\int_{V}d^4x\sqrt{-g} f'\nabla_{\sigma}\nabla_{\beta}\delta g^{\sigma\beta} = \int_{V}d^4x\sqrt{-g} \delta g^{\sigma\beta}\nabla_{\sigma}\nabla_{\beta}f' + \oint_{\partial V}d^3y \varepsilon \sqrt{|h|}n_{\sigma}N^{\sigma}.
\end{equation}
So,
\begin{eqnarray}\label{senlafrontera}
\delta S_{met}&=& \frac{1}{2k}\int_{V}d^4x\sqrt{-g} \left(f' R_{\alpha\beta} + g_{\alpha\beta}\square f' - \nabla_{\alpha} \nabla_{\beta} f' - \frac12 g_{\alpha\beta} f \right)\delta g^{\alpha\beta} + \nonumber\\
&&+ \oint_{\partial V}d^3y \varepsilon \sqrt{h|}n^{\tau}M_{\tau} + \oint_{\partial V}d^3y \varepsilon \sqrt{|h|}n_{\sigma}N^{\sigma}.
\end{eqnarray}
\section{Boundary-Term Evaluation in the Metric Formalism}
We show below that the last two terms in Eq.~(\ref{senlafrontera}) are canceled with the variations of the action $S_{bound}$. First of all it is convenient to express the quantities $M_{\tau}$ and $N^{\sigma}$ depending on the variations $\delta g_{\alpha\beta}$. Let's start replacing Eqs.~(\ref{variaciondeterminanteg}) into (\ref{Mtau}) and (\ref{Nsigma}). We then obtain
\begin{eqnarray}
M_{\tau} &=& -f'g^{\alpha\beta}\nabla_{\tau}\delta g_{\alpha\beta} + g^{\alpha\beta}\delta g_{\alpha\beta}\nabla_{\tau}f' \quad \rm, and\\
N^{\sigma} &=& -f'g^{\sigma\mu}g^{\gamma\nu}\nabla_{\gamma}\delta g_{\mu\nu} + g^{\sigma\mu}g^{\gamma\nu}\delta g_{\mu\nu}\nabla_{\gamma}f'.
\end{eqnarray}
To evaluate the above objects at the border we use that $\delta g_{\alpha\beta}|_{\partial V} = \delta g^{\alpha\beta}|_{\partial V} = 0 $, so that the only non-null terms are the derivatives of $\delta g_{\alpha\beta}$ . Therefore,
\begin{eqnarray}
M_{\tau}|_{\partial V} &=& -f'g^{\alpha\beta}\partial_{\tau}\delta g_{\alpha\beta},\\
N^{\sigma}|_{\partial V} &=& -f'g^{\sigma\mu}g^{\gamma\nu}\partial_{\gamma}\delta g_{\mu\nu}.
\end{eqnarray}
We now calculate the terms $n^{\tau}M_{\tau}|_{\partial V}$ and $n_{\sigma}N^{\sigma}|_{\partial V}$ in the integrals of (\ref{senlafrontera}) on the border:
\begin{eqnarray}
n^{\tau}M_{\tau}|_{\partial V} &=& -f'n^{\tau}(\varepsilon n^{\alpha}n^{\beta} + h^{\alpha\beta})\partial_{\tau}\delta g_{\alpha\beta}\nonumber\\ 
&=& -f'n^{\sigma}h^{\alpha\beta}\partial_{\sigma}\delta g_{\alpha\beta}.
\end{eqnarray}
Analogously,
\begin{eqnarray}
n_{\sigma}N^{\sigma}|_{\partial V} &=& -f'n_{\sigma}(\varepsilon n^{\sigma}n^{\mu} + h^{\sigma\mu})(\varepsilon n^{\gamma}n^{\nu} + h^{\gamma\nu})\partial_{\gamma}\delta g_{\mu\nu}\nonumber\\
&=& -f'n^{\mu}(\varepsilon n^{\gamma}n^{\nu} + h^{\gamma\nu})\partial_{\gamma}\delta g_{\mu\nu}\nonumber\\
&=& -f'n^{\mu}h^{\gamma\nu}\partial_{\gamma}\delta g_{\mu\nu}\nonumber\\
&=& 0,
\end{eqnarray}
where we have used $\varepsilon^2 = 1$ and 
\begin{equation}
g_{\alpha\beta} = h_{\alpha\beta} + n_{\alpha}n_{\beta},
\end{equation}
so that $n^{\alpha}h_{\alpha\beta}=0$ and the tangential derivative  $h^{\gamma\nu}\partial_{\gamma}\delta g_{\mu\nu}$ it is null. 

With these results, the variation of the action $S_{met}$ is
\begin{eqnarray}\label{variacionsmet}
\delta S_{met}&=& \frac{1}{2k}\int_{V}d^4x\sqrt{-g} \left(f' R_{\alpha\beta} + g^{\alpha\beta}\square f' - \nabla_{\alpha} \nabla_{\beta} f' - \frac12 fg^{\alpha\beta} \right)\delta g^{\alpha\beta} + \nonumber\\
&&- \oint_{\partial V}d^3y \, \varepsilon \sqrt{h|}f'n^{\sigma}h^{\alpha\beta}\partial_{\sigma}(\delta g_{\alpha\beta}).
\end{eqnarray}
The variation of the boundary term $S_{bound}$ in the total action is
\begin{eqnarray}\label{variacionsbound}
S_{bound} &=& 2\oint_{\partial V}d^3y \, \varepsilon \sqrt{h|}f''\delta R K + \oint_{\partial V}d^3y \varepsilon \sqrt{|h|}f'n^{\sigma}h^{\alpha\beta}\partial_{\sigma}(\delta g_{\alpha\beta}).
\end{eqnarray}
We note that the second term in (\ref{variacionsbound}) cancels the boundary term in the variation (\ref{variacionsmet}) if additionally we impose $\delta R = 0$ in the border too. Similar argument is given in \cite{Dyer:2008hb}.

Finally, the variation of the total action is $\delta S_{tot} = 0$, and then
\begin{eqnarray}\label{frFieldEqn}
f' R_{\alpha\beta} - \frac12 fg^{\alpha\beta} + g^{\alpha\beta}\square f' - \nabla_{\alpha} \nabla_{\beta} f'  = kT_{\alpha\beta}.
\end{eqnarray}
\section{FLRW Universe with \texorpdfstring{$f(R)$}{fruniverse} Gravity Theory}
Henceforth we take the FLRW metric with $K=0$ 
\begin{equation}\label{metricK0}
{\rm d} s^2 = -{\rm d} t^2 + a^2(t)\sum_{i=1}^3 ({\rm d} x_i)^2 \,,
\end{equation}
where, the non-zero components of the metric are: $g_{00} = -1$ and $g_{ii} = a^2$. The determinant is $g = -a^6$, and then $\sqrt{-g} = a^3$. From the standard relation
\begin{equation}
\Gamma^{\gamma}_{\alpha\beta\gamma} = -\frac12 g^{\alpha}(g_{\alpha\gamma,\beta} + g_{\beta\gamma,\alpha} - g_{\alpha\beta,\gamma}),
\end{equation}
the only non-zero Christoffel symbols are:
\begin{equation}
    \Gamma^0_{ii} = a\dot{a} \quad {\rm and} \quad \Gamma^i_{0i} = \frac{\dot{a}}{a}.
\end{equation}
The non-zero components of the Ricci tensor are
\begin{eqnarray}
R_{00} &=& 3\frac{\ddot{a}}{a},\nonumber\\
R_{ii} &=& a\ddot{a} + 2\dot{a}^2,\nonumber\\
R_{0i} &=& 0 \, \forall \, i,
\end{eqnarray}
and the Ricci scalar is
\begin{equation}
R = 6\left[\frac{\ddot{a}}{a} + \left(\frac{\dot{a}}{a}\right)^2 \right] = 6\left[\dot{H} + 2H^2 \right].
\end{equation}
Using a general $f(R)$ function of the Ricci scalar $R$ the action is written as Eq.~(\ref{totalactionfr}) to obtain the $f(R)$ field equation Eq.~(\ref{frFieldEqn}), whose trace is 
\begin{equation}
    3 \square f' + f' R - 2f = k T,
\end{equation}
where
\begin{equation}
\square f' \equiv \frac{1}{\sqrt{-g}}\partial_{\alpha}\left[\sqrt{-g} g^{\alpha\beta}\partial_{\beta} f' \right]. 
\end{equation}
and $T \equiv g^{\alpha\beta}T_{\alpha\beta}$ \cite{DeFelice:2010aj}. 

Let's define the operator $\mathcal{D}\equiv \nabla_{\alpha}\nabla_{\beta}f' - g_{\alpha\beta} \square$. So, the temporal part of the operator $\mathcal{D}_{\alpha\beta} f'$ results in
 \begin{eqnarray}
 \mathcal{D}_{00} f' &=& \nabla_0\nabla_0 f' - g_{00}\square f'\nonumber\\
 				&=& \partial_0\partial_0 f'- g_{00}\left[\frac{1}{\sqrt{-g}}\partial_0\left(\sqrt{-g} g^{00}\partial_0 f' \right)\right]\nonumber\\
 				&=& \ddot{f'} + \left[\frac{1}{a^3}\partial_0\left(- a^3 \dot{f'} \right)\right]\nonumber\\
 				&=& \ddot{f'} - \left[\frac{1}{a^3}\left(3a^2\dot{a}\dot{f'} + a^3 \ddot{f'} \right)\right],\nonumber\\
 				&=& \ddot{f'} - \frac{3\dot{a}}{a}\dot{f'}- \ddot{f'}\nonumber\\
 				&=&  - 3H\dot{f'}.
\end{eqnarray}
Its $rr$ component corresponds to
\begin{eqnarray}
 \mathcal{D}_{rr} f' &=& \nabla_i \nabla_i f' - g_{rr}\square f'\nonumber\\
&=& \nabla_r \nabla_r f'- g_{rr}\left[\frac{1}{\sqrt{-g}}\partial_{\alpha}\left(\sqrt{-g} g^{\alpha\beta}\partial_\beta f'\right)\right]\nonumber\\
            &=& \partial_r \partial_r f' - a^2\left[\frac{1}{\sqrt{-g}}\partial_{0}\left(\sqrt{-g} g^{00}\partial_0 f'\right) + \frac{1}{\sqrt{-g}}\partial_{r}\left(\sqrt{-g} g^{rr}\partial_r f'\right)\right]\nonumber\\
            &=& \partial_r \partial_r f' - a^2\left[\frac{1}{a^3}\partial_{0}\left(-a^3\dot{f'}\right) + \frac{1}{a^3}\partial_{r}\left(a^3 \frac{1}{a^2}\partial_r f'\right)\right]\nonumber\\
            &=& \partial_r \partial_r f' - a^2\left(-3H\dot{f'}-\ddot{f'}\right) - \frac{1}{a}\left(\partial_{r}a\partial_r f' + a\partial_{r}\partial_{r}f'\right)\nonumber\\
            &=& \partial_r \partial_r f' - a^2\left(-3H\dot{f'}-\ddot{f'}\right) - \frac{1}{a}\left(\partial_{r}a\partial_r f' + a\partial_{r}\partial_{r}f'\right)\nonumber\\
            &=& \partial_r \partial_r f' + a^23H\dot{f'}+a^2\ddot{f'} - \frac{1}{a}\left(\frac{dt}{dr}\dot{a} \frac{dt}{dr}\dot{f'} + a\partial_{r}\partial_{r}f'\right)\nonumber\\
            &=& \partial_r \partial_r f' + a^2 3H\dot{f'}+a^2\ddot{f'} - \frac{1}{a}\left(\frac{dt}{dr}\right)^2\dot{a}\dot{f'}  -\partial_{r}\partial_{r}f'\nonumber\\
             &=& \cancel{\partial_r \partial_r f'} + a^23H\dot{f'}+a^2\ddot{f'} - \frac{1}{a}a^2\dot{a}\dot{f'} - \cancel{\partial_{r}\partial_{r}f'}\nonumber\\
             &=&  a^23H\dot{f'}+a^2\ddot{f'} - a^2 H\dot{f'}\nonumber\\
              &=& a^2 2H\dot{f'} + a^2\ddot{f'},
\end{eqnarray}
where $\left(\frac{dt}{dr}\right)^2 = a^2$ if $ds^2 = 0$. The $0-0$ component of (\ref{frFieldEqn}), assuming that $R=R(r)$ (isotropic and static spacetime)
\begin{equation}
    \partial_r f'= \partial_r f' = \frac{dR}{dr} \frac{df'}{dR}= \frac{dR}{dr}f''.
\end{equation}
If, on the other hand, $R=R(t)$ (homogeneous spacetime), then
\begin{equation}
\dot{f'} \equiv \frac{d}{dt}f'[R(r)]= 0,
\end{equation}
since there is no dependence on time.
\begin{eqnarray} \label{campo00}
f' R_{00} - \frac{1}{2}g_{00}f -\mathcal{D}_{00} f' \square f' &=& \kappa T_{00}\nonumber\\
-3\frac{\ddot{a}}{a}f' + \frac12 f + 3\frac{\dot{a}}{a}\dot{f'} &=& \kappa \rho_m\nonumber\\
-3\left(\dot{H} + H^2\right)f' + \frac12 f + 3H\dot{f'} &=& \kappa \rho_m\nonumber\\
-3\left(\frac{R}{6} - 2H^2 + H^2\right)f' + \frac12 f + 3H\dot{f'} &=& \kappa \rho_m \nonumber\\
-\frac{Rf'}{2} + 3H^2f' + \frac12 f + 3H\dot{f'} &=& \kappa \rho_m \nonumber\\
3H^2 - 3H^2f' - \frac{f- Rf'}{2}  - \frac12 f - 3H\dot{f'} &=& 3H^2 - \kappa \rho_m \nonumber\\
\kappa \rho_m + 3H^2(1 - f') - \frac{f- Rf'}{2}  - \frac12 f - 3H\dot{f'} &=& 3H^2.
\end{eqnarray}
This is the Friedmann equation for $f(R)$ gravity
\begin{equation}\label{friedmanngen}
3H^2  = \kappa \rho_m + 3H^2(1 - f') + \frac{Rf' - f}{2}  - \frac12 f - 3H\dot{f'}.
\end{equation}
In the same way for the component $rr$
\begin{eqnarray} \label{campoij}
f' R_{rr} - \frac{1}{2}g_{rr}f -\mathcal{D}_{rr} f' &=& \kappa T_{rr}\nonumber\\
(a\ddot{a} + 2\dot{a}^2)f' - \frac{1}{2}a^2f -a^2 2H\dot{f'} - a^2\ddot{f'} &=& 
\kappa a^2 p_m\nonumber\\
\left(\frac{\ddot{a}}{a} -\frac{\dot{a}^2}{a^2} +\frac{3\dot{a}^2}{a^2}\right)f'-\frac{1}{2}f -2H\dot{f'}-\ddot{f'} &=& \kappa  p_m\nonumber\\
\left(\dot{H} + 3H^2 \right)f' - \frac{1}{2}f - 2H\dot{f'} - \ddot{f'} &=& \kappa p_m\nonumber\\
\left(3\dot{H} + 6H^2 -(2\dot{H} + 3H^2) \right)f' - \frac{1}{2}f - 2H\dot{f'} - \ddot{f'} &=& \kappa  p_m\nonumber\\
\frac{Rf'}{2} - \left(2\dot{H} + 3H^2 \right)f' - \frac{1}{2}f - 2H\dot{f'} - \ddot{f'} &=& \kappa  p_m\nonumber\\
\left(2\dot{H} + 3H^2 \right) - \left(2\dot{H} + 3H^2 \right)f' - \frac{1}{2}(Rf' - f) - 2H\dot{f'} - \ddot{f'} &=& \kappa p_m + \left(2\dot{H} + 3H^2 \right)\nonumber\\
\left(2\dot{H} + 3H^2 \right) (1-f') - \frac{1}{2}(Rf' - f) - 2H\dot{f'} - \ddot{f'} - \kappa  p_m &=&  \left(2\dot{H} + 3H^2 \right).
\end{eqnarray}
This is the Raychaudhuri equation for $f(R)$ gravity:
\begin{equation}\label{Raygen}
-\left(2\dot{H} + 3H^2 \right) = \kappa p_m   + \frac{1}{2}(Rf' - f) + 2H\dot{f'} + \ddot{f'} - \left(2\dot{H} + 3H^2 \right) (1-f').
\end{equation}
From equations (\ref{friedmanngen}) and (\ref{Raygen}), we can define a density and a pressure associated with the additional terms of curvature in the form
\begin{align}
\kappa \rho_{c} &\equiv 3H^2(1 - f') + \frac{Rf' - f}{2}  - \frac12 f - 3H\dot{f'}.\label{rhoc}\\
\kappa p_{c} &\equiv \frac{1}{2}(Rf' - f) + 2H\dot{f'} + \ddot{f'} - \left(2\dot{H} + 3H^2 \right) (1-f')\label{Pc}.
\end{align}
So the ``curvature'' equation-of-state parameter can be written as
\begin{equation}
    \omega_{c} \equiv \frac{p_{c}}{\rho_{c}} = \frac{\frac{1}{2}(Rf' - f) + 2H\dot{f'} + \ddot{f'} - \left(2\dot{H} + 3H^2 \right) (1-f')}{3H^2(1 - f') + \frac{Rf' - f}{2}  - \frac12 f - 3H\dot{f'}}.
\end{equation}
\section{Continuity equation}
We can obtain the continuity equation for $f(R)$ gravity rewriting the Friedmann equation (\ref{friedmanngen}) using equation (\ref{rhoc}),
\begin{equation}
   3H^2  = \kappa (\rho_m + \rho_c),
\end{equation}
which, upon derivation with respect to cosmic time, yields
\begin{equation}\label{friedmanndot}
6H\dot{H} =\kappa (\dot{\rho}_m + \dot{\rho}_{c}).
\end{equation}
From equation (\ref{Raygen}), we get
\begin{eqnarray}
-2\dot{H}  &=& 3H^2 + \kappa (p_m + p_{c})\nonumber\\
           &=& \kappa (\rho_m + p_m  + \rho_c + p_{c}).
\end{eqnarray}
Replacing this in equation (\ref{friedmanndot}) we obtain
\begin{align}
\kappa \dot{\rho}_m + \kappa \dot{\rho}_{c} + 3H\left(\kappa(\rho_m + p_m) + \kappa(\rho_c + p_{c})\right) &= 0\nonumber\\
 \dot{\rho}_m + 3H \rho_m \left( 1 + \omega_m\right) + \dot{\rho}_c + 3H \rho_c \left( 1 + \omega_c \right)  &= 0.
\end{align}
Since $\dot{\rho}_m + 3H \rho_m \left( 1 + \omega_m \right) = 0$ we get that $\dot{\rho}_c + 3H \rho_c \left( 1 + \omega_c \right) = 0$. Then, the continuity equation is also valid for the curvature terms.

\section{Conclusions}
We can modify the Lagrangian of the Hilbert-Einstein action which yields General Relativity by taking a function of Ricci scalar $f(R)$, thus obtaining an extended field equation. Properly evaluating the terms on the boundary, it is shown that the terms of the perturbed action that appear before the variations with respect to the metric are naturally cancelled out with those in the variation of the boundary term in $S_{bound}$. We stress that this derivation is obtained in the metric formalism.
The dynamical equations in $f(R)$ gravity naturally present extra terms that may account, depending on the chosen function, for the accelerated expansion of the universe without the need to include an exotic component. In this way, the expansion of the universe can be explained as a phenomenon of space-time, as shown in equations (\ref{friedmanngen}) and (\ref{Raygen}). By rewriting the dynamical equations we obtain an equation of state which is valid for both density and a pressure associated with matter and the curvature terms.

\chapter{Legendre Transformation}
\label{Chapter5} 

\lhead{Chapter 4. \emph{Legendre Transformation}}
In this chapter we will start considering a basic Legendre Transformation concept with a geometrical interpretation and its use for fields. Then we will consider a procedure described by Ferrari et. al. in Ref.~\cite{Ferraris_1988} to obtain a new metric which is conformally related to first one, and then recover the Hilbert-Einstein Lagrangian through the Legendre transformation. Additionally, we will consider the inverse problem, following the procedure presented in Ref.~\cite{Magnano:1993bd} we can obtain a $f(R)$ function in the Jordan frame from the Hilbert-Einstein Lagrangian with a scalar field minimally coupled. 
\section{Basic Concepts}
Given a function or a curve --- in 1D, for the sake of the argument --- there are several ways to describe it. For example, the case of a temporal signal it can be seen as a sum of sines or cosines and then it can be expressed in terms of their corresponding amplitudes and frequencies --- this method is known as Fourier Transform. The fact of rewriting the function in terms of new variables is called \textit{a transformation}. 

\begin{figure}
\centering
\includegraphics[width=\textwidth]{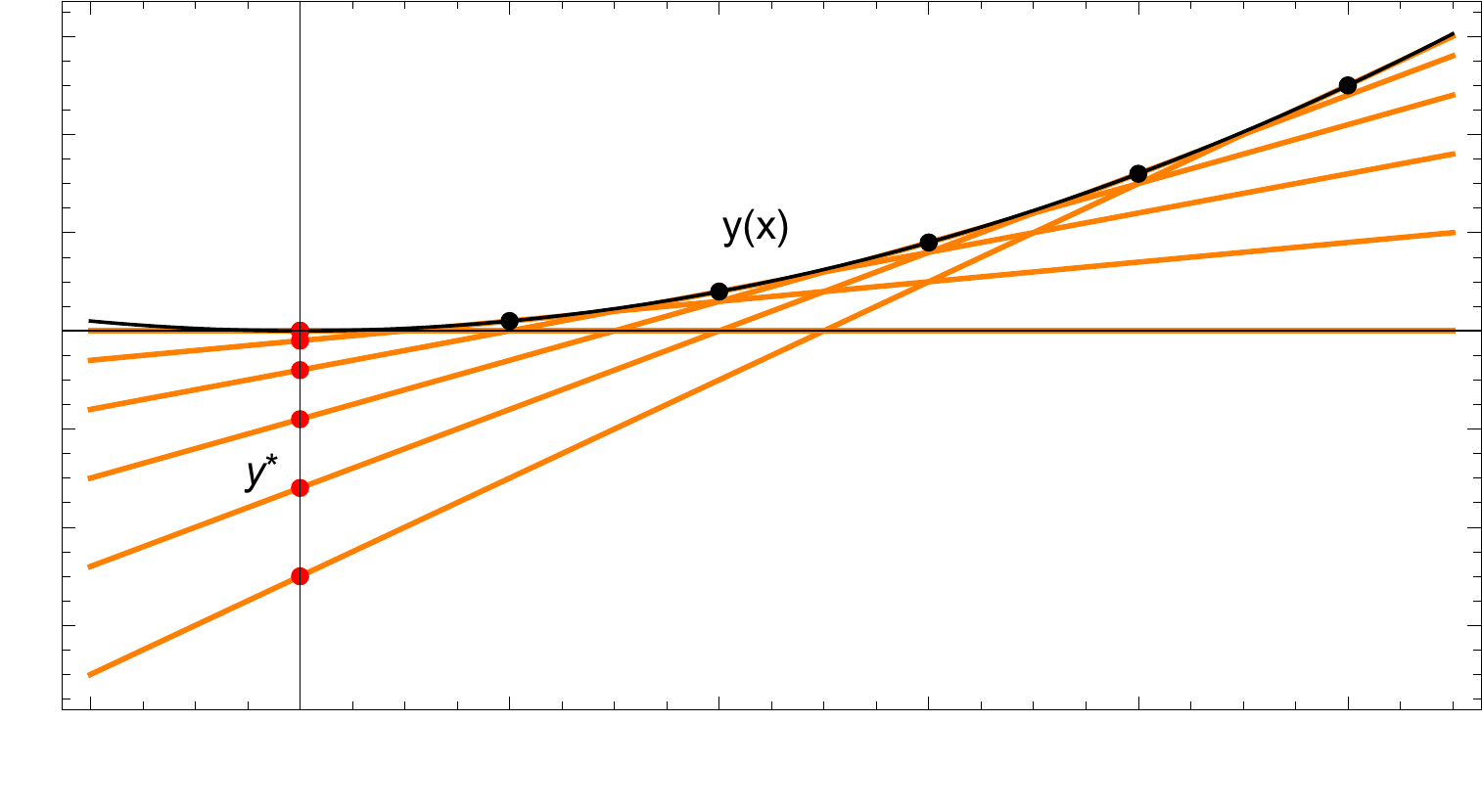}
\caption[Function described by the series of points and by tangent lines.]{$y(x)$ (black/solid) can be described by the series of points (in black) and by the tangent lines (orange/solid). The intersection points of  the tangent line on the $y$ axis $y^*$ (red/point).}
\label{LegendrePlot}
\end{figure}

The intuitive notion of the Legendre transform is the equivalence of describing a curve, instead of as the place of all the points that satisfy the relation $y = y(x)$, as the envelope of a family of tangent lines \cite{callen1985thermodynamics}, as shown in the figure \ref{LegendrePlot}. 

It is possible to build the tangent lines from the slope  $m = m(x)$ at each point and from the intersection of the tangent line on the $y$ axis $y^*$. Finally, using the intersections of the tangent lines according to the slopes, $y^* = y^*(m)$, it is possible to form the family of tangent lines that make up the envelope of the curve. To find the relationship between $y = y(x)$ and $y^* = y^*(m)$, we consider the following argument: the tangent line that goes from the point $(x,y)$ to the intersection $(0,y^*)$, has slope
\begin{equation}
    m = \frac{y(x) - y^*}{x - 0},
\end{equation}
so that
\begin{equation} \label{Y*x}
    y^* = y(x) - m x.
\end{equation}

Given the relationship $y = y(x)$ then the slope is $m = m(x) = y'(x)$ where the prime indicates the derivative with respect to $x$. Finally, it is necessary to invert the relationship between $m$ and $x$, that is, to find $x = x(m)$. It should be considered that only if the second derivative $d^2y/dx^2$ does not change sign within the range of $x$ in which $y(x)$ is defined, there is a single value of the slope for each value of $x$, and vice-verse, that is, there is a 1-to-1 relationship between $m$ and $x$, so the function $m(x)$ it can be inverted to $x(m)$. Replacing the relationship $x = x(m)$ in \ref{Y*x} one obtains
\begin{equation} \label{Y*m}
    y^*(m) = y[x(m)] - mx(m).
\end{equation}
The function $y^*(m)$ is known as the \textit{Legendre Transformation of $y(x)$}.

\section{Legendre transformation for Lagrangians depending on \texorpdfstring{$f(R)$}{frLagrangian}}
%

$f(R)$ gravity theories is presented mainly in three formalisms: Metric, Palatini and Metric-Affine formulations. There is a classical procedure suggested by Einstein and Eddington \cite{einstein1923postcard,eddington1923mathematical} considering the Palatine formalism, i.e. field theories based on Lagrangians of the following kind:  $L_P = L_P(\Gamma, \partial \Gamma)$, to define a metric tensor $g_{\mu\nu}$ obtained from the momentum conjugated (show how this is analogous to momentum and coordinate in Classical Mechanics) to the dynamical connection $\Gamma$ (It should appear in the equation below) as follows:
\begin{equation}
g^{\mu\nu} = \frac{\partial L_P/\partial R_{\mu\nu}}{|det(\partial L_P/ \partial R_{\mu\nu})|^{1/2}}.
\end{equation}
This procedure was extend by Ferraris et al. \cite{Ferraris_1988, Magnano1987} to the case of the metric formalism, with Lagrangians of the form $L = L(R, g_{\mu\nu}, \phi)$, it is defined in this way
\begin{equation}\label{newmetric}
\hat{g}^{\mu\nu} = \frac{\partial L/\partial R_{\mu\nu}}{|\det(\partial L/ \partial R_{\mu\nu})|^{1/2}}.
\end{equation}
Thus, we apply (\ref{newmetric}) to the Lagrangian \ref{Lvacuum} (in vacuum, i.e, no matter nor radiation fields). General Relativity (GR) with a cosmological constant $\Lambda$ would correspond to $f(R) = R - 2\Lambda$. We get
\begin{eqnarray}
\frac{\partial L}{\partial R_{\mu\nu}} &=& \sqrt{-g}\frac{\partial R}{\partial R_{\mu\nu}}\frac{df(R)}{dR},\nonumber\\
                                &=& \sqrt{-g} g^{\mu\nu}f',
\end{eqnarray}
and then
\begin{eqnarray}
\det \left(\frac{\partial L}{\partial R_{\mu\nu}}\right) &=& (\sqrt{-g} f')^4 g^{-1},\nonumber\\
\Rightarrow\left|\det \left(\frac{\partial L}{\partial R_{\mu\nu}}\right)\right|^{1/2} &=& \sqrt{-g} (f')^2.
\end{eqnarray}
Thus
\begin{eqnarray}
 \frac{\partial L/\partial R_{\mu\nu}}{|\det(\partial L/ \partial R_{\rho\sigma})|^{1/2}} &=& g^{\mu\nu} (f')^{-1}.
 \end{eqnarray}
Then
\begin{equation}\label{conformalmetric}
\hat{g}_{\mu\nu} = f'(R) g_{\mu\nu},
\end{equation}
where $\hat{g}_{\mu\nu}$ is a new metric, conformally related to $g_{\mu\nu}$. As Magnano et al. mentioned in \cite{Magnano1987}: 

\blockquote{Accordingly, we can interpret it as a new metric tensor on $M$, which depends on the second-order derivatives of $g_{\mu\nu}$ through the conformal factor $f'(R)$. The set of original variables is called the conformal Jordan frame, while the transformed set, whose dynamics is described by Einstein's equations, is called the conformal Einstein frame.}

We follow the procedure show in \cite{Magnano1987}, i.e., how the Lagrangian (\ref{Lvacuum}) can be described by second-order equations for $\hat{g}_{\mu\nu}$, instead of the fourth-order Eq.~(\ref{frFieldEqn}) in vacuum for $g_{\mu\nu}$

It is convenient to first define a scalar field $p$ by setting
\begin{equation}\label{conformalfac}
    p = p(R) = f'(R).
\end{equation}
Of course, $p$ cannot vanish identically. The following calculations can be carried out in the complement of the (closed) set of stationary points for the functional $f(R)$, i.e., the metrics for which $p$ vanishes. Under this hypothesis we can invert equation (\ref{conformalfac}) to obtain
\begin{equation}
    R_p = R(p).
\end{equation}
Let us now define the functions $f^*(p)$ and $H(p)$ by setting
\begin{eqnarray}
f^*(p) &\equiv& f[R(p)],\\
H(p) &\equiv& p \, R_p - f^*(p).
\end{eqnarray}
Define now a new second-order Lagrangian $\hat{L}$ in the space of variables $\{g_{\mu\nu}, p\}$ by setting
\begin{equation}\label{Lagrangianp}
    \hat{L}(g,p) \equiv \sqrt{-g}\left(p\,R(g) - H(p)\right).
\end{equation}
Of course, inserting equation (\ref{conformalfac}) into equation (\ref{Lagrangianp}) would reproduce the original Lagrangian. Varying equation (\ref{Lagrangianp}) with respect to $g_{\mu\nu}$ and $p$ as independent variables, we obtain the following Euler Lagrange equations:
\begin{align}
\frac{\delta\hat{L}}{\delta g^{\mu\nu}} &= \delta(\sqrt{-g})\left(pR(g) - H(p)\right) +  \sqrt{-g}\delta\left(pR(g) - H(p)\right)\nonumber \\
                         &= \left(-\frac12 \sqrt{-g}g_{\mu\nu} \delta g^{\mu\nu}\right)\bigg(pR(g) - H(p)\bigg
                         ) + \sqrt{-g}\bigg(p \delta R(g) + R \cancelto{0}{\delta} p - \cancelto{0}{\delta} H (p)\bigg)\nonumber\\
                         &=  \sqrt{-g}\left( -\frac12 g_{\mu\nu}(pR(g) - H(p)) + pR_{\mu\nu} + g^{\mu\nu} \square p - \nabla_{\mu}\nabla_{\nu}p \right)\delta g^{\mu\nu} = 0, \label{ecampop}
\end{align}
and
\begin{align}
\frac{\delta\hat{L}}{\delta p} &=  \frac{dR_p}{dp}\frac{\delta\hat{L}}{\delta R_p}\nonumber\\
                                &= \sqrt{-g}\left(- \frac{\delta H(p)}{\delta R_p} \right) \frac{dR_p}{dp}\nonumber\\
                                &=\sqrt{-g}\left(p - f'(R) \right) \frac{dR_p}{dp} = 0. \label{pcampo}
\end{align}
Notice that the last one reproduces definition (\ref{conformalfac}) of $p$, so that equation (\ref{ecampop}) turns out to be equivalent to Eq.~(\ref{frFieldEqn}) in vacuum. In fact, equations (\ref{ecampop}) and (\ref{pcampo}) play the role of ``Hamilton equations'' for the dynamics of the original Lagrangian. Now, it is convenient to replace $p$ by the new auxiliary field $\phi(p)$ defined by
\begin{equation}\label{factorconforme}
    p = e^{\beta\phi},
\end{equation}
where $\beta \equiv \sqrt{2/3}$ so that, replacing (\ref{factorconforme}) in equation (\ref{Rconforme}) and solving for $R$, we find
\begin{eqnarray}\label{RJordan}
 R &=& p\left\{\hat{R} + 3 \beta \hat{g}^{\mu\nu} \nabla_{\mu}\nabla_{\nu}\phi + \hat{g}^{\mu\nu}\nabla_{\mu}\phi \nabla_{\nu}\phi\right\}  \;.
\end{eqnarray}
From (\ref{B22}) we get
\begin{equation}
  \nabla_{\mu}\nabla_{\nu}\phi = \hat{\nabla}_{\mu}\hat{\nabla}_{\nu}\phi + C^{\gamma}_{\mu\nu}\hat{\nabla}_{\gamma}\phi,  
\end{equation}
where
\begin{equation}
     C^{\gamma}_{\mu\nu} = \delta^{\gamma}_{\mu}\hat{\nabla}_{\nu}(\beta\phi/2) + \delta^{\gamma}_{\nu}\hat{\nabla}_{\mu}(\beta\phi/2) - g^{\gamma\sigma}g_{\mu\nu}\hat{\nabla}_{\sigma}(\beta\phi/2).
\end{equation}
 We replace it in (\ref{RJordan}) and obtain
\begin{eqnarray}
 R &=& p\left\{\hat{R} + 3 \beta\,\hat{g}^{\mu\nu}\left[\hat{\nabla}_{\mu}\hat{\nabla}_{\nu}\phi + C^{\gamma}_{\mu\nu}\phi_{,\gamma} \right]  + \hat{g}^{\mu\nu}\phi_{,\mu} \phi_{,\nu}\right\} \nonumber\\
    &=&  p\left\{\hat{R} + 3 \beta\, \hat{g}^{\mu\nu}\hat{\nabla}_{\mu}\hat{\nabla}_{\nu}\phi + \hat{g}^{\mu\nu}\left(\delta^{\gamma}_{\mu}\hat{\nabla}_{\nu}\phi + \delta^{\gamma}_{\nu}\hat{\nabla}_{\mu}\phi - g^{\gamma\sigma}g_{\mu\nu} \hat{\nabla}_{\sigma}\phi \right)\phi_{,\gamma}  +  \hat{g}^{\mu\nu}\phi_{,\mu} \phi_{,\nu}\right\} \nonumber\\
    &=&  p\left\{\hat{R} + 3 \beta\,\hat{g}^{\mu\nu}\hat{\nabla}_{\mu}\hat{\nabla}_{\nu}\phi + 3\hat{g}^{\mu\nu}\phi_{,\mu}\phi_{,\nu} - \hat{g}^{\mu\nu} g^{\gamma\sigma}g_{\mu\nu} \phi_{,\sigma}\phi_{,\gamma} \right\} \nonumber\\
    &=&  p\left\{\hat{R} +  3 \beta\,\hat{g}^{\mu\nu}\hat{\nabla}_{\mu}\hat{\nabla}_{\nu}\phi + 3 \hat{g}^{\mu\nu} \phi_{,\mu}\phi_{,\nu} -   \hat{g}^{\mu\nu} \hat{g}^{\gamma\sigma} \hat{g}_{\mu\nu} \phi_{,\sigma}\phi_{,\gamma}\right\} \nonumber\\
     &=&  p\left\{\hat{R} + 3\beta\, \hat{g}^{\mu\nu}\hat{\nabla}_{\mu}\hat{\nabla}_{\nu}\phi + 3 \hat{g}^{\mu\nu} \phi_{,\mu}\phi_{,\nu} - 4  \hat{g}^{\gamma\sigma} \phi_{,\sigma}\phi_{,\gamma} \right\} \nonumber\\
      &=& p\left\{\hat{R} + 3\beta\, \hat{g}^{\mu\nu}\hat{\nabla}_{\mu}\hat{\nabla}_{\nu}\phi - \hat{g}^{\mu\nu}\phi_{,\mu} \phi_{,\nu}\right\}.
\end{eqnarray}
Then, we replace it in the Lagrangian (\ref{Lagrangianp}) and express it through the new variables $\{\hat{g}^{\mu\nu}, \phi\}$, finding
\begin{eqnarray}
    \hat{L}(\hat{g},\phi) &=& \sqrt{-\hat{g}}p^{-2}\left(p^2\,\left\{\hat{R} + 3\beta\, \hat{g}^{\mu\nu}\hat{\nabla}_{\mu}\hat{\nabla}_{\nu}\phi - \hat{g}^{\mu\nu}\phi_{,\mu} \phi_{,\nu}\right\} - H(p)\right)\nonumber\\
    &=& \sqrt{-\hat{g}} \left(\hat{R} - \hat{g}^{\mu\nu} \phi_{,\mu} \phi_{,\nu} - p^{-2} H\left[p(\phi)\right] \right).
\end{eqnarray}
One can see a divergence term which can be subtracted because it does not affect the dynamics. The Lagrangian (\ref{Lvacuum}) then can be recast in a more familiar form:
\begin{equation} \label{LagrangianE2}
 \hat{L} = \sqrt{-\hat{g}} \left(\hat{R} - \hat{g}^{\mu\nu} \phi_{,\mu} \phi_{,\nu} - 2V(\phi)\right),
\end{equation}
where $\hat{R}$ is the Ricci scalar obtained from $\hat{g}_{\mu\nu}$. In other words, in the Einstein frame, the gravitational dynamics is set by a GR-like term ($\hat{R}$) and a the scalar field $\phi$, which satisfies a set of nonlinear Klein-Gordon equations, as an ordinary minimally-coupled massive scalar field subject to the potential
\begin{equation}
V(\phi) \equiv \frac{1}{2p^2}\Big\{p(\phi) R[p(\phi)] - f[R(p(\phi))] \Big\},
\end{equation}
which is completely determined by the particular $f(R)$ chosen.
\section{The Inverse Problem}\label{inverseproblem}
From now on, the super(sub)scripts ``$^E$" and  ``$^J$" indicate the frame (Einstein and Jordan, respectively) where the quantity is defined. We drop the subscript in $R_J$ (and in $\phi_E$ --- see below) to avoid excessive cluttering of the equations. 
We write the modified gravitation Lagrangian in JF (in the vacuum, i.e, no matter/radiation fields) as
\begin{equation}\label{Lvacuum}
L_J = \sqrt{-g^J} f(R),
\end{equation}
where $g^J\equiv \det (g^J_{\mu\nu})$. General Relativity (GR) with a cosmological constant $\Lambda$ would correspond to $f(R) = R - 2\Lambda$. 
We saw in Chapter \ref{Chapter3} that a variational procedure in the metric formalism yields fourth-order equations for the metric \cite{Sotiriou:2008rp}, namely
\begin{equation}\label{EqfieldfR}
 R_{\mu\nu} f' -\frac12  g^J_{\mu\nu} f + g^J_{\mu\nu}\, \Box f' - \nabla_{\mu}\nabla_{\nu} f' = 0,
\end{equation}
where $f' \equiv {\rm d} f/{\rm d} R$.  

In the present work we start by examining the \textbf{inverse problem}: from a scalar field $\phi$ and its potential $V_E(\phi)$, we map $L_E$ in Eq.~(\ref{LagrangianE2}) onto the corresponding $L_J$ in Eq.~(\ref{Lvacuum}). 
From equation (\ref{LagrangianE2}) we see that potential is
\begin{equation}\label{potencialVJ}
V_J(p) = \frac{1}{2p^2}\bigg(p \, R(p) - f[R(p)] \bigg),
\end{equation}
From which we obtain
\begin{equation} \label{fdif}
2 p^2 V_J(p) = R\, p - f.
\end{equation}
We can also isolate R in (\ref{potencialVJ}) to arrive at
\begin{equation}\label{R(p)}
    R(p) = 2 p V_J(p) + \frac{f}{p}
\end{equation}
To ensure the consistency of the Legendre transformation, a solution to (\ref{fdif}) should also meet the R-Regularity condition $d^2f/dR^{2} \neq 0$.

Deriving (\ref{fdif}) with respect to $R$, and using the condition above, yields
\begin{eqnarray}
2 p^2 \frac{dp}{dR}\frac{dV_J(p)}{dp} + 4p\frac{dp}{dR} V_J(p) &=& p + R\frac{dp}{dR} - p,\nonumber\\
 2 p^2 \frac{dV_J(p)}{dp}+ 4 p V_J(p)&=& R.
\end{eqnarray}
Using (\ref{R(p)}) we get
\begin{equation}\label{f(p)}
f[R(p)] = p^2 \left[2V_J(p) + 2 p \frac{dV_J(p)}{dp} \right],
\end{equation}
or, in terms of the $\phi$ field, using that $p \equiv \exp{(\beta \,\phi)}$, we can rewrite (\ref{f(p)}) and (\ref{R(p)}) as \cite{Motohashi:2017vdc}
\begin{equation}\label{fphi}
f(\phi) = {\rm e}^{2 \beta\, \phi} \left[2V_E(\phi) + 2 \beta^{-1}  \frac{dV_E(\phi)}{d\phi} \right],
\end{equation}
\begin{equation}\label{Rphi}
R(\phi) = {\rm e}^{\beta\, \phi} \left[4 V_E(\phi) + 2 \beta^{-1} \frac{dV_E(\phi)}{d\phi} \right].
\end{equation}
We obtain a parametric solution for the $f(R)$ function that depends completely on the potential $V_E(\phi)$. A constant potential in (\ref{LagrangianE2}) is interpreted as a cosmological constant $V_E(\phi) = \Lambda$, in which case we get the function $f(R) = \frac{1}{8\Lambda}R$ --- this result is also obtained as a particular (``singular'') solution of (\ref{fdif}), which in this case is a Clairaut \cite{clairaut1934solution}. For a non constant potentials the solutions do exist, but this are practically inaccessible, since one is usually unable to solve (\ref{R(p)}) analytically \cite{Magnano:1993bd}.

\begin{figure}
\begin{center}
\includegraphics[width=\textwidth]{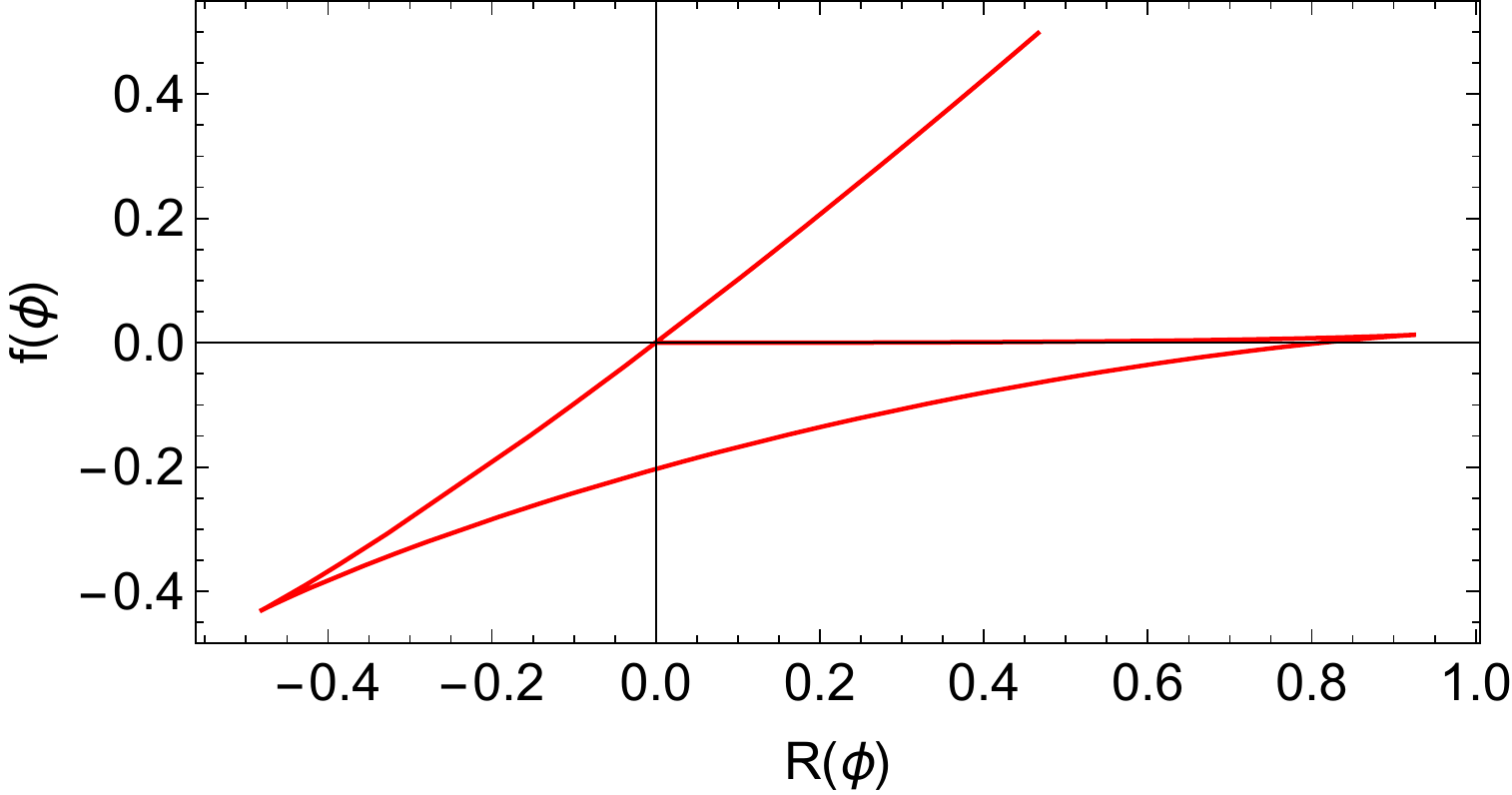}
\end{center}
\caption[Parametric plot of $f(R)$]{Parametric plot of $f(R)$ given by Eqs. (\ref{fVE}), (\ref{RVE}). Note that $f'>0$ for all $R$.}
\label{swallowtail}
\end{figure}

If we apply the above equations to the simplest possible (nontrivial) potential for a scalar field, namely
\begin{eqnarray}\label{VE}
V_E(\phi) = \frac{1}{2}m^2_\phi \, \phi^2,
\end{eqnarray}
we then obtain the corresponding parametric form of $f(R)$:
\begin{align}
\label{fVE}
f(\phi) &= m_\phi^2   \,\frac{\phi( \beta \, \phi + 2)}{\beta} \, {\rm e}^{2 \beta \phi },\\
\label{RVE}
R(\phi) &=2\, m_\phi^2  \frac{\phi  (\beta \, \phi +1)}{\beta} \, {\rm e}^{ \beta \phi },
\end{align}
which we plot in Fig.~\ref{swallowtail}.  Throughout this thesis, we will refer to the three stages of this plot as branches of the system.

One can easily see that
\begin{equation}
f'\equiv \frac{df}{dR} = e^{\beta\, \phi} > 0 \quad \forall \phi,
\end{equation}
and
\begin{equation}
\frac{d^2f}{dR^2}= \frac{d\phi}{dR}\frac{d}{d\phi}\left( \frac{df}{dR}\right)= 
\frac{\beta^2/2}{2 \beta^2 V_E + 3\beta V_E'+ V_E''},
\label{d2fdR2}
\end{equation}
where $V_E'\equiv dV_E/d\phi$ and $V_E''\equiv d^2V_E/d\phi^2$. The sign of $\frac{d^2f}{dR^2}$ is given by the sign of the denominator of (\ref{d2fdR2}).

\section{Conclusions}
We followed and reproduced the procedure in \cite{Ferraris_1988, Magnano1987} to obtain a new metric from (\ref{newmetric}) using the Lagrangian vacuum case $L = \sqrt{-g}\,f(R)$. Using Legendre's transformation with the new metric, a Hilbert-Einstein-like Lagrangian is recovered with an effective scalar field. It can be said that the fourth-order field equations in the Jordan frame include in themselves the gravitational interaction that is presented by the nonlinear Lagrangian in the Einstein frame. We considered also the inverse problem for the $f(R)$ gravity. It is presented a parametric solution that are completely determined by the choice of a particular potential in the Einstein frame and, ultimately, by $f(R)$ itself.
 
\chapter{Thermodynamics: van der Waals Theory}
\label{Chapter6}

\lhead{Chapter 5. \emph{Thermodynamics: van der Waals Theory}}

In this chapter we briefly describe the thermodynamics for the van der Waals model (vdW) following the calculations in \cite{kittel1970thermal,johnston2014thermodynamic}. We will show the most relevant thermodynamic quantities that describe the system and we study the phase transitions of the system, the equilibrium conditions under the Maxwell construction to define the regions of coexistence and metastability regions of gas and liquid phases.

\section{van der Waals Equation}

The ideal-gas model is based on the approximation of a system of many free non interacting particles in the classical regimen. ``Free'' means that the particles are confined in a container with volume $V$ with no other restrictions or external forces, each with an associated kinetic energy (proportional to the temperature $T$) and momentum, and able to transfer it during elastic collisions with the walls of the container. These interactions, in turn, translate into the pressure $P$. The equation of state for this case is the well known $P V=N k_B T$, where $N$ is the number of particles and $k_B$ is the Boltzmann constant. 

In 1873 \cite{van1873over}, van der Waals pointed out that this so-called ``particles'' actually have a volume $b/N_A$ where $N_A$ is Avogadro's number. Hence, the volume $V$ in the ideal equation is replaced by $V-Nb$. He also considered the interaction between particles: the pressure on the wall is reduced proportional to the numbers of the interacting pairs of particles, or upon the square of the number of particles per unit volume $aN^2/V^2$, where $a$ and $b$ are constants of the particular gas. Accordingly, he suggested the following state equation:
\begin{equation}\label{vdWeq}
    P = \frac{N k_B T}{V-Nb} - \frac{aN^2}{V^2}.
\end{equation}

In the classical limit, the Fermi-Dirac and Bose-Einstein distribution functions lead to an identical result for the average number of atoms in an orbital\footnote{The energy of a system is the total energy of all particles, kinetic plus potential, with account taken of interactions between particles. A quantum state of the system is a state of all particles. Quantum states of a one-particle system are
called orbitals \cite{kittel1970thermal}.}. Thus, let $\varepsilon$ be the energy of an orbital occupied by one particle. The Fermi-Dirac and Bose-Einstein distribution functions $f(\varepsilon)$ for the average occupancy of and orbital at energy $\varepsilon$ are \cite{kittel1970thermal}
\begin{equation}
    f(\varepsilon) = \frac{1}{e^{(\varepsilon - \mu)/K_B T } \pm 1},
\end{equation}
where $\mu$ is the chemical potential, the plus sign is for the Fermi-Dirac distribution, and minus for the Bose-Einstein distribution. For the classical regime, $e^{(\varepsilon - \mu)/K_B T } \gg 1 \,\forall \varepsilon$. Then the average occupancy of and orbital of energy $\varepsilon$ simplifies to
\begin{equation}\label{averagefun}
    f(\varepsilon) \approx e^{(\mu - \varepsilon)/K_B T } = \lambda e^{- \varepsilon/K_B T },
\end{equation}
where $\lambda \equiv e^{\mu / K_B T }$. 

The thermal average of the total number of atoms equals the number of atoms known to be present. This number must be the sum over all orbitals of the distribution function $N =$ \textless$N$\textgreater $\; = \sum_i f(\varepsilon_i)$ where $i$ is the index of an orbital of energy $\varepsilon_i$. So, replacing equation (\ref{averagefun}) in the sum we obtain
\begin{equation}
N = \lambda \sum_i e^{- \varepsilon/K_B T }.
\end{equation}
For orbitals of free particles the sum is the partition function $Z_1 = n_Q V$, where $n_Q \equiv \left(m k_B T/2\pi \hslash^2 \right)^{3/2}$ is the quantum concentration for a single free atom in volume V, whereby $N=\lambda Z_1$. As a result,
\begin{equation}
N = \lambda Z_1 = \lambda n_Q V,
\end{equation}
so,
\begin{equation}
    \lambda = \frac{N}{n_Q V} =  e^{\mu / K_B T }.
\end{equation}
We will calculate the background thermodynamic quantities for the ideal gas in order to obtain an extension for the van der Waals gas. The chemical potential $\mu$ for the ideal gas is \cite{kittel1970thermal,johnston2014thermodynamic}
\begin{equation}
    \mu \equiv k_B T \ln\left(\frac{N}{V n_Q}\right).
\end{equation}
We can calculate the free energy function $F$, also known as Helmholtz free energy, from its relation with the chemical potential through 
\begin{equation}
    \mu(T,V,N) \equiv \left(\frac{\partial F}{\partial N}\right)_{T,V}.
\end{equation}
Then,
\begin{eqnarray}\label{freeEnergy}
    F(T,V,N) &=& \int_0^N  \mu(T,V,N) \, dN\nonumber\\ 
    &=& k_B T \int_0^N  (\ln N - \ln(V n_Q)) dN\nonumber\\
    &=&  k_B T N \bigg(\ln N - 1 - \ln(V n_Q) \bigg)\nonumber\\
    &=&  -k_B T N \left[\ln\left(\frac{V n_Q}{N}\right) + 1\right].
\end{eqnarray}
Using the relation with the free energy, the pressure $P$ is
\begin{equation}
    P \equiv - \left(\frac{\partial F}{\partial V}\right)_{T,N} = \frac{N T}{V},
\end{equation}
which is the equation of state of the ideal gas.

We also obtain the entropy $S$ from the free energy:
\begin{equation}
S(T,V,N) \equiv - \left(\frac{\partial F}{\partial T}\right)_{V,N} = N \left[\ln\left(\frac{V n_Q}{N}\right) + \frac52\right].
\end{equation}
Then we can calculate the thermal energy $U$ from the definition
\begin{equation}
U\equiv F + k_B T S = \frac32 N k_B T.
\end{equation}
The heat capacity at constant volume $C_V$ is
\begin{equation}
C_V \equiv \left(\frac{\partial U}{\partial T}\right)_{V,N} = \frac32 N k_B.
\end{equation}
The enthalpy $H$ is given by
\begin{equation}
H  \equiv U + P V = \frac52 N k_B T.
\end{equation}

Therefore, the heat capacity at constant pressure $C_P$ is
\begin{equation}
C_P \equiv \left(\frac{\partial H}{\partial T}\right)_{P} = \frac52 N k_B.
\end{equation}

Now, considering the interaction of particles via the Lennard-Jones potential, the average potential energy per particle is $\phi_{ave} = - Na/V$, where the parameter $a \geq 0$ is an average value of the potential energy per unit. Consequently, the change in the internal energy $U$ due to the attractive part is given by $\Delta U = N \phi_{ave} = -N^2a/V$. On the other hand, the change in the free energy is $\Delta F \approx \Delta U$ since the parameter $a$ introduces no entropy at the system. Then, replacing $V$ by $V- Nb$ in the free-energy expression for the ideal gas (\ref{freeEnergy}) and add $\Delta F$ we obtain \cite{johnston2014thermodynamic,kittel1970thermal}
\begin{equation}\label{freevdW}
 F = - K_B T N \left(\ln\left[\frac{(V-N b) n_Q}{N}\right] + 1\right) - \frac{N^2a}{V}.
\end{equation}

Proceeding as before we can obtain the other thermodynamic function: the entropy for the vdW fluid is 
\begin{eqnarray}
S(T,V,N) &=&  - \left(\frac{\partial F}{\partial T}\right)_{V,N} = -K_B N \left(\ln\left[\frac{(V-N b) n_Q}{N}\right] + \frac52\right).
\end{eqnarray}
The internal energy is
\begin{eqnarray}\label{UvdW}
U &=& F + TS = \frac32NK_B T - \frac{N^2a}{V}.
\end{eqnarray}
The heat capacity at constant volume is $C_V = \left(\partial U/\partial T \right)_{V,N} = 3/2 N K_B T$; is the same as the ideal gas because the second term of the internal energy does not depend on the temperature. From (\ref{freevdW}) we obtain the pressure function
\begin{eqnarray}\label{pressvdW}
P &=& - \left(\frac{\partial F}{\partial V}\right)_{T,N} = \frac{N T}{V- Nb} - \frac{N^2a}{V}.
\end{eqnarray}
which is, again, the van der Waals equation of state (\ref{vdWeq}). Now, using (\ref{UvdW}) and (\ref{pressvdW}) we can obtain the enthalpy
\begin{eqnarray}\label{EntalphyvdW}
H  &=& U + P V = N\left(\frac{3 K_B T}{2} + \frac{ K_B T V}{V- Nb} - \frac{N^2a}{V} \right).
\end{eqnarray}
But, in this case, we cannot obtain an analytical expression of the heat capacity at constant pressure because it is impossible to obtain the volume in terms of pressure for the vdW model. We will make this calculation below with reduced quantities formulation for simplification.%

\subsection{Critical Values of the vdW Gas}
\begin{figure}
\begin{center}
\includegraphics[width=\textwidth]{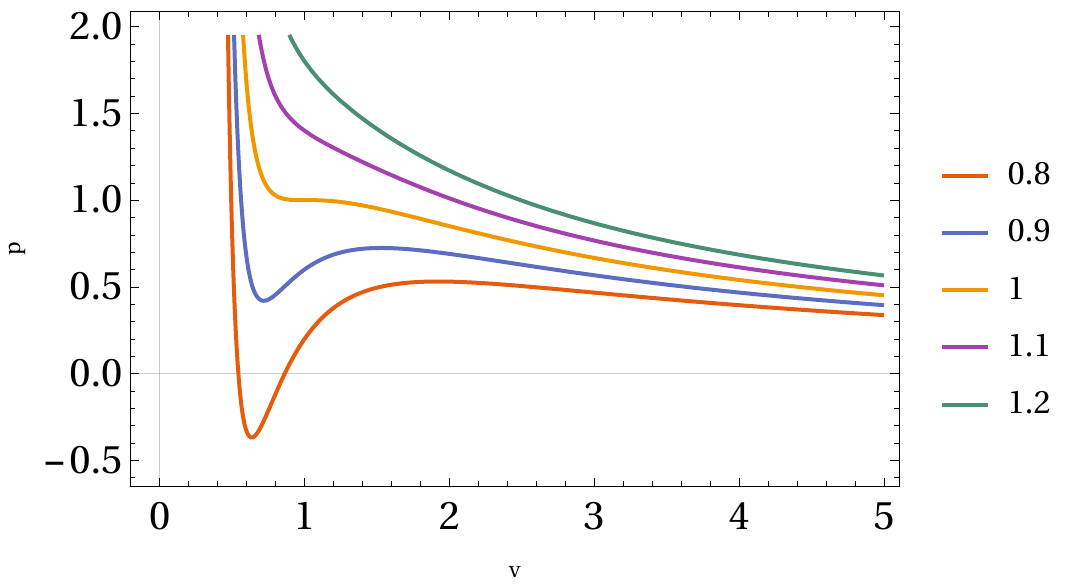}
\end{center}
\caption[vdW equation]{vdW equation of state for the reduced form (\ref{vdWequ2}) with isothermal values $\tau = 0.8, 0.9, 1, 1.1, 1.2$. Critical curve (orange/solid) for $\tau = 1$.}
\label{Plotpv}
\end{figure}

The so-called critical quantities
\begin{equation}
    P_c = \frac{a}{27b^2}, \,\,\, T_c = \frac{8a}{27b}, \,\,\, V_c = 3Nb.
\end{equation}
are the critical pressure, critical temperature and critical volume, respectively, that define the critical point of the van der Waals fluid. Solving this system for $a$, $b$ and $N$ we obtain
\begin{equation}\label{reduce}
    a = \frac{27 T_c^2}{64 P_c}, \,\,\, b = \frac{T_c}{8 P_c}, \,\,\, N = \frac{8P_cV_c}{3T_c}. 
\end{equation}
We can rewrite the vdW state equation (\ref{vdWeq}) in terms of the critical values using the following normalized variables
\begin{equation}
    p \equiv \frac{P}{P_c}, \,\,\, \tau \equiv \frac{T}{T_c}, \,\,\, v \equiv \frac{V}{V_c}.
\end{equation}
Then, using (\ref{reduce}), we can rewrite (\ref{vdWeq}) as
\begin{equation}\label{vdWequ2}
    p = \frac{8\tau}{3v -1}- \frac{3}{v^2}. 
\end{equation}
In figure (\ref{Plotpv}), we plot the $p-v$ plane for several isothermals values.
Values of $a$ and $b$ are usually obtained by fitting to the observed $P_c$ and $T_c$. At the critical point, the curve of $p \times v$ at constant $\tau$ has a point of inflection. Here the local maximum and minimum of the $p-v$ curve coincide, and there is no separation between the vapor and liquid phases. At a horizontal point of inflection the conditions \cite{callen1985thermodynamics, kittel1970thermal}
\begin{equation}
\left(\frac{\partial p}{\partial v}\right)_{\tau} = 0, \quad {\rm and} \quad \left(\frac{\partial^2 p}{\partial v^2}\right)_{\tau} = 0,
\end{equation}
are satisfied by the vdW state equation (\ref{vdWequ2}). No phase separation exists above $T_c$.

\begin{figure}
\centering
\includegraphics[width=0.45\textwidth]{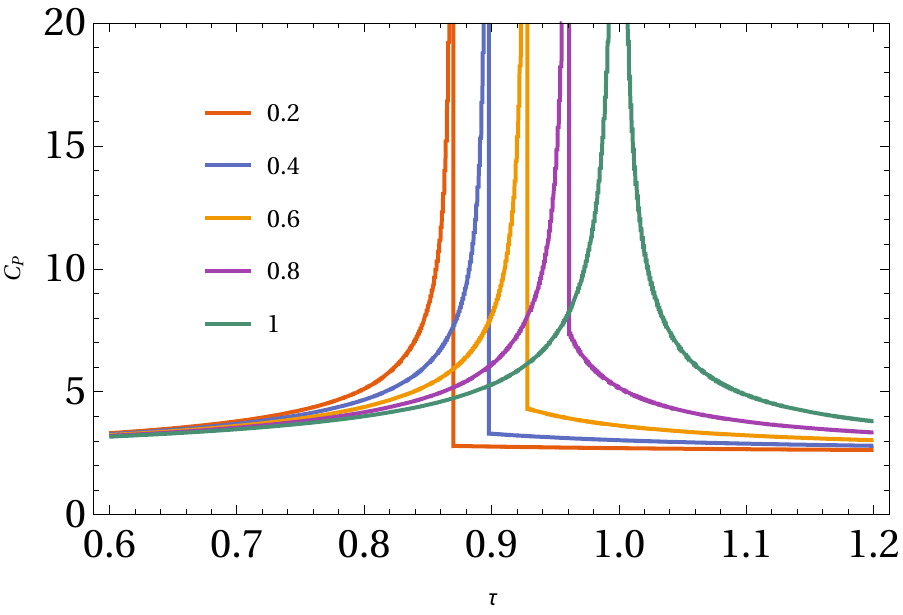}
\includegraphics[width=0.45\textwidth]{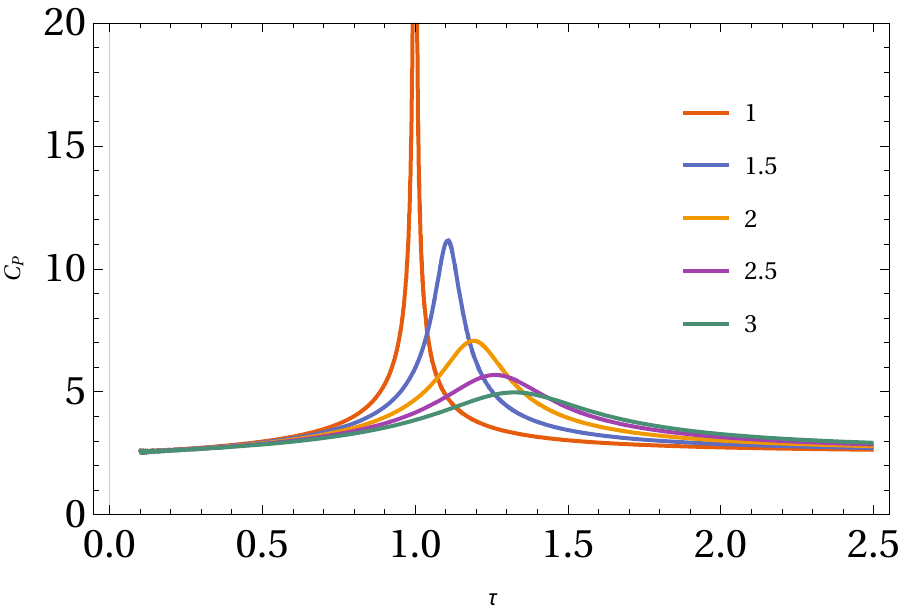}
\caption[Heat capacity at constant pressure versus reduced temperature.]{Heat capacity at constant pressure versus reduced temperature. {\bf Left panel:} reduced pressures from 0.2 to 1 in 0.2 increments. {\bf Right panel:} reduced pressures from 1 to 3 in 0.5 increments. Labels indicate the pressure for each curve.}
\label{cpvdw}
\end{figure}

We also can write the free energy in terms of the reduced/normalized variables. First of all, we write the quantum concentration as $n_Q = n_{Qc} \tau^{3/2}$, where $n_{Qc} \equiv \left(m k_B T_c/2\pi \hslash^2 \right)^{3/2}$. Then, we can rewrite
\begin{equation}
    \frac{F}{P_c V_c} = - \frac83 \tau \left[\ln\left(x_c \tau^{3/2}(3 v - 1)\right) +1 \right] - \frac{3}{v},
\end{equation}
where $x_c =  n_{Qc} T_c / 8 P_c $. In this way, the reduced form of entropy $s$ is
\begin{equation}
    \frac{s}{N k_B} = \ln\bigg[x_c \tau^{3/2}(3 v - 1)\bigg] + \frac52,
\end{equation}
and the enthalpy is
\begin{equation}
    \frac{H}{P_c V_c} = \frac{4\tau (5v -1 )}{3v - 1} - \frac6v.
\end{equation}
The heat capacity at constant pressure is calculated from the relation 
\begin{equation} \label{cpequation}
    C_p = \tau \frac{\partial s(p,\tau)}{\partial \tau}.
\end{equation}
In order to obtain $s(p,\tau)$, we solve numerically $v(p)$ from (\ref{vdWequ2}) using \verb|NSolve| function from the \verb|Mathematica| software --- see figure (\ref{cpvdw}).

\subsection{The Gibbs Function and Chemical Potential of the vdW}
The Gibbs function of the van der Waals fluid is obtained from $G = F + P V$. Therefore, from (\ref{freevdW}) and (\ref{pressvdW}), this results in
\begin{equation}
    G(V,T,N) =  -K_B T N \left(\ln\left[\frac{(V-N b) n_Q}{N}\right] + 1\right) - \frac{2N^2a}{V} + \frac{N T}{V- Nb}.
\end{equation}
As we can see, the Gibbs function is written in terms of $V$, $T$, $N$ instead $P$, $T$, $N$. Although it is not possible to put it in terms of these later variables explicitly, it is possible to do so parametrically using the sets $G(V,T,N)$ and $P(V,T,N)$. We want $G(P,T,N)$ because we can then obtain the chemical potential $\mu(P,T)$ as $G(P,T,N)/N$, since $\mu$ determines the phase coexistence relation. When the liquid ($\ell$) and gas (g) chemical potentials satisfy $\mu_\ell < \mu_g$ the liquid is more stable than the gas phase, or equivalently, when the Gibbs function satisfies $G_\ell < G_g$ and vice versa. The two phases can coexist if  $\mu_\ell = \mu_g$ or $G_\ell = G_g$. Also, we can obtain the chemical potential from the free energy in equation (\ref{freevdW}):
\begin{eqnarray}
    \mu(T,V,N) &=& \left(\frac{\partial F}{\partial N} \right)_{T,N} \nonumber\\
                &=&  -K_B T \ln\left[\frac{(V-N b) n_Q}{N}\right]  - \frac{2Na}{V} + \frac{N T b}{V- Nb} \nonumber\\
                &=&  -K_B T \ln\left[\frac{V-N b}{N}\right] - \frac{2Na}{V} + \frac{N T b}{V- Nb} -K_B T \ln n_Q.
\end{eqnarray}
 which can be rewritten in term of the reduced variables using (\ref{reduce}) as
\begin{equation}\label{chemicalpotreduce}
    \frac{\mu}{T_c K_B} = - \tau \ln\left[3v -1 \right] + \frac{\tau}{3v -1} - \frac{9}{4v K_B} - \tau \log \left[\frac{n_Q T_c}{8P_c}\right].
\end{equation}
\begin{figure}[t]
\begin{center}
\includegraphics[width=0.8\textwidth]{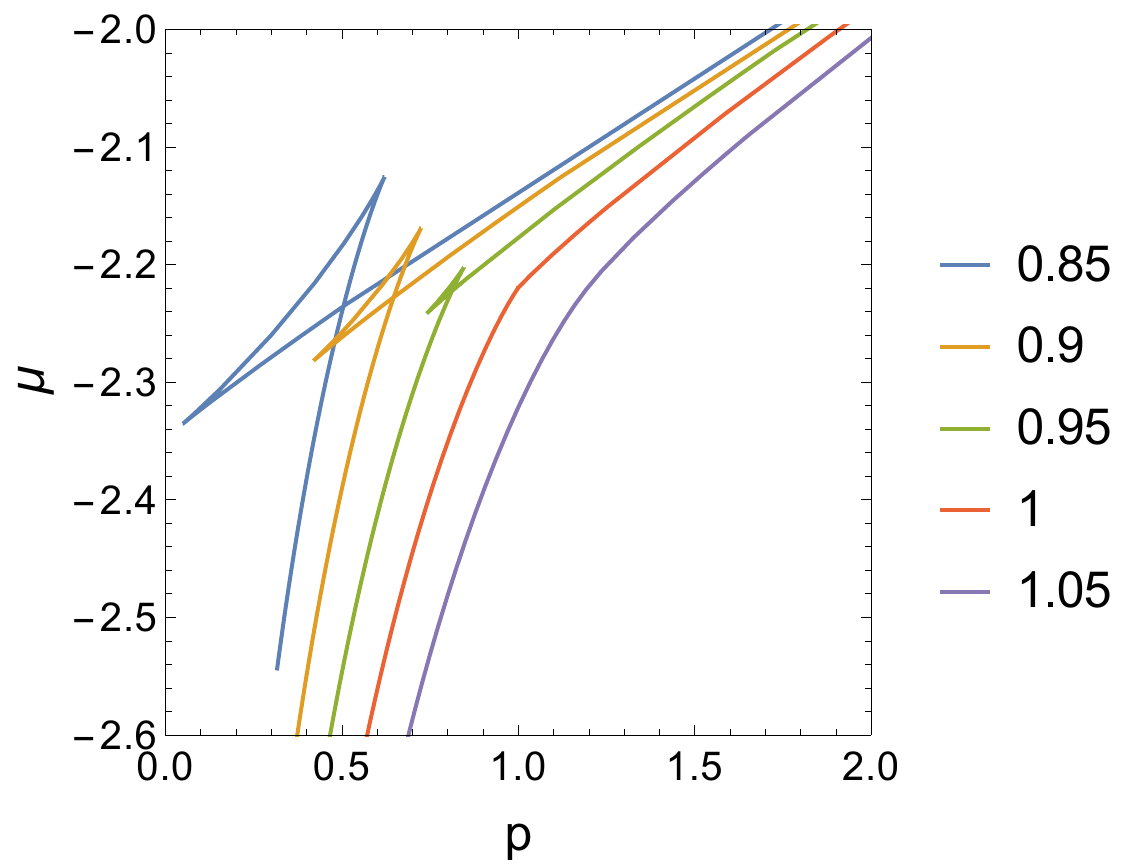}
\end{center}
\caption[ParametricPlot of reduced pressure vs chemical potential.]{ParametricPlot of reduced pressure (\ref{vdWequ2}) vs chemical potential (\ref{chemicalpotreduce}) for the isothermals $\tau = 0.85, 0.9, 0.95, 1, 1.05$. It is show a critical curve (in orange/solid) for $\tau = 1$.}
\label{Plotpmu}
\end{figure}
Figure \ref{Plotpmu} shows the parametric plot of the chemical potential versus pressure for several temperatures; the lowest branch represents the stable phase, the other branches represent metastable phases. The pressure at which the branches cross determines the transition between gas and liquid; this pressure is called \textit{the equilibrium vapor pressure} \cite{kittel1970thermal}.

\begin{figure}
\center
{\includegraphics[width=0.8\textwidth]{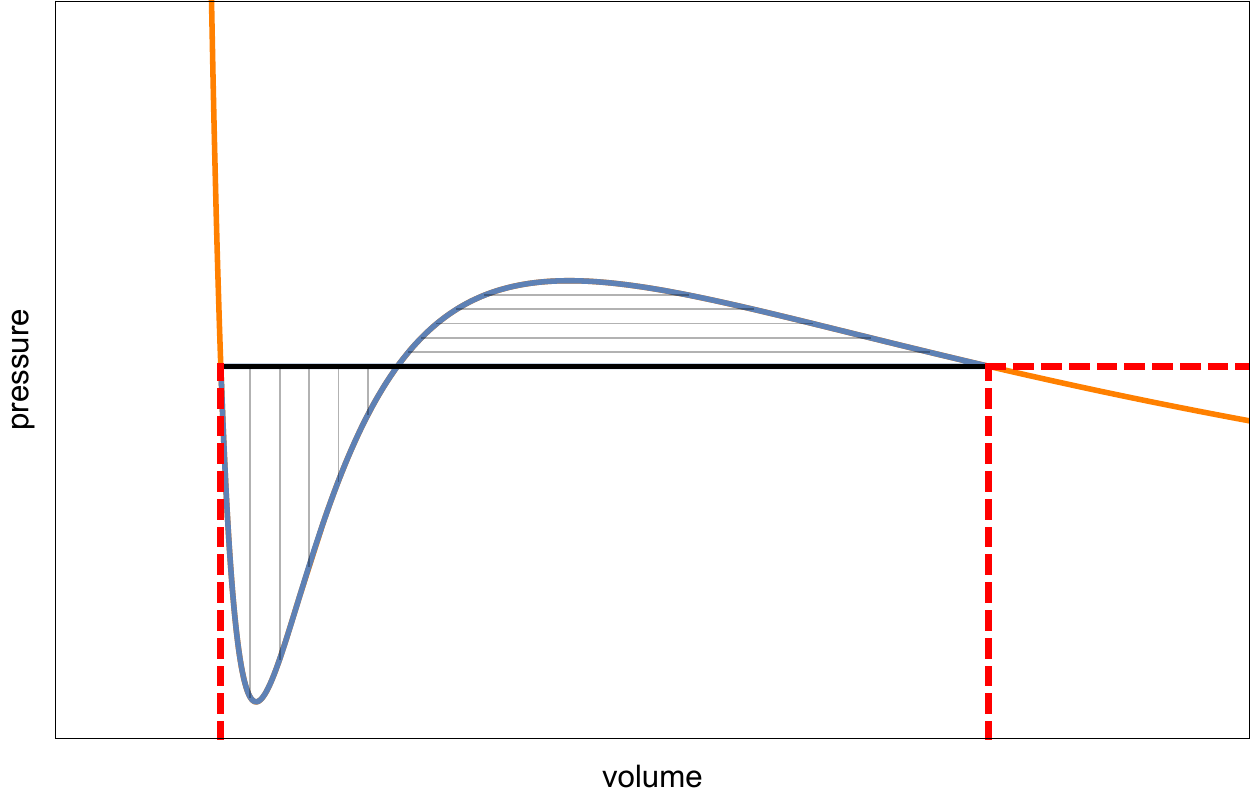}
\put(-270,170){Liquid}
\put(-35,80){Gas}
\put(-200,100){Coexistence line}
\put(-280,0){$v_1$}
\put(-71,0){$v_2$}
\put(10,113){$p_X$}}\medskip
\caption[Pressure vs volume at reduced temperature.]{Plot of $p \times v$ at reduced temperature $\tau = 0.85$.  In the region $v < v_1$, only the liquid phase exists and in the region $v > v_2$, only the gas phase exists. The phases coexist between $v_1$ and $v_2$. The value of $v_1$ or $v_2$ is determined by the condition that $\mu_l(\tau,p) = \mu_g(\tau,p)$ along the horizontal line between $v_1$ and $v_2$. This will occur if the shaded area below the line is equal to the shaded area above the line. This is Maxwell’s construction. The pressure $p_X$ is a constant pressure part of the $p-v$ isotherm at which the gas and liquid coexist as indicated by the horizontal line.}
\label{plotpvmaxwell}
\end{figure}

In terms of the numerical integral of $p \times v$ isotherm over the two-phase region at temperature $\tau$, Maxwell construction states that
\begin{equation}
    \int_{v_1}^{v_2} \bigg[p(\tau,v) - p_X (\tau) \bigg] dv = p_X(\tau)\left(v_1 - v_2 \right) = 0.
\end{equation}
The limit of the local stability of the system under small fluctuation occurs at conditions where the phase separation into liquid and vapour phases should take place. This limit defines a curve known as spinodal curve defined by the condition that the second derivative of Gibbs energy function (with respect to concentration) is zero, that is the locus of points of the inflection points on the Gibbs energy function. On the other hand, the locus where two phases may coexist in metastable equilibrium define a curve known as binodal curve, which is determined by the condition that the first derivative of Gibbs energy function is zero. 
\begin{figure}
\begin{center}
\includegraphics[width=0.8\textwidth]{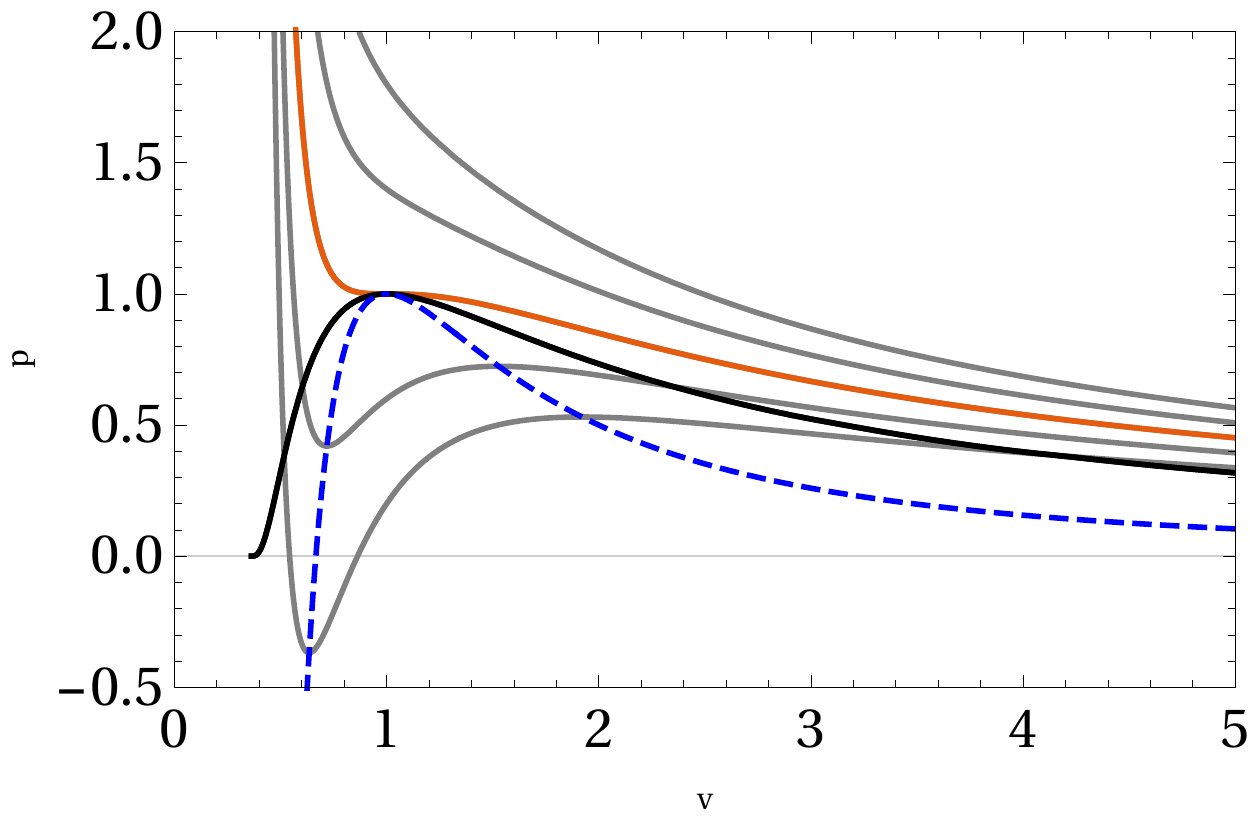}
\end{center}
\caption[Spinodal and binodal curves for the vdW.]{Spinodal (blue/dashed) and binodal (black/solid) curves for the vdW. Spinodal curve is the (inner) boundary between the regions of the metastable and the unstable region. There is a intermediate zone between the binodal and the spinodal curve known as metastable region.}
\label{PlotBS}
\end{figure}
\newpage
\section{Conclusions}
We presented the thermodynamic properties of the van der Waals model which is an improvement of the ideal model that considers the interactions and the volume of the ``particles'' of the system. The van der Waals model is the simplest model that presents a phase transition between liquid and gas phases. Under Maxwell construction, it is possible to obtain the spinodal curve, which defines a stability region of the system. Furthermore, it is possible to find a curve using the extrema points of the Gibbs function, known as binodal curve, which defines a metastable region of the system.
 
\part{Results}

\chapter{Exploring a Stability Criterion in \texorpdfstring{$f(R)$}{frstab} Gravity Theory}
\label{Chapter7}
\lhead{Chapter 6. \emph{Exploring a Stability Criterion in \texorpdfstring{$f(R)$}{frstable} Gravity Theory}}

In this chapter we analyze the stability criteria in $f(R)$ Gravity Theory under three different approach they are: the mass criteria, the catastrophe theory approximation and the thermodynamics analogy.

\section{The Stability Mass Criteria}\label{stable}
 In $f(R)$ theories, the effective potential for $R$ is obtained from the trace of Eq.~(\ref{EqfieldfR}):
\begin{eqnarray} \label{tracefieldfR}
 3\square f' + f' R - 2f =  0,
\end{eqnarray}
which can be rewritten as \cite{Nojiri:2004dw}
\begin{equation}\label{traceR}
\square R + \frac{f'''}{f''} (\nabla_{\alpha}R) \,(\nabla^{\alpha}R) + \frac{f'R-2 f}{3 f''} = 0.
\end{equation}
We will address only homogeneous spacetime, where spatial derivatives vanish, and thus
\begin{equation}\label{frtrace}
\ddot R +  3 H \dot R + \frac{f'''}{f''} \dot R^2   + \frac{dV_J}{dR}  = 0 ,
\end{equation}
where
\begin{equation}
\frac{dV_J}{dR}\equiv  \frac{2 f - f'R}{3 f''}. 
\label{dVdR}
\end{equation}

\begin{figure}
\begin{center}
\includegraphics[width=0.8\textwidth]{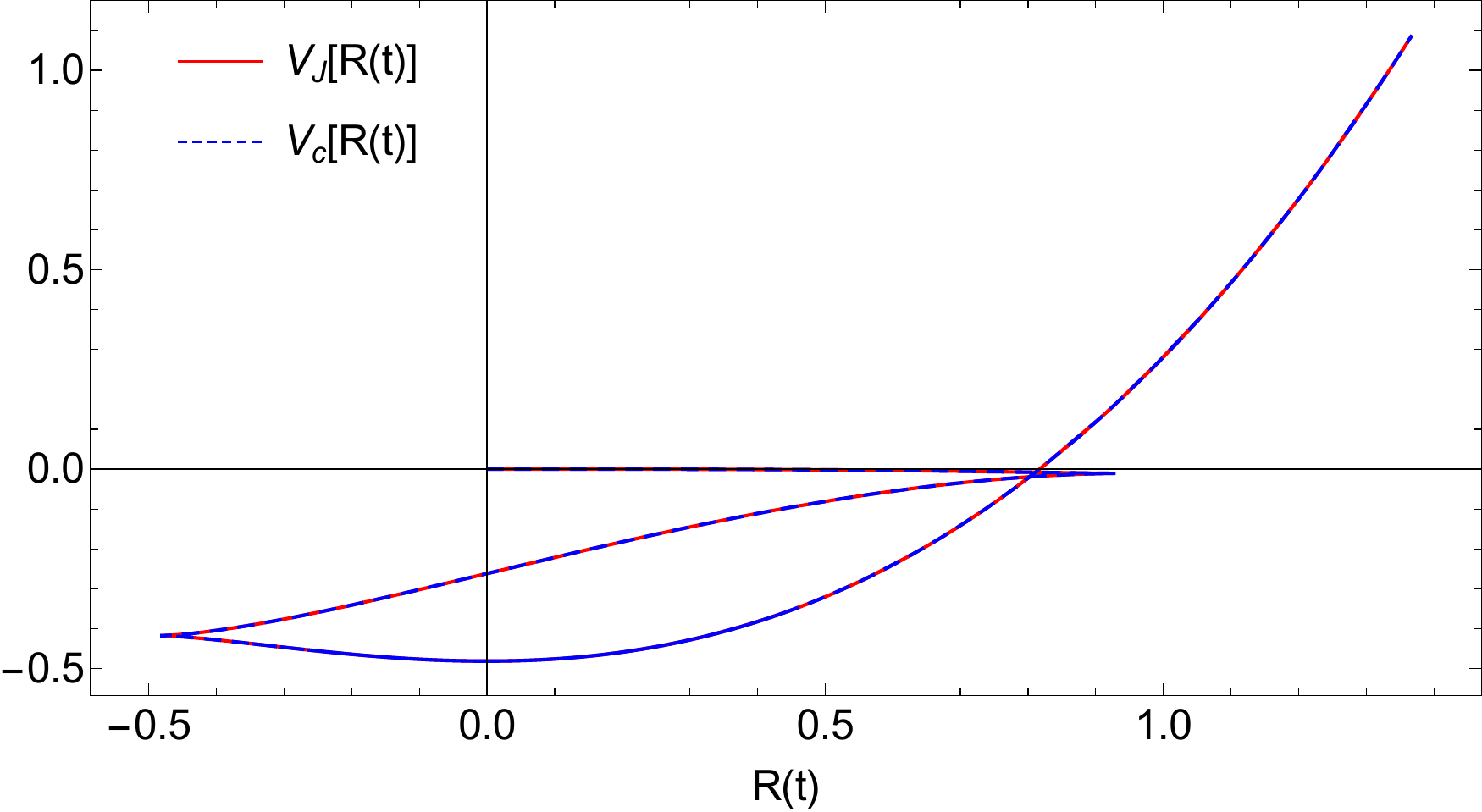}
\end{center}
\caption[Potential $V_J(R)$ given by the numerical integration.]{Potential $V_J(R)$ given by the numerical integration in Eq.~(\ref{VJt}), blue/solid line, and its fit by the cusp catastrophe $V(R;u,v,c)$, defined by Eq.~(\ref{cusp}), red/dashed line. The last branch, which presents a minimum at $R=0$, seems thicker (and not dashed) because it is swept more times.}
\label{Vcusp_fit}
\end{figure}

It is revealing to know the potential $V_J(R)$ itself, which can be readily obtained from
\begin{equation}\label{VJt}
V_J[R(t)] = \int^{R(t)} \frac{dV_J}{dR} \frac{dR}{dt'} dt',
\end{equation}
once we know $R(t)$ and assuming $V_J(0)= 0$. From Eqs.~(\ref{Rphi}), (\ref{VJt}), the numerical solution of Eq.~(\ref{eqphisq}) and the initial conditions defined above, we plot $V_J(R)$ in Fig.~\ref{Vcusp_fit}. As one would expect, it is a multivalued function, whose branches correspond to those of $f(R)$. Its changing shape indicates that the potential is an essential dynamical ingredient --- its extrema and their corresponding effective masses --- changes as time passes by. 

The stability of any given configuration is determined by the signal of the squared-mass term of the corresponding (effective) potential at each one of its equilibrium points. It is well known that a given equilibrium point is stable if the corresponding squared mass is positive. In the present case, the (effective) squared mass in the JF is then given by the second derivative of Eq.~(\ref{dVdR}):
\begin{align}
\label{mJ}
m^2_J(R) \equiv \frac{d^2 V_J}{{\rm d} R^2} &= -\frac{R}{3} + \frac{f'}{3 f''} +  \frac{ f'''}{3 (f'')^2}\left( -2 f +  f' R\right), \\
 \label{mJ1}
 & = -\frac{R}{3} + \frac{f'}{3 f''}
\end{align}
where we have used that ${\rm d} V_J/{\rm d} R=0$ (i.e, at an extremum of $V_J(R)$) in the last step. The same expression was obtained by Baghram \cite{Baghram:2007df}:
\begin{equation}
\label{mB}
m_{B}^{2}(R) \equiv - \frac{R}{3} + \frac{f'}{3 f''},
\end{equation}
when we calculate $m_J^2(R)$ at one of the extrema of $V_J(R)$. This simply points out the obvious equivalence between our approach and the perturbative one.

\begin{figure}[h!]
\center
\noindent\stackinset{l}{25pt}{d}{140pt}
{\includegraphics[width=0.25\textwidth]{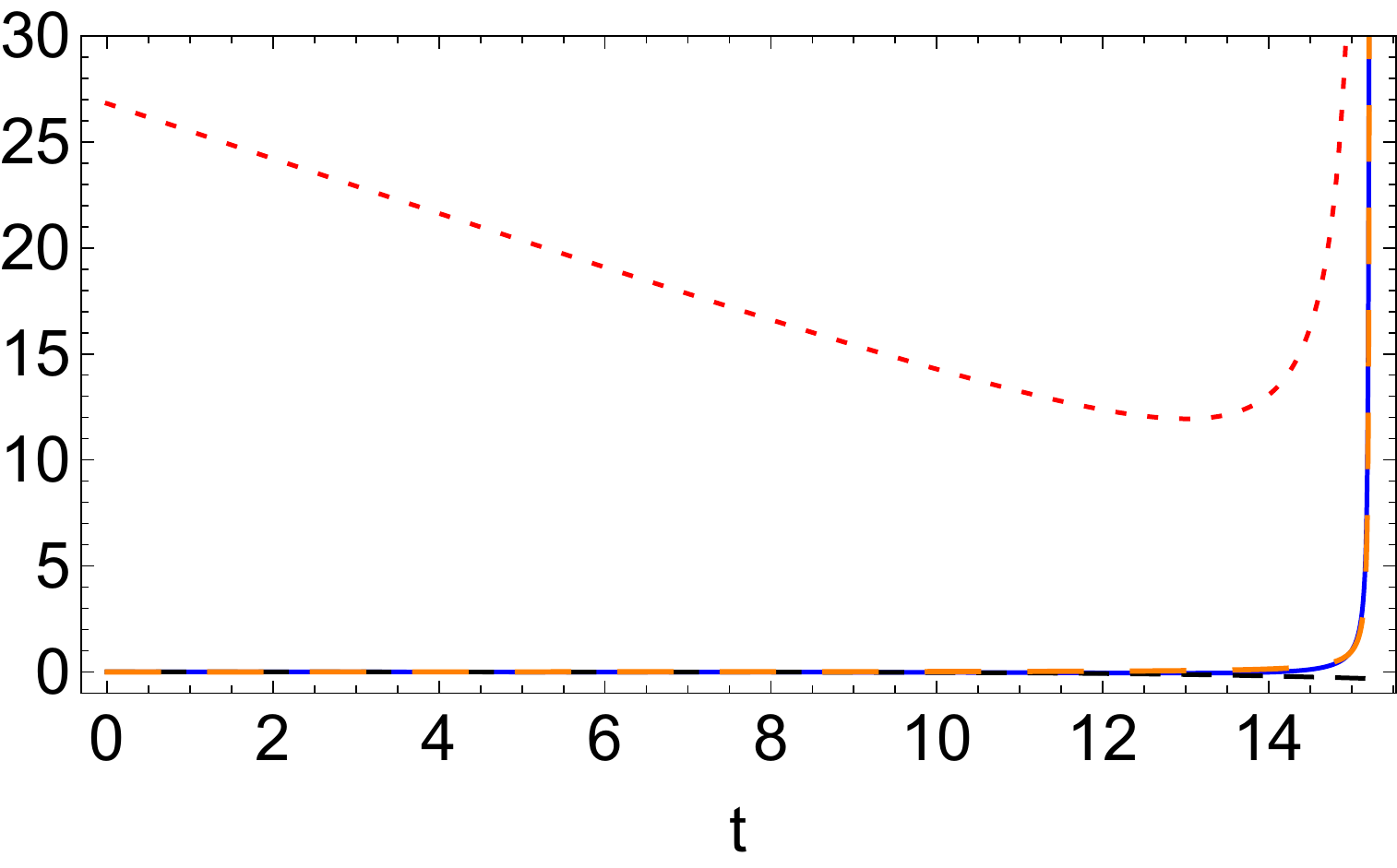}}
\noindent\stackinset{r}{20pt}{d}{40pt}
{\includegraphics[width=0.37\textwidth]{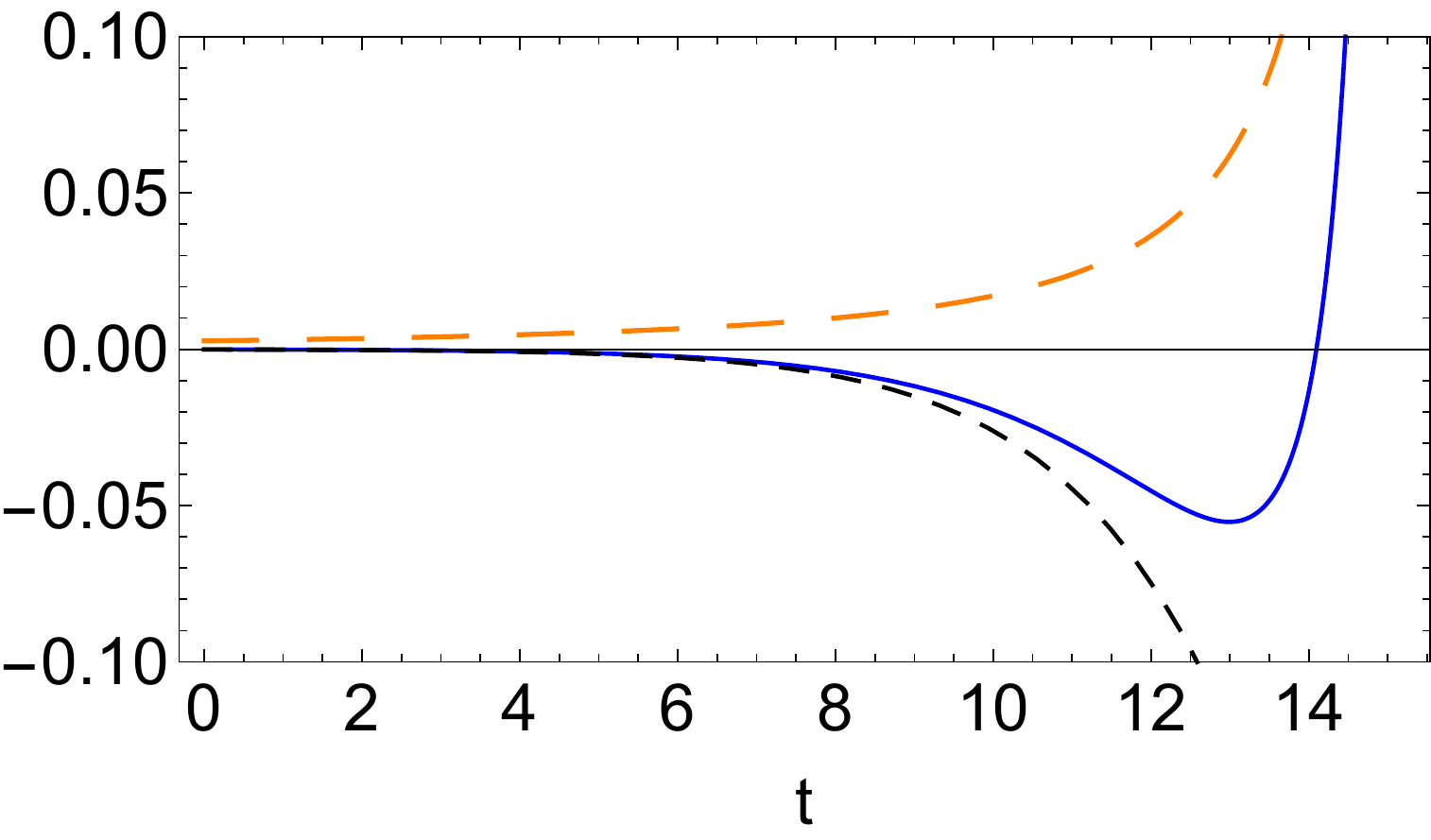}}
{\includegraphics[width=\textwidth]{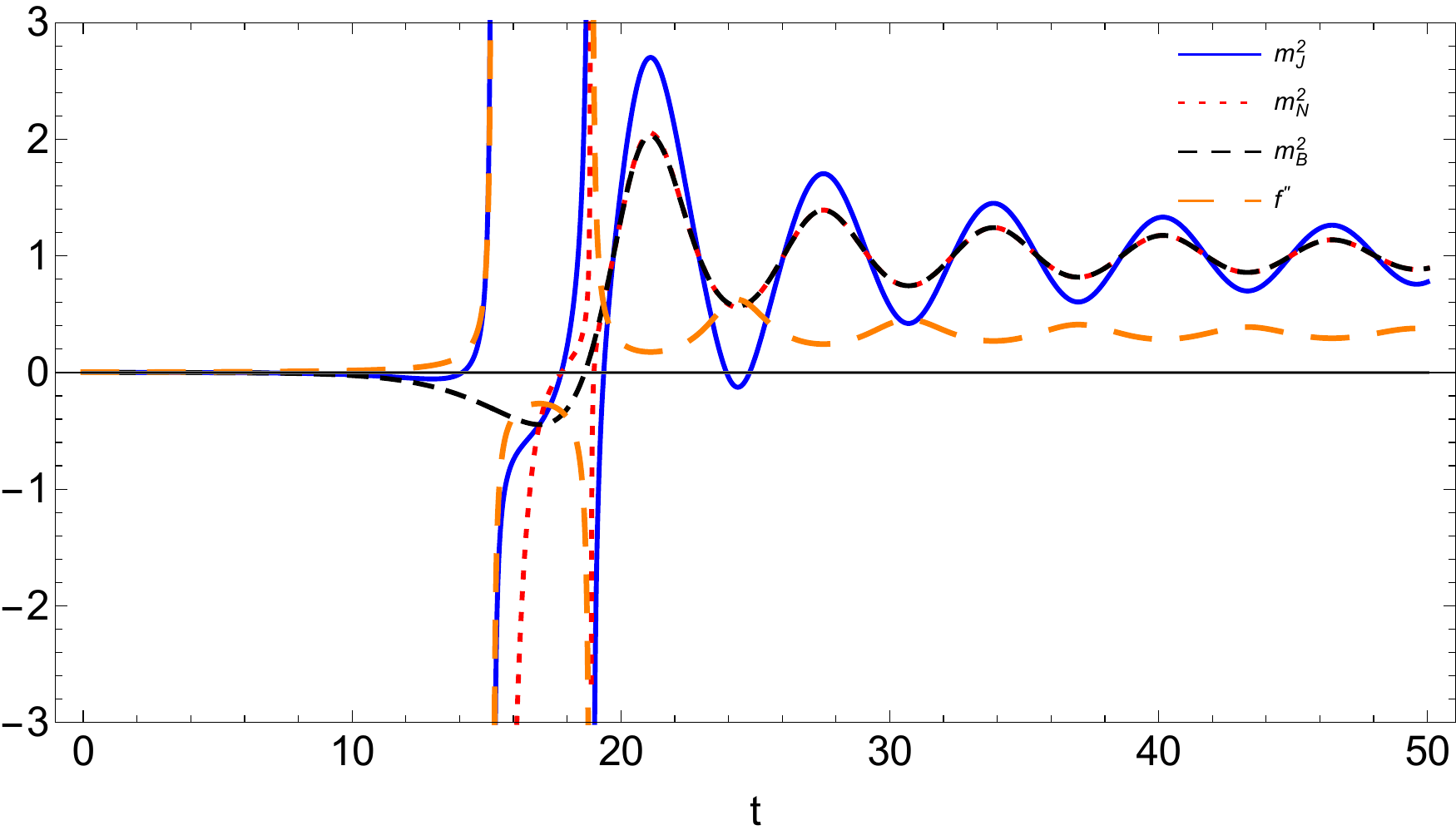}}
\caption[Mass definitions]{Different masses: $m^2_J$ (\ref{mJ}) in blue/solid, $m^2_N$ (\ref{mN}) in red/dotted, $m^2_B$ (\ref{mB}) in black/dashed and $f''$ in orange/large-dashed as functions of time. We have used $N = 60$ efolds and $m_\phi = 1$. The insets show the behavior of the curves in the first branch.}
\label{MAll}
\end{figure}

It is important to compare Eq.~(\ref{mJ}) to other mass definitions previously presented in the literature, namely Nojiri's \cite{Nojiri:2004dw} and Sotiriou and Faraoni's \cite{Sotiriou:2008rp}, respectively:
\begin{align}
\label{mN}
m_{N}^2(R) &\equiv -\frac{R}{3} + \frac{f'}{3 f''} +  \frac{f'''}{3 (f'')^2}( -2 f + f' R+ R),\\
\label{mSF}
m_{SF}^2(R) &\equiv   \frac{1}{3f''} \left(\frac1\epsilon -	f' \right),
\end{align}
all of which rely on an expansion around the (homogeneous) background. 
The difference between Eqs.~(\ref{mJ}) and (\ref{mN}) is the very last term in the latter (namely, $R$). Such term was obtained \cite{Nojiri:2004dw} assuming GR in the background and in the presence of matter: $- 8 \pi G_N T \approx R $, where $T$ is the trace of the matter energy-momentum tensor. Therefore, according to that reasoning, Nojiri's expression (\ref{mN}) will also agree with ours when matter is absent --- which is precisely the case studied here --- and when calculated at one of the extrema of $V_J(R)$. We compare the mass definitions --- Eqs.~(\ref{mJ}), (\ref{mB}) and (\ref{mN}) --- can be seen in Fig.~\ref{MAll} as function of time for the $f(R)$ presented here.



The authors of the latter mass expression, Eq.~(\ref{mSF}), use a perturbative approach around GR (and this is the crucial point here), having written $f(R) = R + \epsilon \, \Delta(R)$, where $\epsilon\ll 1$. 
The equation of the corresponding mass $m^2_{SF}$ obviously depends on $\epsilon$, but
since it can be as small as wanted, requiring the mass to be positive is equivalent to demanding $f''(R)>0$ (see Fig.~\ref{MAll})  --- which became the standard criterion for stability in the current literature.


\section{The Stability Criteria from Catastrophe Theory Approximation}\label{catastrophesection}

As we mentioned in the previous section, the potential $V_J(R)$ has a dynamical character, changing over time. With that piece of information in mind, we notice that it can be written, across different branches, as different expressions  of a $4^{\rm th}$-order polynomial: 
\begin{equation}\label{cusp}
V_c(R; u,v,c) = \frac{R^4}{4} + u \, \frac{R^2}{2} + v \, R + c.
\end{equation}
It is important to point out that the above expression is {\bf not} a series expansion. Actually, it is the simplest one that features the necessary characteristics to explain the behaviour of $R(t)$ --- the most important of which is the unique minimum in the final stage, around which the system oscillates; it also explains the initial-phase evolution.  

\begin{figure}[t]
\centering
\noindent\stackinset{l}{80pt}{b}{130pt}
{\includegraphics[width=0.15\textwidth]{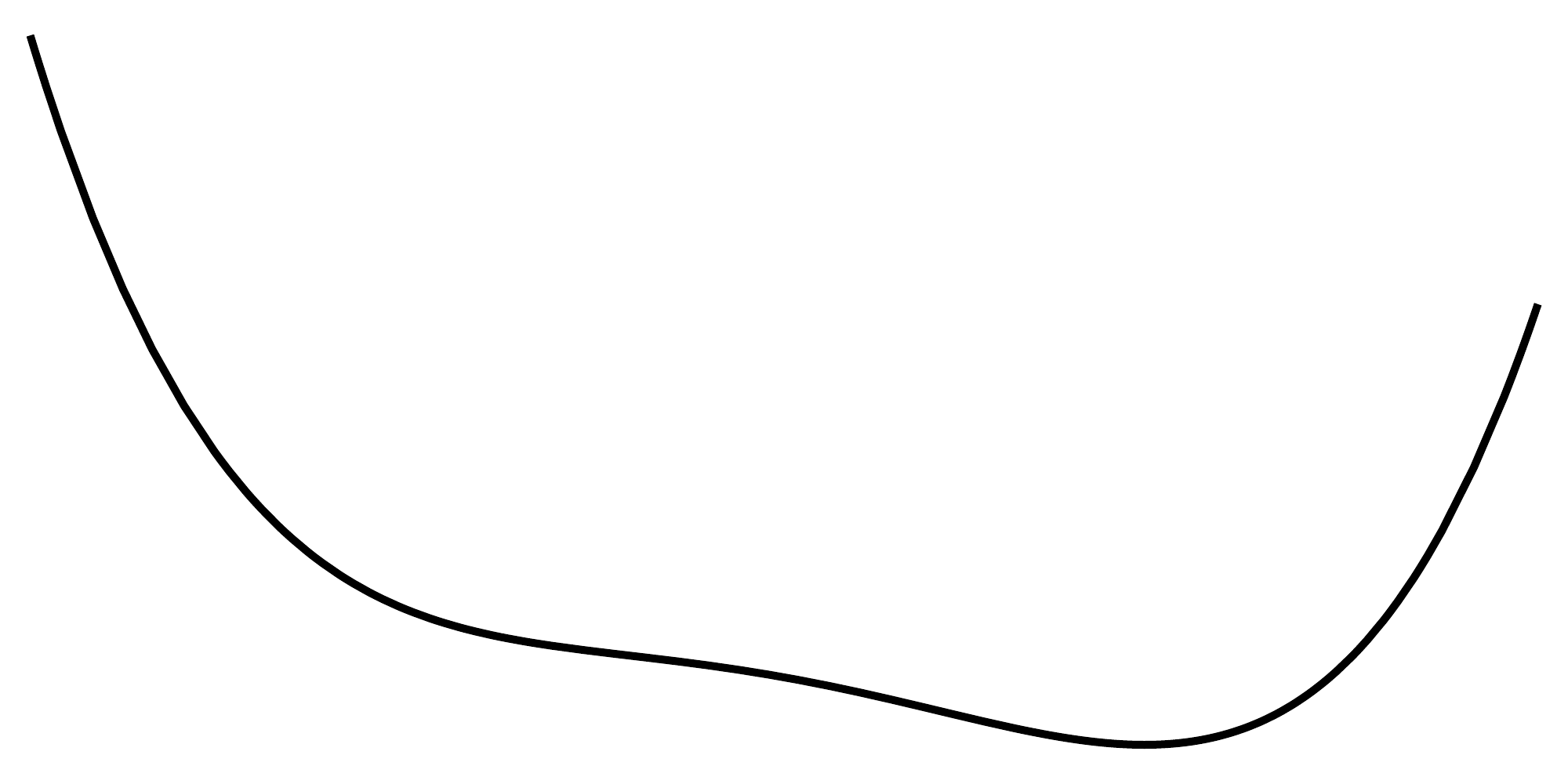}}
\noindent\stackinset{l}{170pt}{b}{150pt}
{\includegraphics[width=0.15\textwidth]{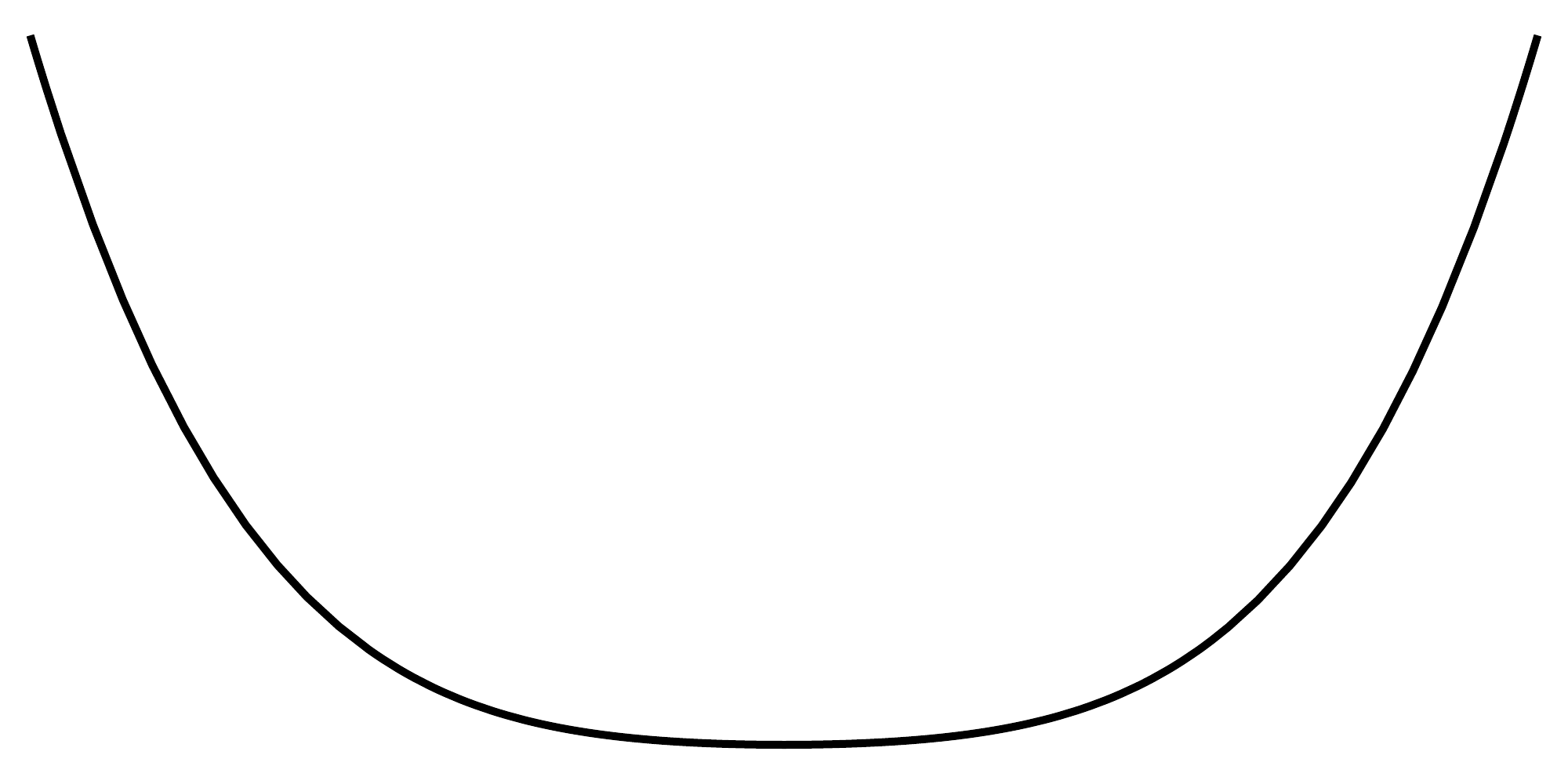}}
\noindent\stackinset{l}{260pt}{b}{130pt}
{\includegraphics[width=0.15\textwidth]{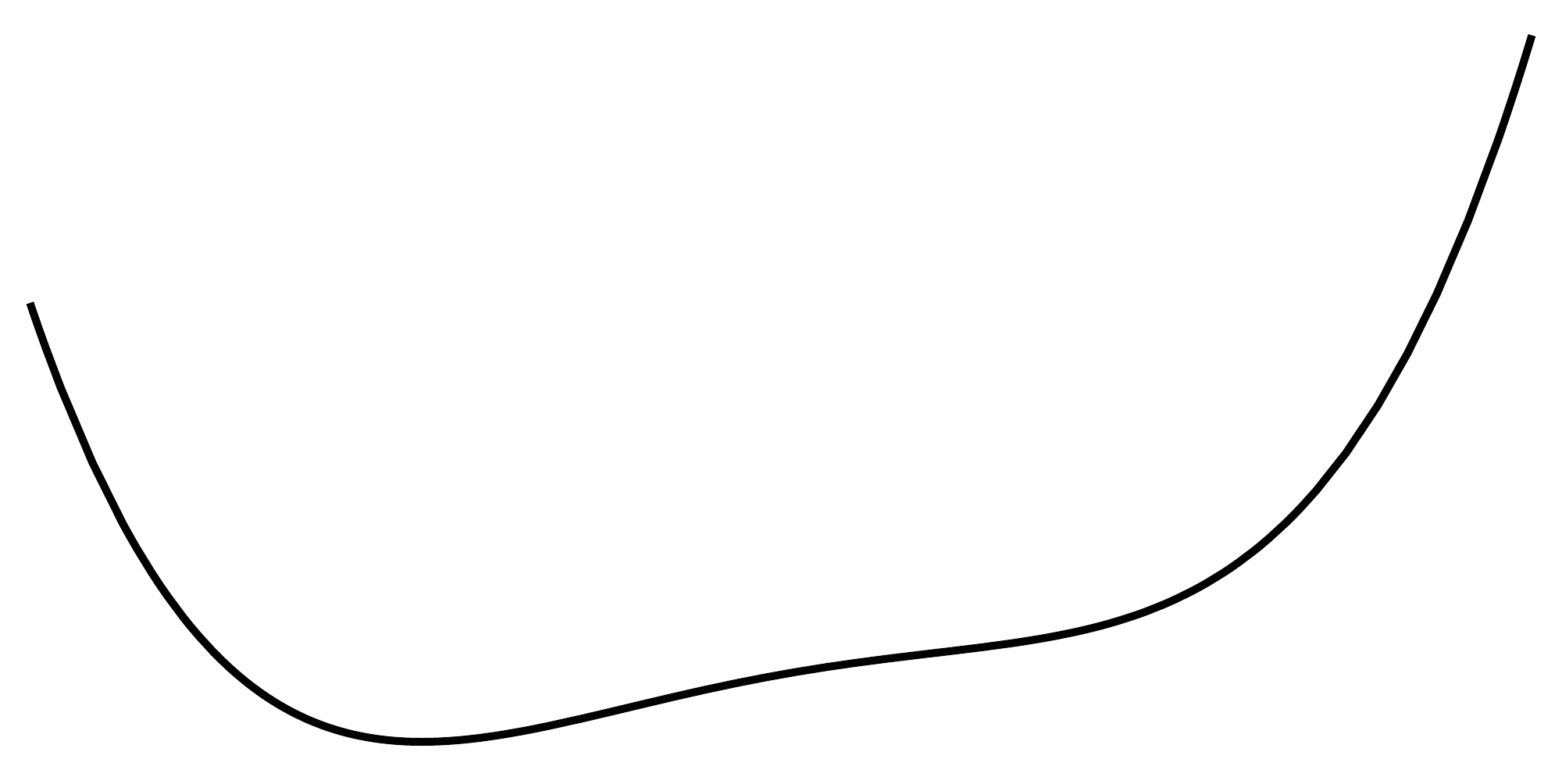}}
\noindent\stackinset{l}{80pt}{b}{35pt}
{\includegraphics[width=0.15\textwidth]{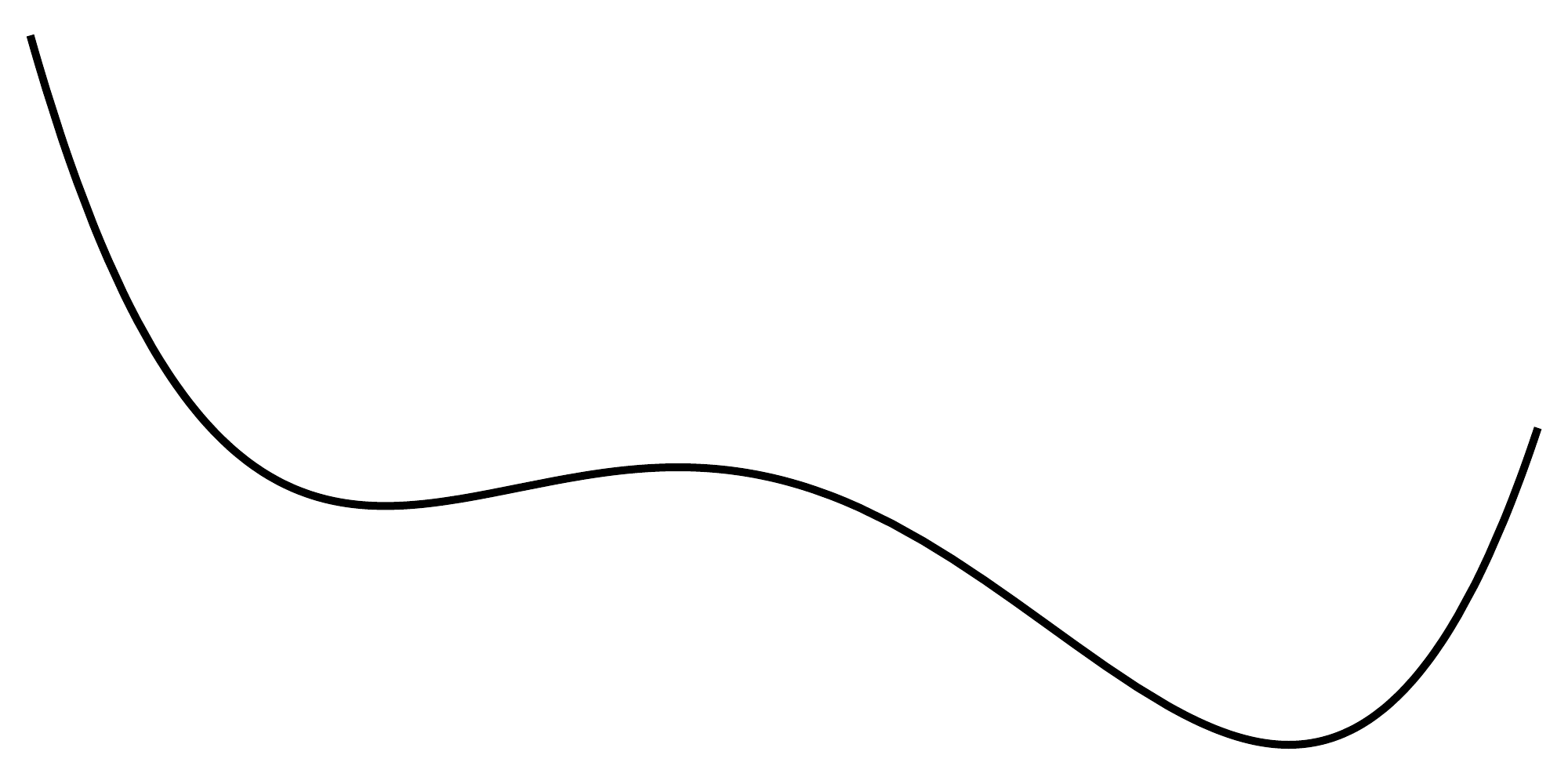}}
\noindent\stackinset{l}{170pt}{b}{35pt}
{\includegraphics[width=0.15\textwidth]{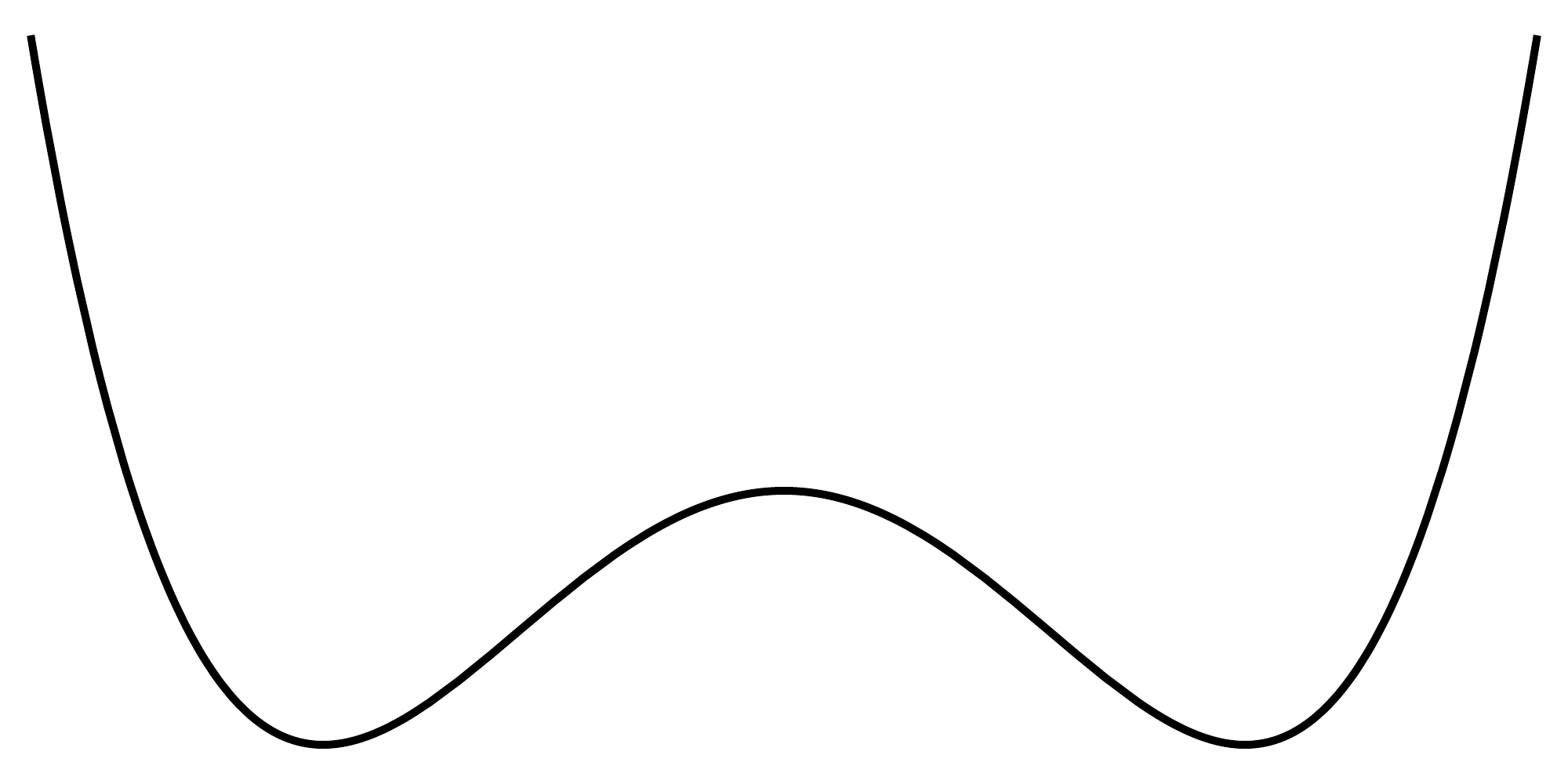}}
\noindent\stackinset{l}{260pt}{b}{35pt}
{\includegraphics[width=0.15\textwidth]{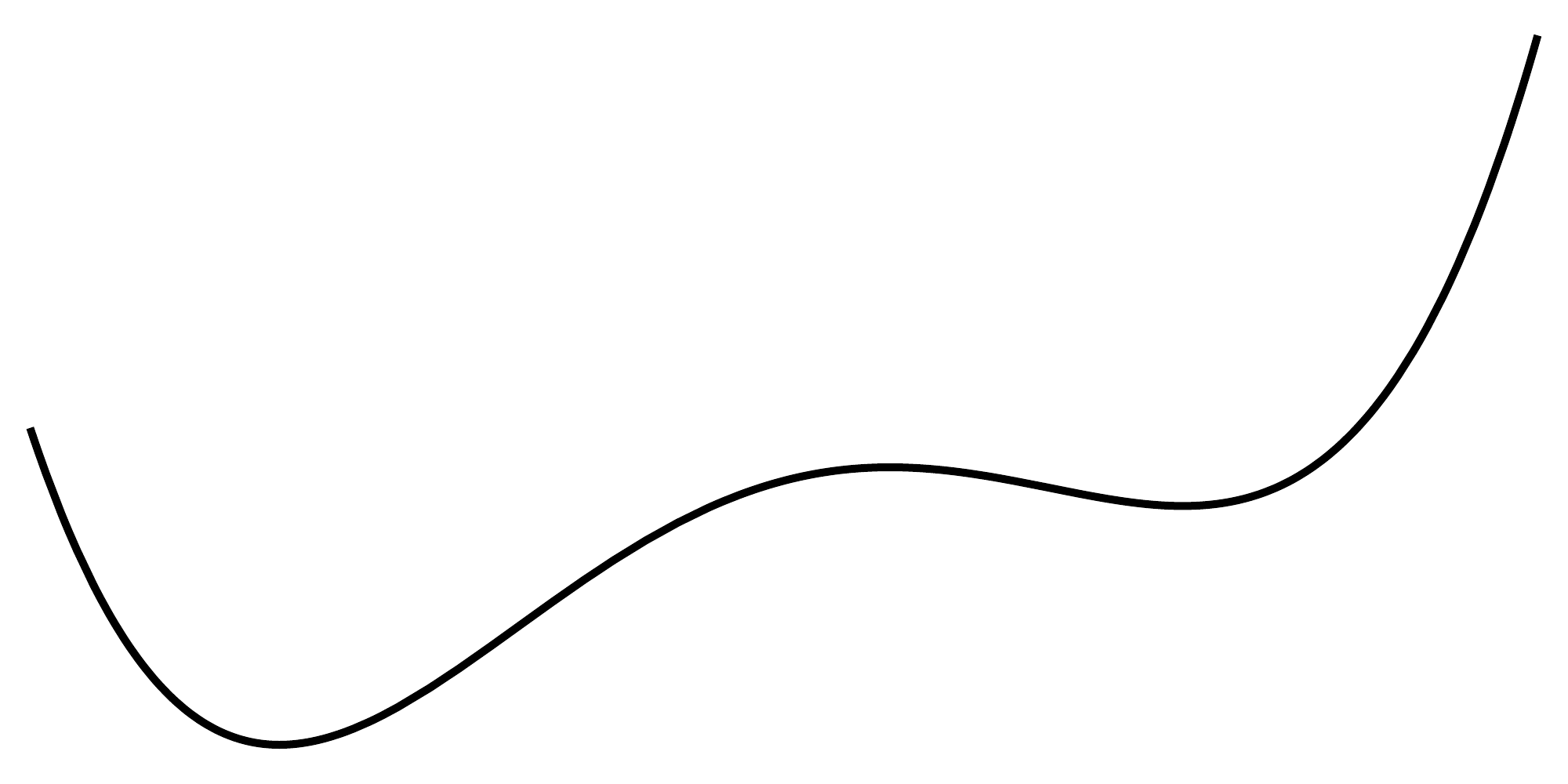}}
{\includegraphics[width=0.9\textwidth]{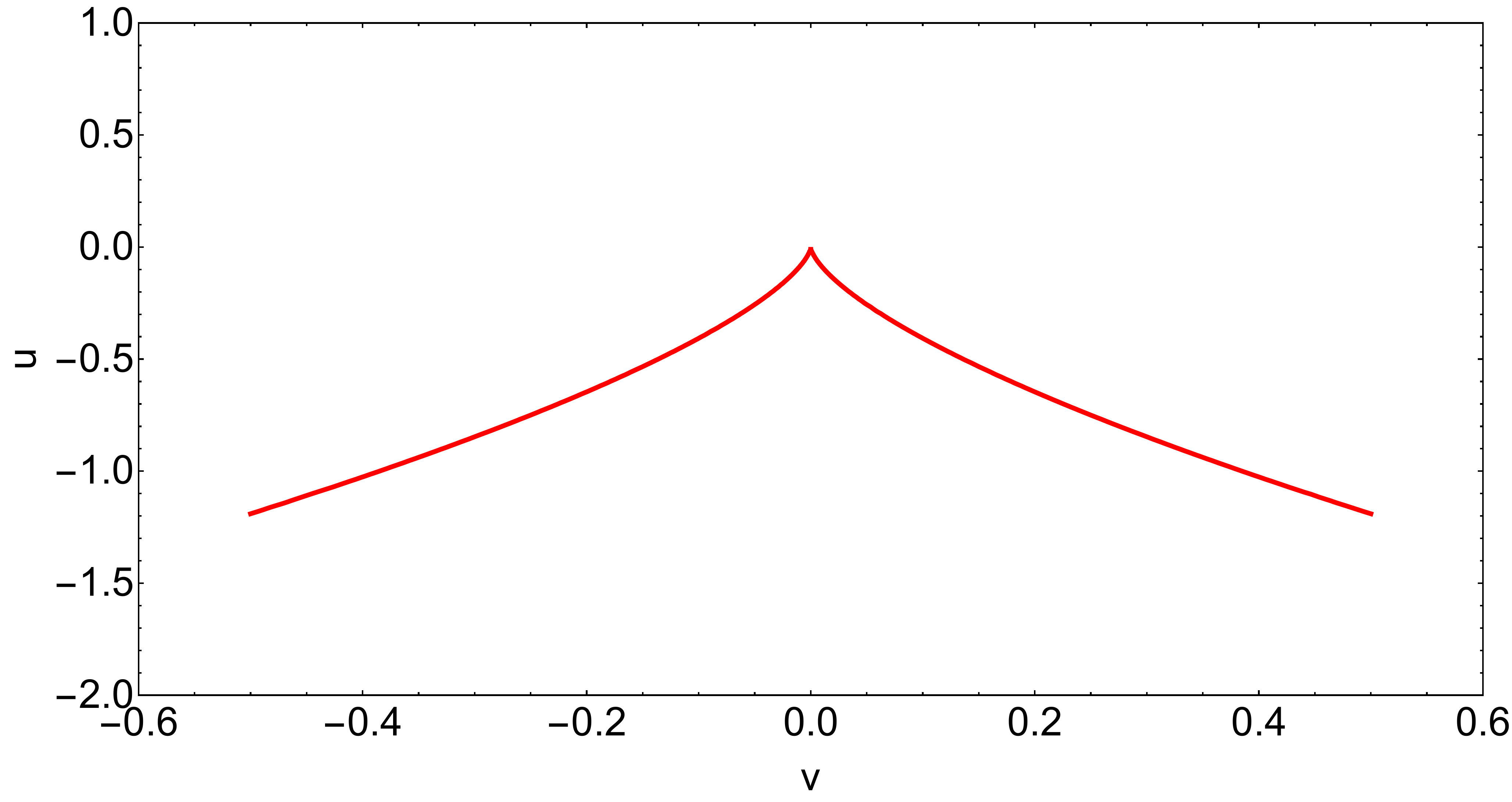}}\medskip
\caption[Control space.]{Control space $\{u \times v\}$ and the corresponding qualitative shape of $V_c(R;u,v,c)$ in each region.}
\label{cusp2D}
\end{figure}

The number of equilibrium states (i.e, extrema of $V_c$) obviously depends on the {\it control parameters} $u$ and $v$ --- see Fig.~\ref{cusp2D}. 
The coalescence of extrema and the change of dominance are studied by Catastrophe Theory \cite{poston1996catastrophe}. The expression (\ref{cusp}) for $V_c(R)$ is an elementary or normal form known as {\it cusp catastrophe}. The parameters 
$\{u,v,c\}$ span the so-called {\it control space} and $R$ defines the 1-D {\it behavior space}. 
The solutions of the system of equations $\{ \partial_R V_c(R;u,v,c) = 0 \, , \partial^2_R V_c(R;u,v,c) = 0\} $ generate the {\it bifurcation set} of the catastrophe. They represent the generation (or the annihilation) of a stable state and an unstable one (a minimum and a maximum of $V_c$, respectively). For Eq.~(\ref{cusp}), the bifurcation set is known as a {\it cusp} and it is given by
\begin{equation}\label{cuspequation}
3^3 \, v^2= - 2^2 \, u^3,
\end{equation}
and it is plotted in Fig.~\ref{cusp2D}.

On the cusp itself, two extrema (one minimum, one maximum) coalesce. The sign of $v$ defines the tilt of the potential, i.e, for $v>0$, the global minimum will be the one at $R<0$. For larger $u$, there is only one minimum of $V_c(R)$.

\begin{figure}
\centering
\includegraphics[width=0.9\textwidth]{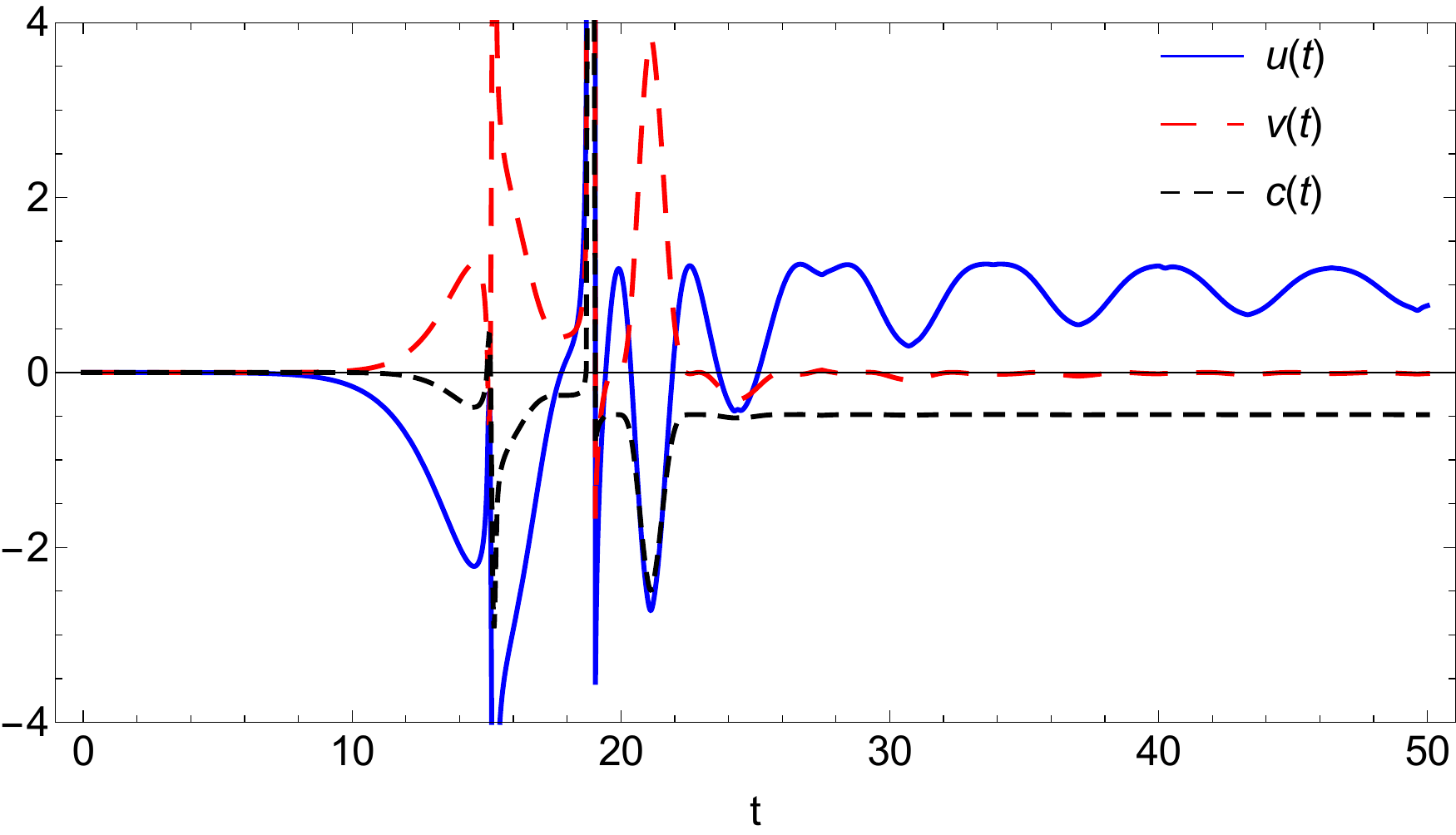}
\caption[Plot of parameters $u(t)$, $v(t)$ and $c(t)$.]{Plot of parameters $u(t)$, $v(t)$ and $c(t)$. Notice the divergences at the first two turnarounds of $R(t)$ (at $t_1 \approx 15.2$ and $t_2 \approx  18.9$). } 
\label{uvc}
\end{figure}

In order to obtain the time evolution of the control parameters for the case at hand, we have fitted $V_c[R(t);u(t),v(t),c(t)]$ to the potential $V_J[R(t)]$ for consecutive time ranges, using smaller intervals whenever close to one of the two sideways spikes in Fig.~\ref{swallowtail}. That returned the functions $u(t)$, $v(t)$ and $c(t)$,  plotted in Fig.~\ref{uvc}. They diverge at the first two turnarounds of $R(t)$ (at $t_1 \approx 15.2$ and $t_2 \approx  18.9$) but never again, even though $R(t)$ does oscillate endlessly. That happens because one can write
\begin{equation}
 \frac{d^2V_J}{dR^2}= \frac{d^2 V_J}{d\phi^2}\frac{1}{\left(\frac{dR}{d\phi} \right)^2}  , 
\end{equation}
whenever $dV_J/dR=0$, i.e, at the equilibrium points. Therefore, when the system is at such a point {\bf and} $dR/d\phi = 0$ --- which happens  {\it only} at the first two turnarounds of $R(t)$ --- one then gets $d^2V_J/dR^2\to \infty$. Both the potential $V_J(R)$ and its fit $V_c(R; u,v,c)$ are plotted in Fig.~\ref{Vcusp_fit}, showing an almost perfect agreement.

\begin{figure}
\includegraphics[width=0.47\textwidth]{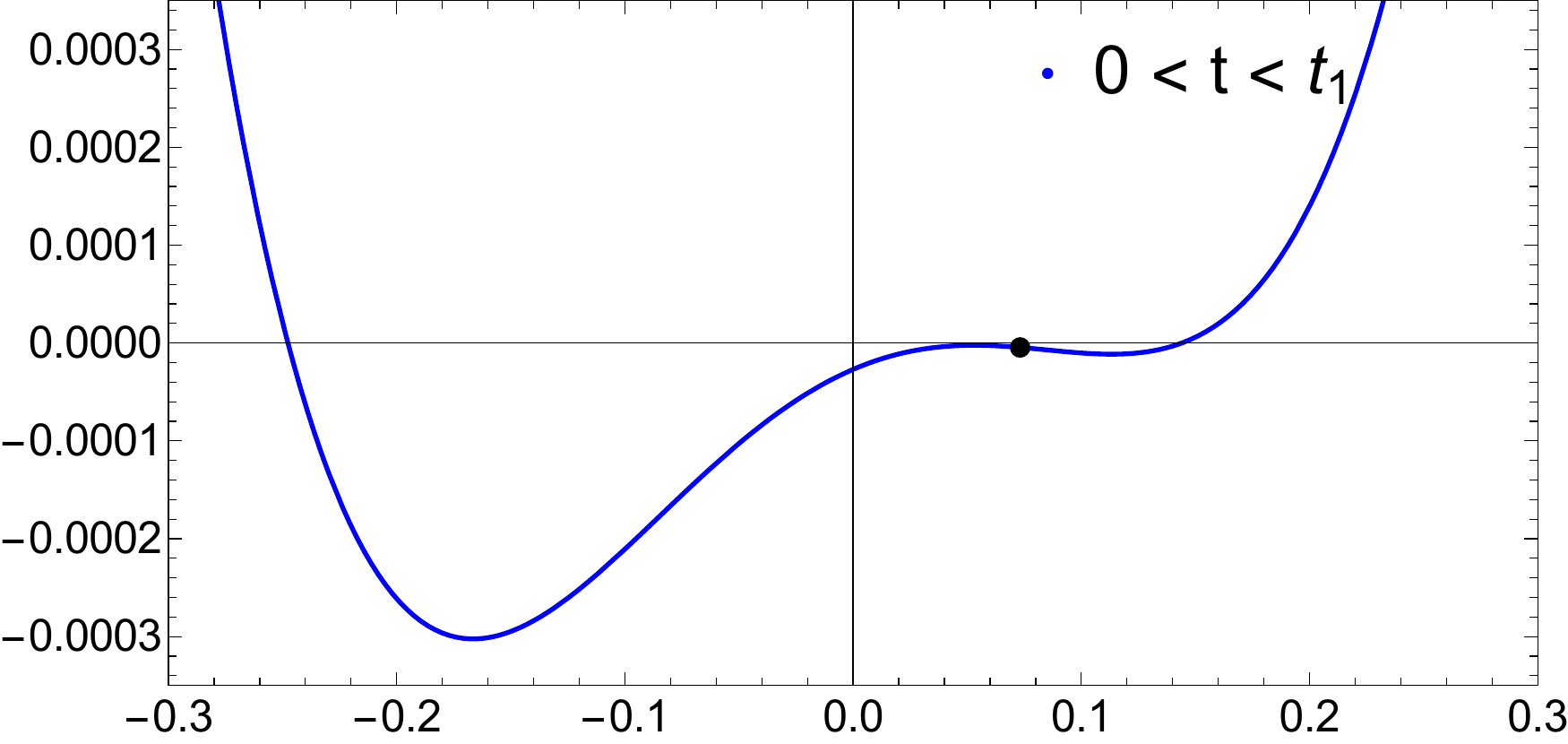}
\includegraphics[width=0.45\textwidth]{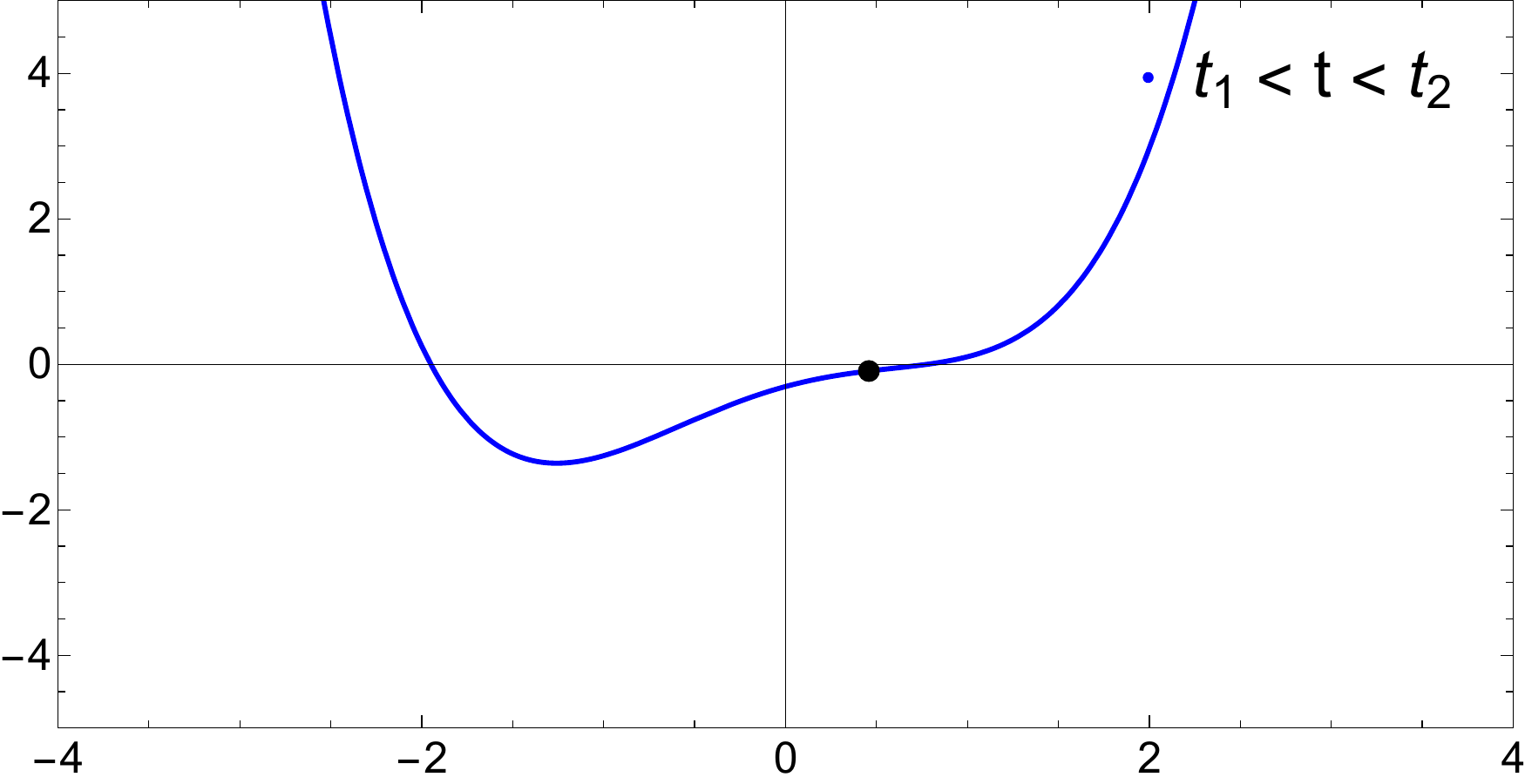}\\
\includegraphics[width=0.46\textwidth]{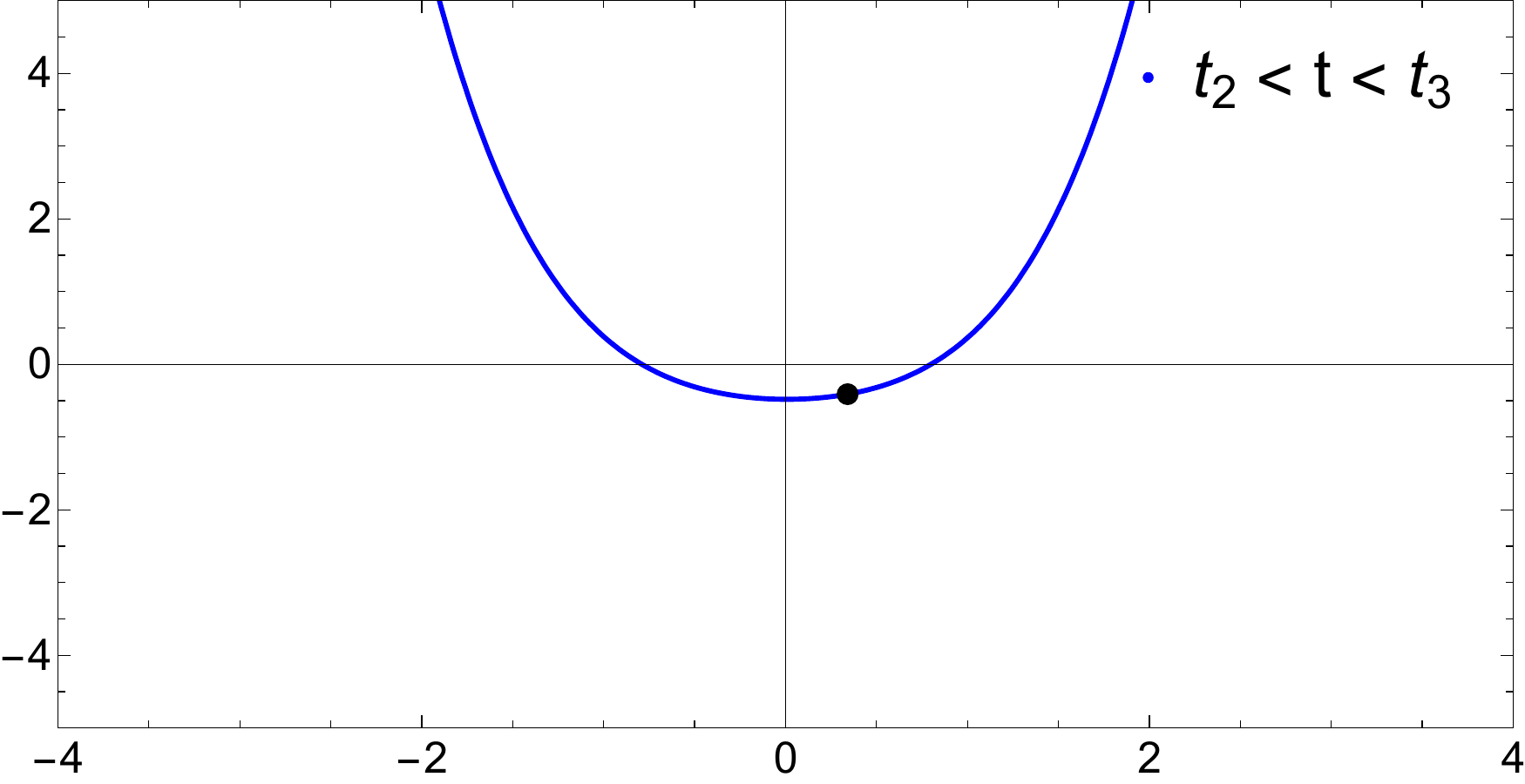}
\includegraphics[width=0.455\textwidth]{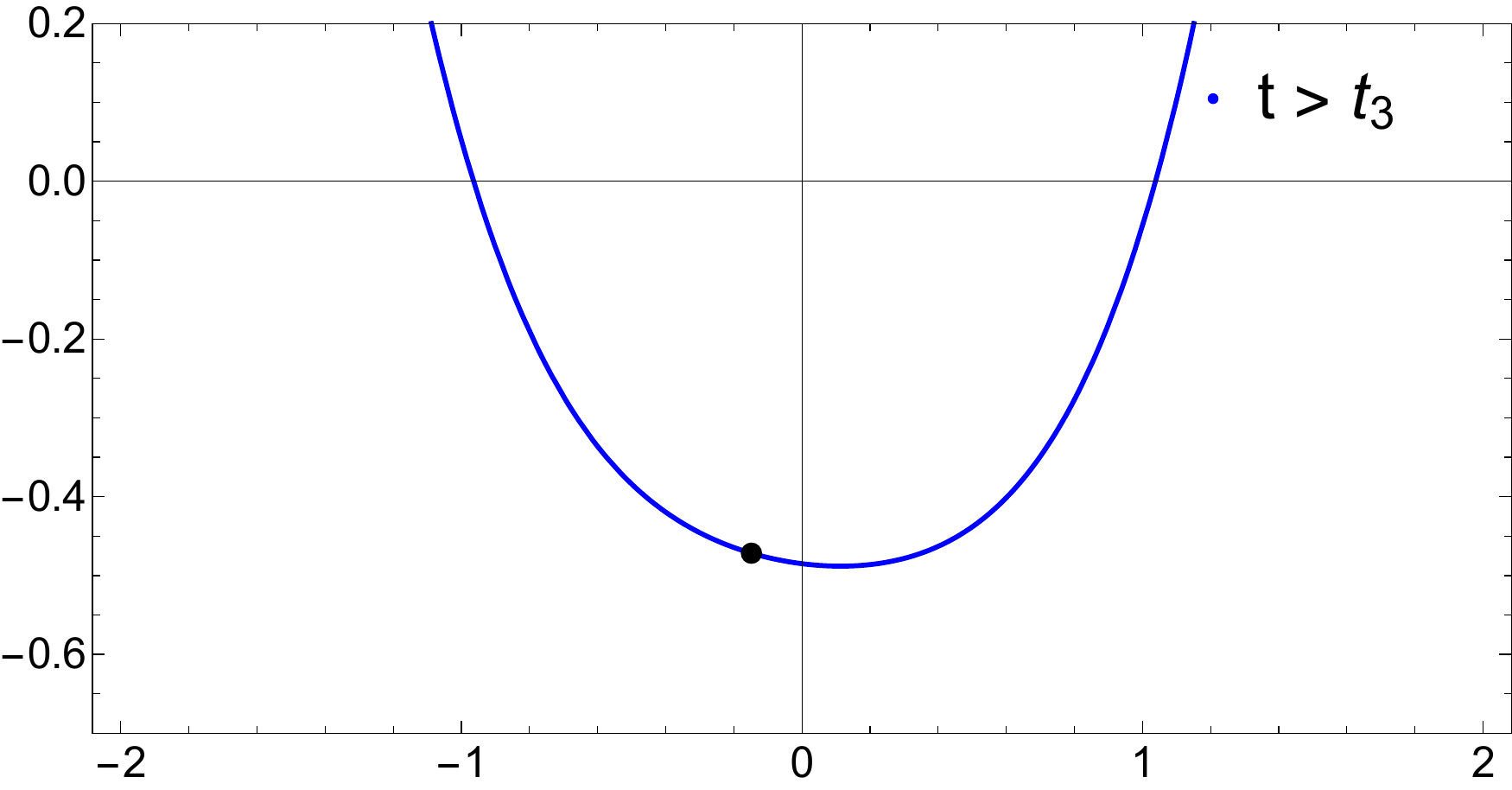}
\caption[Snapshots of the potential $V_c(R;u(t),v(t),c(t))$ as a function of $R$]{Snapshots of the potential $V_c(R;u(t),v(t),c(t))$ as a function of $R$ (see labels for time ranges), where $t_1 \approx 15.2$, $t_2 \approx  18.9$, $t_3 \approx  21.1$. The changes of branch occur at $t=t_1, t_2$. The black dot marks the value of $R(t)$ in the middle of the corresponding time range.}
\label{multi}
\end{figure}

Figure~\ref{multi} shows four snapshots of $V_c(R; u,v,c)$; the black dot indicates the value of $R(t)$ for a particular value of $t$ in the time range indicated in each panel. One can clearly see that the system is continuously trying to reach the equilibrium, but the positions of the minima (and even their number) change across the panels, as times passes. For $t>t_3$, there is only one minimum, which is will be reached by the system only in the asymptotic future.

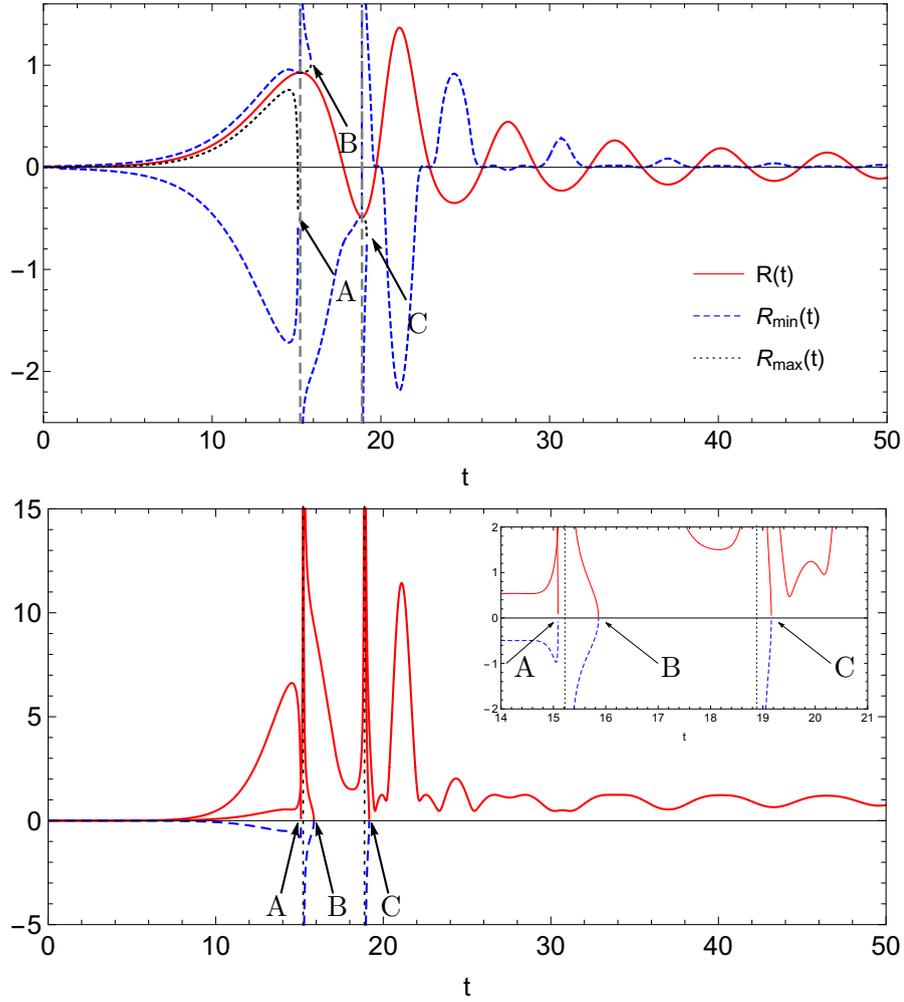
\begin{figure}
\centering
{\includegraphics[width=0.8\textwidth]{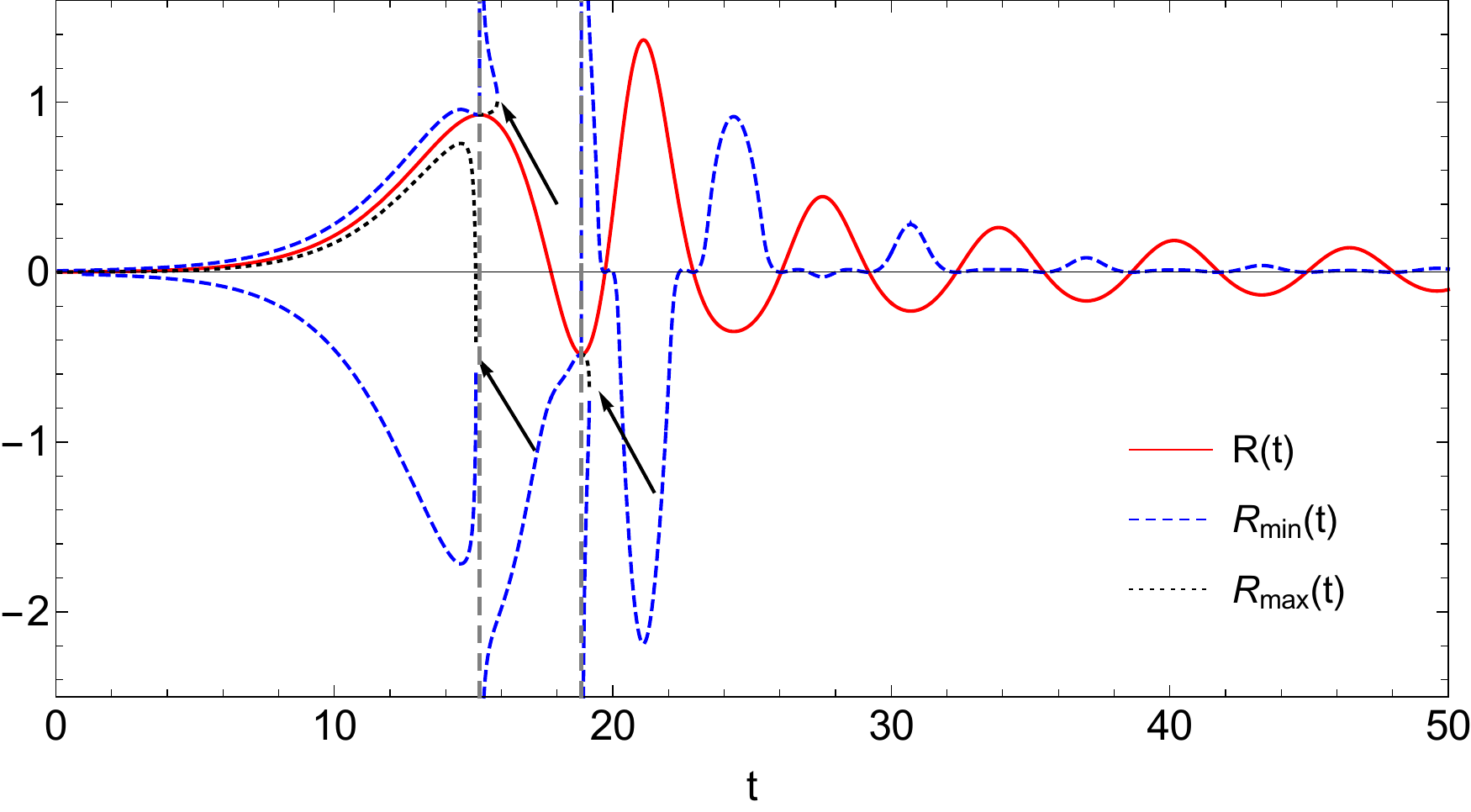}
\put(-212,72){A}
\put(-211,128){B}
\put(-185,60){C}}
\noindent\stackinset{r}{10pt}{t}{10pt}
{\includegraphics[width=0.35\textwidth]{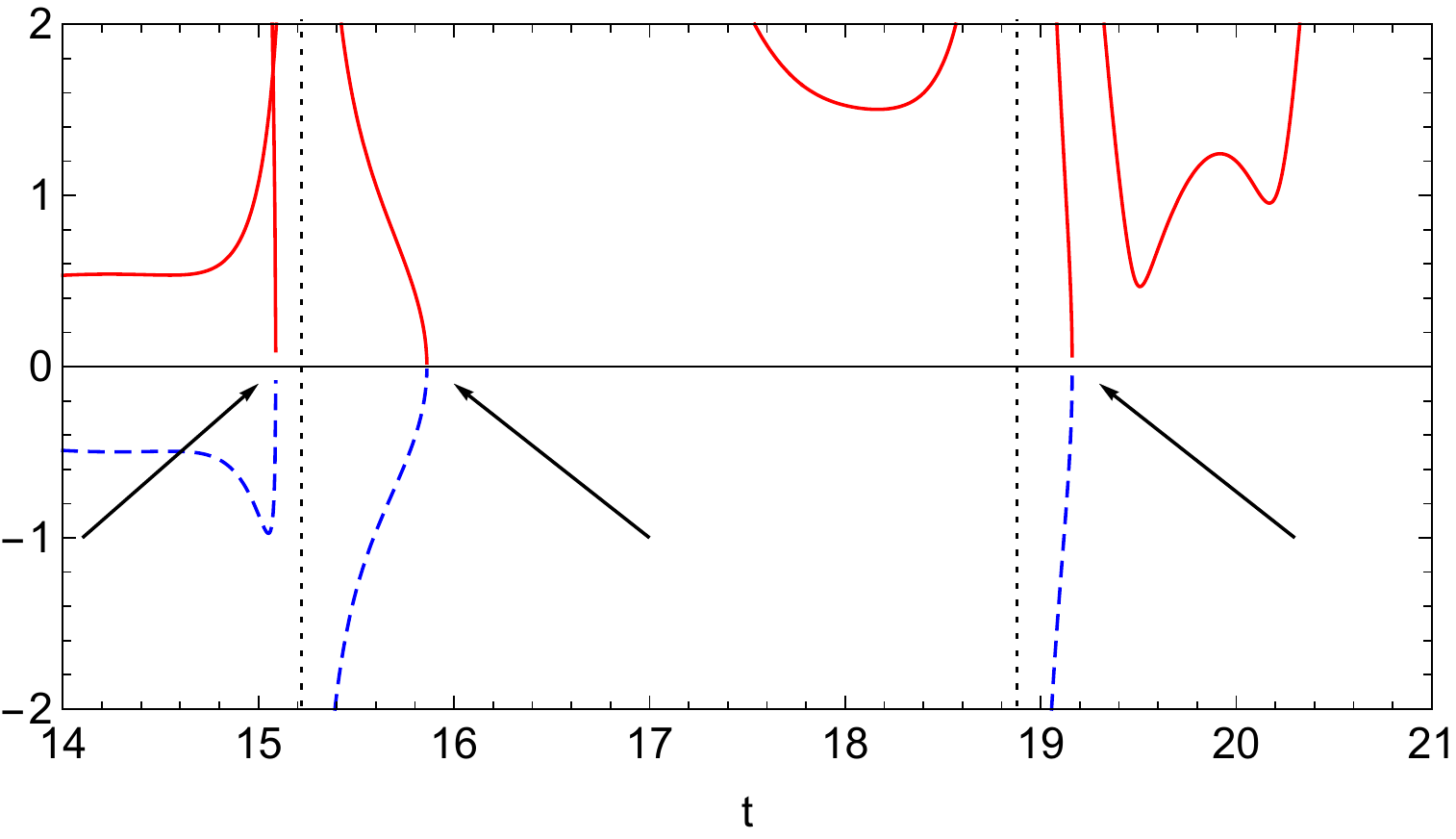}
\put(-135,25){A}
\put(-80,25){B}
\put(-15,25){C}}
{\includegraphics[width=0.8\textwidth]{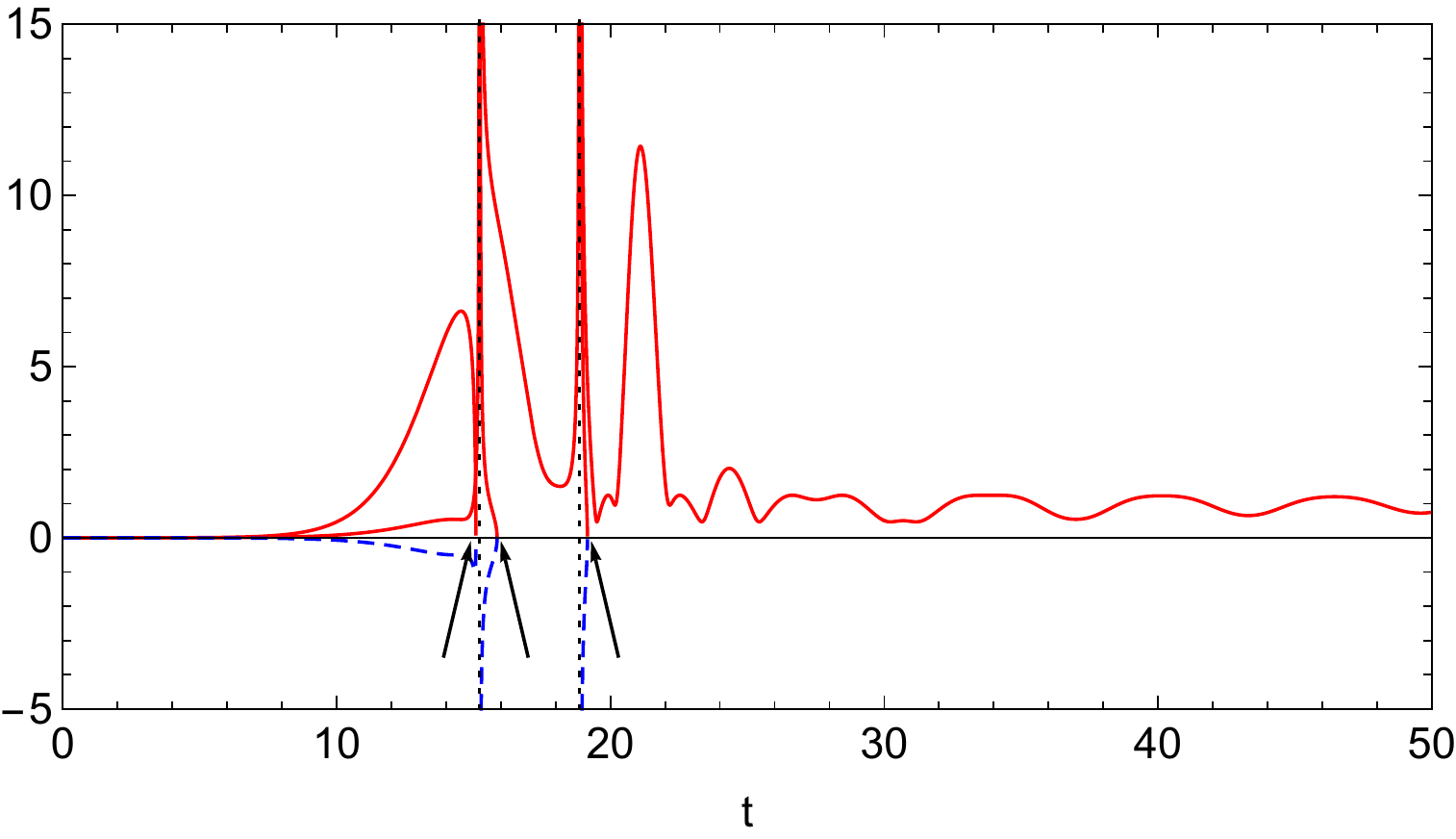}
\put(-238,33){A}
\put(-215,33){B}
\put(-195,33){C}}\medskip
\caption[Plot of $R(t)$ and the extrema and the effective mass squared.]{{\bf Upper panel:} Plot of $R(t)$ (red/solid line) and the extrema of $V_c[R; u(t), v(t), c(t)]$ as times passes. One can see that the system is always tying to catch up with one of the minima of the potential, $R_{\rm min}(t)$ (blue/dashed line). In particular, in the first branch, it is moving away from the local maximum $R_{\rm max}(t)$ (black/dotted line). {\bf Lower panel:} Effective mass squared $m^2_J[R(t)]$, calculated at each of the extrema (minimum in red/solid lines; maximum in blue/dashed line) of $V_c[R; u,v,c]$ plotted as a function of time. See inset for zooming in the coalescence region. {\bf Both panels:} Arrows point out the three coalescences (labeled $A$, $B$ and $C$) of one maximum and one minimum, when the cusp is crossed --- see Fig.~\ref{cusp2D}.}
\label{Rmins-mJ_fig}
\end{figure}

Indeed, the system keeps trying to reach one of the minima of the potential, as we can see in Fig.~\ref{Rmins-mJ_fig} (upper panel). The squared mass $m_J^2[R(t)]$ calculated at each extremum is plotted as a function of time in the lower panel of Fig.~\ref{Rmins-mJ_fig}. One can clearly see the coalescence of one maximum ($m_J^2<0$) and one minimum ($m_J^2>0$) at the points indicated by the arrows --- in both panels. 

There are clearly three distinct regimes, each one corresponding to a different branch of the evolution:

{\it Initial branch} ($t<t_1$):
$R$ slowly drifts away from its initial value, i.e, the initial configuration is (mildly) unstable as the system slowly drifts away from the local maximum ($R\sim 0$) towards a local minimum at $R>0$ (see Fig.~\ref{multi}). When it is about to reach it, at $t=t_1$, it coalesces with the local maximum (arrow $A$ in Fig.~\ref{Rmins-mJ_fig}, upper panel). A slowly-varying $R$ is a clear sign of an almost-de Sitter universe, i.e, there is inflation in the JF, even in the absence of a cosmological constant and with an approximately linear Lagrangian (since $f''\,R\ll 1$). Inflation is automatically terminated when, in the EF, $\frac{1}{2}{\dot\phi}^2 \sim V_E(\phi)$ and the slow-roll approximation breaks down.
Nevertheless, one can clearly see from Fig.~\ref{swallowtail} that $f''>0$ in the first branch. The standard criterion for stability ($f''>0$) fails here because, in spite of the almost-linear Lagrangian, it cannot be considered a perturbation of GR, where $f'\equiv 1$ as opposed to $f'\ll 1$, here. One might interpret this feature as a modified effective Gravitational Constant $\tilde G_N\equiv G_N/f'\gg G_N$.

{\it Intermediate branch} ($t_1<t<t_2$): 
The system is moving between two different stable configurations.  This system starts close to the local minimum ($m^2_J>0$), which annihilates with the local maximum --- see arrow $B$ in Fig.~\ref{Rmins-mJ_fig}, lower panel. Then it moves towards the only minimum left, at $R<0$, which coalesces with a local maximum that is created at the divergence --- see  arrow $C$ in Fig.~\ref{Rmins-mJ_fig}. One can see from Fig.~\ref{swallowtail} that $f''<0$ in this time range.

{\it Final branch} ($t>t_2$):
From $t=t_2$ on, there is only one minimum at $R=0 \, \forall t$ (where $m_J^2> 0$), around which the system is oscillating with a damping amplitude. At $t=t_3 > t_2$, the system reaches its furthest point away from equilibrium, which corresponds to the highest value of $R$. In this branch, and only here, $f''>0$ gives the expected answer on the stability of the system on this branch because the perturbative approach around GR does hold. For $t>t_3$, both $\phi$ and $R$ are in phase with each other and a plain chain-rule calculation of $dR/{\rm d} t = (dR/d\phi)\cdot(d\phi/{\rm d} t)$ yields a finite result for all $t$.

\begin{figure}[t]
\begin{center}
\includegraphics[width=0.6\textwidth]{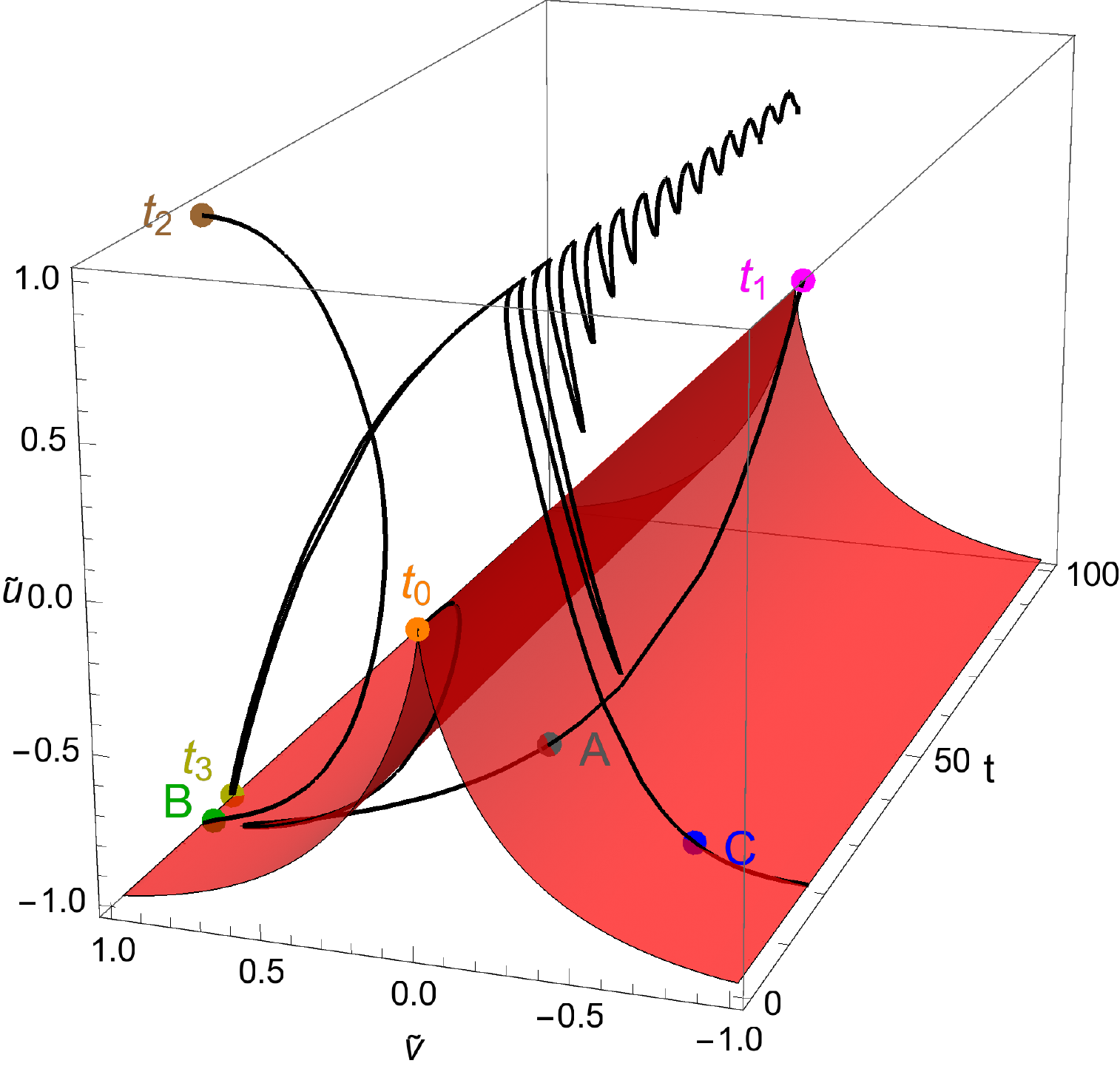}
\end{center}
\caption[Compactified control space.]{Compactified control space $\{ \tilde u  \times \tilde v \}$ (and an extra dimension for time $t$). The cusp surface is given by Eq.~(\ref{cuspequation}). The curve indicates the trajectory of the system, given by the parametric plot $\{\tilde u(t) \times \tilde v(t)\}$. The position of the system at $t=t_0, t_1, t_2$ and $t_3$ are  labeled. The circles labeled $A$, $B$ and $C$ mark the moments when the trajectory crosses the bifurcation set.}
\label{fitcusp}
\end{figure}

In order to circumvent the aforementioned infinities in the control parameters $u(t)$ and $v(t)$ (explicitly shown in Fig.~\ref{uvc}), we  compactify the parameter space by defining
\begin{equation}
\left\{
\begin{array}{l}
\tilde u(t) \equiv \tanh[u(t)], \quad {\rm and} \\
\tilde v(t) \equiv \tanh[v(t)] .
\end{array}
\right.
\end{equation}
In Fig.~\ref{fitcusp} we have plotted both the surface that corresponds to the bifurcation set (generated by Eq.~(\ref{cuspequation}) and slided along the time direction) and the path followed by the system in the compactified control space $\{ \tilde u(t) \times \tilde v(t) \}$ with an extra dimension for the time $t$. 
One can see the system starts at the cusp ($\{\tilde u,\tilde v\}=\{0,0\}$) and spends most of the time above it, i.e, in the 1-solution region (recall Fig.~\ref{cusp2D}). As we mentioned before, the system crosses the bifurcation set three times (labeled $A$, $B$ and $C$), producing corresponding changes in the structure of the minima of the effective potential $V_J(R)$. On the other hand, their dominance, i.e, the global minimum, is set by the tilt of the potential, i.e, by the sign of the parameter $\tilde v$ (or $v$). As we will see in the next section, it follows a characteristic behavior of phase transitions\cite{poston1996catastrophe}. 

The different criteria (positive squared masses and positive $f''$) may seem equivalent at usual cosmological conditions, but, as we will see in the next section, this is not so.  In particular, as we will show in the next section, the initial inflationary phase is {\bf unstable} only according to the masses defined above --- and, therefore, it is fated to end --- and, at the same time, {\bf stable} according to  a naive application of the $f''$ rule.  

In order to achieve that goal, we need first to determine the solution(s) of our equation of motion (\ref{frtrace}) and, to do that, we have to determine our effective potential $V_J(R)$, Eq.~(\ref{VJt})

Before skipping to the next Section, we generalize the potential $V_E(\phi)$, to include a shift in the $\phi$-vacuum value and a cosmological constant --- both, defined, obviously, in the Einstein frame.

\begin{figure}[t]
\center
\includegraphics[width=0.42\textwidth]{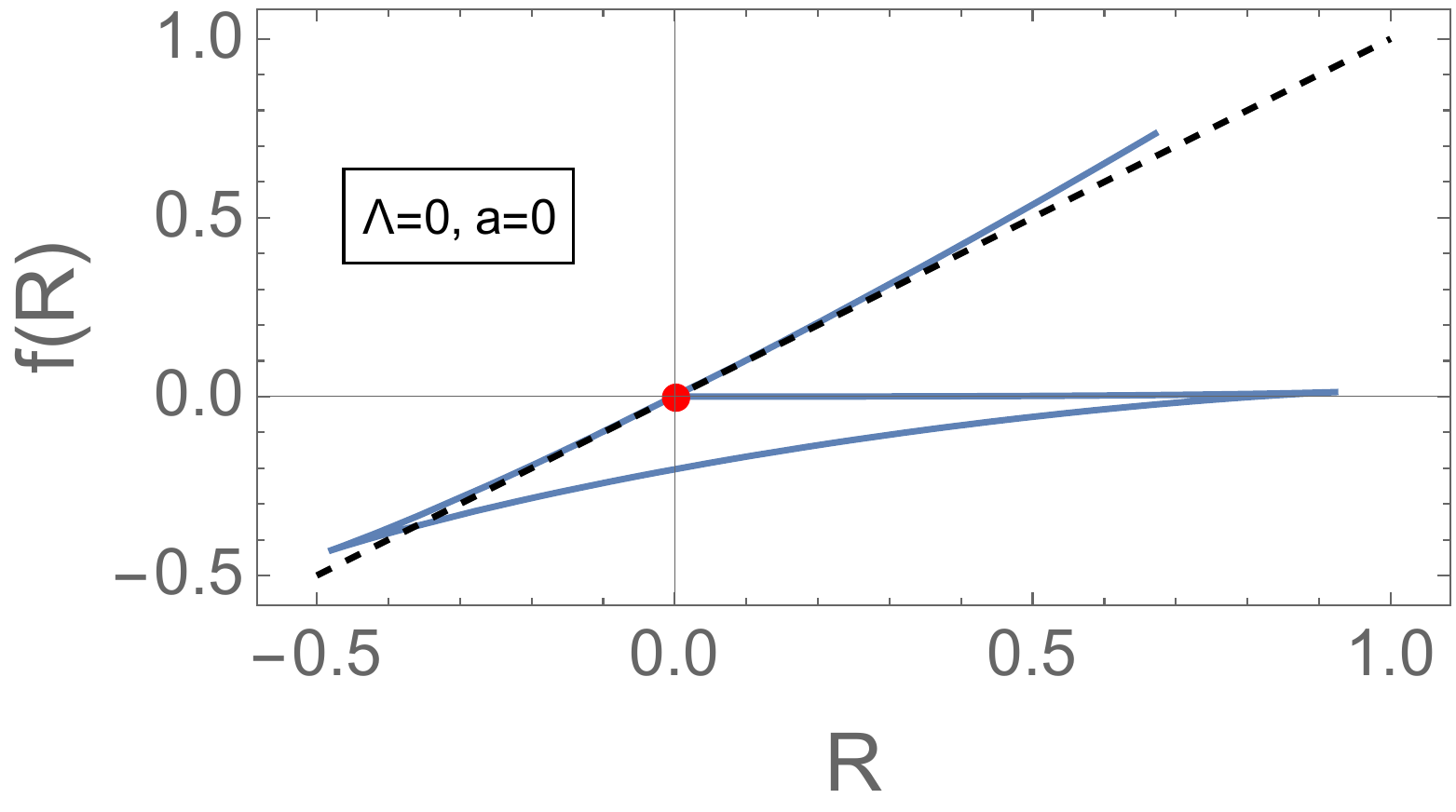}
\includegraphics[width=0.42\textwidth]{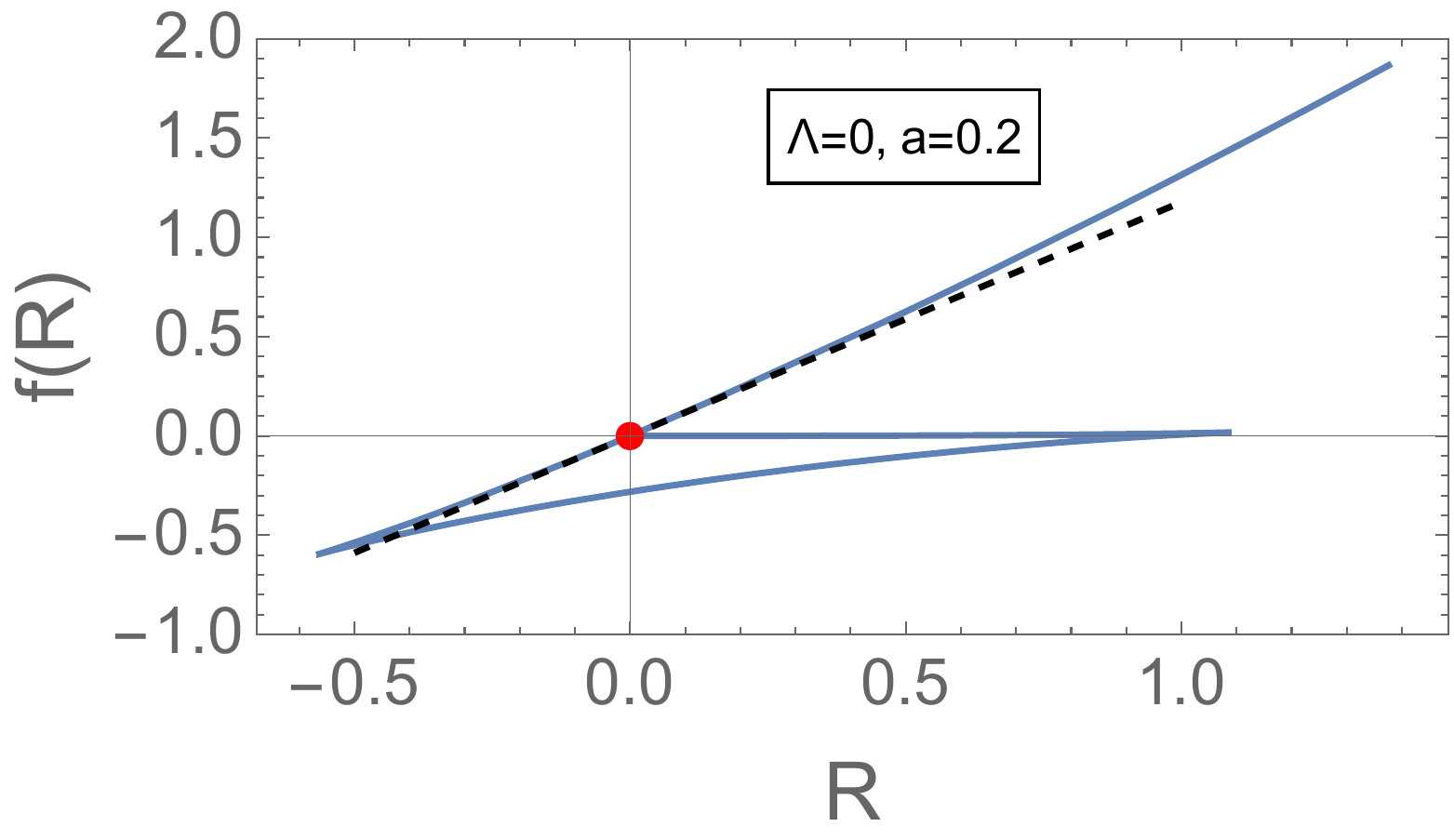}
\includegraphics[width=0.42\textwidth]{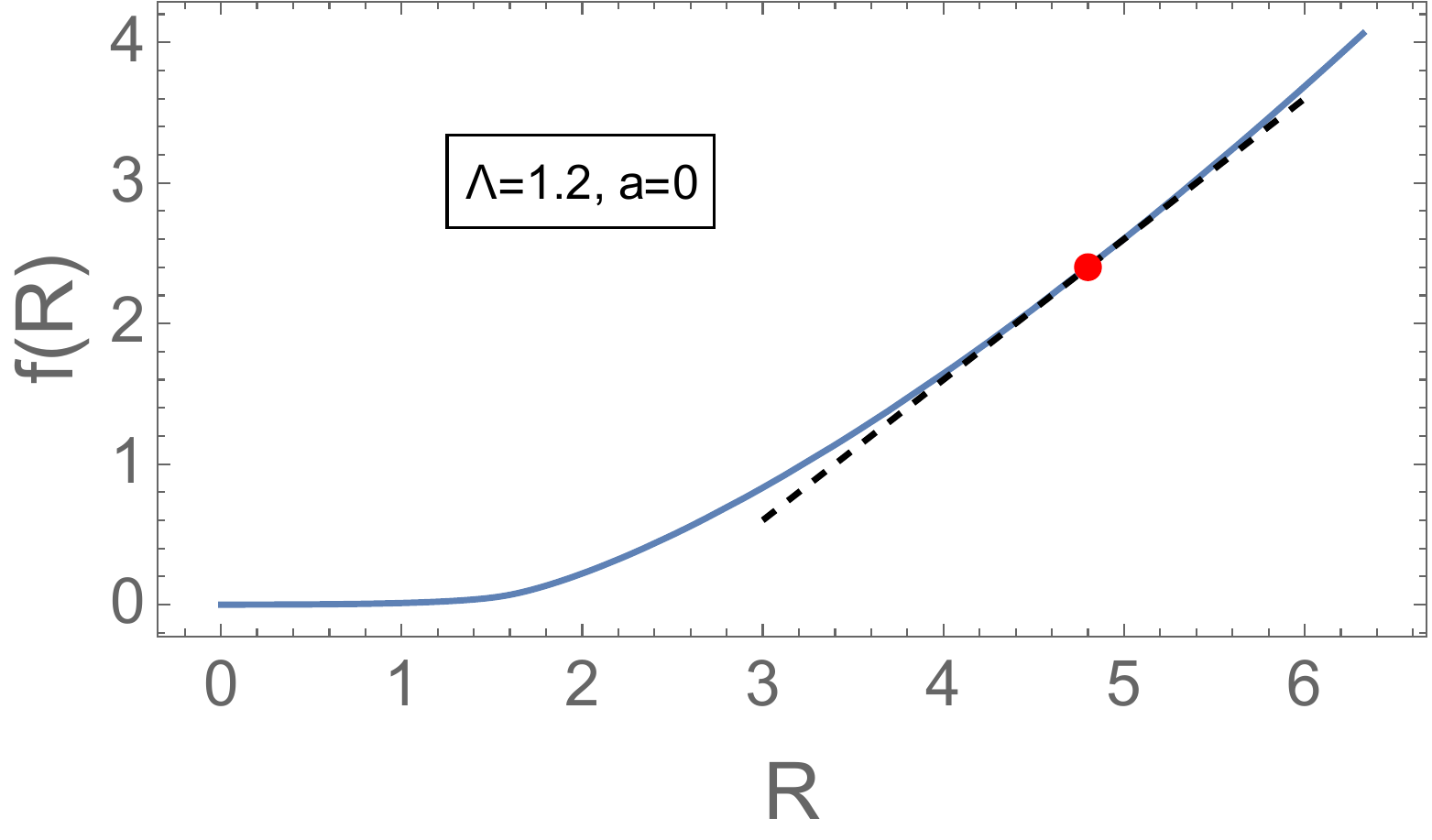}
\includegraphics[width=0.42\textwidth]{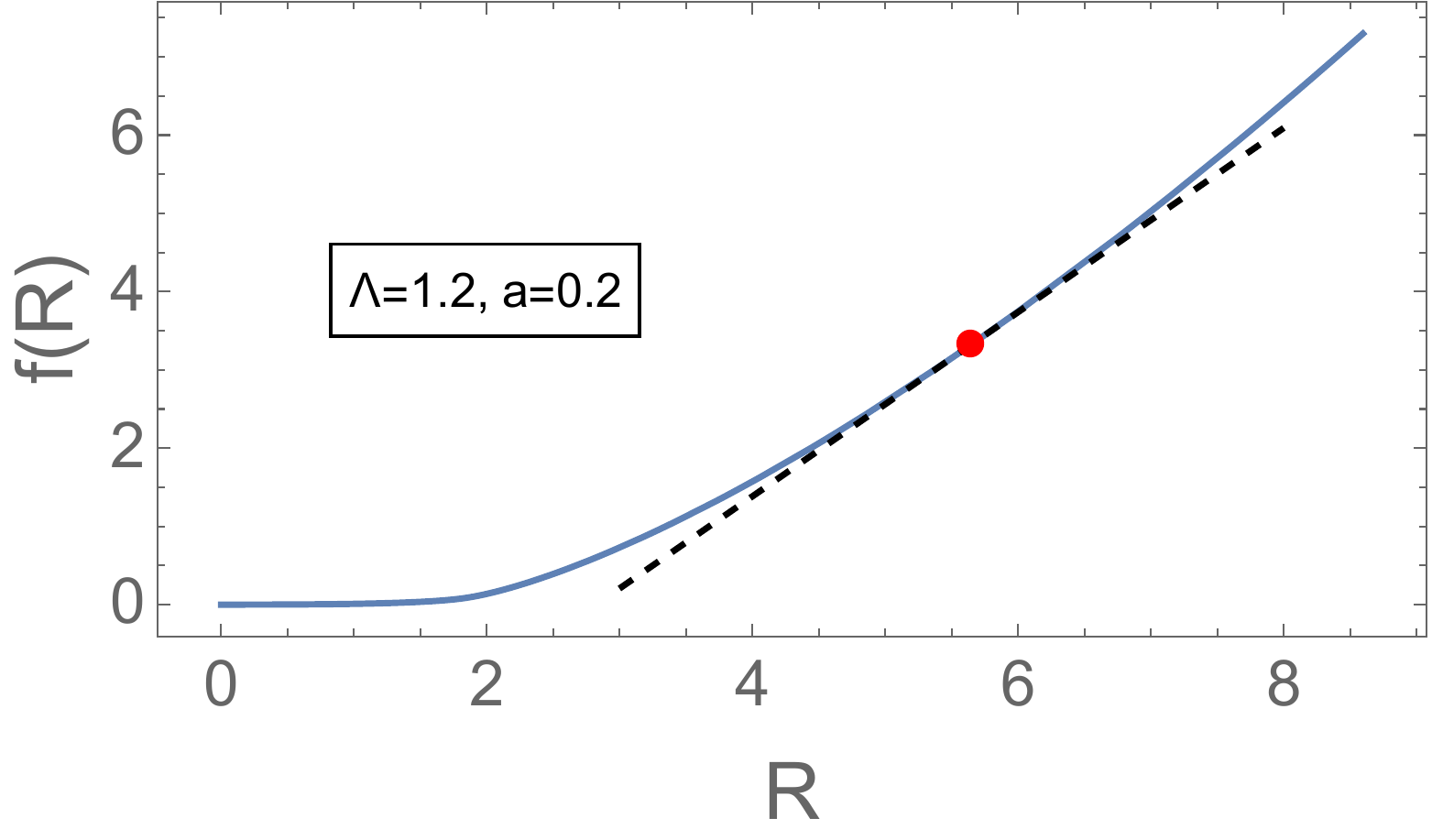}
\caption[Parametric plots of $f(R)$.]{Parametric plots of $f(R)$ given by Eqs. (\ref{fVEL}, \ref{RVEL}) for $\phi \in [-15,0.5]$ and for the parameters $\{\Lambda,a\}$ shown in the respective insets. In all panels, $f'>0 \, \forall R$.  The change in $a$ only re-scales the plot. Note that the panels with low $\Lambda$ present a 3-branch structure. In all of them, the dashed line is given by $f(R) = \exp(\beta a) R - 2 \Lambda_J$, the linear behavior of $f$ at the red dot, which  indicates the final de Sitter solution (Minkowski, if $\Lambda=0$), reached when $\phi=a$. The field $\phi$ and $a$ are given in Planck-Mass ($M_P$) units, $R$ and $\Lambda$ are given in $M_P^4$. We used $m_\phi=1 M_P$. 
}
\label{swallowtailLambda}
\end{figure}

Following an established procedure \cite{Magnano:1993bd} in section \ref{inverseproblem}, one arrives at the parametric expressions (\ref{Rphi}) and (\ref{fphi}), we apply this equations to the simplest possible (nontrivial) potential for a scalar field, to which we add an {\it ad hoc} Cosmological Constant $\Lambda$, namely
\begin{equation}\label{VELambda}
V_E(\phi) = \frac{1}{2}m^2_\phi \, (\phi-a)^2 + \Lambda.
\end{equation}
For now, $\Lambda$ and $a$ are written just for the sake of completeness, but they will turn out to be key ingredients later on. 
We then obtain the corresponding parametric form of $f(R)$:
\begin{align}
\label{fVEL}
f(\phi) &= e^{2 \beta  \phi } \bigg[m_\phi^2 (a-\phi ) \Big(a  -  \phi -\frac{2}{\beta}\Big)+2 \Lambda \bigg],\\
\label{RVEL}
R(\phi) &= 2 e^{\beta  \phi } \bigg[ m_\phi^2 (a-\phi ) \Big(a  -  \phi -\frac{1}{\beta}\Big)+2 \Lambda \bigg],
\end{align}
which we plot in Fig.~\ref{swallowtailLambda}. If $\Lambda<\Lambda_c$ (to be defined later on), the curve features a 3-branch structure.  In all of them, from the above expressions, one has $df/dR\equiv f' = \exp(\beta \phi)$. In particular, on the final branch, when the field $\phi$ oscillates around its potential minimum ($\phi=a$), one recovers GR only if $f'=\exp(\beta a)=1$, i.e, if $a=0$.
Regardless of $a$, the system does reach a de Sitter state with a non vanishing $R= R_{\rm dS}\equiv 4 \Lambda \exp(\beta a)$ and, therefore, a corresponding effective cosmological constant in the JF, given by $\Lambda_J \equiv \Lambda \exp(2 \beta a)$, and an effective gravitational constant  $G_{\rm eff}\equiv G_N \exp(-\beta a)$, where $G_N$ is the standard gravitational constant. In other words, at the final stage ($\phi \approx a$), the modified Lagrangian given by Eqs.~(\ref{fVEL}) and (\ref{RVEL}) can be written as the linear function
$f(R) = \exp(\beta a) R - 2 \Lambda_J$.


\begin{figure}
\center
\includegraphics[width=0.42\textwidth]{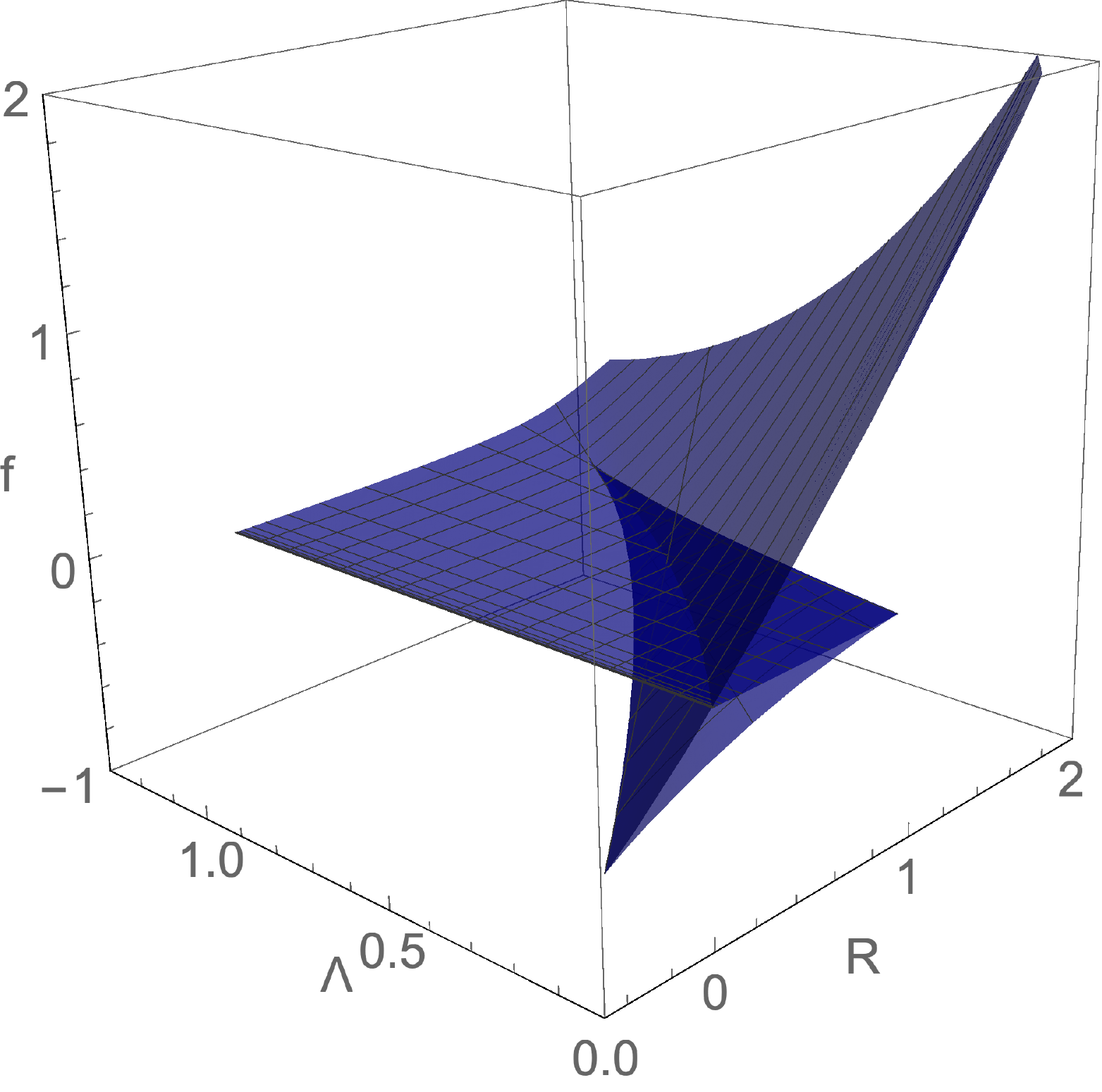}
\includegraphics[width=0.42\textwidth]{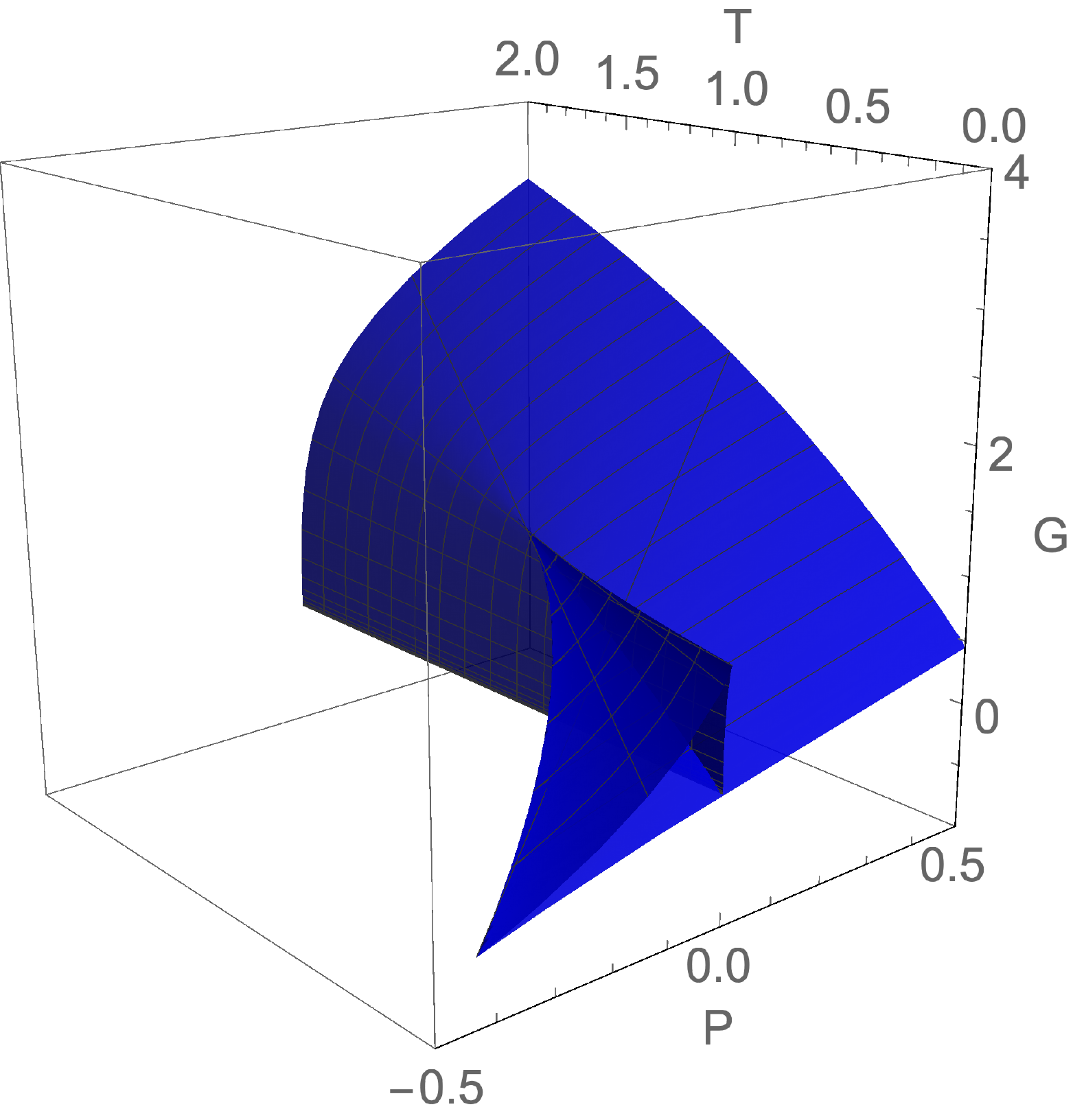}
\caption[Plots of $f(R,\Lambda)$, and $G(P,T)$]{Plots of $f(R,\Lambda)$, given by Eqs.~(\ref{fVEL}, \ref{RVEL}), and $G(P,T)$, given by Eqs.~(\ref{Gphi}) and (\ref{Pphi}) with $\beta=\sqrt{2/3}$ and $a=0$. The latter panel can be accessed on line (for a turnable picture) at \url{https://tinyurl.com/s7am2px}.}
\label{swallow3D}
\end{figure}

The behaviour of $f(R)$ for different values of $\Lambda$ is shown in Fig.~\ref{swallow3D} --- slices are shown in Fig.~\ref{swallowtailLambda}. The attentive reader may recognize a similiar surface for the van der Waals gas \cite{callen} (vdW, from now on). Indeed, Fig.~\ref{swallow3D} bears strong resemblance to the Gibbs potential $G$ for the vdW gas as a function of its temperature $T$ and its pressure $P$. The self-intersecting line indicates the coexistence curve of two phases and the pair of sideways peaks correspond to the metastable states. 

\section{A Numerical Example}\label{numerical}

From now on, we will investigate the potential given in Eq.~(\ref{VE}) as a standard toy-model inflationary potential in the EF --- initially, we will keep $a=\Lambda=0$, except when necessary for a cleaner picture and noted so. 

First of all, we have to determine the time evolution of $R(t)$ and $\phi(t)$. We recall that throughout this study there is no matter nor radiation; the $\phi$ field is pure gravity. In GR, that would imply $R=0 \, \forall \, t$. In $f(R)$ theories, on the other hand, $R$ has a dynamical behavior of its own. Here, it suffices to use $R[\phi(t)]$ (defined in the JF) from Eq.~(\ref{RVEL}) and  $\phi(t)$ (in the EF) from the standard equation of motion for a scalar field in an expanding homogeneous spacetime Eq.~(\ref{kleingordon}), where the initial conditions for the numerical solution of Eq.~(\ref{eqphisq}) are the standard ones in the slow-roll approximation \cite{Linde:2007fr}: 
\begin{equation}
\phi(0) = - \sqrt{2(1 + 2 N)}\approx - 15.5 \, , \quad
\dot \phi(0) = \sqrt{\frac23}\approx 0.81,
\end{equation}
which correspond to 
\begin{equation}
R(0)  \approx 
3.4 \times 10^{-3}\, ,\quad \dot R(0) \approx 
1.8 \times 10^{-3}.\footnote{Where $\phi$ is given in Planck-Mass ($M_P$) units, $R$ is given in $M_P^4$,
and $N=60$ is the number of efolds.} 
\end{equation}

We point out that in standard $\phi^2$ inflation, the slow roll is an attractor \cite{Grain_2017} so that the initial conditions do not need to be fine tuned. In the corresponding phase in the JF, where we fit $f(R) \approx R^{2.2}$, the same happens.
\begin{figure}[t]
\center
\noindent\stackinset{r}{15pt}{t}{60pt}
{\includegraphics[width=0.45\textwidth]{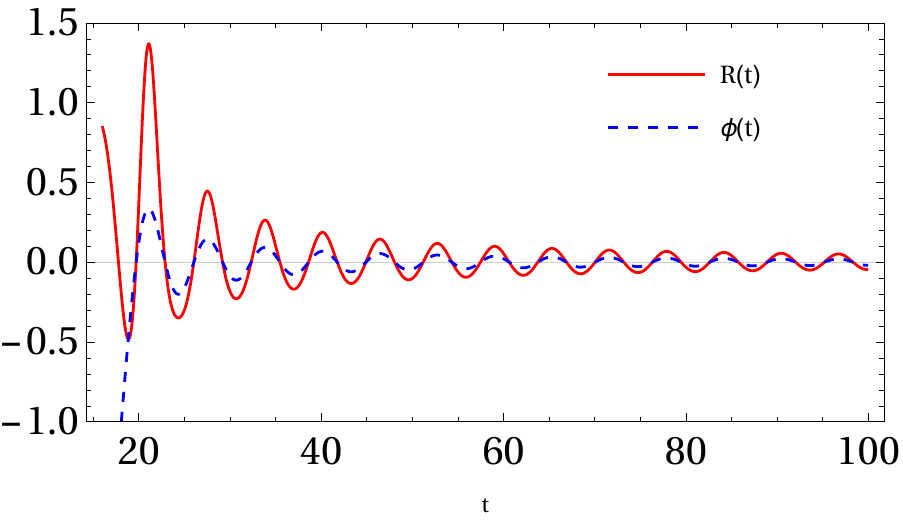}}
{\includegraphics[width=.9\textwidth]{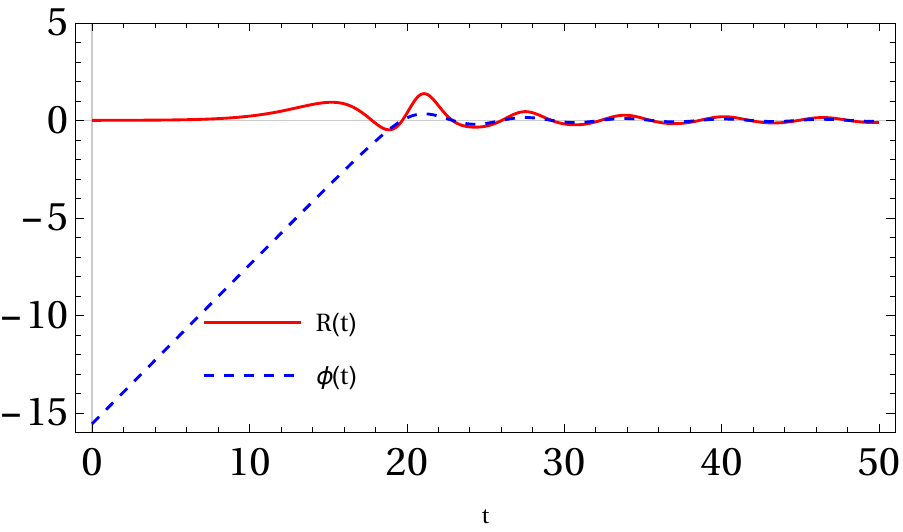}}
\caption[Numerical solution for $R(t)$ and $\phi(t)$]{Numerical solution for $R(t)$ (red/solid) and $\phi(t)$ (blue/dashed), given by Eq.~(\ref{RVEL}) and the numerical solution of Eq.~(\ref{eqphisq}), respectively, with $N = 60$ efolds, using the potential defined in Eq.~(\ref{VELambda}) and $m_\phi = 1$.}
\label{Rb050}
\end{figure}

One can follow the evolution of the system along Fig.~\ref{swallowtailLambda} (top panels): The system starts close to the origin and slowly moves along the first branch (close to the horizontal axis), generating an initially inflationary phase (since $R\approx {\rm const}$). It then quickly sweeps through the second branch (where $f''<0$) and then oscillates around the origin along the almost-linear third branch (where GR is recovered for $a=0$). 

Accordingly, in Fig.~\ref{Ufigs}, the system starts 
 on a stable (asymptotic) solution $\phi\to-\infty$ ($V\to+\infty$), but the slow roll  drives the field towards the origin. Eventually, it settles down at the minimum $\phi=a$ ($V=\exp(-\beta a)$).

The same behavior can be seen in Fig.~\ref{sbin}, as follows: The system  starts at $V\to\infty$, which is a stable configuration only if the temperature is above the binodal curve, i.e, either slightly below or above $T_c$ (thick black curve). On the other hand, if T is low enough (like the lowest gray curve, which corresponds to $T=0$), the system starts at a metastable phase (the binodal region) --- the initial inflationary solution is indeed momentary.
Either way, the effective fluid quickly crosses the spinodal curve (the unstable region) and then oscillates around 
$P=2 T \exp(2 \beta a)$ and $V=\exp(-\beta a)$, indicated by a gray circle for $T=0$ in Fig.~\ref{sbin}. At this temperature, the system ends exactly on the binodal curve. For higher temperatures, though, the system settles down above the binodal line, i.e, in a stable configuration.

Each description above explains the same evolution from a different point of view; each one uses a different  --- but equivalent --- fluid, as we shall see now. 
\subsection{Einstein Frame}
\begin{figure}[t]
\center
\includegraphics[width=0.8\textwidth]{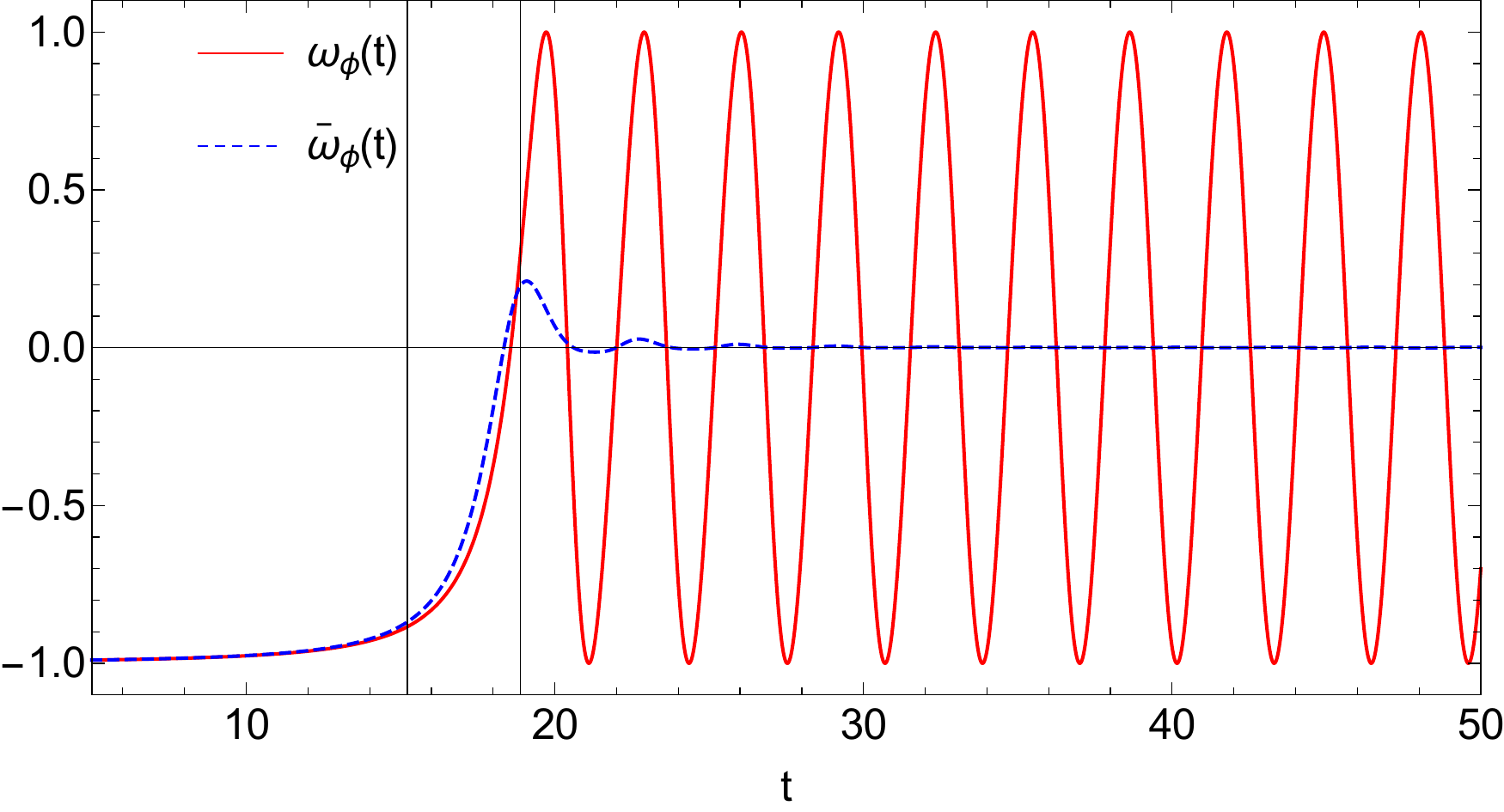}
\caption[Equation-of-state parameter in the EF.]{Equation-of-state parameter ($w_\phi$ and its time average $\bar{w}_\phi$) for the $\phi$ field, defined in the EF, as functions of time, for $\Lambda=0$. The vertical lines correspond to $t=t_1$ and $t=t_2$, when $R'(t_1)=R'(t_2)=0$, i.e, at the sideways peaks in Fig.~\ref{swallowtailLambda} (top panels).}
\label{wphi}
 \end{figure}

We plot in Fig.~\ref{wphi}, along each of such aforementioned periods, the corresponding equation-of-state parameter for the $\phi$ field (defined in the EF):
\begin{equation}
    w_\phi(t) \equiv \frac{p_\phi(t)}{\rho_\phi(t)} \equiv 
    \frac{\frac{1}{2}{\dot\phi}^2-V_E[\phi(t)]}{\frac{1}{2}{\dot\phi}^2+V_E[\phi(t)]},
\end{equation}
and its average over one period $T$ (defined in the final oscillatory phase).
%
There are clearly two distinct phases: the early inflationary period, characterized by $w_\phi \approx \bar{w}_\phi \approx -1$, and the dust-like phase, when $w_\phi$ oscillates between $\pm 1$ and $\bar{w}_\phi=0$, as for the traditional inflaton field in the JF \footnote{At some point, the inflaton field should couple to matter (which is absent in our model from the beginning) to start (p)reheating --- the study of such phase is beyond the scope of the present paper.}. The sideways peaks, at $t_{1,2}$, indicate the transition between the aforementioned phases. 

\subsection{Jordan Frame}
\begin{figure}[t]
\center
\includegraphics[width=0.45\textwidth]{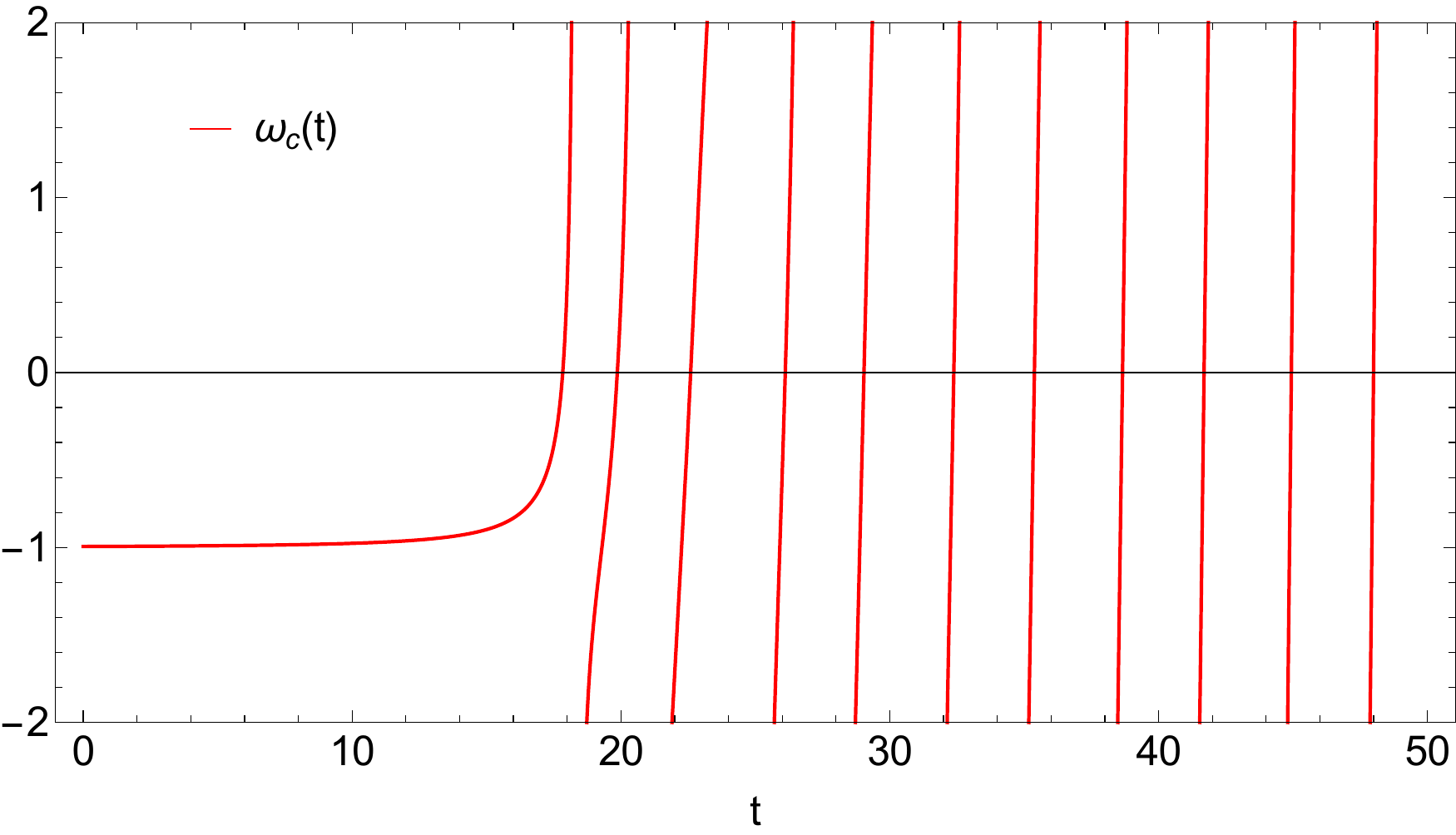}
\includegraphics[width=0.45\textwidth]{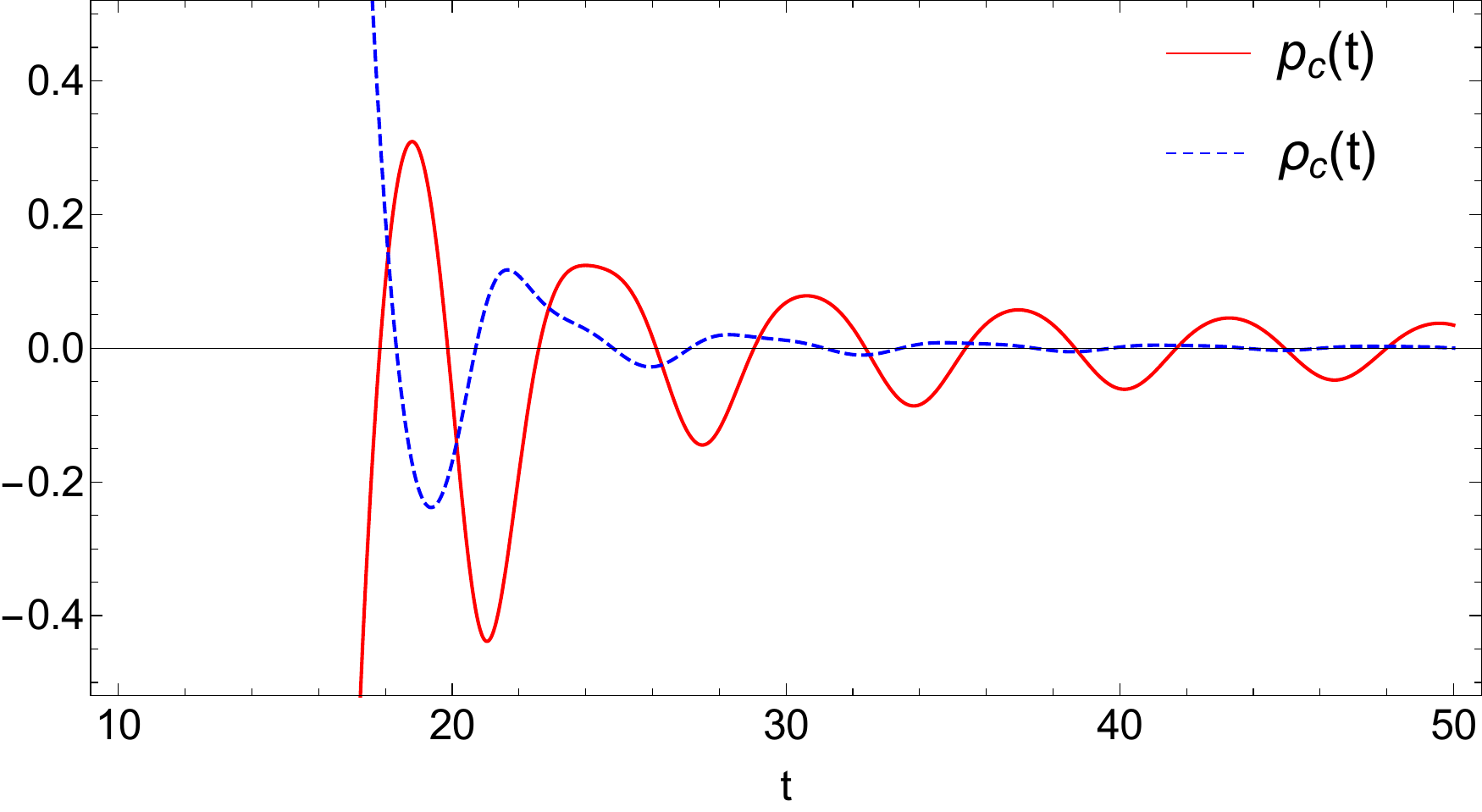}
\caption[Equation-of-state parameter in the JF.]{{\bf Left panel:} Equation-of-state parameter $\omega_c$ for the ``curvature fluid'' in the JF, for $\Lambda=0$. The divergences, all of them non-physical, correspond to $\rho_c=0$, which happens periodically while the field $\phi$ oscillates around the minimum of its potential $V_E(\phi)$. {\bf Right panel:} Corresponding pressure $p_c$ (red solid curve) and density $\rho_c$ (blue dashed line) for the  ``curvature fluid''. 
}
\label{wJF}
\end{figure}

There is a corresponding behavior in the JF, of course. One can define a ``curvature fluid'' whose energy density and pressure are defined as, respectively:
\begin{align}
8\pi G \rho _{c} &\equiv \left( f'R - f\right)/2 - 3H\dot{f'} + 3H^{2}(1-f') \label{rhocurv}\\
8\pi G p_{c}  &\equiv \ddot{f'} + 2H\dot{f'} - (2\dot{H}+3H^{2})(1-f') +  (f-f' R)/2 \label{pcurv}.
\end{align}
In Fig.~\ref{wJF} we plot the corresponding equation-of-state parameter  $\omega_c \equiv p_c/\rho_c$ (left panel), $\rho_c(t)$, $p_c(t)$ (right panel), all of them defined in the JF, for $\Lambda=0$. In the inflationary phase, the  curvature fluid behaves as a cosmological constant ($\omega_{c}\approx -1$), as expected, since it is responsible for the accelerated quasi-de Sitter expansion. In the oscillatory phase, on the other hand, the behaviour of $\omega_{c}$ is  not usual just because $\rho_{c}$ vanishes periodically, whenever $\phi(t)=0$ at the bottom of its potential $V_E(\phi)$ --- see Fig.~\ref{wJF}, right-hand panel. Nevertheless, there are no divergences of {\it physical} quantities. If $\Lambda\neq 0$, then $\omega_c = \omega_\phi = -1$ in the final stages, as expected. 

\section{The Stability Criteria from Thermodynamics Analogy}\label{ThermoAnalogySection}

The mere similarity between Fig.~\ref{swallow3D} and the Gibbs potential might be just a coincidence. Nevertheless, there is indeed a deeper connection: the whole system --- its equilibrium points, stability and evolution --- is determined by the Internal Energy $U$, the Gibbs potential $G$ and its critical points, as we will now see. 

For now, let us associate the Cosmological Constant $\Lambda$ to an effective temperature $T\equiv\Lambda$. On the other hand, we do not directly identify $G$ to $f$ and neither $P$ to $R$. We rather use an slightly more general {\it Ansatz}: we define a new pair of coordinates $\{-G,P\}$ as a rotation of the original one $\{f,R\}$:
\begin{equation}\label{ansatz}
\left(
\begin{tabular}{c}
$-G$ \\
$P$
\end{tabular}
\right)
\equiv 
\left(
\begin{tabular}{cc}
$\cos\theta$ &  $- \sin\theta$ \\
$\sin\theta$ & $\cos\theta$
\end{tabular}
\right)
\left(
\begin{tabular}{c}
$f$ \\
$R$
\end{tabular}
\right),
\end{equation}
%
%
%
%
which yields
\begin{align}
%
 G (\phi,T) &=
 e^{\beta  \phi } \sin (\theta ) \left(\frac{2 (\phi -a) (\beta  (\phi -a)+1)}{\beta }+4 T\right) + \nonumber \\
  &-e^{2 \beta  \phi } \cos (\theta ) \left(\frac{2 (\phi -a)}{\beta }+(\phi -a)^2+2 T\right),
 \label{Gphitheta}  \\
%
P (\phi,T) &=e^{2 \beta  \phi } \sin (\theta ) \left(\frac{2 (\phi -a)}{\beta }+(\phi -a)^2+2 T\right)+\nonumber \\
&+e^{\beta  \phi } \cos (\theta ) \left(\frac{2 (\phi -a) (\beta  (\phi -a)-1)}{\beta }+4 T\right).
\label{Pphitheta}
\end{align}

The effective volume $V$ is the variable ``canonically conjugated" to the effective pressure $P$, i.e, since
\begin{equation}
dG(P,T) = V \cdot  dP - S \cdot dT,
\label{dG}
\end{equation}
one can define an effective volume
\begin{equation}
V \equiv \left.\frac{\partial G}{\partial P}\right|_T = \left.\frac{\partial G/\partial\phi}{\partial P/\partial\phi}\right|_T = \frac{1-e^{\beta  \phi } \cot (\theta )}{e^{\beta  \phi }+\cot (\theta )},
\label{Vphitheta}
\end{equation}
which can be inverted and yield
\begin{equation}
\phi = \frac{1}{\beta}\log \left(\frac{1-V \cot (\theta )}{\cot (\theta )+V}\right).
\end{equation}

\begin{figure}[t]
\center
\includegraphics[width=0.8\textwidth]{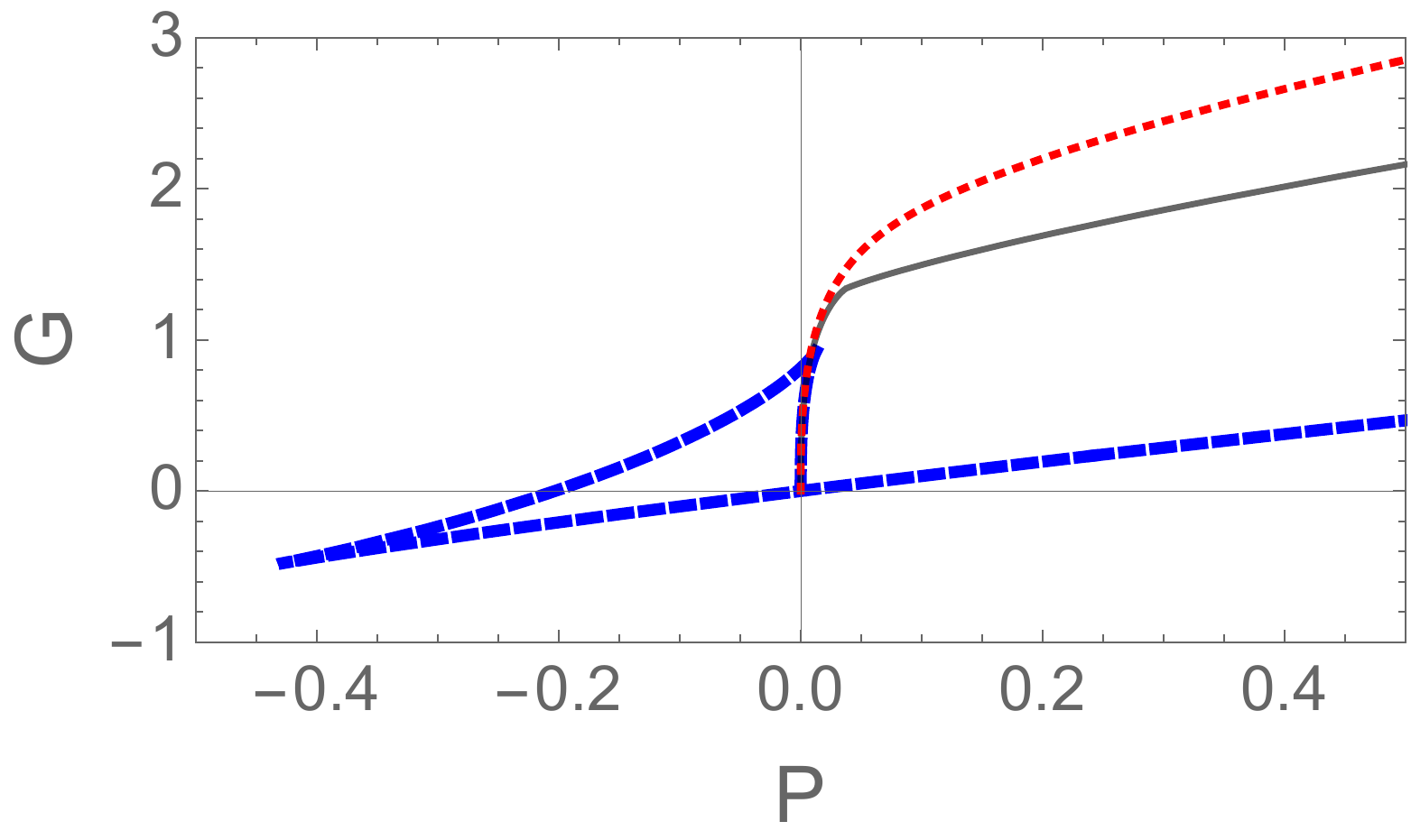}
\caption[Plot of the Gibbs Potential $G$ as a function of the pressure $P$]{Plot of the Gibbs Potential $G$ as a function of the pressure $P$, for $\beta=\sqrt{2/3}$, $\theta=\theta_*$ and $T=0$ (dashed blue), $T=T_c=15/16$ (solid black) and $T=1.5$ (dotted red).}
\label{Gphi_fig}
\end{figure}

In order to define the exact correspondence, i.e, the value of $\theta$, we only require that the volume is positive and unlimited from below.
Indeed, such procedure yields $\theta = \theta_* \equiv  \pi/2$ and simpler parametric expressions for the previously defined thermodynamic quantities: 
\begin{align}
G &= e^{\beta  \phi } \left(\frac{2 (\phi -a) [\beta  (\phi -a)+1]}{\beta }+4 T\right) \label{Gphi},\\
P &= e^{2 \beta  \phi } \left(\frac{2 (\phi -a)}{\beta }+(\phi -a)^2+2 T\right),
\label{Pphi}\\
V &= \exp(-\beta \phi) \quad \Leftrightarrow \quad \phi = -\frac{1}{\beta}\log(V),
\label{Vphi}
\end{align}
and the corresponding  plot in Fig.~\ref{swallow3D} (right panel). In Fig.~\ref{Gphi_fig}, we plot the curve $G(P)$ for different temperatures $T$, each one corresponding to a different vertical section of the previous 3D figure. 

One can also calculate the Helmholtz energy 
\begin{align}
F(T,V) &\equiv G - P \cdot V   = \frac{e^{2 \beta  \phi } \csc (\theta ) \big[(a-\phi )^2+2 T\big]}{e^{\beta  \phi }+\cot (\theta )} \\
& = \frac{1}{V}\bigg[\left(a+\frac{1}{\beta }\log V\right)^2+2 T\bigg] \qquad {\rm if} \quad \theta=\pi/2
\end{align}
from which one can define the entropy as
\begin{align}
S (T,V) &\equiv - \left.\frac{\partial F}{\partial T}\right|_V = -\frac{2 \sin (\theta ) (V \cot (\theta )-1)^2}{\cot (\theta )+V} \\
& = -\frac{2}{V},  \qquad {\rm if} \quad \theta=\pi/2.
\end{align}
One can then realize that the specific heat at constant volume vanishes, since $C_V \equiv T \cdot \partial S/\partial T|_V  = 0 \, \forall T$. Such feature is not unusual: it has been already found in studies of  thermodynamics and phase transitions of black holes \cite{Dolan_2011}.

\begin{figure}[t]
\center
\includegraphics[width=0.45\textwidth]{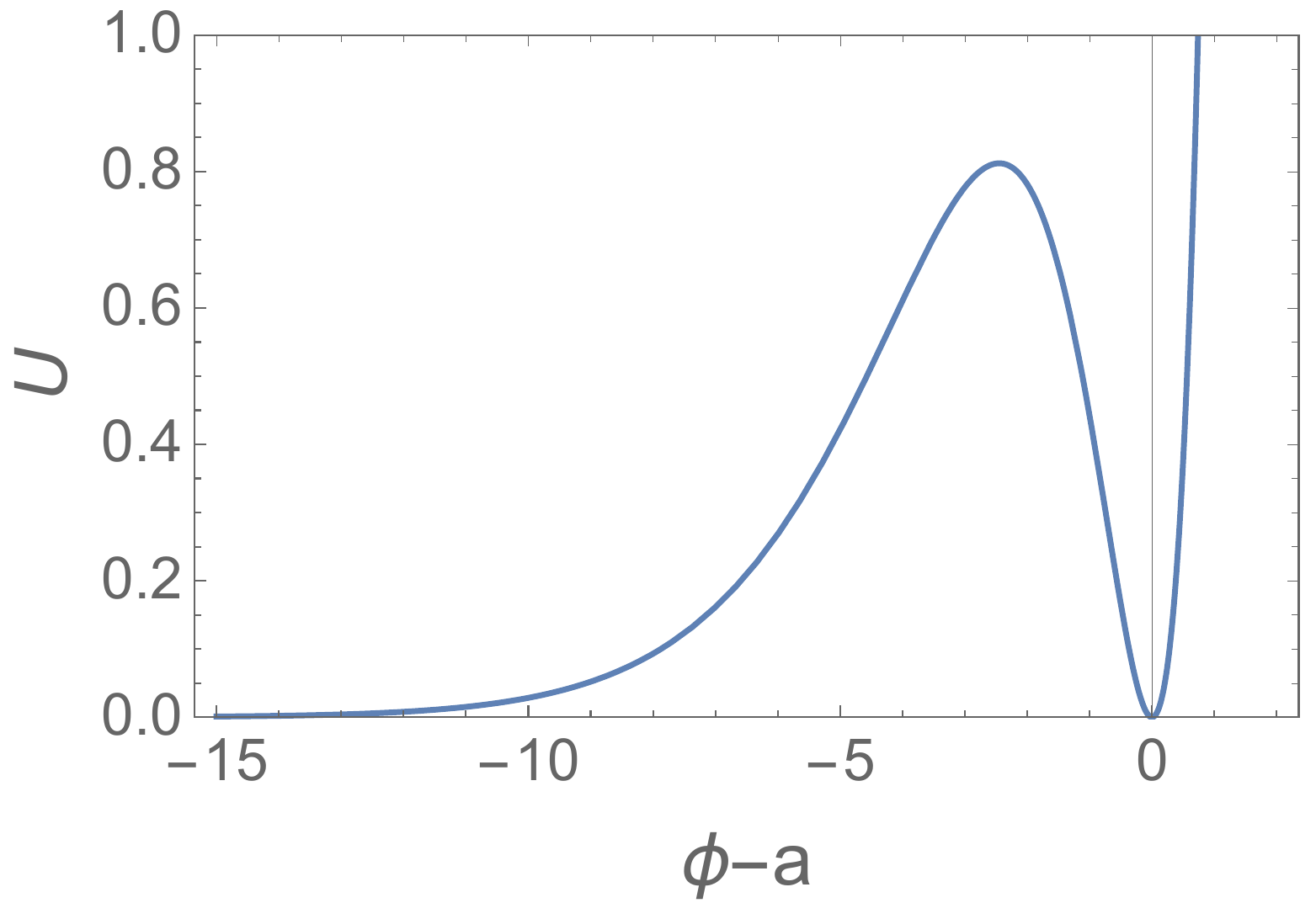}
\includegraphics[width=0.45\textwidth]{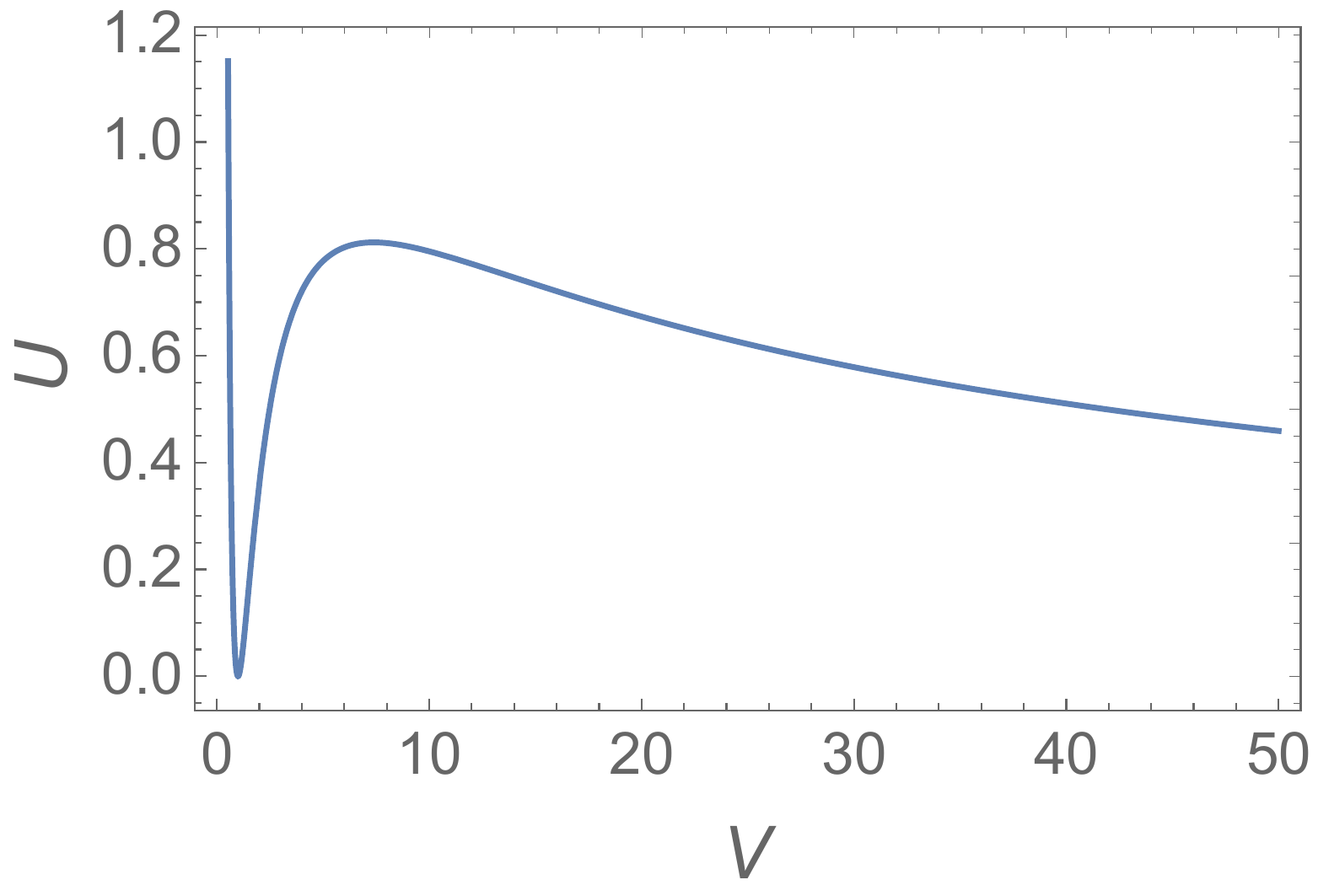}
\caption[Plot of the internal energy $U$]{Plot of the internal energy $U$ with $\theta=\theta_*=\pi/2$ as a function of $\phi-a$ (left panel) and of the volume $V$ with $a=0$ (right panel). We recall that $V = \exp(-\beta \phi)$. }
\label{Ufigs}
\end{figure}

The internal energy $U(T,V)$ is given by its standard definition:
\begin{align}
U &\equiv G - P \cdot V + T \cdot S = \frac{(a-\phi )^2 e^{2 \beta  \phi } \csc (\theta )}{e^{\beta  \phi }+\cot (\theta )}  \\
&= \frac{1}{V} \left(a+\frac{1}{\beta }\log V\right)^2, \qquad {\rm if} \quad \theta=\pi/2,
\end{align}
for which $\phi=a$ (accordingly, $V=\exp(-\beta a)$) is always a minimum. It turns out that also $U$ is only a function of the volume $V$ and {\it not} of the temperature $T$. One might acknowledge the existence of another two equilibrium points: an asymptotic one (a local minimum at $\phi\to-\infty$) and a local maximum (whose position depends on $\theta$) --- see Fig.~\ref{Ufigs}.

\begin{figure}[t]
\center
\includegraphics[width=0.7\textwidth]{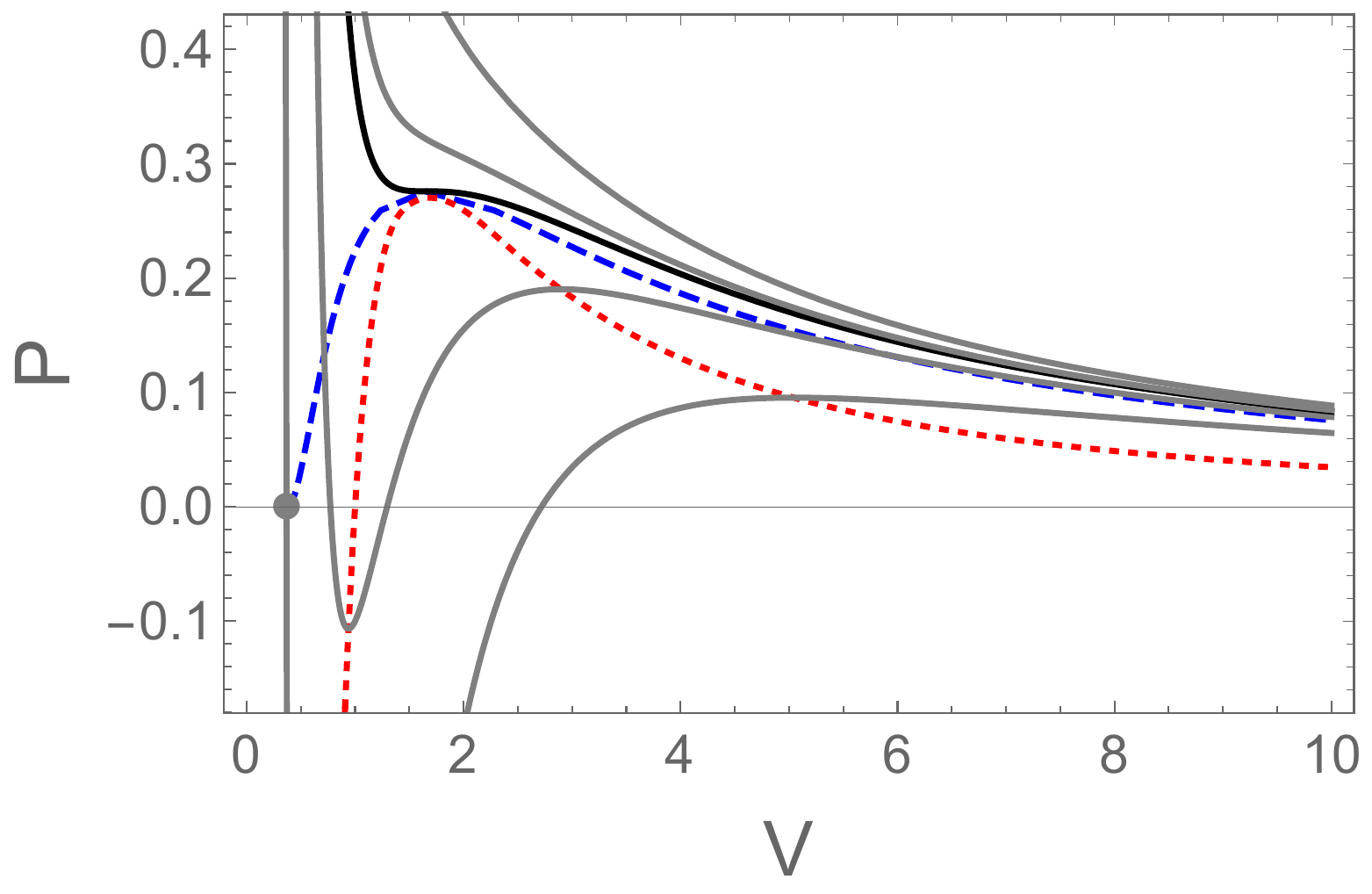}\\ 
\hspace{1cm}
\includegraphics[width=0.3\textwidth]{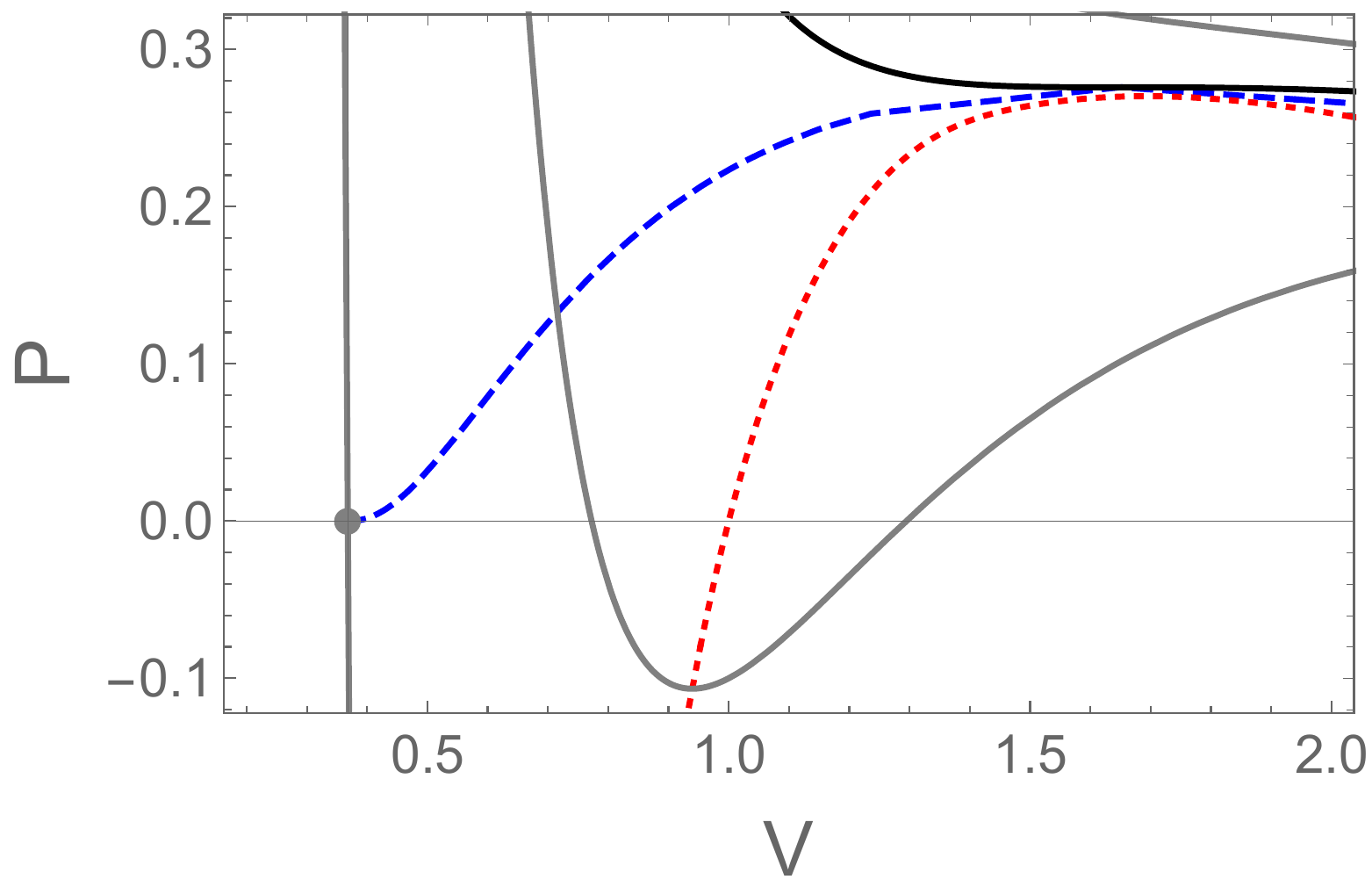}
\includegraphics[width=0.3\textwidth]{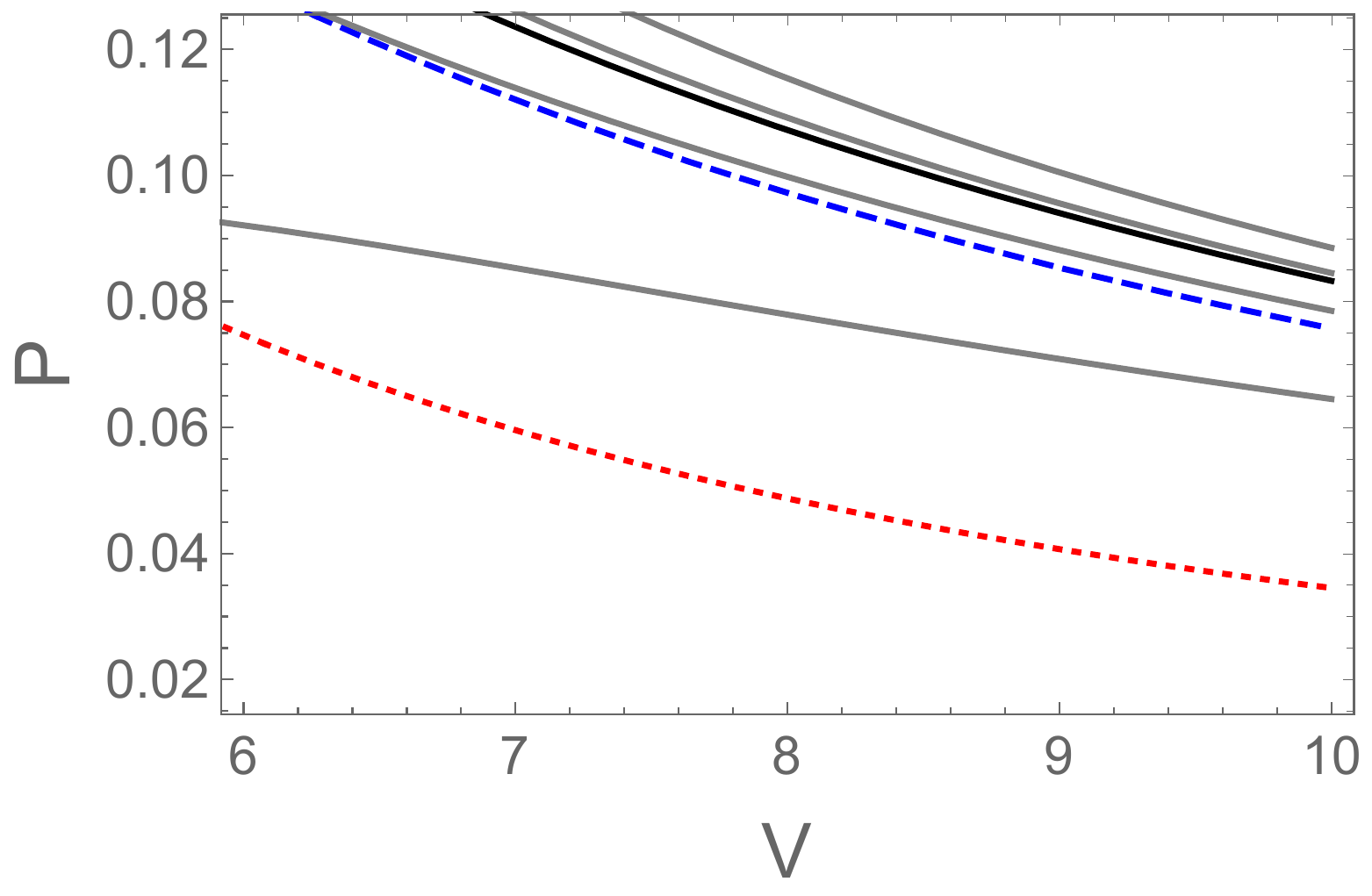}
\caption[Plot of the effective pressure $P$ as a function of the effective volume $V$]{Plot of the effective pressure $P$ as a function of the effective volume $V$, for $a=a_*\equiv 1/\beta$ and different values of temperature: $T=T_c\equiv 15/16$ (solid thick black); lower (higher) curves, in solid thin gray, correspond to lower (higher) temperatures. The {\bf spinodal curve} is plotted in dotted red. The {\bf binodal curve} is plotted in dashed blue. The gray circle indicates the final configuration ($\phi=a$) for the $T=0$ case (higher temperatures correspond to  higher final pressures). {\bf Lower panels:} zooming into the left and right-hand ends of the volume axis to show the behavior of the same curves. Note that the binodal curve does end at the gray circle.}
\label{sbin}
\end{figure}

 From now on, we shall always use $\theta=\theta_*=\pi/2$.
Equations~(\ref{Pphi}) and (\ref{Vphi}) yield the equation of state for our vdW-like  ``efective gas'', i.e, an expression that relates $P$, $V$ and $T$:
\begin{equation}
P = \frac{\beta  \left(a^2 \beta -2 a+2 \beta  T\right)+(2 a \beta -2 +\log V )\log V }{\beta ^2 V^2}.
\label{PV}
\end{equation}
%
The behaviour of $P(V)$ for four different values of $T$ is shown in Fig.~\ref{sbin}, which bears strong resemblance to a vdW gas\footnote{Nevertheless, here one obtains $P\propto T V^{-2}$ in the high-temperature limit, instead of the standard ideal-gas behavior $P \propto T V^{-1}$.}. Even though the equations of state are not exactly the same, they do describe the same phenomena, as we will now see. 

For instance, we can define the binodal and spinodal curves, that indicate, respectively, the regions of metastability and instability of the system --- see Fig.~\ref{sbin}. The former can be obtained using two equivalent calculations --- from the self-intersecting points of the Gibbs function and from the Maxwell construction --- supporting the results from each other. The latter curve is obtained from the extrema of the Gibbs function (see Fig.~\ref{swallowtail}), i.e, the first two turning points (extrema) of $R(t)$. The {\it critical point} $\{P_c,T_c,V_c\}$, defined at the crossing of those curves, indicates the end of the coexistence line. We will come back to those curves in the next section.

\begin{figure}[t]
\center
\includegraphics[width=0.8\textwidth]{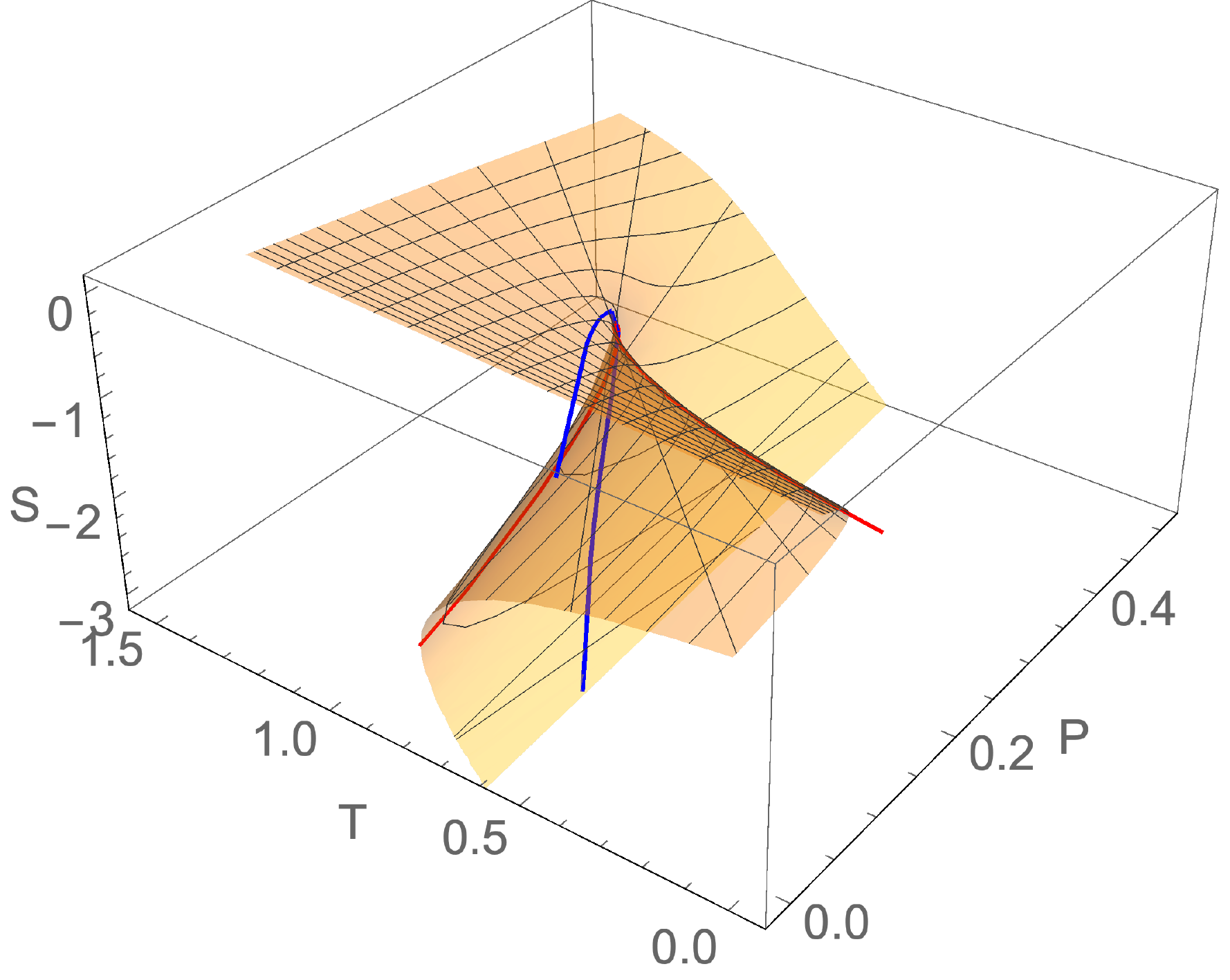}
\caption[Surface given by $S(P,T)$ ]{Surface given by $S(P,T)$ for $a=a_*$.  The spinodal and binodal curves are indicated in red (horizontal cusp shape) and blue (vertical ``$\subset$" shape), respectively. A turnable version is available at \url{https://tinyurl.com/wopt7fq}.}
\label{spt}
\end{figure}

The entropy as a function of pressure and temperature provides another very important piece of information.  $S(P,T)$ is depicted in Fig.~\ref{spt}, which also shows the spinodal and binodal curves.  The region where the entropy is multi-valued is known in Catastrophe Theory \cite{saunders1980introduction} as a cusp and indicates the existence of a first-order phase transition and unstable configurations.

\begin{figure}[t]
\center
\includegraphics[width=0.8\textwidth]{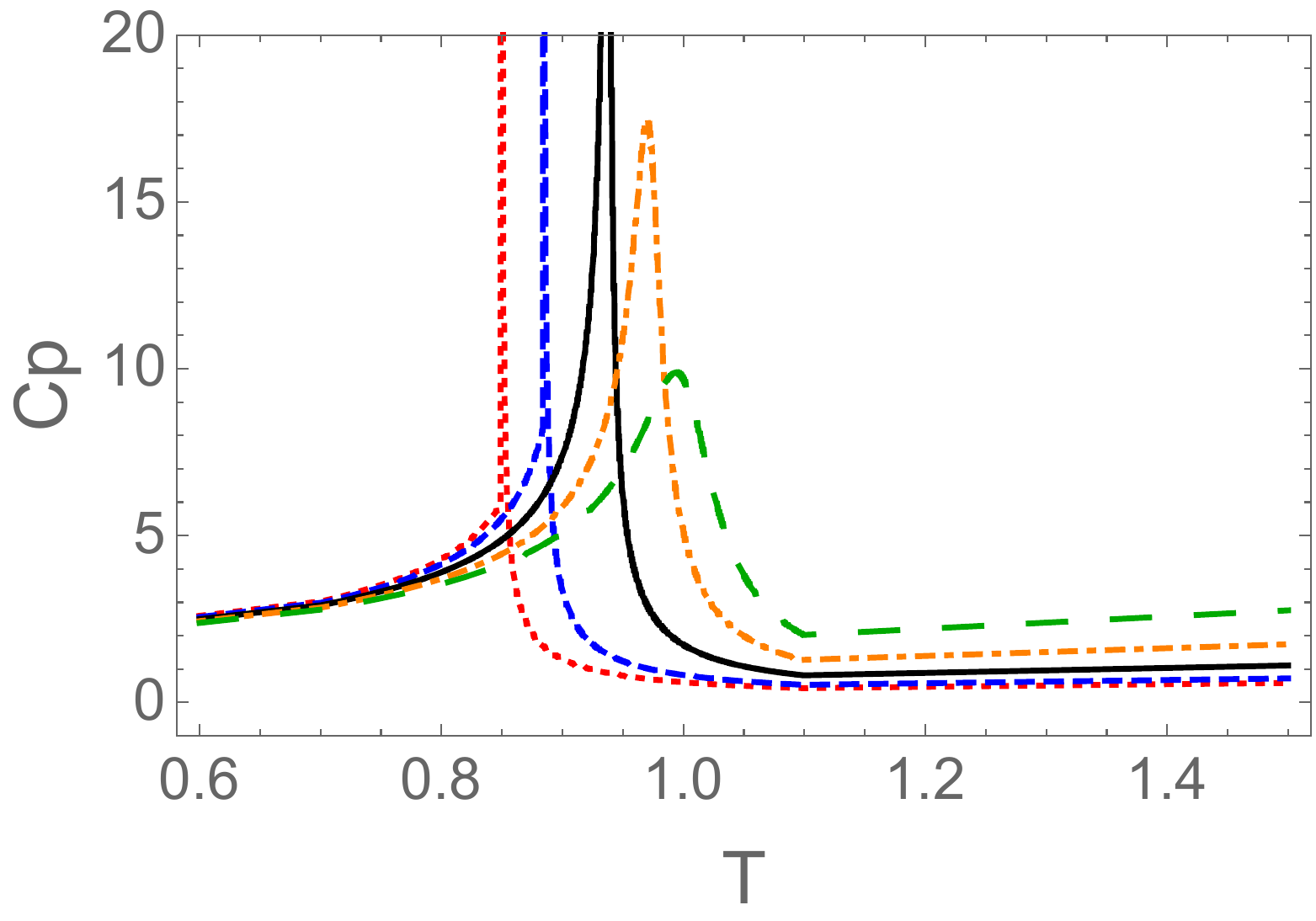}
\caption[Specific heat at constant pressure $C_P$]{Behavior of the specific heat at constant pressure $C_P$ as a function of the temperature $T$ close to its transition value ($T_c=15/16\approx 0.94$ if $P=P_c$), for different values of pressure (from left to right): $0.85 P_c$ (dotted red), $0.85 P_c$ (dashed blue), $P_c$ (solid black), $1.1 P_c$ (dot-dashed orange) and $1.2 P_c$ (long-dashed green). In all curves, $a=a_*$, for which $P_c\approx 1.51$.}
\label{Cp}
\end{figure}

From $S(P,T)$ we can get the specific heat at constant pressure, $ C_P \equiv T \cdot \partial S/\partial T |_P$, shown in Fig.~\ref{Cp}. We obtain the expected behavior for temperatures around the coexistence curve, for pressures both below (finite jump) and above (smooth behavior) the critical value $P_c$. We also obtain the usual divergence at the critical point $\{T_c,P_c\}$ (solid black line in Fig.~\ref{Cp}) as given by $C_P|_{P_c} \sim [(T_c-T)/T_c]^\alpha$, with $\alpha\approx 1.00$. 

\subsection{vdW fluid}

\begin{figure}[!t]
\center
\includegraphics[width=0.8\textwidth]{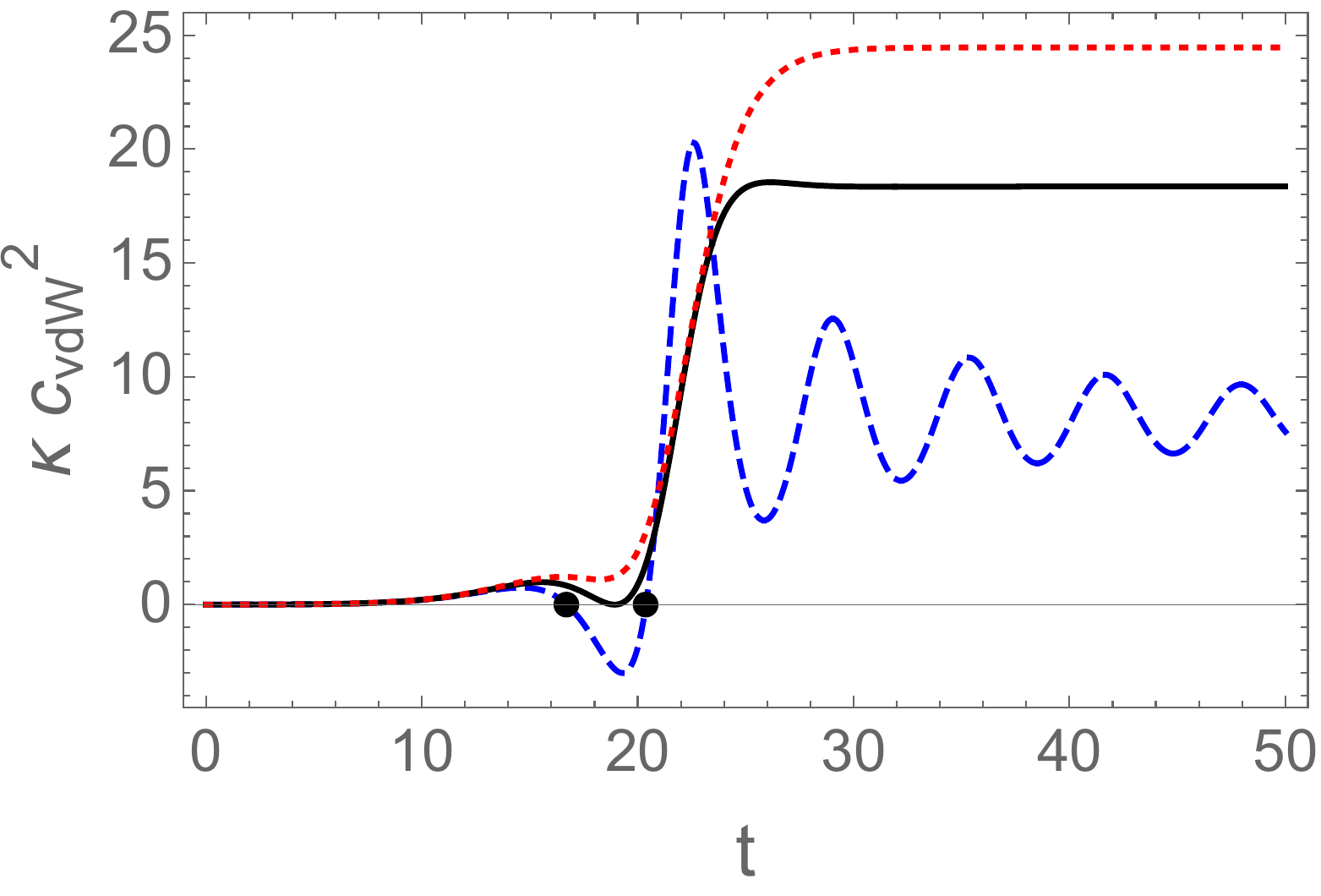}
\caption[Plot of $\kappa \cdot c_{\rm vdW}^2 \equiv \kappa \cdot \dot P/ \dot \rho$ for the effective vdW gas]{Plot of $\kappa \cdot c_{\rm vdW}^2 \equiv \kappa \cdot \dot P/ \dot \rho$ for the effective vdW gas, for $a=a_*$ and $T=0$ (dashed blue), $T=15/16$ (black) and $T=1.5$ (dotted red), as functions of time. The black dots indicate when $\dot R(t)=0$, i.e, at the sideway peaks in Fig.~\ref{swallowtail}, between which $f''(R)<0$.}
\label{cs2}
\end{figure}

Two important pieces of information are available only from the vdW gas and {\it not} from either the curvature fluid or the $\phi$ field. 

One of them is its sound speed squared, defined as $c_{\rm vdW}^2 \equiv \dot P/\dot \rho = -(V^2/\kappa) \dot P / \dot V$ (where we define $\kappa>0$ by $\rho =: \kappa /V$) and  plotted in Fig.~\ref{cs2}. We can see that $c_{\rm vdW}^2<0$ {\it only} between the first two extrema of $R(t)$, i.e, in the second branch (see Fig.~\ref{swallowtail}), when $f''<0$, as expected from the usual {\it perturbative} argument on stability of $f(R)$ theories \cite{Sotiriou:2008rp}.  Obviously, for $T>T_c$, the second branch is suppressed and one obtains $c_{\rm vdW}^2>0 \, \forall t$. 
With an imaginary sound speed, fluctuations grow exponentially fast, but, during the spinodal decomposition process, only a given range of wavelength do so \cite{CHOMAZ2004263}. This is similar to a feature that has already been proposed in the preheating scenario \cite{preheating}. Further details will be the subject of future work. 

Another important feature is the sudden change in the entropy, from $S(\phi\to-\infty)=0$ to $S(\phi=a)=-2 e^{2 \beta a}$, marking the release of latent heat, just as expected in an ordinary first-order phase transition, which has already been pointed out by the $C_P$ behavior, shown in the previous section. The relation with (p)reheating will also be the subject of future work.

\chapter*{Conclusions and Outlook}

Let us present briefly our main results and some prospects of this works.

In this thesis we investigate the Ricci scalar function $f(R)$ in the Jordan frame, for the vacuum case, obtained from a inflationary potential $V(\phi)$ in the Einstein frame. We used, in the beginning, a simpler $\phi^2$ potential as a toy model. We were faced with a wide selection of possibilities, questions and hypotheses when we saw the result of the $f(R)$ function and its particular form showed in fig.~\ref{swallowtail}, some of which we could prove and publish.

In our first analysis, we investigated the standard stability criteria of $f(R)$, based on the fact that this criteria did not properly describe the behavior of the stability during the phase of inflation, since the stability of any given configuration is determined by the signal of the squared-mass term of the corresponding potential. In this particular case, the mass and the standard $f''>0$ stability criteria only agree in the final stage, when the system oscillates around the global minimum $R=0$. The standard benchmark $f''>0$ does indicate stability only when the modified theory is close to GR. Although this condition is clearly stated at the derivation of this criterion --- see, for instance, Ref.~\cite{Sotiriou:2008rp} --- such requirement is usually taken for granted. We also stress that it is not sufficient for the Lagrangian to be a linear function of $R$, as in the first branch of the present model. The $m^2_J$ criterion, on the other hand, holds in any phase of the evolution.
 
The reader may point out that the system defined in the EF ---  GR with a scalar field with a standard $\phi^2$ potential --- is always stable and therefore only a singular mapping to JF could introduce an instability. Nevertheless, we recall that the initial inflationary phase is not permanent. It is this instability --- one may say transition --- from inflation to a power-law expansion that $m_J^2<0$ refers to.
 
In our second exploration of the stability system, we used Catastrophe Theory in order to analyze the dynamical character of the potential $V_J(R)$ over the time. We made an approximation using a $4^{\rm th}$-order polynomial (\ref{cusp}) known as {\it cusp catastrophe}, with which we were able to characterize the evolution of the system during its three different branches. We found that, the initial configuration is (mildly) unstable as the system slowly drifts away from the local maximum ($R\sim 0$) towards a local minimum at $R>0$, so the standard criterion for stability ($f''>0$) fails here because, in spite of the almost-linear Lagrangian, it cannot be considered a perturbation of GR, where $f'\equiv 1$ as opposed to $f'\ll 1$, here. We also point out that, this feature might be interpreted as a modified effective Gravitational Constant $\tilde G_N\equiv G_N/f'\gg G_N$. Then, in the second branch, the system is moving between two different stable configurations. Finally, the system reaches its furthest point away from equilibrium, and the standard criterion for stability $f''>0$ gives the expected answer on the stability of the system on this branch because the perturbative approach around GR does hold.

While this thesis was written, we became aware of Ref.~\cite{mielke2008inflation} whose authors also apply Catastrophe Theory to the problem discussed here. The results are, nevertheless, completely different. In particular, we characterize the existence of a {\it cusp} catastrophe, rather than a {\it swallowtail}. Although the shape of Fig.~\ref{swallowtail} is indeed similar to a swallowtail, the resemblance is only geometrical: the structure of the maxima and minima of the effective potential would have to change only at the edges, which does not happen here. 

In our third approach, we extended our study including two parameters a trivial shift ($a$) in the vacuum expectation value of the field and the Cosmological Constant $\Lambda$ in the potential of our toy model. This two ingredients were indispensable pieces to make the thermodynamical analogy with the van der Waals gas. Especially, we were able to recognize the closeness behaviour of $f(R)$ for different values of $\Lambda$ to the Gibbs potential $G$ for the vdW gas as a function of its temperature $T$ and its pressure $P$ as it is shown in Fig.~\ref{swallow3D}. In order to match the corresponding variables, we associated the Cosmological Constant $\Lambda$ to an effective temperature $T\equiv\Lambda$ and identify $G$ and $P$ with $f$ and $R$ defining a new pair of coordinates $\{-G,P\}$ as a rotation of the original one $\{f,R\}$ through the {\it Ansatz} (\ref{ansatz}). With this unexpected piece of information it was brought to light a third ``frame'', where the system is described by a vdW-like gas. The whole thermodynamics picture then follows: binodal and spinodal curves, phase transition, critical quantities (pressure, volume and temperature), entropy jumps, specific-heat divergence (and the corresponding critical exponent). It is the first time that we are able to distinguish the coexistence curve of two phases for this model and characterize the unstable, metastable and stable states. 

The toy model here presented is able to generate inflation in the early universe even if $\Lambda=0$, as expected from the standard $\phi^2$ potential in the EF. The mechanism in the JF, on the other hand, is a modification of GR: $f(R)\sim R^{2.2}$ --- similar to the already known Starobinsky's $R+R^2$ model  \cite{Starobinsky:2007hu}. 
 
We are currently investigating other physical consequences of the approach here presented, that may indicate that either the potential we assume is a simple toy model, as expected, or that there might be some compatibility with observable quantities from inflation, for instance. In particular, during the spinodal decomposition process, only a given range of wavelength is exponentially amplified \cite{CHOMAZ2004263}. A similar feature has already been proposed in the preheating scenario \cite{preheating}. 

We also recall that a non vanishing $a>0$ reduces the value of the effective Newton's constant and, at the same time, generates a large effective cosmological constant in the JF (see discussion in section \ref{inverseproblem}). In a more speculative note, we hypothesize that such a mechanism could be used to (almost) cancel out a bare $\Lambda_0$ in the JF, if $\Lambda<0$. Moreover, the cosmological constant in the EF is {\it not} a dynamical quantity in the present work, but it may become so if it is actually the vacuum energy of another field which happens to go through a phase transition of its own.

In any case, the inverse mapping from EF to JF and the phase transition still stand and may be a key feature in a more detailed model. We are currently examining other potentials $V_E(\phi)$ and further generalizations (see appendix \ref{AppendixC} and  \ref{AppendixD}) --- Indeed, non-trivial potentials have been investigated before \cite{mielke2008inflation} but with no mention to the thermodynamics we develop here.

\section*{Publication and Participation in Events}
The following publication have been derived from this thesis:

\begin{itemize}
    \item[$\cdot$] The thermodynamic approach for the stability criteria in $f(R)$ theories of gravity presented in chapter \ref{Chapter7} was published in the \textit{Journal of Cosmology and Astroparticle Physics} in June 2020 \cite{Peralta_2020}.
\end{itemize}

The main results derived from this thesis have been presented in the following events:

\begin{itemize}

\item[$\cdot$] CoCo2o2o: Cosmology in Colombia, Bogot\'a, Colombia\\
\textbf{Talk} - \textit{Thermodynamics of $f(R)$ theories of gravity.}\\
From September 23 to 25, 2020.

\item[$\cdot$] IWARA2020 Video Conference - 9th International Workshop on Astronomy and Relativistic Astrophysics, Mexico City, Mexico. \\ \textbf{Talk} - \textit{Thermodynamics of $f(R)$ theories of gravity.}\\
From September 6 to 12, 2020.

\item[$\cdot$] 2nd Workshop on Current Challenges in Cosmology. Centro de Convenciones Universidad Antonio Nari\~no, Bogot\'a, Colombia.\\ \textbf{Poster} - \textit{$f(R)$ Gravity: From Einstein to Jordan Frames.}\\
 From October 29 to November 2, 2018.	

\item[$\cdot$] BSCG XVII - Brazilian School of Cosmology and Gravitation, Centro Brasiliero de Pesquisas F\'isicas (CBPF), Rio de Janeiro, Brazil.\\ \textbf{Poster} and \textbf{Talk} - \textit{$f(R)$ Gravity: From Einstein to Jordan Frames.}\\
 From July 16 to July 21, 2018.
 
 \item[$\cdot$] XIII Workshop on New Physics in Space\\ International Center for Theorical Physics (ICTP), South American Institute for Fundamental Research (SAIRF), S\~ao Paulo, Brazil.\\ \textbf{Poster} - \textit{Inflation Driven by the Inflaton or by $f(R)$ Modified Gravity: Einstein versus Jordan Frames.}\\
From november 28 to december 1, 2016.	
\end{itemize}


\addtocontents{toc}{\vspace{2em}} 

\appendix 



\chapter{Term evaluation
\texorpdfstring{$g^{\alpha\beta}(\delta\Gamma^{\sigma}_{\alpha\beta}) - g^{\alpha\sigma} (\delta\Gamma^{\gamma}_{\alpha\gamma})$}{termeva}} 

\label{AppendixA} 

\lhead{Appendix A. \emph{Term evaluation
$g^{\alpha\beta}(\delta\Gamma^{\sigma}_{\alpha\beta}) - g^{\alpha\sigma} (\delta\Gamma^{\gamma}_{\alpha\gamma})$}} 


We calculated the evaluation of the term below
\begin{eqnarray}\label{Evaluation}
g^{\alpha\beta}(\delta\Gamma^{\sigma}_{\alpha\beta}) - g^{\alpha\sigma} (\delta\Gamma^{\gamma}_{\alpha\gamma}).
\end{eqnarray}
We know that
\begin{eqnarray} \label{varcristofelB}
\delta \Gamma^{\sigma}_{\alpha\beta} &=& \frac12 \delta g^{\sigma\gamma} \left[ \partial_{\alpha} g_{\gamma\beta} +\partial_{\beta}g_{\gamma\alpha} - \partial_{\gamma}g_{\alpha\beta} \right] + \frac12 g^{\sigma\gamma} \left[\partial_{\beta}\delta g_{\gamma\alpha} + \partial_{\alpha} \delta g_{\gamma\beta} - \partial_{\gamma} \delta g_{\alpha\beta} \right],
\end{eqnarray}
also
\begin{eqnarray}
\nabla_{\gamma} \delta g_{\alpha\beta} = \partial_{\gamma} \delta g_{\alpha\beta} - \Gamma^{\sigma}_{\gamma\alpha}\delta g_{\sigma\beta} - \Gamma^{\sigma}_{\gamma\beta}\delta g_{\alpha\sigma}.
\end{eqnarray}
Then, we replace the covariant derivatives in the equation (\ref{varcristofelB})
\begin{eqnarray}
\delta\Gamma^{\sigma}_{\alpha\beta} &=& \delta g^{\sigma\gamma}g_{\lambda\gamma}\Gamma^{\lambda}_{\alpha\beta}  + \frac12 \delta g^{\sigma\gamma} \Big[\nabla_{\alpha}\delta g_{\beta\gamma} + \Gamma^{\lambda}_{\alpha\beta} \delta g_{\lambda\gamma} + \Gamma^{\lambda}_{\alpha\gamma}\delta g_{\lambda\beta}\nonumber\\ 
 && + \left(\nabla_{\beta}\delta g_{\alpha\gamma} + \Gamma^{\lambda}_{\alpha\beta} \delta g_{\lambda\gamma} + \Gamma^{\lambda}_{\beta\gamma}\delta g_{\lambda\alpha} \right) - \left(\nabla_{\gamma}\delta g_{\alpha\beta} + \Gamma^{\lambda}_{\alpha\gamma} \delta g_{\lambda\beta} + \Gamma^{\lambda}_{\beta\gamma}\delta g_{\lambda\alpha}\right) \Big] \nonumber\\
&=& \delta g^{\sigma\gamma}g_{\lambda\gamma}\Gamma^{\lambda}_{\alpha\beta}  + \frac12 \delta g^{\sigma\gamma} \Big[\nabla_{\alpha}\delta g_{\beta\gamma} + \nabla_{\beta}\delta g_{\alpha\gamma} - \nabla_{\gamma}\delta g_{\alpha\beta} \Big] + g^{\sigma\gamma}\Gamma^{\lambda}_{\alpha\beta}\delta g_{\lambda\gamma},
\end{eqnarray}
then, we replace $\delta g_{\alpha\beta} = - g_{\alpha\mu}g_{\beta\nu}\delta g^{\mu\nu}$
\begin{eqnarray} \label{gammauno}
\delta\Gamma^{\sigma}_{\alpha\beta} &=&  \delta g^{\sigma\gamma}g_{\lambda\gamma}\Gamma^{\lambda}_{\alpha\beta} - \delta g^{\mu\nu} g^{\sigma\nu}g_{\gamma\mu}g_{\lambda\nu}\Gamma^{\lambda}_{\beta\alpha}\nonumber\\
    && + \frac12 g^{\sigma\gamma} \Big[ \nabla_{\alpha}\delta g_{\gamma\beta} +\nabla_{\beta}\delta g_{\gamma\alpha} - \nabla_{\gamma}\delta g_{\beta\alpha} \Big]\nonumber\\
&=& \underline{\delta g^{\sigma\gamma} g_{\lambda\gamma}\Gamma^{\lambda}_{\beta\alpha} } - \underline{\delta g^{\mu\nu}\delta^{\sigma}_{\mu}g_{\lambda\nu}\Gamma^{\lambda}_{\beta\alpha} } + \frac12 g^{\sigma\gamma} \Big[\nabla_{\beta}\delta g_{\gamma\alpha} + \nabla_{\alpha}\delta g_{\gamma\beta} - \nabla_{\gamma}\delta g_{\beta\alpha} \Big]\nonumber\\
&=&\frac12 g^{\sigma\gamma} \Big[\nabla_{\beta}\delta g_{\gamma\alpha} + \nabla_{\alpha}\delta g_{\gamma\beta} - \nabla_{\gamma}\delta g_{\beta\alpha} \Big].
\end{eqnarray}
Same for the term
\begin{eqnarray} \label{gammados}
\delta\Gamma^{\gamma}_{\alpha\gamma} &=& \frac12 \delta g^{\sigma\gamma} \left[\partial_{\alpha}g_{\sigma\gamma} + \partial_{\gamma} g_{\sigma\alpha} - \partial_{\sigma}g_{\alpha\gamma} \right] + \frac12 g^{\sigma\gamma} \left[\partial_{\alpha}\delta g_{\sigma\gamma} + \partial_{\gamma} \delta g_{\sigma\alpha} - \partial_{\sigma} \delta g_{\alpha\gamma} \right]\nonumber\\
&=& \delta g^{\sigma\gamma}g_{\sigma\gamma}\Gamma^{\lambda}_{\alpha\gamma} + \frac12 g^{\sigma\gamma} \Big[\nabla_{\alpha}\delta g_{\sigma\gamma} + \underline{\nabla_{\gamma}\delta g_{\sigma\alpha} } - \underline{\nabla_{\sigma}\delta g_{\alpha\gamma}} \Big] + g^{\sigma\gamma} \Gamma^{\lambda}_{\alpha\gamma} \delta g_{\lambda\sigma}\nonumber\\
&=& \delta g^{\sigma\gamma}g_{\sigma\gamma}\Gamma^{\lambda}_{\alpha\gamma} - g^{\sigma\gamma}g_{\lambda\mu}g_{\sigma\nu}\delta g^{\mu\nu}\Gamma^{\lambda}_{\alpha\gamma} + \frac12 g^{\sigma\gamma}\nabla_{\alpha}\delta g_{\sigma\gamma}\nonumber\\
&=& \frac12 g^{\sigma\gamma}\nabla_{\alpha}\delta g_{\sigma\gamma}.
\end{eqnarray}
We replace (\ref{gammauno}) and (\ref{gammados}) in (\ref{Evaluation}) and finally we get
\begin{eqnarray} \label{difgamma}
g^{\alpha\beta}(\delta\Gamma^{\sigma}_{\beta\alpha}) - g^{\alpha\sigma} (\delta\Gamma^{\gamma}_{\alpha\gamma}) &=& g^{\alpha\beta} \, \frac12 g^{\sigma\gamma} \Big[\nabla_{\beta}\delta g_{\gamma\alpha} + \nabla_{\alpha}\delta g_{\gamma\beta} - \nabla_{\gamma}\delta g_{\beta\alpha} \Big] - g^{\alpha\sigma}\, \frac12 g^{\sigma\gamma}\nabla_{\alpha}\delta g_{\sigma\gamma}\nonumber\\
&=& g_{\mu\nu}\nabla_{\sigma}\delta g^{\mu\nu} + \nabla_{\gamma}\delta g^{\sigma\gamma}.
\end{eqnarray}

\chapter{Conformal Transformation} 

\label{AppendixB} 

\lhead{Appendix B. \emph{Conformal Transformation}} 

We define the conformal transformation and the mapping from the quantities in one frame to their corresponding {\it Doppelg\"angers} in the other frame. We show the explicit calculations follow the arguments in \cite{Wald:1984rg}

\section{Previous Notions: Derivative Operators and Parallel Transport}

The \textit{derivate operator} $\nabla$ (sometimes called a \textit{covariant derivate}) on a manifold $M$ it is a mapping that takes each tensor field of the type $(k,l)$ to a tensor field of the type $(k,l+1)$ and satisfies the following five conditions:
\begin{itemize} 
\item[1.] Linearity: For all $A, B \in \mathcal{T}(k,l)$ and $a, b \in \mathbb{R}$,
\small
\begin{align}\label{B1}
\nabla_{\gamma} \left( \alpha A^{a_{1} \cdots a_{k}}\,_{b_{1}\cdots b_{l}} + \beta B^{a_{1} \cdots a_{k}}\,_{b_{1}\cdots b_{l}}\right ) =& \nonumber\\ &\, \alpha \nabla_{\gamma} \left (A^{a_{1} \cdots a_{k}}\,_{b_{1}\cdots b_{l}}\right ) + \beta \nabla_{\gamma}\left (B^{a_{1} \cdots a_{k}}\,_{b_{1}\cdots b_{l}}\right).
\end{align} 
\normalsize
\item[2.] The Leibnitz rule: For all $A \in \mathcal{T}(k,l), B \in \mathcal{T}(k',l')$,
\begin{align} \label{B2}
\nabla_{\gamma} \left(A^{a_{1} \cdots a_{k}}\,_{b_{1}\cdots b_{l}}B^{a_{1} \cdots a_{k}}\,_{b_{1}\cdots b_{l}}\right ) 
=& \nonumber\\ & \nabla_{\gamma} \left(A^{a_{1} \cdots a_{k}}\,_{b_{1}\cdots b_{l}}\right )B^{a_{1} \cdots a_{k}}\,_{b_{1}\cdots b_{l}} + \nonumber\\ 
& + A^{a_{1} \cdots a_{k}}\,_{b_{1}\cdots b_{l}}\nabla_{\gamma} \left(B^{a_{1} \cdots a_{k}}\,_{b_{1}\cdots b_{l}}\right ).
\end{align}
\item[3.] Commutativity with contraction: For all $A \in \mathcal{T}(k,l),$
\begin{equation} \label{B3}
\nabla_{\gamma} \left(A^{a_{1} \cdots c \cdots a_{k}}\,_{b_{1}\cdots c \cdots b_{l}}\right ) = \nabla_{\gamma} A^{a_{1} \cdots a_{k}}\,_{b_{1}\cdots b_{l}} \,.
\end{equation}
\item[4.] Consistency with the notion of tangent vectors as directional derivatives on scalar fields: For all $f \in \mathcal{\lambda}$ and for all $t^{\alpha} \in V_{p}$,
\begin{equation} \label{B4}
t(f) = t^{\alpha}\nabla_{\alpha} .
\end{equation}
\item[5.] Torsion free: For all $f \in \mathcal{\lambda}$.
\begin{equation} \label{B5}
\nabla_{\alpha}\nabla_{\beta} f = \nabla_{\beta}\nabla_{\alpha} f.
\end{equation}
\end{itemize}

As is shown in \cite{Wald:1984rg} $\hat{\nabla}_{\alpha} - \nabla_{\alpha}$ (where $\hat \nabla$ is the derivative operator associated with a different choice of coordinate system.) defines a map of dual vectors at the point $p$ (as opposed to dual vector fields defined in a neighborhood of $p$) to tensors of type $(0,2)$ at $p$. By property (1.), this map is linear. Consequently $\hat{\nabla}_{\alpha} - \nabla_{\alpha}$ defines a tensor of a type (1,2) at $p$, which is denoted as $C^{\gamma}_{\alpha\beta}$,
\begin{equation} \label{B10}
\nabla_{\alpha} w_{\beta} = \hat{\nabla}_{\alpha}w_{\beta} - C^{\gamma}_{\alpha\beta}w_{\beta}.
\end{equation}
This displays the possible disagreements of the actions of $\nabla_{\alpha}$ and $\hat{\nabla}_{\alpha}$ on dual vectors fields. A symmetry property of $C^{\gamma}_{\alpha\beta}$ follows immediately from condition (5.). If we let $w_{\beta} = \nabla_{\beta}f = \hat{\nabla}_{\beta}f$, we find
\begin{equation}\label{divf}
    \nabla_{\alpha} \nabla_{\beta} f = \hat{\nabla}_{\alpha}\hat{\nabla}_{\beta}f - C^{\gamma}_{\alpha\beta}\hat{\nabla}_{\gamma}f.
\end{equation}
Since both $\nabla_{\alpha} \nabla_{\beta} f$ and $\hat{\nabla}_{\alpha}\hat{\nabla}_{\beta}$ are symmetric in $\alpha$ and $\beta$, it follows that $C^{\gamma}_{\alpha\beta}$ must be also have this property $C^{\gamma}_{\alpha\beta} = C^{\gamma}_{\beta\alpha}$. 

Given a derivative operator $\nabla_{\alpha}$ we can define the notion of parallel transport of a vector along a curve $C$ with a tangent $t^{\alpha}$. A vector $v^{\alpha}$ given at each point on the curve is said to be \textit{parallelly transported} (as one moves along) the curve if the equation
\begin{equation} \label{B11}
t^{\alpha}\nabla_{\alpha} v^{\beta} = 0,
\end{equation}
is satisfied along the curve. Given two vectors $v^{\alpha}$ and $w^{\alpha}$, we demand that their inner product $g_{\alpha\beta}v^{\alpha}w^{\beta}$ remain unchanged if we parallel-transport them along any curve. Thus we require
\begin{equation} \label{B12}
t^{\alpha}\nabla_{\alpha}(g_{\beta\gamma}v^{\beta}w^{\gamma}) = 0,
\end{equation}
for $v^{\beta}$ and $w^{\gamma}$ satisfying equation (\ref{B11}). We use the Leibniz rule and we obtain
\begin{equation} \label{B13}
t^{\alpha}v^{\beta}w^{\gamma}\nabla_{\alpha} g_{\beta\gamma} = 0.
\end{equation}
Equation (\ref{B13}) will hold for all curves and parallelly transported vectors if and only if
\begin{equation} \label{B14}
\nabla_{\alpha} g_{\beta\gamma} = 0,
\end{equation}
which is the additional condition imposed on $\nabla_{\alpha}$. That this equation  uniquely determines $\nabla_{\alpha}$ is shown by the following theorem
\begin{theorem}
Let $g_{\alpha\beta}$ a metric. Then there exists a unique derivative operator $\nabla_{\alpha}$ satisfying $\nabla_{\alpha} g_{\beta\gamma} = 0$.
\end{theorem}
\textit{Prof}. Let $\hat{\nabla}_{\alpha}$ any derivative operator, e.g., an ordinary derived operator associated with a coordinate system. we attempt to solve for $C^{\alpha}_{\beta\gamma}$ so that the derivative operator determined by $C^{\alpha}_{\beta\gamma}$  will satisfy the required property. We will prove this theorem by showing that a unique solution for $C^{\alpha}_{\beta\gamma}$ exists. we have
\begin{equation} \label{B15}
0 = \nabla_{\alpha} g_{\beta\gamma} = \hat{\nabla}_{\alpha}g_{\beta\gamma} - C^{\delta}_{\alpha\beta}g_{\delta\gamma} - C^{\delta}_{\alpha\gamma}g_{\beta\delta},
\end{equation}
that is,
\begin{equation} \label{B16}
 C_{\gamma\alpha\beta} + C_{\beta\alpha\gamma} = \hat{\nabla}_{\alpha}g_{\beta\gamma}.
\end{equation}
By index substitution, we also have
\begin{align}
C_{\gamma\beta\alpha} + C_{\alpha\beta\gamma} =&\, \hat{\nabla}_{\beta}g_{\alpha\gamma}, \label{B17}\\ 
C_{\beta\gamma\alpha} + C_{\alpha\gamma\beta} =&\, \hat{\nabla}_{\gamma}g_{\alpha\beta}. \label{B18}
\end{align}
We added equations (\ref{B16}), (\ref{B17}), and then sustract equation (\ref{B18}). Using the symmetry property  of $C^{\alpha}_{\beta\gamma}$, we find
\begin{equation} \label{B19}
2 C_{\gamma\beta\alpha} = \hat{\nabla}_{\alpha}g_{\beta\gamma} + \hat{\nabla}_{\beta}g_{\alpha\gamma} - \hat{\nabla}_{\gamma}g_{\alpha\beta},
\end{equation}
that is,
\begin{equation} \label{B20}
C^{\gamma}_{\alpha\beta} = \frac12g^{\gamma\delta}\left \{\hat{\nabla}_{\alpha}g_{\beta\gamma} + \hat{\nabla}_{\beta}g_{\alpha\gamma} - \hat{\nabla}_{\gamma}g_{\alpha\beta}\right \}.
\end{equation}
This choice of $C^{\gamma}_{\alpha\beta}$ solves equation (\ref{B14}) and it's unique, which completes the proof \cite{Wald:1984rg}.
\section{Conformal Transformation}
Be $M$ a n-dimensional manifold with metric $g_{\alpha\beta}$. It is possible to rescale the metric tensor through
\begin{equation} \label{B21}
\hat{g}_{\alpha\beta} \equiv \Omega^{2}(x^\alpha) g_{\alpha\beta},
\end{equation}
where $\Omega$ is a smooth, strictly positive function. The metric $\hat{g}_{\alpha\beta}$ is said to arise from $g_{\alpha\beta}$ via a \textit{conformal transformation}. The length of timelike and spacelike intervals and the norm of timelike and spacelike vectors change because the rescaling, but the null vectors and null intervals of the metric $g_{\alpha\beta}$ remain null in $\hat{g}_{\alpha\beta}$. Therefore, the spacetime $(M, g_{\alpha\beta})$ and $(M, \hat{g}_{\alpha\beta})$ have the same causal structure (this is obviously true). We also have $\hat{g}^{\alpha\beta} = \Omega^{-2}g^{\alpha\beta}$, since then $\hat{g}^{\alpha\beta}\hat{g}_{\beta\gamma} = g^{\alpha\beta}g_{\beta\gamma} = \delta^{\alpha}_{\gamma}$. 
\begin{figure}
\begin{center}
\includegraphics[width=\textwidth]{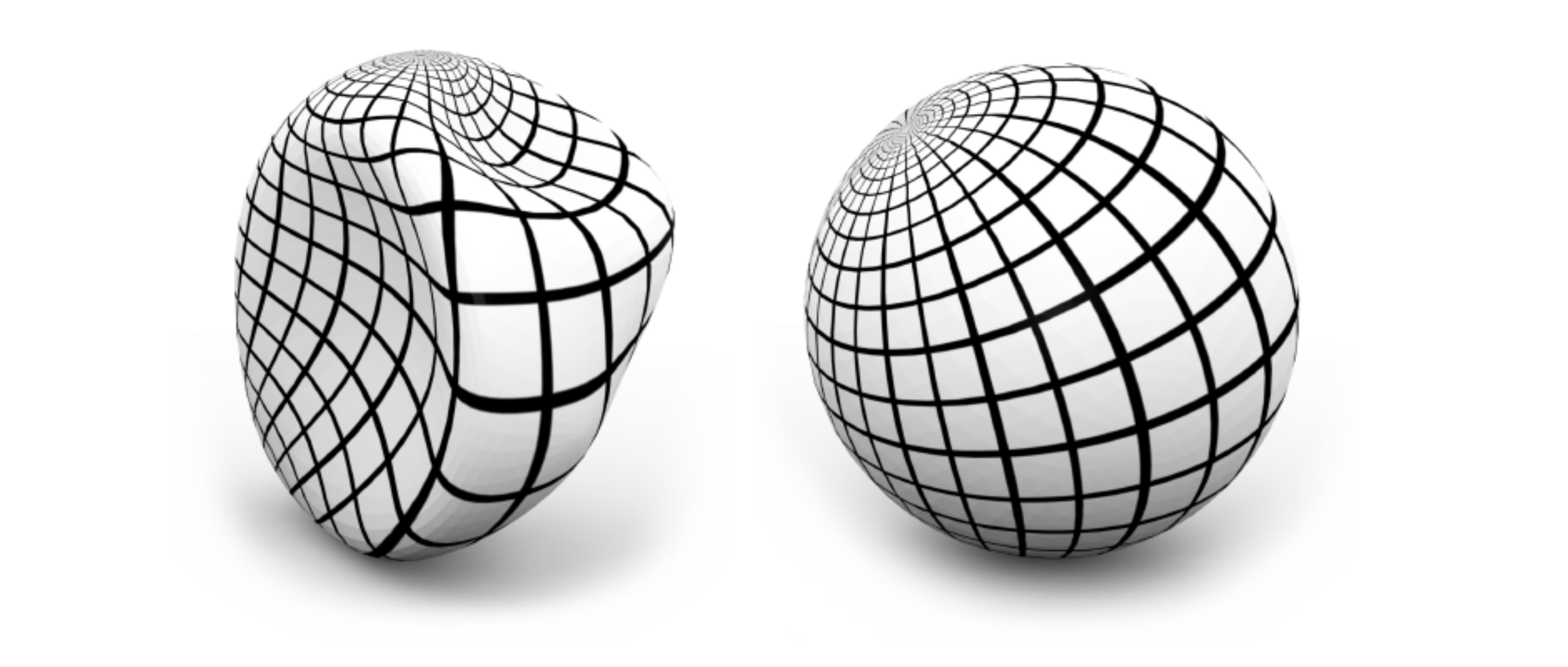}
\end{center}
\caption[Conformal map between two closed surfaces.]{Conformal map between two closed surfaces. Credits: \textsuperscript{\textcopyright} DGD - Discretization in Geometry and Dynamics - SFB Transregion 109.}
\label{conformalmap}
\end{figure}

Let $\nabla_{\alpha}$ denote the derivative operator associated with $g_{\alpha\beta}$, and let $\hat{\nabla}_{\alpha}$ denote the derivative operator associated with $\hat{g}_{\alpha\beta}$. The relation between $\hat{\nabla}_{\alpha}$ and $\nabla_{\alpha}$ is given by equation (\ref{B10}). Reversing the roles of $\nabla_{\alpha}$  and $\hat{\nabla}_{\alpha}$:
\begin{equation}\label{B22}
\hat{\nabla}_{\alpha}w_{\beta} =  \nabla_{\alpha} w_{\beta} - C^{\gamma}_{\alpha\beta}w_{\gamma},
\end{equation}
where
\begin{equation} \label{B23}
C^{\gamma}_{\alpha\beta} = \frac12 \hat{g}^{\gamma\delta}\bigg[ \nabla_{\alpha}\hat{g}_{\beta\delta} +  \nabla_{\beta}\hat{g}_{\alpha\delta} -  \nabla_{\delta}\hat{g}_{\alpha\beta} \bigg].
\end{equation}
However, since $\nabla_{\alpha} g_{\beta\gamma} =0$, we have
\begin{equation} \label{B24}
 \nabla_{\alpha} \hat{g}_{\beta\gamma} =  \nabla_{\alpha}\left (\Omega^{2}g_{\beta\gamma}\right ) = 2\Omega g_{\beta\gamma} \nabla_{\alpha} \Omega.
\end{equation}
Hence, we obtain
\begin{align} \label{B25}
C^{\gamma}_{\alpha\beta} =&\, \Omega^{-1}g^{\gamma\delta}\left \{g_{\beta\delta} \nabla_{\alpha}\Omega + g_{\alpha\delta} \nabla_{\beta}\Omega - g_{\alpha\delta} \nabla_{\delta}\Omega \right \}\nonumber\\
		 =&\, \delta^{\gamma}_{\beta}  \nabla_{\alpha}\ln\Omega + \delta^{\alpha}_{\gamma} \nabla_{\beta}\ln\Omega - g^{\gamma\delta}g_{\alpha\beta} \nabla_{\delta}\ln\Omega\nonumber\\
		 =&\, 2\delta^{\gamma}\,_{(\alpha}\nabla_{\beta)} \ln \Omega - g^{\gamma\delta}g_{\alpha\beta} \nabla_{\delta} \ln \Omega. 
\end{align}
The subindice notation for totally simmetric and totally antisymmetic parts of tensors is defined as\cite{Wald:1984rg}: $T_{(\alpha\beta)} = \frac12 \left( T_{\alpha\beta} + T_{\beta\alpha} \right) $, and $T_{[\alpha\beta]} = \frac12 \left( T_{\alpha\beta} - T_{\beta\alpha} \right) $

The curvature $R_{\alpha\beta\gamma}\,^{\delta}$ associated to $ \nabla_{\alpha}$ is given by
\begin{equation} \label{B26}
{R_{\alpha\beta\gamma}}^{\delta} \, w_{\delta} \equiv \nabla_{\alpha} \nabla_{\beta} w_{\gamma} -  \nabla_{\beta} \nabla_{\alpha} w_{\gamma}.
\end{equation}
Similarly, the curvature $\hat{R}_{\alpha\beta\gamma}\,^{\delta}$ associated to $\hat{\nabla}_{\alpha}$ is given by
\begin{equation}
{\hat R_{\alpha\beta\gamma}}\,^d\,\, w_{\delta} = \hat{\nabla}_{\alpha}\hat{\nabla}_{\beta} w_{\gamma} - \hat{\nabla}_{\beta}\hat{\nabla}_{\alpha} w_{\gamma}.
\end{equation}
Replacing $\hat{\nabla}_{\alpha}$ for its equivalent in the equation (\ref{B22}), we obtain
\begin{align*}
{\hat{R}_{\alpha\beta\gamma}}\,^{\delta}\,\,  w_{\delta} =&\, \hat{\nabla}_{\alpha}\hat{\nabla}_{\beta} w_{\gamma} - \hat{\nabla}_{\beta}\hat{\nabla}_{\alpha} w_{\gamma} \\
=&\, \hat{\nabla}_{\alpha}\left (\nabla_{\beta} w_{\gamma} - C^{\delta}_{\beta\gamma}w_{\delta}\right) - \hat{\nabla}_{\beta}\left ( \nabla_{\alpha} w_{\gamma} - C^{\delta}_{\alpha\gamma}w_{\delta}\right),\nonumber\\
=&\,  \nabla_{\alpha} \left (\nabla_{\beta} w_{\gamma}  - C^{\delta}_{\beta\gamma}w_{\delta}\right) -  C^{\epsilon}_{\alpha\gamma}\left ( \nabla_{\beta} w_{\epsilon}  - C^{\delta}_{\beta\epsilon}w_{\delta}\right ) - 		C^{\epsilon}_{\alpha\beta}\left ( \nabla_{\epsilon} w_{\gamma}  - C^{\delta}_{\epsilon\gamma}w_{\delta}\right ) - \\
&\, - \nabla_{\beta} \left ( \nabla_{\alpha} w_{\gamma}  - C^{\delta}_{\alpha\gamma}w_{\delta}\right ) +  C^{\epsilon}_{\beta\gamma}\left ( \nabla_{\alpha} w_{\epsilon}  - C^{\delta}_{\alpha\epsilon}w_{\delta}\right ) + 						 C^{\epsilon}_{\alpha\beta}\left ( \nabla_{\epsilon} w_{\gamma}  - C^{\delta}_{\epsilon\gamma}w_{\delta}\right )\\
=&\, \nabla_{\alpha} \nabla_{\beta} w_{\gamma} -  \nabla_{\beta} \nabla_{\alpha} w_{\gamma} -  \nabla_{\alpha}\left (C^{\delta}_{\beta\gamma} w_{\delta}\right ) + \nabla_{\beta}\left (C^{\delta}_{\alpha\gamma} w_{\delta}\right)-\\
	&\, - C^{\epsilon}_{\alpha\gamma} \nabla_{\beta} w_{\epsilon} +  C^{\epsilon}_{\beta\gamma}  \nabla_{\alpha} w_{\epsilon} +  C^{\epsilon}_{\alpha\gamma} C^{\delta}_{\beta\epsilon} w_{\delta} -  C^{\epsilon}_{\beta\gamma}C^{\delta}_{\alpha\epsilon}w_{\delta}, \\
=&\, R_{\alpha\beta\gamma}\,^{\delta} w_{\delta} - C^{\delta}_{\beta\gamma} \nabla_{\alpha} w_{\delta} - \left (\nabla_{\alpha} C^{\delta}_{\beta\gamma}\right)w_{\delta} + 										C^{\delta}_{\alpha\gamma}\nabla_{\beta} w_{\delta} + \left (\nabla_{\beta} C^{\delta}_{\alpha\gamma}\right)w_{\delta} - \\
&\, - C^{\epsilon}_{\alpha\gamma}\nabla_{\beta}w_{\epsilon} + C^{\epsilon}_{\beta\gamma}\nabla_{\alpha} w_{\epsilon} + \left(C^{\epsilon}_{\alpha\gamma} C^{\delta}_{\beta\epsilon} - C^{\epsilon}_{\beta\gamma}C^{\delta}_{\alpha\epsilon}\right)w_{\delta}\\
=&\, R_{\alpha\beta\gamma}\,^{\delta} w_{\delta} - \left (2\nabla_{[\alpha} C^{\delta}\,_{\beta]\gamma} + 2 C^{\epsilon}\,_{\gamma[\alpha} C^{\delta}\,_{\beta]\epsilon} \right )w_{\delta}.
\end{align*}

Thus, the relation between the curvature, $\hat{R}_{\alpha\beta\gamma}\,^{\delta}$, associated to $\hat{\nabla}_{\alpha}$ and the curvature $R_{\alpha\beta\gamma}\,^{\delta}$ associated to $\nabla_{\alpha}$ It is given by the following expression
\begin{equation}
\hat{R}_{\alpha\beta\gamma}\,^{\delta} = R_{\alpha\beta\gamma}\,^{\delta} - 2\nabla_{[\alpha} C^{\delta}\,_{\beta]\gamma} + 2 C^{\epsilon}\,_{\gamma[\alpha} C^{\delta}\,_{\beta]\epsilon}.
\end{equation}
Therefore, using the formula (\ref{B25}) for $C^{\gamma}_{\alpha\beta}$, we find
\small
\begin{align*}
\hat{R}_{\alpha\beta\gamma}\,^{\delta} =&\, R_{\alpha\beta\gamma}\,^{\delta} - 2\nabla_{[\alpha} C^{\delta}\,_{\beta]\gamma} + 2 C^{\epsilon}\,_{\gamma[\alpha} C^{\delta}\,_{\beta]\epsilon} \\
=&\, R_{\alpha\beta\gamma}\,^{\delta} - \nabla_{\alpha} C^{\delta}_{\beta\gamma} +  \nabla_{\beta} C^{\delta}_{\alpha\gamma} +  C^{\epsilon}_{\alpha\gamma} C^{\delta}_{\beta\epsilon} -  C^{\epsilon}_{\beta\gamma} C^{\delta}_{\alpha\epsilon}\\
=&\, R_{\alpha\beta\gamma}\,^{\delta} - \nabla_{\alpha}\left\{\delta^{\delta}_{\beta} \nabla_{\gamma}\ln\Omega + \delta^{\delta}_{\gamma} \nabla_{\beta}\ln\Omega - g^{\delta\lambda}g_{\beta\gamma}\nabla_{\lambda}\ln\Omega\right\} + \\
&\,  +  \nabla_{\beta}\left \{\delta^{\delta}_{\alpha} \nabla_{\gamma}\ln\Omega + \delta^{\delta}_{\gamma} \nabla_{\alpha}\ln\Omega -g^{\delta\lambda}g_{\alpha\gamma} \nabla_{\lambda}\ln\Omega\right \} +\\
&\, + \left\{\delta^{\epsilon}_{\gamma} \nabla_{\alpha}\ln\Omega +\delta^{\epsilon}_{\alpha} \nabla_{\gamma}\ln\Omega -g^{\epsilon\lambda}g_{\alpha\gamma} \nabla_{\lambda}\ln\Omega\right\}\cdot \left\{\delta^{\delta}_{\beta} \nabla_{\epsilon}\ln\Omega + \delta^{\delta}_{\beta} \nabla_{\gamma}\ln\Omega -g^{\delta\lambda}g_{\beta\epsilon} \nabla_{\lambda}\ln\Omega\right\} - \\
&\, - \left\{\delta^{\epsilon}_{\gamma} \nabla_{\beta}\ln\Omega +\delta^{\epsilon}_{\beta} \nabla_{\gamma}\ln\Omega -g^{\epsilon\lambda}g_{\beta\epsilon} \nabla_{\lambda}\ln\Omega\right\} \cdot \left\{\delta^{\delta}_{\alpha} \nabla_{\epsilon}\ln\Omega + \delta^{\delta}_{\epsilon}\nabla_{\alpha}\ln\Omega - g^{\delta\lambda}g_{\alpha\epsilon} \nabla_{\lambda}\ln\Omega\right\},
\end{align*}
\normalsize
then
\small
\begin{align*}
\hat{R}_{\alpha\beta\gamma}\,^{\delta} =&\, R_{\alpha\beta\gamma}\,^{\delta} - \delta^{\delta}_{\gamma}\nabla_{\alpha} \nabla_{\gamma}\ln\Omega - \delta^{\delta}_{\gamma}\nabla_{\alpha} \nabla_{\beta}\ln\Omega + g^{\delta\lambda}g_{\beta\gamma}\nabla_{\alpha} \nabla_{\lambda}\ln\Omega +\\
&\, +\delta^{\delta}_{\alpha} \nabla_{\beta} \nabla_{\gamma}\ln\Omega + \delta^{\delta}_{\gamma} \nabla_{\beta}\nabla_{\alpha}\ln\Omega - g^{\delta\lambda}g_{\alpha\gamma} \nabla_{\beta} \nabla_{\lambda}\ln\Omega +\\
&\, +\nabla_{\alpha}\ln\Omega\left \{\delta^{\delta}_{\beta} \nabla_{\gamma}\ln\Omega + \delta^{\delta}_{\gamma} \nabla_{\beta}\ln\Omega - g^{\delta\lambda}g_{\beta\gamma} \nabla_{\lambda}\ln\Omega\right\}+\\
&\, + \nabla_{\gamma}\ln\Omega\left \{\delta^{\delta}_{\beta} \nabla_{\alpha}\ln\Omega + \delta^{\delta}_{\alpha} \nabla_{\beta}\ln\Omega - g^{\delta\lambda}g_{\beta\alpha}\nabla_{\lambda}\ln\Omega\right\}-\\
&\, -\delta^{\delta}_{\beta}g^{\epsilon\lambda}g_{\gamma\alpha} \nabla_{\epsilon}\ln\Omega\cdot \nabla_{\lambda}\ln\Omega -g^{\delta\lambda}g_{\gamma\alpha} \nabla_{\lambda}\ln\Omega\cdot \nabla_{\beta}\ln\Omega + g^{\delta\lambda}g_{\gamma\alpha} \nabla_{\lambda}\ln\Omega\cdot \nabla_{\beta}\ln\Omega-\\
&\, -\nabla_{\beta}\ln\Omega\left \{\delta^{\delta}_{\alpha} \nabla_{\gamma}\ln\Omega + \delta^{\delta}_{\gamma} \nabla_{\alpha}\ln\Omega - g^{\delta\lambda}g_{\alpha\gamma} \nabla_{\lambda}\ln\Omega\right\}-\\
&\, - \nabla_{\gamma}\ln\Omega\left \{\delta^{\delta}_{\alpha} \nabla_{\beta}\ln\Omega + \delta^{\delta}_{\beta} \nabla_{\alpha}\ln\Omega - g^{\delta\lambda}g_{\alpha\beta}\nabla_{\lambda}\ln\Omega\right\}+\\
&\, + \delta^{\delta}_{\alpha}g^{\epsilon\lambda}g_{\gamma\beta} \nabla_{\epsilon}\ln\Omega\cdot \nabla_{\lambda}\ln\Omega + g^{\delta\lambda}g_{\gamma\beta} \nabla_{\lambda}\ln\Omega\cdot \nabla_{\alpha}\ln\Omega -g^{\delta\lambda}g_{\gamma\beta} \nabla_{\lambda}\ln\Omega\cdot \nabla_{\alpha}\ln\Omega\\
=&\, R_{\alpha\beta\gamma}\,^{\delta} - \delta^{\delta}_{\gamma} \nabla_{\alpha} \nabla_{\gamma}\ln\Omega +\delta^{\delta}_{\alpha} \nabla_{\beta} \nabla_{\gamma}\ln\Omega + g^{\delta\lambda}g_{\beta\gamma} \nabla_{\alpha} \nabla_{\lambda}\ln\Omega -g^{\delta\lambda}g_{\alpha\gamma} \nabla_{\alpha} \nabla_{\lambda}\ln\Omega +\\
&\, + \delta^{\delta}_{\beta} \nabla_{\gamma}\ln\Omega\cdot \nabla_{\alpha}\ln\Omega - \delta^{\delta}_{\alpha} \nabla_{\beta}\ln\Omega\cdot \nabla_{\gamma}\ln\Omega-\\ &\,-\delta^{\delta}_{\beta}g^{\epsilon\lambda}g_{\gamma\alpha}\nabla_{\epsilon}\ln\Omega\cdot \nabla_{\lambda}\ln\Omega +\delta^{\delta}_{\alpha}g^{\epsilon\lambda}g_{\gamma\beta}\nabla_{\epsilon}\ln\Omega\cdot \nabla_{\lambda}\ln\Omega-\\
&\, -g^{\delta\lambda}g_{\gamma\beta} \nabla_{\alpha}\ln\Omega\cdot \nabla_{\lambda}\ln\Omega + g^{\delta\lambda}g_{\gamma\alpha} \nabla_{\beta}\ln\Omega\cdot \nabla_{\lambda}\ln\Omega,
\end{align*}
\normalsize
so
\begin{align}
\hat{R}_{\alpha\beta\gamma}\,^{\delta} =& \, R_{\alpha\beta\gamma}\,^{\delta} + 2\delta^{\delta}\,_{[\alpha}\nabla_{\beta]} \nabla_{\gamma}\ln\Omega - 2g^{\delta\lambda}g_{\gamma}\,_{[\alpha}\nabla_{\beta]} \nabla_{\lambda}\ln\Omega + 2\left(\nabla_{[\alpha}\ln\Omega\right )\delta^{\delta}\,_{\beta]} \nabla_{\gamma}\ln\Omega-\nonumber\\
&\, -2g_{\gamma[\alpha}\delta^{\delta}\,_{\beta]}g^{\epsilon\lambda} \nabla_{\epsilon}\ln\Omega\cdot \nabla_{\lambda}\ln\Omega - 2\left(\nabla_{[\alpha}\ln\Omega\right )g_{\beta]\gamma} g^{\delta\lambda}\nabla_{\lambda}\ln\Omega.
\end{align} 
Contracting over $\beta$ and $\delta$ we obtain
\begin{align*}
\hat{R}_{\alpha\gamma} =&\, R_{\alpha\gamma} - n \nabla_{\alpha} \nabla_{\gamma}\ln\Omega + 2 \nabla_{\alpha} \nabla_{\gamma}\ln\Omega - g^{\delta\lambda}g_{\alpha\gamma} \nabla_{\delta} \nabla_{\lambda}\ln\Omega + n \nabla_{\gamma}\ln\Omega\cdot \nabla_{\alpha}\ln\Omega-\\
&\, -  \nabla_{\alpha}\ln\Omega\cdot \nabla_{\gamma}\ln\Omega -ng^{\epsilon\lambda}g_{\gamma\alpha} \nabla_{\epsilon}\ln\Omega\cdot \nabla_{\lambda}\ln\Omega + g^{\epsilon\lambda}g_{\gamma\alpha} \nabla_{\epsilon}\ln\Omega\cdot \nabla_{\lambda}\ln\Omega- \\ 
&\, - \nabla_{\alpha}\ln\Omega\cdot \nabla_{\gamma}\ln\Omega +g^{\epsilon\lambda}g_{\gamma\alpha} \nabla_{\epsilon}\ln\Omega\cdot \nabla_{\lambda}\ln\Omega,
\end{align*}
then
\begin{align} 
\hat{R}_{\alpha\gamma} =&\, R_{\alpha\gamma} - (n-2) \nabla_{\alpha} \nabla_{\gamma}\ln\Omega + (n-2) \nabla_{\alpha}\ln\Omega \nabla_{\gamma}\ln\Omega - \nonumber\\
&\, - (n-2)g^{\delta\lambda}g_{\alpha\gamma} \nabla_{\delta}\ln\Omega\cdot \nabla_{\lambda}\ln\Omega - g^{\delta\lambda}g_{\alpha\gamma} \nabla_{\delta} \nabla_{\lambda}\ln\Omega.
\end{align}
Finally contracting with $\hat{g}^{\alpha\gamma} = \Omega^{-2}g^{\alpha\gamma}$
\begin{align*}
\hat{g}^{\alpha\gamma}\hat{R}_{\alpha\gamma} =&\, \Omega^{-2}\Big\{ g^{\alpha\gamma}R_{\alpha\gamma} - (n-2)g^{\alpha\gamma} \nabla_{\alpha} \nabla_{\gamma}\ln\Omega + (n-2)g^{\alpha\gamma} \nabla_{\alpha}\ln\Omega \nabla_{\gamma}\ln\Omega - \\
&\, - n(n-2)g^{\delta\lambda} \nabla_{\delta}\ln\Omega\cdot \nabla_{\lambda}\ln\Omega - ng^{\delta\lambda} \nabla_{\delta} \nabla_{\lambda}\ln\Omega\Big\},
\end{align*}
thus
\begin{equation} \label{B31}
 \hat{R} = \Omega^{-2}\left \{ R - 2(n-1)g^{\alpha\gamma} \nabla_{\alpha} \nabla_{\gamma}\ln\Omega - (n-2)(n-1)g^{\alpha\gamma} \nabla_{\alpha}\ln\Omega\cdot \nabla_{\gamma}\ln\Omega \right \}.
\end{equation}
For the case in which $ \nabla_{\alpha}$ is the ordinary (covariant) derivative operator, then $C^{\gamma}_{\alpha\beta}$ it is denoted as $\Gamma^{\gamma}_{\alpha\beta}$, the Christoffel symbol. For the purposes of the study, the manifold taken is 4-dimensional, then, replacing $n=4$ in (\ref{B31}) we obtain
\begin{equation}\label{Rconforme}
\hat{R} =\Omega^{-2}\left \{ R - 6g^{\alpha\gamma} \nabla_{\alpha} \nabla_{\gamma}\ln\Omega - 6g^{\alpha\gamma} \nabla_{\alpha}\ln\Omega\cdot \nabla_{\gamma}\ln\Omega \right \}.
\end{equation}

\chapter{The \texorpdfstring{$\phi^{4}$}{phi4} Potential Case} 

\label{AppendixC} 

\lhead{Appendix C. \emph{The \texorpdfstring{$\phi^{4}$}{phi4} Potential Case}} 
In this appendix we will present the main results of the stability criteria for the $\phi^{4}$ potential using a trivial shift ($a$) in the vacuum expectation value of the field and the Cosmological Constant $\Lambda$. We briefly present the analysis of the slow-roll approximation and the numerical case. Finally, we will show the main result of the stability mass criteria and the corresponding thermodynamics analogy. 
\section{Slow-Roll Analysis}
As we show in chapter \ref{Chapter2} the modification of the potential includes a free parameter $a$ as $\phi$-field Vacuum Expectation Value -- (VEV) and the cosmological constant $\Lambda$ 
\begin{equation}\label{quarticpot}
V_{q}(\phi) = \frac14 m^4_{\phi} (\phi-a)^{4} + \Lambda.
\end{equation}
As we showed before, we turn off the value of $\Lambda$ in the potential in the slow-roll analysis. The slow-roll parameters (\ref{parametrosr1}) and (\ref{parametrosr2}) can be written as
\begin{eqnarray}\label{epsilonq}
\epsilon &=&  \frac{M^{2}_{P}}{2}\left (\frac{V_{q}'(\phi)}{V_{q}(\phi)}\right )^{2} = \frac{8 M_P^2}{(\phi -a)^2},\\
\eta &=& M^{2}_{P}\left |\frac{V_{q}''(\phi)}{V_{q}(\phi)} \right | = \frac{12 M_P^2}{(\phi -a)^2}\label{etaq},
\end{eqnarray}
where ($'$) indicates the derivative with respect to $\phi$, and the subscript ``$q$" means quartic case. Slow-roll ends when $\epsilon \simeq 1$, so the scalar field value at the end of inflation $\phi_{\mathrm{end}}$, according to (\ref{epsilonq}), is
\begin{equation} \label{phiendq}
\phi_\mathrm{end} \simeq a-2 \sqrt{2} M_P.
\end{equation}
In order to obtain the initial value of the scalar field ($\phi_{\mathrm{*}}$), equation (\ref{efolds}) can be used; once the integral is solved and replaced in (\ref{phiendq}), one obtains
\begin{eqnarray} \label{Nq}
N   &=& \frac{1}{M^2_{P}}\int^{\phi_\mathrm{*}}_{\phi_\mathrm{end}}\frac{V_{q}(\phi)}{V'_{q}(\phi)}d\phi = \frac{1}{M_P^2}\left(\frac{a \phi_\mathrm{end}}{4}-\frac{a \phi_{\mathrm{*}}}{4}-\frac{\phi_\mathrm{end}^2}{8}+\frac{\phi_{\mathrm{*}}^2}{8}\right).
\end{eqnarray}
Solving for $\phi_{\mathrm{*}}$ from (\ref{Nq}) we obtain
\begin{eqnarray}\label{phiestrellaq}
\phi_\mathrm{*} = a-2 \sqrt{2} \sqrt{M_P^2 (\text{N}+1)}\;.
\end{eqnarray}
An analytical expression of the field $\phi_{q} = \phi_{q}(t)$ can be obtained by solving the system (\ref{KGsr}) and (\ref{friedmannsr}) for the value of the field (\ref{phiestrellaq}), with $N = 60$ efolds, is
\begin{eqnarray}\label{phitq}
\phi_{q}(t) = a-2 \sqrt{122}M_P e^{-\frac{2 m_{\phi} ^2 t}{\sqrt{3}}},
\end{eqnarray}
assuming the slow-roll evolution.

\section{Numerical Solution}
\begin{figure}
\begin{center}
\includegraphics[width=0.6\textwidth]{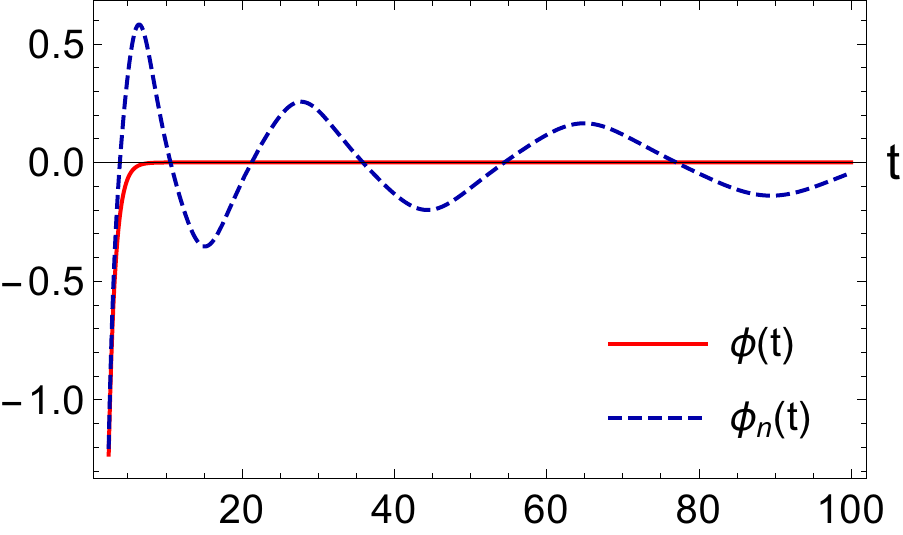}    
\end{center}
\caption[Numerical and analytic solutions for $\phi_{q}(t)$]{Analityc (red/solid) and numerical (blue/dashed) solutions for $\phi_{sq}(t)$ given by Eq.~(\ref{phitsq}), with $N = 60$ efolds, using $m_\phi = 1$ and $M_P = 1$. Note that, before $t\sim 4$, the curves are very close.}
\label{phi4plot}
\end{figure}

We used \verb|Mathematica|\textsuperscript{\textcopyright} software \cite{Mathematica} to obtain a numerical solution of the full system equation of movement for the scalar field (\ref{kleingordon}), and the initial conditions are the standard ones from the analytic slow-roll solution set. For $N=60$ efolds we obtain:
\begin{equation}
\phi_{q}(0) = a-2 \sqrt{122}\, M_P \approx a- 22.1 M_P \, , \quad
\dot \phi_{q}(0) = 4 \sqrt{\frac{122}{3}}\, m_{\phi}^2\approx 25.51 \, m_{\phi}^2.
\end{equation}
The numerical and analytical solutions are shown in Figure \ref{phi4plot}.

\begin{figure}
\center
\includegraphics[width=0.8\textwidth]{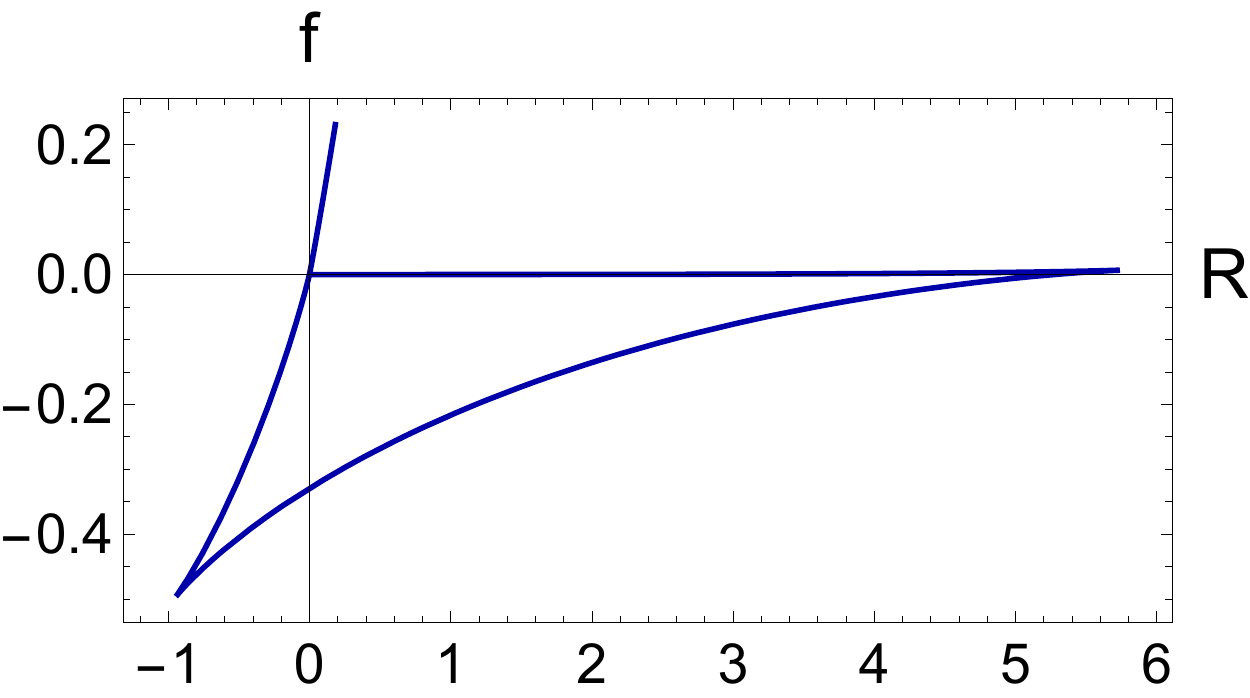}
\caption[Parametric plots of $f(R)$ Quartic case]{Parametric plots of $f(R)$ given by Eqs. (\ref{fq}, \ref{Rq}) for $\phi \in [-20,0.4]$ and for the parameters $\Lambda=0$, and $a=0$. The field $\phi$ and $a$ are given in Planck-Mass ($M_P$) units, $R$ and $\Lambda$ are given in $M_P^4$. We used $m_\phi=1 M_P$.}
\label{fRq}
\end{figure}
The spectral index (\ref{ns}), and the tensor-to-scalar ratio (\ref{ratiots}) calculated for $N=60$ (corresponding to the scale $k=0.002/\mathrm{Mpc}$), are
\begin{eqnarray}
n_s &=&1+\frac{48 M_P^4 (N+1)}{\Lambda +16 M_P^4 (N+1)^2}-\frac{512 M_P^8 (N+1)^3}{\left(\Lambda +16 M_P^4 (N+1)^2\right)^2},\\
r &=& \frac{4096 M_P^8 (N+1)^3}{\left(\Lambda +16 M_P^4 (N+1)^2\right)^2}.
\end{eqnarray}
The values of the spectral index and the tensor-to-scalar ratio obtained by the quartic inflationary potential are not favorable according to the last PLANCK's report \cite{Planck2018}.    
\section{\texorpdfstring{$f(R)$}{frq} function from quartic potential}
We then obtain the corresponding parametric form of $f(R)$ given by Eqs. (\ref{fphi}, \ref{Rphi}):
\begin{align}
\label{fq}
f(\phi) &= e^{\beta  \phi } \left(\frac{(a-\phi )^3 (a \beta -\beta  \phi -2)}{\beta }+4 \Lambda \right),\\
\label{Rq}
R(\phi) &= 2 e^{2 \beta  \phi } \left(\frac{(\phi -a)^3}{\beta }+\frac{1}{4} (a-\phi )^4+\Lambda \right).
\end{align}
\newpage
\section{The Stability Mass Criteria}\label{stableq}
\begin{figure}[t]
\center
{\includegraphics[width=0.8\textwidth]{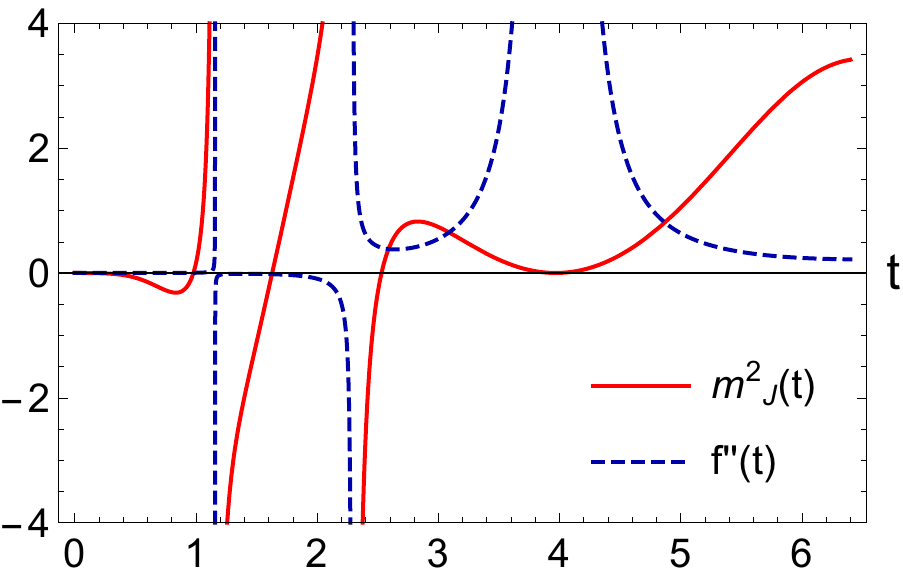}}
\caption[Mass of Quartic Potential]{Quartic potential mass: $m^2_J$ (\ref{mJ}) in red/solid, and $f''$ in blue/dashed as functions of time. We have used $N = 60$ efolds and $m_\phi = 1$,$\Lambda = 0$, and $a=0$.}
\label{Mqpot}
\end{figure}

In chapter \ref{Chapter7} we point out that the stability of any given configuration is determined by the signal of the squared-mass term of the corresponding (effective) potential at each one of its equilibrium points. In Fig~\ref{Mqpot} we compare the behavior of the second derivative of $f(R)$ and the $m^2_J$ for the quartic potential (\ref{quarticpot}).
\section{The Stability Criteria from Thermodynamics Analogy}\label{ThermoAnalogySectionq}
For this quartic potential case, the thermodinamic association (\ref{ansatz}) of the new pair of coordinates $\{-G,P\}$ as a rotation of the original one $\{f,R\}$ yields
\small
\begin{align}
 G (\phi,T) &=
e^{\beta  \phi } \sin (\theta ) \left(\frac{(a-\phi )^3 (a \beta -\beta  \phi -2)}{\beta }+4 T\right)-2 e^{2 \beta  \phi } \cos (\theta ) \left(\frac{(\phi -a)^3}{\beta }+\frac{1}{4} (a-\phi )^4+T\right) ,
 \label{Gphithetaq}  \\
P (\phi,T) &= 2 e^{2 \beta  \phi } \sin (\theta ) \left(\frac{(\phi -a)^3}{\beta }+\frac{1}{4} (a-\phi )^4+T\right)+e^{\beta  \phi } \cos (\theta ) \left(\frac{(a-\phi )^3 (a \beta -\beta  \phi -2)}{\beta }+4 T\right),
\label{Pphithetaq}\\
V &= \left.\frac{\partial G/\partial\phi}{\partial P/\partial\phi}\right|_T = \frac{1-e^{\beta  \phi } \cot (\theta )}{e^{\beta  \phi }+\cot (\theta )}.
\label{Vphithetaq}
\end{align}
\normalsize
Eq.~(\ref{Vphithetaq}) can be inverted and yield
\begin{equation}
\phi = \frac{1}{\beta}\log \left(\frac{1-V \cot (\theta )}{\cot (\theta )+V}\right).
\end{equation}

\begin{figure}
\center
\includegraphics[width=0.7\textwidth]{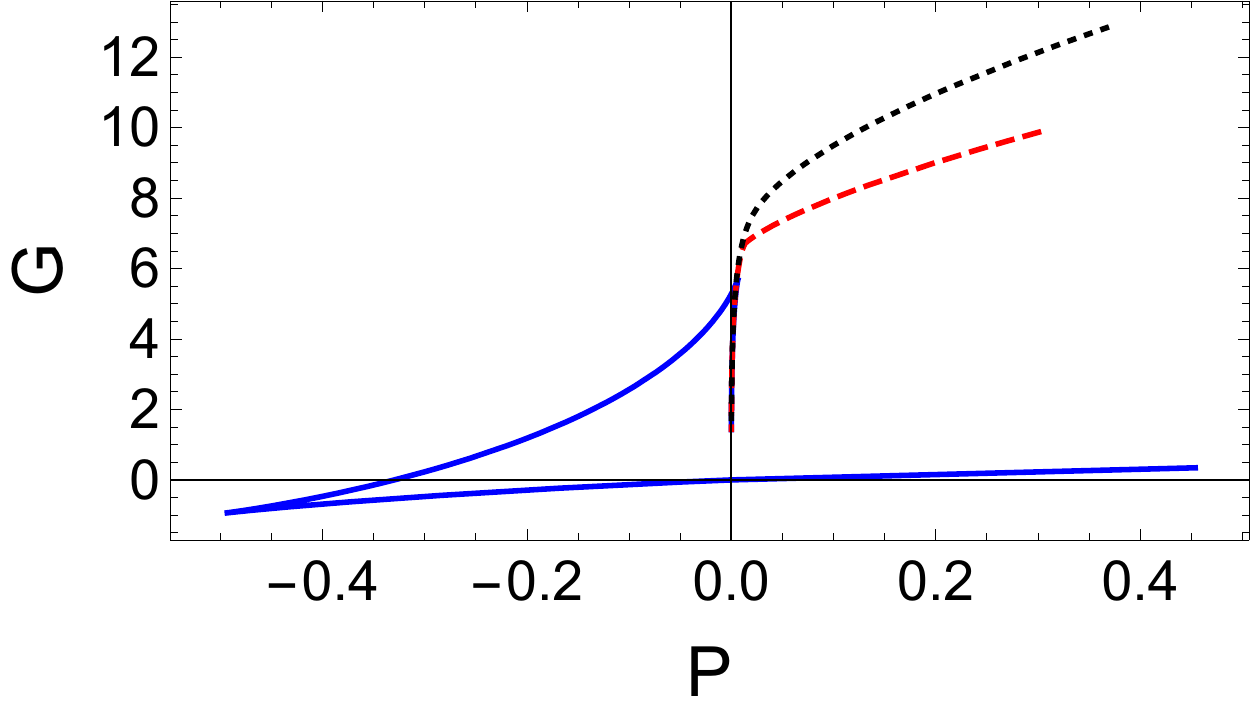}
\caption[Plot of the Gibbs for quartic potential]{Parametricplot of the Gibbs Potential $G$ Eq.~(\ref{Gphiq}) as a function of the pressure $P$ (Eq.~\ref{Pphiq}), for $\beta=\sqrt{2/3}$, $a=0$, $\theta=\theta_*$ and $T=0$ (solid blue), $T=T_c \simeq 19.331$ (dashed red) and $T=30$ (dotted black).}
\label{Gphiq_fig}
\end{figure}

In order to define the exact correspondence, i.e, the value of $\theta$, we only require that the volume  is positive and unlimited from below. Such procedure yields $\theta = \theta_* \equiv  \pi/2$ and simpler parametric expressions for the previously defined thermodynamic quantities: 
\begin{align}
G &= e^{\beta  \phi } \left(\frac{(a-\phi )^3 (a \beta -\beta  \phi -2)}{\beta }+4 T\right),\label{Gphiq}\\
P &= 2 e^{2 \beta  \phi } \left(\frac{(\phi -a)^3}{\beta }+\frac{1}{4} (a-\phi )^4+T\right),\label{Pphiq}\\
F &= \frac{1}{2} e^{\beta  \phi } \left(a^4-4 a^3 \phi +6 a^2 \phi ^2-4 a \phi ^3+4 T+\phi ^4\right),\label{Fphiq}\\
S &= -2 e^{\beta  \phi }\label{Sphiq},\\
U &= \frac{1}{2} (a-\phi )^4 e^{\beta  \phi },\label{Uphiq}\\
V &= \exp(-\beta \phi) \quad \Leftrightarrow \quad \phi = -\frac{1}{\beta}\log(V).\label{Vphiq}
\end{align}
In Fig.~\ref{Gphiq_fig}, we plot the curve $G(P)$ for different temperatures $T$.

\begin{figure}
\center
\includegraphics[width=0.7\textwidth]{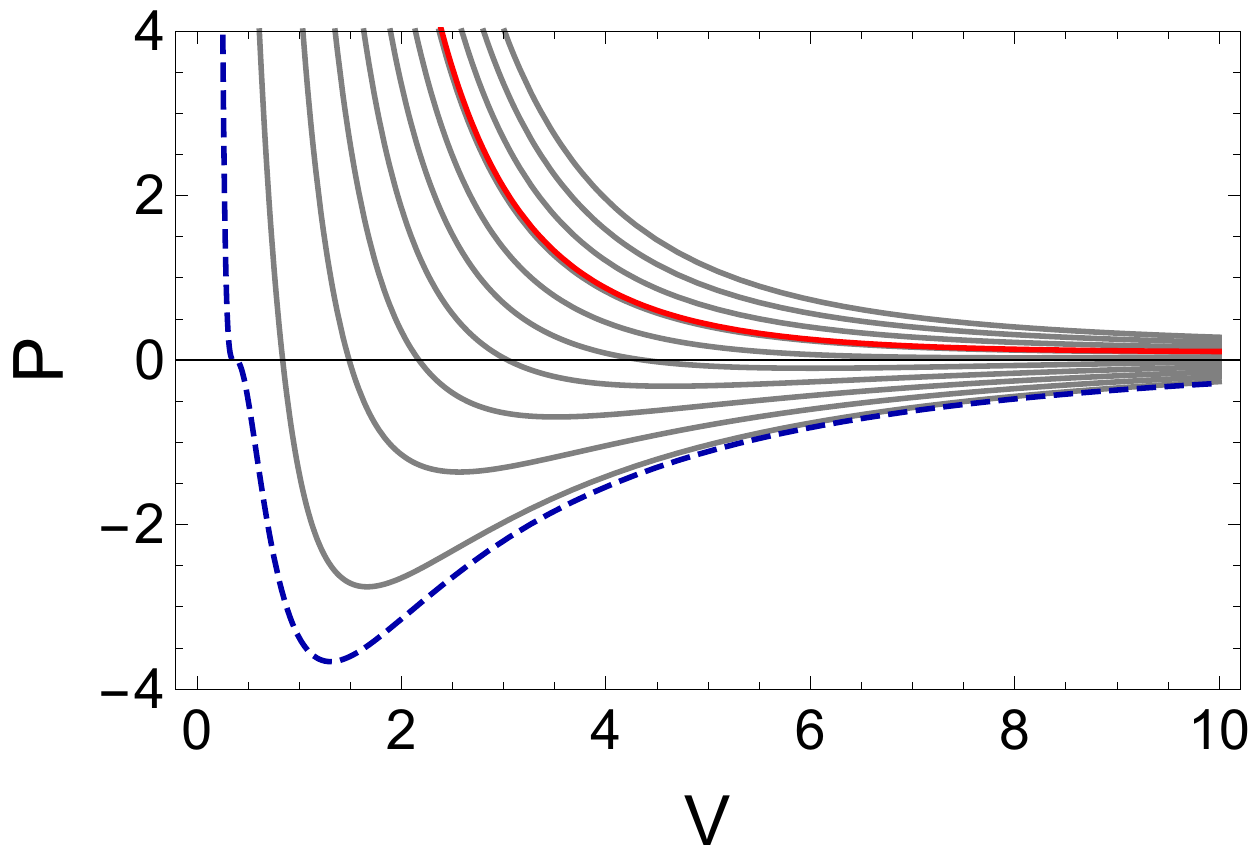}
\includegraphics[width=0.7\textwidth]{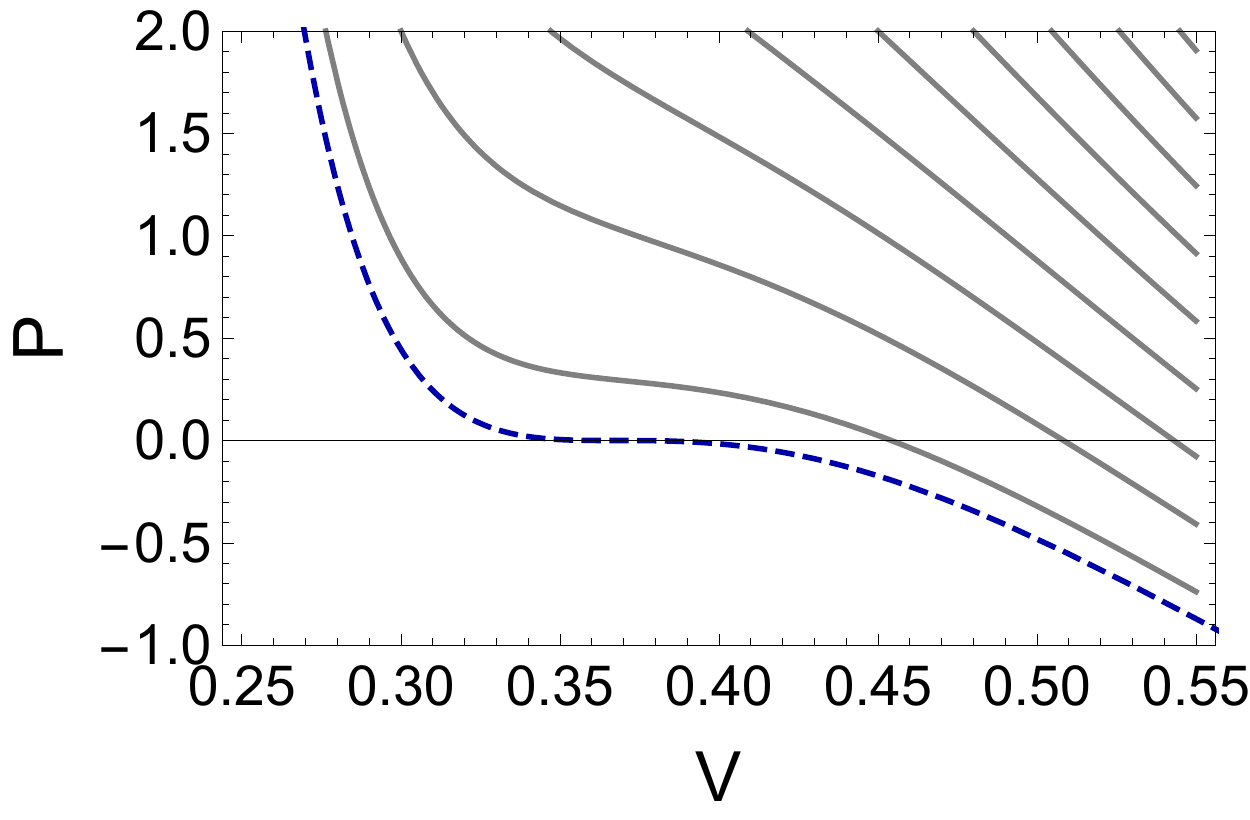}
\caption[Plot of the Preassure for quartic potential]{\textbf{Upper panel:} Parametricplot of the Pressure Plot of the effective pressure $P$ as a function of the effective volume $V$ Eq~(\ref{PVq}), for $a=0$ and different values of temperature: $T=0$ (dashed blue) and $T=T_c \simeq 19.331$ (solid thick red) and higher curves (solid thin gray) correspond to higher temperatures $T>0$. \textbf{Lower panel:} zooming on the region where $V \in (0.25,0.55)$ for different values of temperature: $T=0$ (dashed blue) and higher curves (solid thin gray) correspond to higher temperatures $T>0$.}
\label{Pphiq_fig}
\end{figure}

We can obtain an explicit equation of state for our vdW-like  ``efective gas'' for this case, for $\theta=\theta_*=\pi/2$, using equations~(\ref{Pphiq}) and (\ref{Vphiq}). The behaviour of $P(V)$ for four different values of $T$ is shown in Fig.~\ref{Pphiq_fig}.
\begin{equation}
P = \frac{(a \beta +\log (V))^4-4 (a \beta +\log (V))^3+4 \beta ^4 T}{2 \beta ^4 V^2}.
\label{PVq}
\end{equation}

\chapter{The Double Well Potential Case} 

\label{AppendixD} 

\lhead{Appendix D. \emph{The Double Well Potential Case}} 
In this appendix we will present the main results of the stability criteria for the double well potential adding \textit{ad hoc} the Cosmological Constant $\Lambda$. We briefly present the analysis of the slow-roll approximation and the numerical solution for the field. Finally, we will show the main result of the stability mass criteria and the corresponding thermodynamics analogy. 
\section{Slow-Roll Analysis}
The double well potential depends on one parameter model of inflation $a$ as $\phi$-field Vacuum Expectation Value -- (VEV), but we add the cosmological constant $\Lambda$ which, as we showed in chapter \ref{Chapter7}, play an important role in the thermodynamics interpretation 
\begin{equation}\label{dwpot}
V_{dw}(\phi) = \frac14 m^4_{\phi} (\phi^2-a^2)^{2} + \Lambda.
\end{equation}
For this analysis we turn off the value of $\Lambda$ will not significantly affect the initial value of the field, therefore it can be ignored. The slow-roll parameters (\ref{parametrosr1}) and (\ref{parametrosr2}) can be written as
\begin{eqnarray}
\epsilon &=&  \frac{M^{2}_{P}}{2}\left (\frac{V_{dw}'(\phi)}{V_{dw}(\phi)}\right)^{2} = \frac{8 M_P^2 \phi ^2}{\left(\phi ^2-a^2\right)^2}\label{epsilondw},\\
\eta &=& M^{2}_{P}\left |\frac{V_{dw}''(\phi)}{V_{dw}(\phi)} \right | = \frac{8 M_P^2 \left(a^2+\phi ^2\right)}{\left(a^2-\phi ^2\right)^2}\label{etadw},
\end{eqnarray}
where ($'$) indicates the derivative with respect to $\phi$, and the subscript ``$dw$" means double well. The parameter $\eta$  is equal to $\eta= 8 M_P^2/a^2$ when $\phi = 0$. In order for slow-roll to be valid, this last value should be less than one which amounts to $a/M_P > 2 \sqrt{2}$. This constraint on the parameter a shows that the symmetry breaking scale needs to be superPlanckian \cite{Martin:2013tda}. Slow-roll ends when $\epsilon \simeq 1$, so the scalar field value at the end of inflation $\phi_{\mathrm{end}}$, according to (\ref{epsilondw}), is
\begin{equation} \label{phienddw}
\phi_\mathrm{end} \simeq-\sqrt{2}M_P-\sqrt{a^2+2 M_P^2}.
\end{equation}
In order to obtain the initial value of the scalar field ($\phi_{\mathrm{*}}$), equation (\ref{efolds}) can be used; once the integral is solved and replaced in (\ref{phienddw}), one obtains
\begin{eqnarray} \label{Ndw}
N   &=& \frac{1}{M^2_{P}}\int^{\phi_\mathrm{*}}_{\phi_\mathrm{end}}\frac{V_{dw}(\phi)}{V'_{dw}(\phi)}d\phi = \frac{1}{8 M_P^2}\left(2 a^2 (\log(\phi_{\mathrm{*}})-\log (\phi_\mathrm{end}))+\phi_\mathrm{end}^2-\phi_\mathrm{*}^2\right).
\end{eqnarray}
Solving for $\phi_{\mathrm{*}}$ from (\ref{Ndw}) we obtain
\footnotesize
\begin{eqnarray}\label{phiestrelladw}
\phi_\mathrm{*} = -a \sqrt{-W_0\left(-\frac{\left(\sqrt{a^2+2 M_P^2}+\sqrt{2} M_P\right)^2 \exp \left(-\frac{2 M_P \left(\sqrt{2} \sqrt{a^2+2 M_P^2}+M_P (4 N+2)\right)}{a^2}-1\right)}{a^2}\right)}\;.
\end{eqnarray}
\normalsize
An analytical expression of the field $\phi_{dw} = \phi_{dw}(t)$ can be obtained by solving the system (\ref{KGsr}) and (\ref{friedmannsr}) for the value of the field (\ref{phiestrelladw}), with $N = 60$ efolds, is
\begin{eqnarray}\label{phitdw}
\phi_{dw}(t) = -a e^{\frac{4 t}{\sqrt{3}}} \sqrt{-W_0},
\end{eqnarray}
where
\begin{equation}
    W_0= W_0\left[-\frac{1}{a^2}\left(e^{-\frac{2 M_P \left(\sqrt{2} \sqrt{a^2+2 M_P^2}+242 M_P\right)}{a^2}-1} \left(\sqrt{a^2+2 M_P^2}+\sqrt{2} M_P\right)^2\right)\right],
\end{equation}
assuming the slow-roll approximation.
\begin{figure}
\begin{center}
\includegraphics[width=0.6\textwidth]{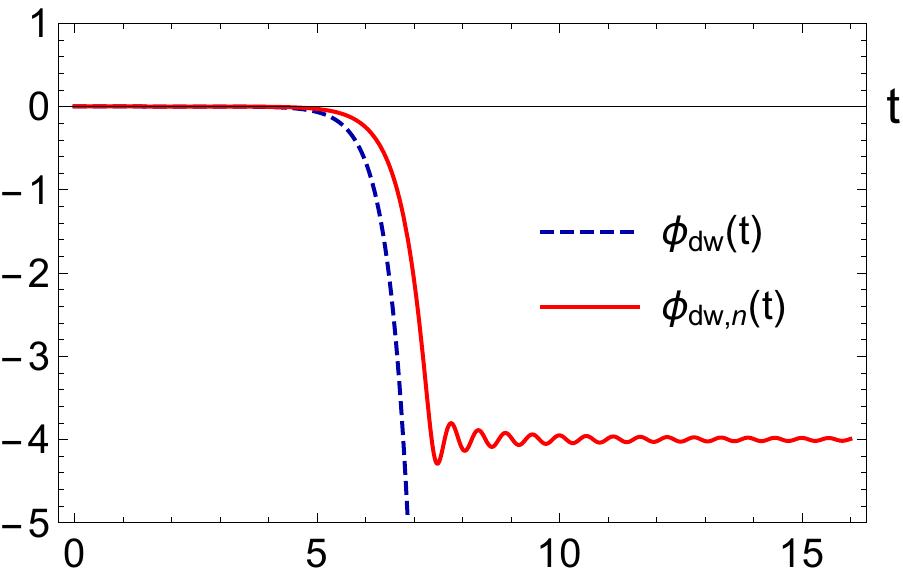}    
\end{center}
\caption[Numerical and analytic solutions for $\phi_{dw}(t)$]{Numerical solution (red/solid) and analytic solutions (blue/dashed) for $\phi_{dw}(t)$ given by Eq.~(\ref{phitdw}), with $N = 60$ efolds, using $m_\phi = 1$, $M_P = 1$, $a=4$ and $\Lambda = 0$. Before $t\sim 1$, the curves are very close.}
\label{phidwplot}
\end{figure}
\section{Numerical Solution}
\begin{figure}
\center
\includegraphics[width=0.8\textwidth]{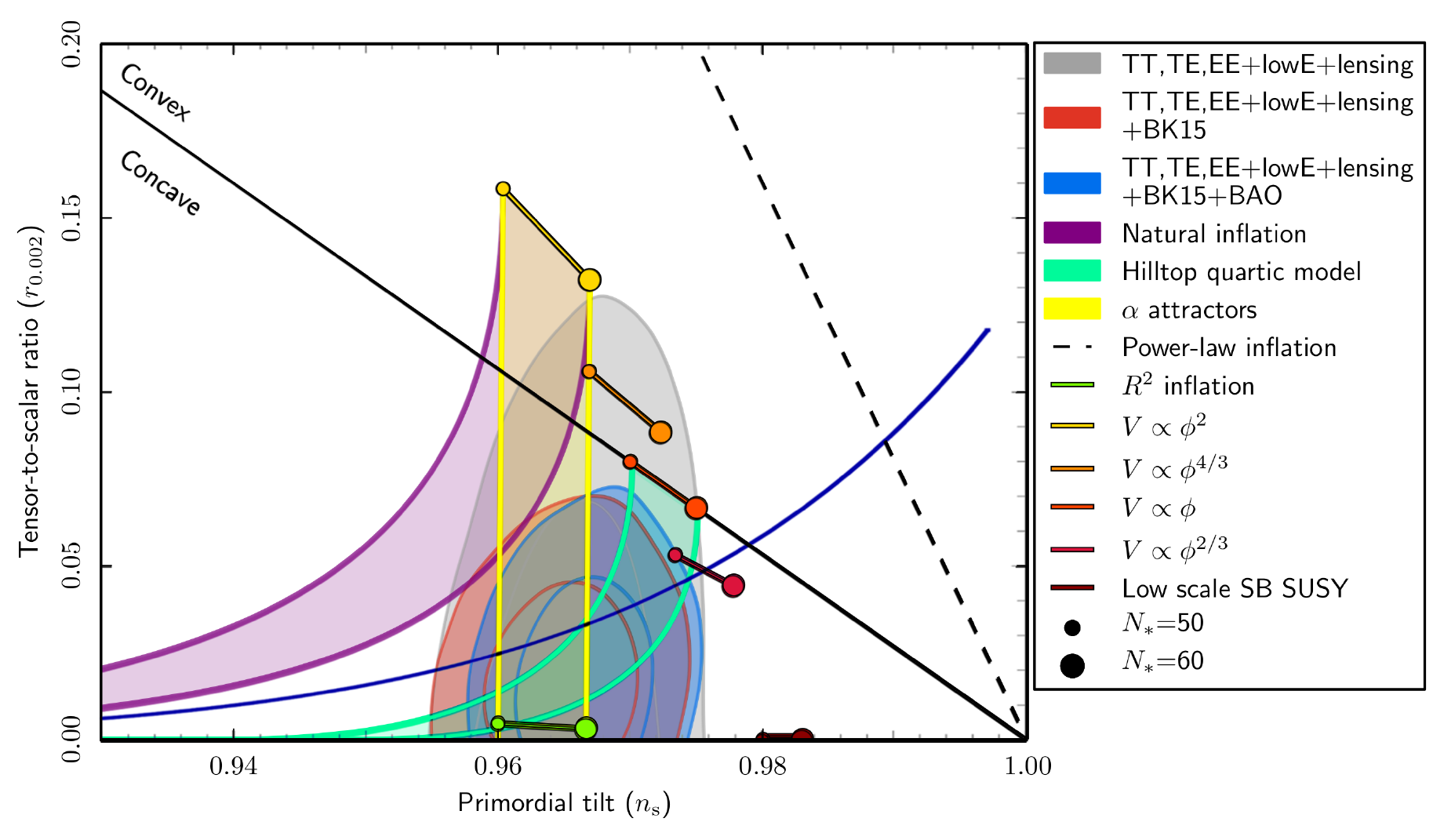} \caption[Planck data 2018 for dw potential]{Marginalized joint $68\%$ and $95\%$ CL regions for $n_s$ and $r$ at $k = 0.002 \text{Mpc}^{-1}$ from Planck alone and in combination with BK15 or BK15+BAO data, compared to the theoretical predictions of selected inflationary models. The double well potential model (blue/solid) was made for $\Lambda =0$, $N=60$ efolds, $m_{\phi}=1M_P$ and varying the parameter $a$ in the range $(5,100)M_P$, noticing that in range between $(13.5,19)$ $M_P$ the region for $n_s$ and $r$ is in the $95\%$ CL region from Planck alone data. Image adapted from: ESA and the Planck Collaboration \cite{Planck2018}.} 
\label{fig:planck2018dw} 
\end{figure}

We used \verb|Mathematica|\textsuperscript{\textcopyright} software \cite{Mathematica} to obtain a numerical solution of the full system equation of movement for the scalar field (\ref{kleingordon}), and the initial conditions are the standard ones from the analytic slow-roll solution set, for $N=60$ efolds we obtain:
\begin{equation}
\phi_{dw}(0) = -a \sqrt{-W_0}, \quad {\rm and} \quad \dot \phi_{dw}(0) = -\frac{4 a\, m_{\phi}^2\sqrt{-W_0}}{\sqrt{3}}.
\end{equation}
The numerical and analytical solutions are shown in Figure \ref{phidwplot}.
The spectral index (\ref{ns} and the tensor-to-scalar ratio (\ref{ratiots})), calculated for $N=60$ (corresponding to the scale $k=0.002/\mathrm{Mpc}$), are
\footnotesize
\begin{eqnarray}
n_s &=& -\frac{8 a^2 m_{\phi}^4 M_P^2 \left(3 W_0+1\right)}{a^4 m_{\phi}^4+a^4 m_{\phi}^4 W_0^2+2 a^4 m_{\phi}^4 W_0+4 \Lambda }+\frac{a^6 m_{\phi}^8 M_P^2 W_0 \left(W_0+1\right)^2}{\left(\frac{1}{4} a^4 m_{\phi}^4 \left(W_0+1\right)^2+\Lambda \right)^2}+1,\\
r &=& -\frac{8 a^2 m_{\phi}^8 M_P^2 W_0 \left(a^2 W_0 + a^2\right)^2}{\left(\frac{1}{4} m_{\phi}^4 \left(a^2 W_0 + a^2\right)^2+\Lambda \right)^2}.
\end{eqnarray}
\normalsize
The values of the spectral index and the tensor-to-scalar ratio obtained by the double well potential are favorable according to the last PLANCK's report \cite{Planck2018} see fig.~\ref{fig:planck2018dw}. The $\Lambda$ value does not significantly affect the results of these parameters.    

\section{\texorpdfstring{$f(R)$}{frdw} function from double well potential}
\begin{figure}[t]
\centering
\includegraphics[width=0.425\textwidth]{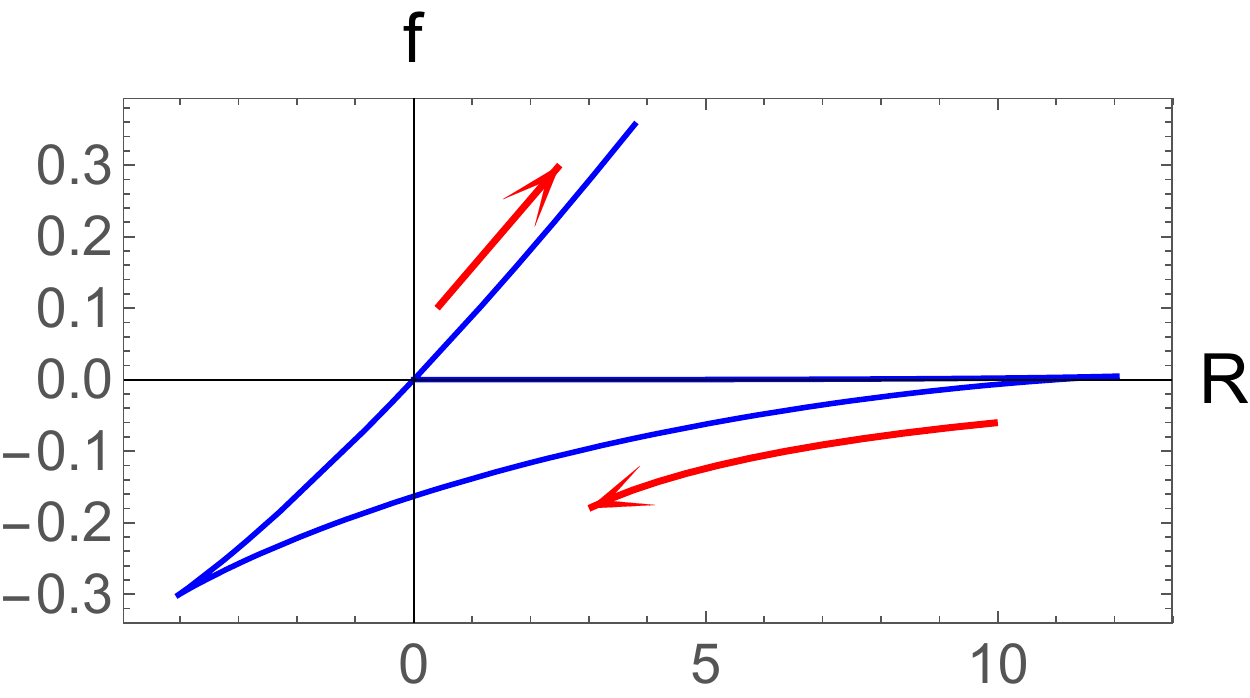}
\includegraphics[width=0.425\textwidth]{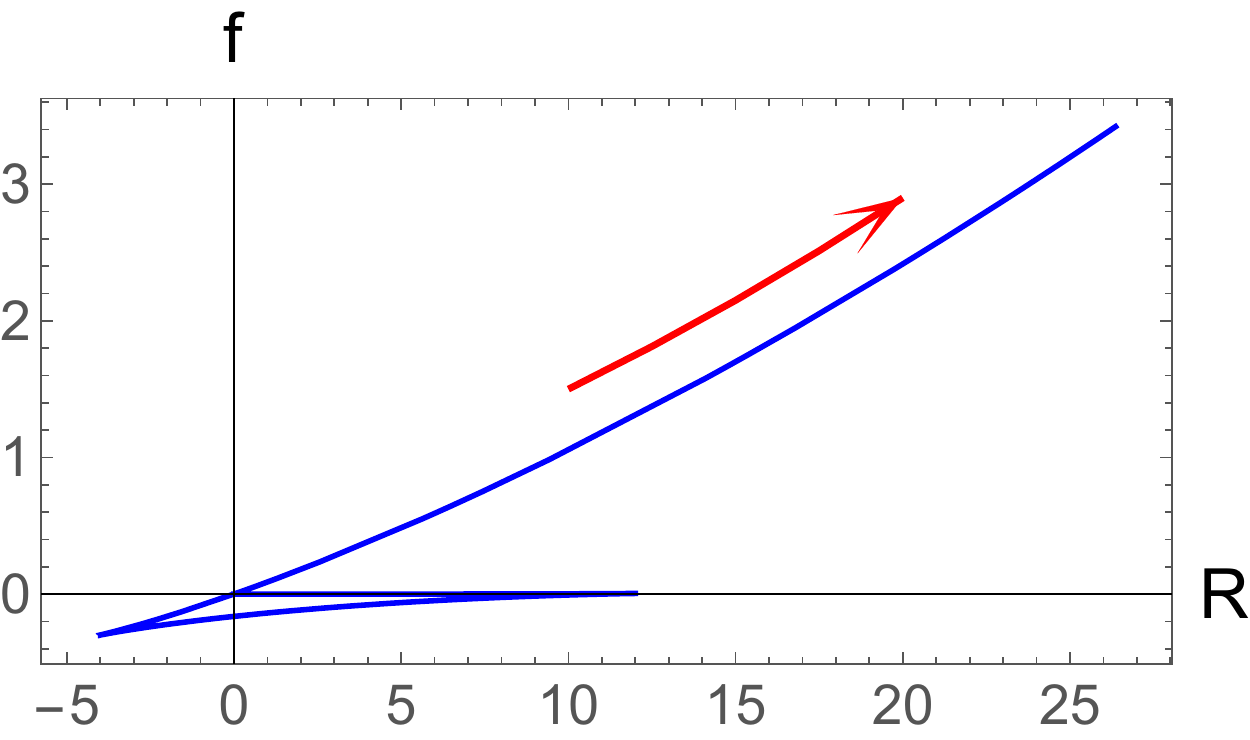}
\includegraphics[width=0.425\textwidth]{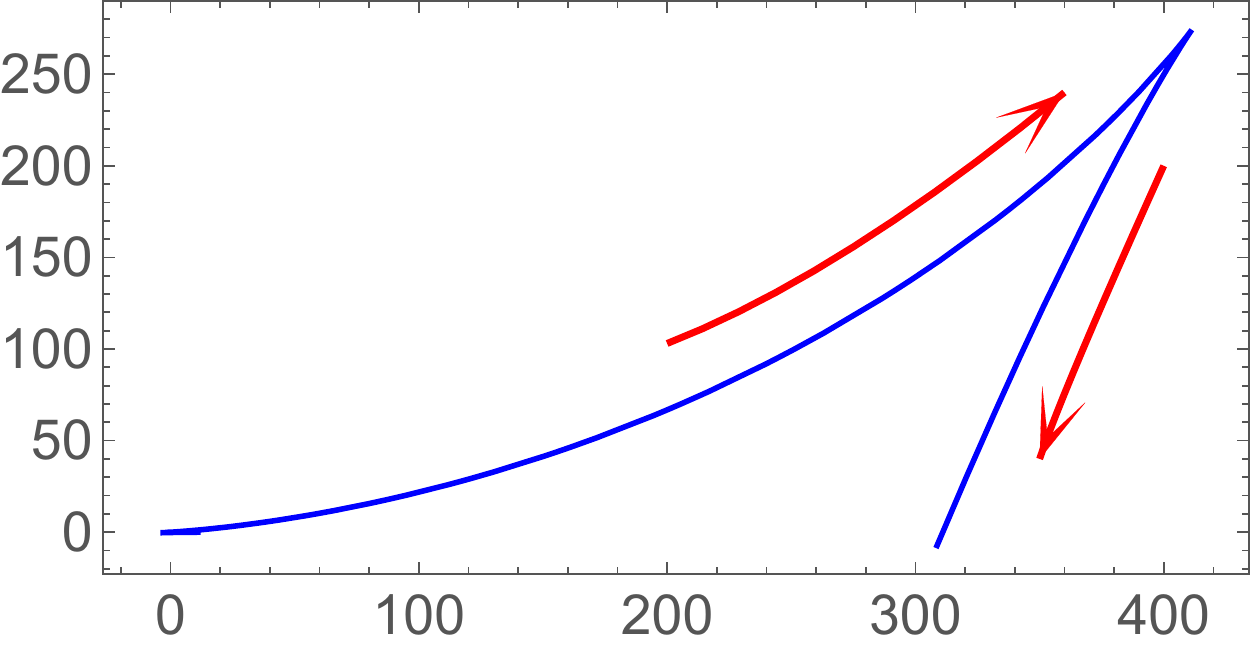}
\includegraphics[width=0.425\textwidth]{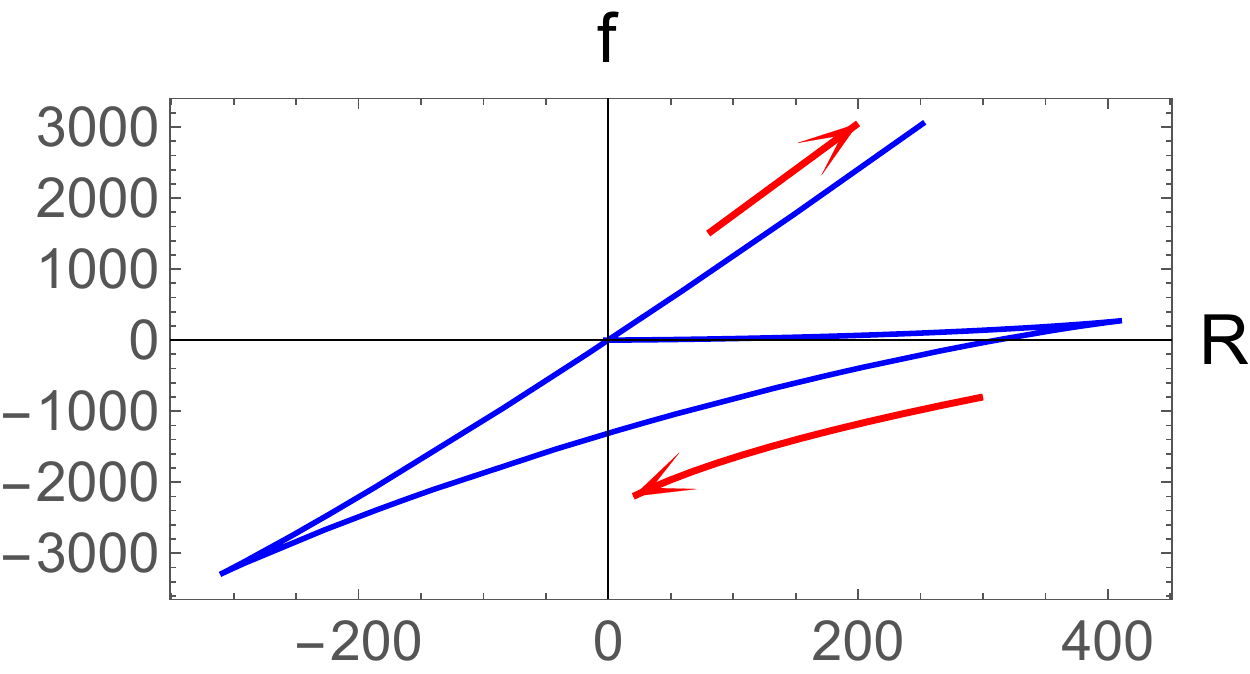}
\caption[Parametric plots of $f(R)$ Double Well Potential]{Parametric plots of $f(R)$ given by Eqs. (\ref{fdw}, \ref{Rdw}) for the parameters $m_\phi=1 M_{P}$, $\Lambda=0$ and $a=3$ in different ranges of $\phi$. \textbf{Upper left panel:} $\phi \in [-50,-2.8]$, a first catastrophic behavior is shown. \textbf{Upper right panel:} $\phi \in [-50,-2.2]$. \textbf{Lower left panel:} $\phi \in [-50,1.43]$, begins a second catastrophic behavior. \textbf{Lower right panel:} $\phi \in [-50,3.1]$, shows the complete second catastrophe. \textbf{All panels:} Arrows indicate the path of $f(R)$ as the field $\phi$ evolves. The field $\phi$ and $a$ are given in Planck-Mass ($M_{P}$) units, $R$ is given in $M_{P}^4$.}
\label{fRdw}
\end{figure}
We then obtain the corresponding parametric form of $f(R)$ given by Eqs. (\ref{fphi}, \ref{Rphi}) 
\begin{align}
\label{fdw}
f(\phi) &= 2 e^{2 \beta  \phi } \left(\frac{m_{\phi}^4 (a-\phi ) (a+\phi ) \left(a^2 \beta -\phi  (\beta  \phi +4)\right)}{4 \beta }+\Lambda \right),\\
\label{Rdw}
R(\phi) &= e^{\beta  \phi } \left(\frac{m_{\phi}^4 (a-\phi ) (a+\phi ) \left(a^2 \beta -\phi  (\beta  \phi +2)\right)}{\beta }+4 \Lambda \right).
\end{align}
In fig.~\ref{fRdw}, we show an interesting double catastrophic behavior in two different scales.
\section{The Stability Mass Criteria}\label{stabledw}
\begin{figure}[t]
\center
{
\noindent\stackinset{r}{30pt}{t}{85pt}{\includegraphics[width=0.4\textwidth]{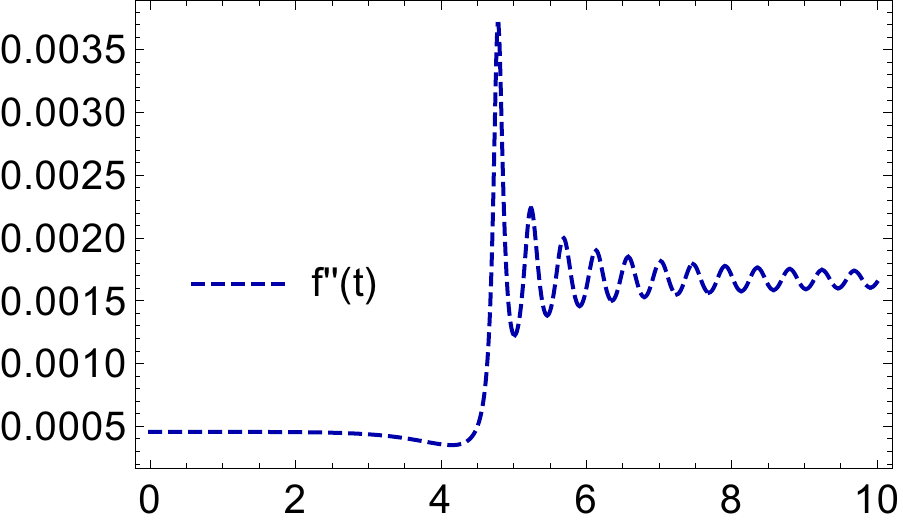}}
{\includegraphics[width=0.85\textwidth]{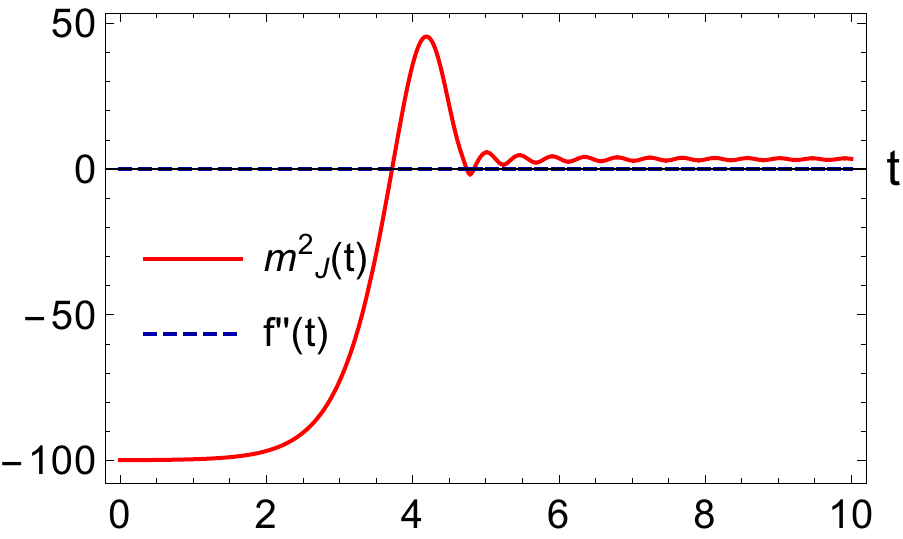}}}
\caption[Mass of Double Well Potential]{Double well potential mass: $m^2_J(t)$ (\ref{mJ}) in red/solid, and $f''(t)$ in blue/dashed as functions of time. We have used $N = 60$ efolds and $m_\phi = 1$, $a=5$, and $\Lambda = 0$. \textbf{Inset panel:} Behavior of $f''(t)$ between $t \in (0,10)$.}
\label{Mdwpot}
\end{figure}

In Fig~\ref{Mdwpot} we compare the behavior of the second derivative of $f(R)$ and the mass term $m^2_J$ (\ref{mJ}) for the double well potential (\ref{dwpot}). We can note again that the two criteria coincide only at the end of inflation.

\section{The Stability Criteria from Thermodynamics Analogy}\label{ThermoAnalogySectiondw}
\begin{figure}[t]
\center
\includegraphics[width=0.45\textwidth]{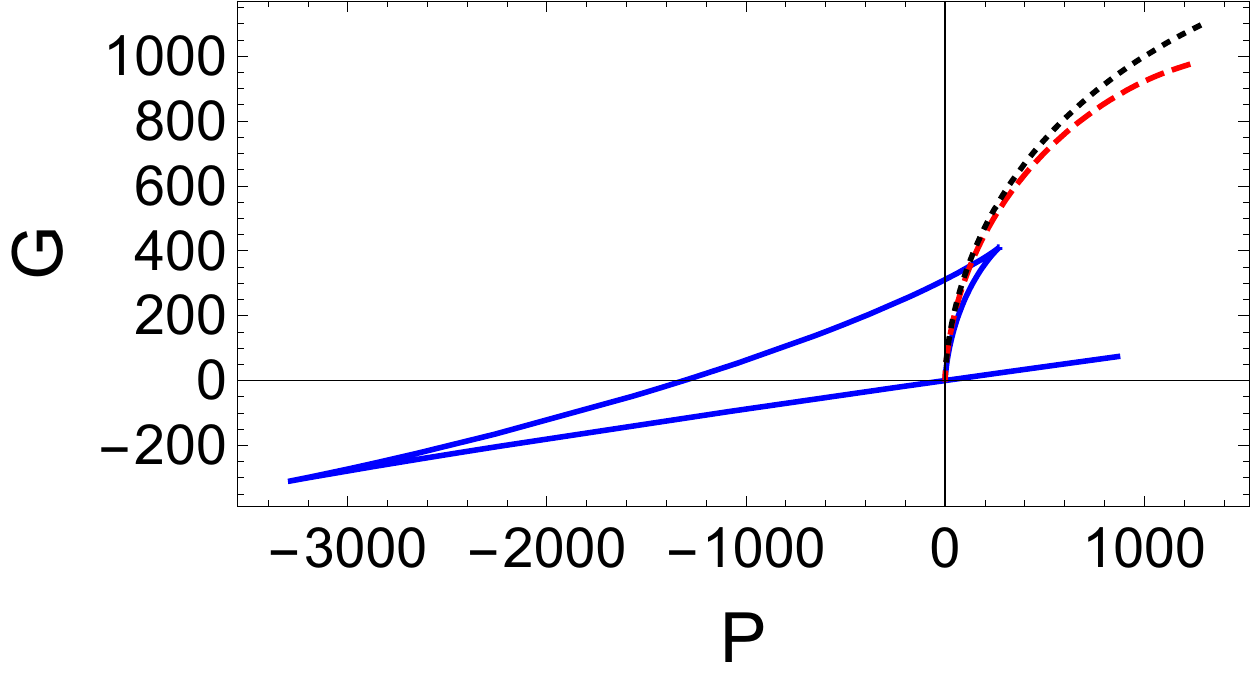}
\includegraphics[width=0.45\textwidth]{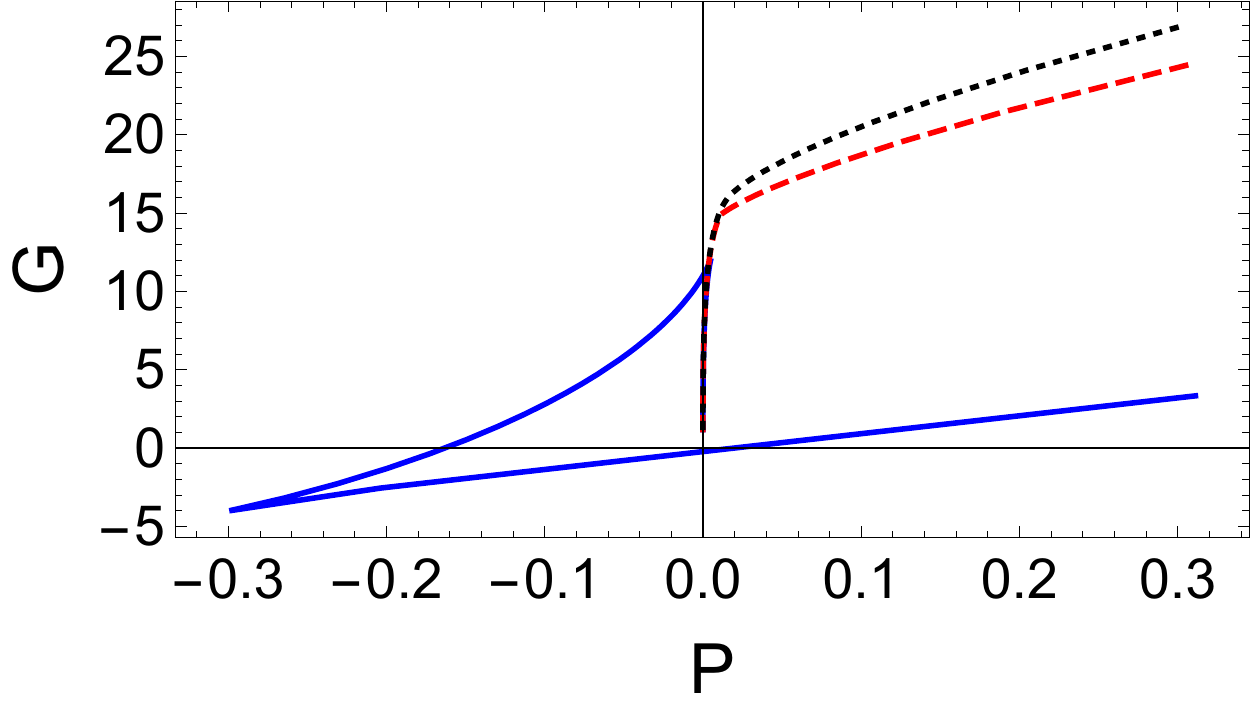}
\caption[Plot of the Gibbs Potential for DW Potential]{Parametricplot of the Gibbs Potential $G$ Eq.~(\ref{Gphidw}) as a function of the pressure $P$ Eq.~(\ref{Pphidw}), for $\beta=\sqrt{2/3}$, $a=3$, $\theta=\theta_*$ and different values of temperature. \textbf{On the left side:} The Gibbs Potential in the scale of $V=[0.084, 20]$ which could be associated a critical temperature $T=T_{c_1} \simeq 46.796$ (dashed red), $T=0$ (solid blue), and $T=60$ (dotted black). \textbf{On the right side:} The Gibbs Potential in the scale of $V=[10,1000]$ which could be associated a critical temperature $T=T_{c_2} \simeq 167.303$ (dashed red), $T=0$ (solid blue), and $T=200$ (dotted black).}
\label{Gphidw_fig}
\end{figure}

For the double well potential case, the thermodinamic association (\ref{ansatz}) of the new pair of coordinates $\{-G,P\}$ as a rotation of the original one $\{f,R\}$ yields
\begin{align}
 G (\phi,T) &=
4 e^{\beta  \phi } \sin (\theta ) \left(\frac{2 \phi  \left(\phi ^2-a^2\right)}{\beta }+\left(a^2-\phi ^2\right)^2+T\right)+\nonumber\\
&-2 e^{2 \beta  \phi } \cos (\theta ) \left(\frac{4 \phi  \left(\phi ^2-a^2\right)}{\beta }+\left(a^2-\phi ^2\right)^2+T\right),
 \label{Gphithetadw}  \\
P (\phi,T) &= 2 e^{2 \beta  \phi } \sin (\theta ) \left(\frac{4 \phi  \left(\phi ^2-a^2\right)}{\beta }+\left(a^2-\phi ^2\right)^2+T\right)+\nonumber\\
&+4 e^{\beta  \phi } \cos (\theta ) \left(\frac{2 \phi  \left(\phi ^2-a^2\right)}{\beta }+\left(a^2-\phi ^2\right)^2+T\right),
\label{Pphithetadw} \\
V &= \left.\frac{\partial G/\partial\phi}{\partial P/\partial\phi}\right|_T = \frac{1-e^{\beta  \phi } \cot (\theta )}{e^{\beta  \phi }+\cot (\theta )}. \label{Vphithetadw}
\end{align}
Eq.~(\ref{Vphithetadw}) can be inverted and yield
\begin{equation}
\phi = \frac{1}{\beta}\log \left(\frac{1-V \cot (\theta )}{\cot (\theta )+V}\right).
\end{equation}
In order to obtain a positive volume and unlimited from below, we require that $\theta = \theta_* \equiv \pi/2$, with this value we get simpler parametric expressions for the previously defined thermodynamic quantities: 
\begin{align}
G &= 4 e^{\beta  \phi } \left(\frac{2 \phi  \left(\phi ^2-a^2\right)}{\beta }+\left(a^2-\phi ^2\right)^2+T\right),\label{Gphidw}\\
P &= 2 e^{2 \beta  \phi } \left(\frac{4 \phi  \left(\phi ^2-a^2\right)}{\beta }+\left(a^2-\phi ^2\right)^2+T\right),\label{Pphidw}\\
F &= 2 e^{\beta  \phi } \left(a^4-2 a^2 \phi ^2+T+\phi ^4\right),\label{Fphidw}\\
S &= -2 e^{\beta  \phi }\label{Sphidw},\\
U &= 2 \left(a^2-\phi ^2\right)^2 e^{\beta  \phi },\label{Uphidw}\\
V &= \exp(-\beta \phi) \quad \Leftrightarrow \quad \phi = -\frac{1}{\beta}\log(V)\label{Vphidw}.
\end{align}
In Fig.~\ref{Gphidw_fig}, we show the the Gibbs Potential $G(P)$ for different temperatures $T$.

\begin{figure}[t]
\center
\includegraphics[width=0.8\textwidth]{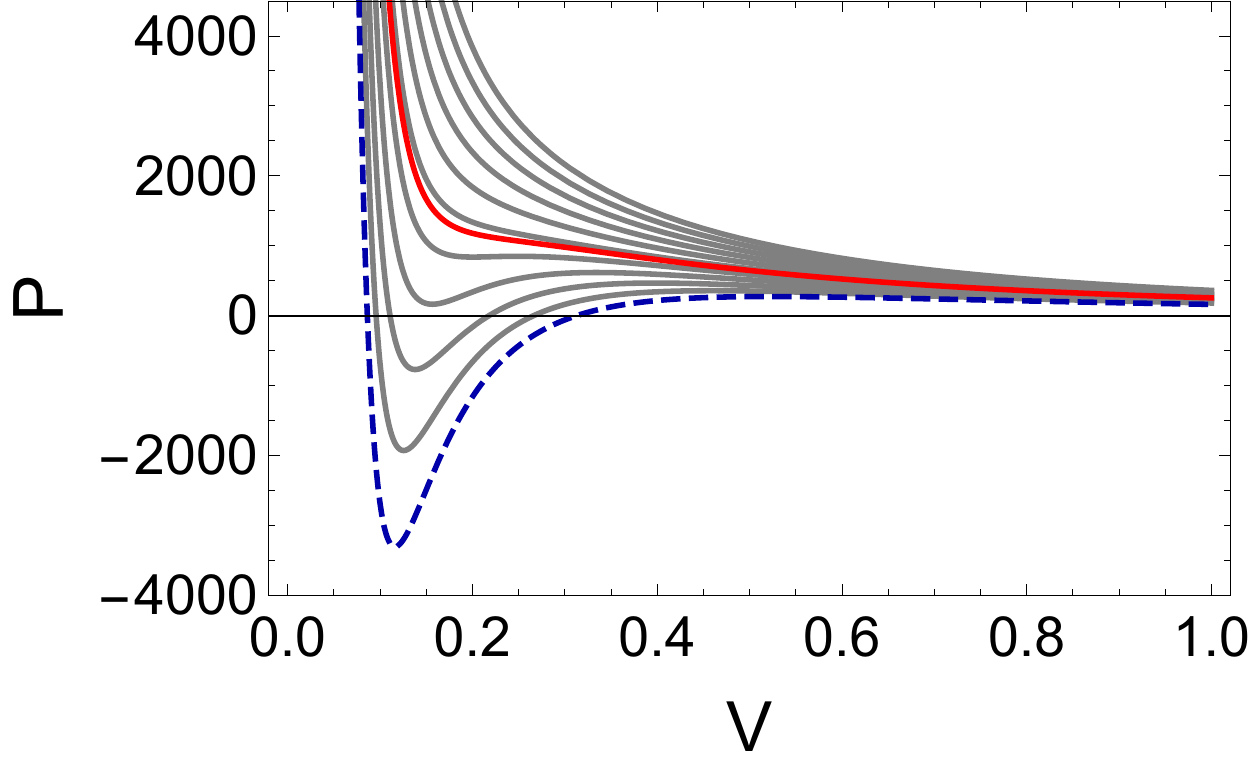}
\includegraphics[width=0.8\textwidth]{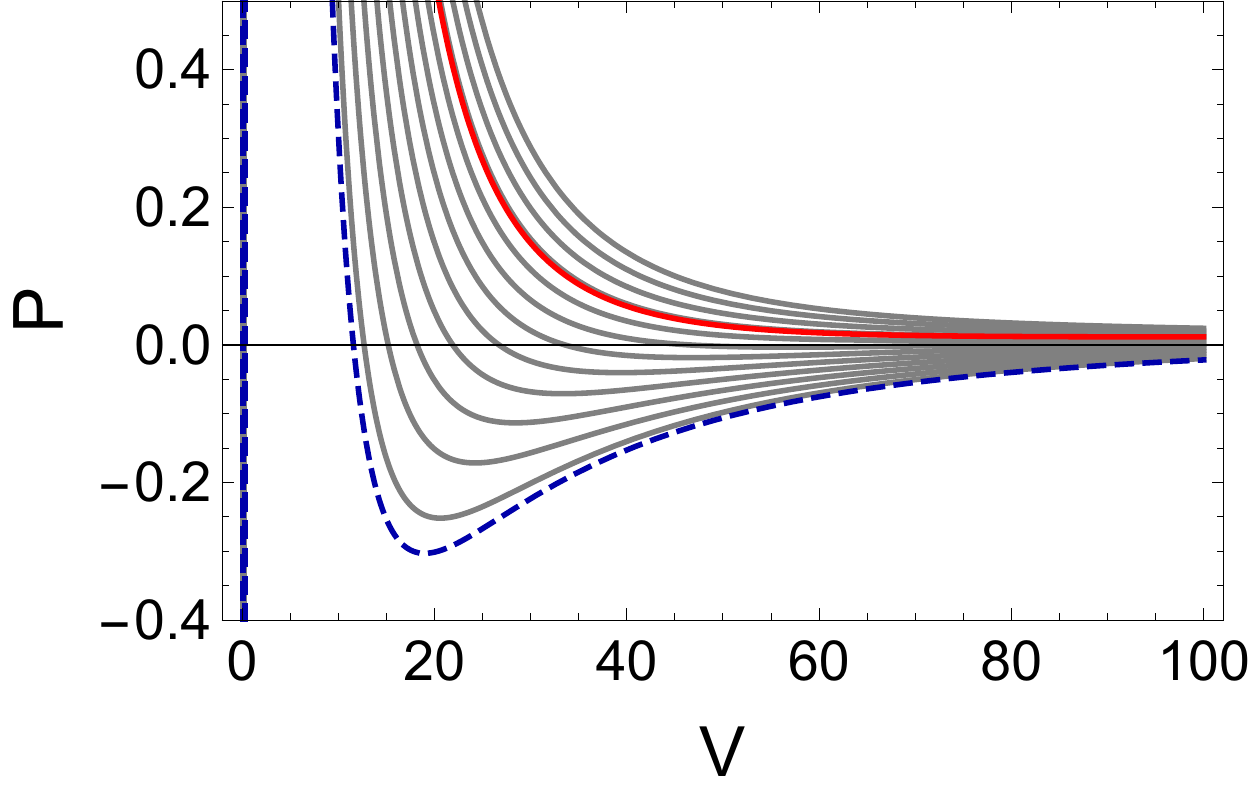}
\caption[Plot of the effective pressure $P$ for DW Potential]{Parametricplot of the effective pressure $P$ as a function of the effective volume $V$ Eq.~(\ref{PVdw}), for $\beta=\sqrt{2/3}$, $a=3$, $\theta=\theta_*$, and different values of temperature. \textbf{Upper panel:} The vdW-like  ``efective gas'' behavior in the scale of $V=[0,1]$ which could be associated a critical temperature $T=T_{c_1} \simeq 46.796$ (solid thick red)  and $T=0$ (dashed blue). \textbf{Lower panel:} Another vdW-like  ``efective gas'' behavior in the scale of $V=[0,100]$ which could be associated a critical temperature $T=T_{c_2} \simeq 167.303$ (solid thick red)  and $T=0$ (dashed blue).}
\label{pvdwpot}
\end{figure}

For $\theta=\theta_*=\pi/2$, equations~(\ref{Pphidw}) and (\ref{Vphidw}) yield the equation of state for our vdW-like  ``efective gas'', i.e, an expression that relates $P$, $V$ and $T$. We show the behaviour of this equation of state for four different values of $T$ in Fig.~\ref{pvdwpot}. 
\begin{equation}
P = \frac{2 \left(\left(a^2-\frac{\log ^2(V)}{\beta ^2}\right)^2+\frac{4 a^2 \beta ^2 \log (V)-4 \log ^3(V)}{\beta ^4}+T\right)}{V^2}.
\label{PVdw}
\end{equation}

\addtocontents{toc}{\vspace{2em}} 

\backmatter


\printbibliography

\end{document}